\documentclass[11pt,oneside]{book}

\usepackage[inline]{enumitem}
\usepackage[numbers,merge]{natbib}
\usepackage{marginfix}
\usepackage{tocloft}
\usepackage[numbib]{tocbibind}
\usepackage[fleqn]{amsmath}
\usepackage{amssymb}
\usepackage{IEEEtrantools}
\usepackage{imakeidx}
\usepackage[table,dvipsnames]{xcolor}
\usepackage[breakable,skins]{tcolorbox}
\usepackage{soul} 
\usepackage[normalem]{ulem}
\usepackage{multicol}
\usepackage{bold-extra}
\usepackage[top=3cm,bottom=3cm,outer=3cm,inner=3cm, heightrounded,marginparwidth=2cm,marginparsep=10pt]{geometry}
\usepackage{longtable}
\usepackage{makeidx}
\usepackage{hyperref}
\usepackage[pdftex,color]{changebar}
\usepackage{framed} 
\definecolor{background}{rgb}{0.9,0.9,0.9}
\colorlet{shadecolor}{background}
\sethlcolor{background}
\usepackage{calc}
\usepackage{pifont}
\newcommand{\cmark}{\ding{51}}
\newcommand{\xmark}{\ding{55}}
\usepackage{pdflscape}
\usepackage{pdflscape}
\usepackage{tikz}
\usetikzlibrary{calc}
\reversemarginpar
\definecolor{RED}{RGB}{255, 0, 0}
\definecolor{PURPLE}{RGB}{128, 0, 128}
\definecolor{BLUE}{RGB}{0, 0, 255}
\definecolor{CYAN}{RGB}{0, 128, 255}

\colorlet{crls}{orange}
\colorlet{mdf}{Green}

\newcommand\MOLSCAT{{{\scshape molscat}}}
\newcommand\BOUND{{\scshape bound}}
\newcommand\FIELD{{\scshape field}}
\newcommand\lastversion{2022.0}
\newcommand\currentversion{2025.0}
\newcommand\file[1]{{\tt #1}}
\newcommand\mylabel[1]{\label{#1}
}
\newcommand\etal{{\it et al.}}
\newcommand\inpitem[1]{{\tt #1}\index{{\tt #1} in \namelist{\&INPUT}}}
\newcommand\basisitem[1]{{\tt #1}\index{{\tt #1} in \namelist{\&BASIS}}}
\newcommand\potlitem[1]{{\tt #1}\index{{\tt #1} in \namelist{\&POTL}}}
\newcommand\common[1]{{\tt #1}}
\newcommand\module[1]{{\tt #1}}
\newcommand\iounit[1]{{\tt #1}}
\newcommand\namelist[1]{{\tt #1}}
\newcommand\code[1]{{\tt #1}}
\newcommand\var[1]{{\tt #1}\index{{\tt #1} (internal)}}
\newcommand\prog[1]{{\tt #1}}
\newcommand\bcalV{\boldsymbol{\cal V}}
\newlength\myitembox
\newlength\myitemlength
\newlength\mythirdlength
\newlength\myfourthlength
\def\msa{m_{s{A}}}
\def\msb{m_{s{B}}}
\def\mia{m_{i{A}}}
\def\mib{m_{i{B}}}
\def\mfa{m_{f{A}}}
\def\mfb{m_{f{B}}}
\def\sa{s_{A}}
\def\sb{s_{B}}

\def\fa{f_{A}}
\def\fb{f_{B}}
\def\Pab{P_{AB}}

\def\sixj#1#2#3#4#5#6{\left\{\begin{matrix}#1&#2&#3\\#4&#5&#6\end{matrix}\right\}}

\SetEnumitemKey{twocol}{
nosep,
before=\raggedcolumns\begin{multicols}{2}\raggedright,
after=\end{multicols}}
\SetEnumitemKey{threecol}{
nosep,
before=\raggedcolumns\begin{multicols}{3}\raggedright,
after=\end{multicols}}
\SetEnumitemKey{fourcol}{
nosep,
before=\raggedcolumns\begin{multicols}{4}\raggedright,
after=\end{multicols}}
\SetEnumitemKey{fivecol}{
nosep,
before=\raggedcolumns\begin{multicols}{5}\raggedright,
after=\end{multicols}}
\renewcommand{\d}{{d}}
\def\mcol{red}
\def\bcol{blue}
\def\fcol{green}
\def\mbcol{purple}
\def\mfcol{brown}
\def\bfcol{cyan}
\hypersetup{
   colorlinks=true,linkcolor = purple,urlcolor=magenta,citecolor=purple}
\makeindex[title=Index of key variables and arrays]
\begin{document}
\title{\MOLSCAT, \BOUND\ and \FIELD\ \\ \ \\ Version \currentversion\\ \ \\ User Manual}
\author{Jeremy M. Hutson and C. Ruth Le Sueur}
\date{\today}
\maketitle

\pagenumbering{roman} \tableofcontents \mainmatter \pagenumbering{arabic}
\chapter{\texorpdfstring{Preamble}{\ref{preamble}: Preamble}}\mylabel{preamble}

\section{Citing the programs}\sectionmark{Citing the programs}\mylabel{citing}

Any publication that uses \MOLSCAT, \BOUND\ or \FIELD\ should cite both the
version of the program used:

Jeremy M. Hutson and C. Ruth Le Sueur, \MOLSCAT: a program for non-reactive quantum scattering
calculation on atomic and molecular collisions, Version \currentversion, \hfil\break
\url{https://github.com/molscat/molscat}.

Jeremy M. Hutson and C. Ruth Le Sueur, \BOUND: a program for bound states of interacting pairs of
atoms and molecules, Version \currentversion, \url{https://github.com/molscat/molscat}.

Jeremy M. Hutson and C. Ruth Le Sueur, \FIELD: a program for bound states of interacting pairs of
atoms and molecules as a function of external field, Version \currentversion, \hfil\break
\url{https://github.com/molscat/molscat}.

and the published paper(s):

Jeremy M. Hutson and C. Ruth Le Sueur, `\MOLSCAT: a program for non-reactive
quantum scattering calculations on atomic and molecular collisions',
\hfil\break\noindent Computer Physics Communications 241, 9-18 (2019):
\hfil\break\noindent \url{https://doi.org/10.1016/j.cpc.2019.02.014}.

Jeremy M. Hutson and C. Ruth Le Sueur, `\BOUND\ and \FIELD: programs for
calculating bound states of interacting pairs of atoms and molecules",
\hfil\break\noindent Computer Physics Communications 241, 1-8 (2019):
\hfil\break\noindent \url{https://doi.org/10.1016/j.cpc.2019.02.017}.

The current and previous versions of this documentation are available from
\hfil\break\noindent \url{https://arxiv.org/abs/1903.06755}.

\section{Licensing}

These programs are free software: they can be redistributed and/or modified under the terms of the GNU General Public License, version 3, as published by
the Free Software Foundation. The full text of the license is available from
\url{https://www.gnu.org/licenses/} and is included in the file \file{COPYING}
included in the distribution.

\section{Updates}\mylabel{updates}

Program updates are made available via github at
\url{https://github.com/molscat/molscat}. No github user account is needed. If
using a Linux machine with git installed, the commands to create a
directory containing the program source code and associated files are

\code{mkdir molscat}

\code{git clone https://github.com/molscat/molscat molscat}

On subsequent occasions the source code can be updated to the latest
distributed version simply by navigating to the molscat directory and issuing
the command

\code{git pull}

\section{Scope of the programs}\mylabel{scope}

\MOLSCAT\ is a general-purpose package for performing non-reactive quantum
scattering calculations for atomic and molecular collisions using
coupled-channel methods. Simple atom-molecule and molecule-molecule collision
types are coded internally and additional ones may be handled with plug-in
routines. Plug-in routines may include external magnetic, electric or photon
fields (and combinations of them). Simple interaction potentials are coded
internally and more complicated ones may be handled with plug-in routines.

\BOUND\ is a general-purpose package for performing calculations of bound-state
energies in weakly bound atomic and molecular systems using coupled-channel
methods. It solves the same sets of coupled equations as \MOLSCAT, and can use the
same plug-in routines if desired, but with different boundary conditions.

\FIELD\ is a development of \BOUND\ that locates external fields at which a
bound state exists with a specified energy.  One important use is to locate the
positions of magnetically tunable Feshbach resonance positions in ultracold
collisions.

Versions of these programs before 2019.0 were released separately. However,
there is a significant degree of overlap between their internal structures and
usage specifications. This manual therefore describes all three, with careful
identification of parts that are specific to one or two of the programs.

The authors would be grateful to know of any bugs encountered in any of the
programs or errors in this documentation.

\section{Program capabilities}\mylabel{capabilities}

\MOLSCAT, \BOUND\ and \FIELD\ all construct sets of coupled differential
equations that represent the Hamiltonian for atom-atom, atom-molecule or
molecule-molecule interactions. They have built-in capabilities to generate the
coupled equations for
\begin{itemize}[nosep]
\item Atom + linear rigid rotor;
\item Atom + vibrating diatom;
\item Linear rigid rotor + linear rigid rotor;
\item Atom + symmetric top;
\item Atom + asymmetric top;
\item Asymmetric top + linear molecule;
\item Atom + rigid corrugated surface: diffractive (elastic) scattering and
    band structure.
\end{itemize}
The programs can set up the coupled equations for
\begin{itemize}[nosep]
\item Full close-coupling calculations, with no dynamical approximations;
\item Effective potential approximation;
\item CS (coupled-states/centrifugal-sudden) approximation;
\item Decoupled $L$-dominant approximation;
\item Infinite-order sudden approximation (\MOLSCAT\ only).
\end{itemize}
The programs also include an interface for a plug-in basis-set suite to set up
other sets of coupled equations. Basis-set suites can be programmed to take
account of one or more external fields, such as electric, magnetic and photon
fields. Two such suites are included in this distribution, for
\begin{itemize}[nosep]
\item $^1$S atom + $^3\Sigma$ diatom in a magnetic field;
\item Alkali-metal atom + alkali-metal atom in a magnetic field, including
    hyperfine structure.
\end{itemize}

Input energies may be specified with respect to a particular scattering
threshold if desired. For basis sets that take account of external field(s),
the reference energy is often field-dependent.

The interaction potential between the interacting species may be supplied in a
variety of ways. Very simple potentials may be supplied as explicit input data.
More sophisticated potentials may be supplied as external routines that provide
either:
\begin{itemize}[nosep]
\item the coefficients for an expansion of the potential in suitable
    internal coordinates at a supplied value of the interspecies distance; or
\item the values of the potential for a supplied set of internal
    coordinates and the interspecies distance; the programs then integrate
    over the internal coefficients to obtain the expansion coefficients.
\end{itemize}

The programs loop over good quantum numbers such as the total angular momentum
and parity of the interacting pair, and construct a separate set of coupled
equations for each set of values of the good quantum numbers.

The programs solve the coupled equations using a variety of numerical methods,
all of which propagate the wavefunction matrix or its log-derivative across a
range of values of the interspecies separation.

\MOLSCAT\ performs scattering calculations at a list or grid of energies where
there is at least one channel that is asymptotically open (energetically
accessible at long range). It propagates the solutions outwards from short
range to a point at long range where the interaction potential is
insignificant. It matches the solution at long range to analytic functions that
describe the solutions in the absence of the interaction potential and obtains
the scattering S matrix. It then has options to
\begin{itemize}[nosep]
\item Combine S matrices from different values of total angular momentum
    and other symmetries to calculate state-to-state integral cross
    sections;
\item Combine S matrices from different values of total angular momentum
    and parity to calculate spectroscopic line-shape cross sections;
\item Output S matrices to files for subsequent processing to calculate differential cross
    sections or generalised transport, relaxation and Senftleben-Beenakker cross sections;
\item Converge on and characterise scattering resonances or predissociating / quasibound states
    of the collision complex through their signature in the eigenphase sum (multichannel phase
    shift);
\item Output K matrices to files for subsequent processing to characterise
    scattering resonances;
\item Calculate low-energy scattering properties such as scattering
    lengths, scattering volumes and effective ranges;
\item Converge on and characterise low-energy Feshbach resonances in the
    scattering length as a function of external field(s).
\end{itemize}

\BOUND\ seeks bound states in a specified range of energies, usually at
energies where all channels are asymptotically closed (energetically
inaccessible at long range). It propagates the solutions outwards from short
range and inwards from long range to a matching point in the classically
allowed region. It compares the outwards and inwards solutions and attempts to
converge on energies where the wavefunction is both continuous and continuously
differentiable at the matching point. These are the bound-state energies of the
system.

\BOUND\ has options to
\begin{itemize}[nosep]
\item Calculate expectation values without explicit wavefunctions;
\item Calculate bound-state wavefunctions and output them to files for
    subsequent processing.
\end{itemize}
For basis sets that take account of external field(s), \BOUND\ can loop over a
list or grid of values of the external fields and locate bound states at each
field.

\FIELD\ operates in a similar way to \BOUND\, but seeks values of the external
field(s) at which bound states exist at a specified list or grid of energies.

\section{Program limitations}\mylabel{limitations}

The programs propagate coupled equations with respect to a single integration
variable $R$ (usually the interparticle distance). They require that the
kinetic energy operator in the variable $R$ can be written in the form
$[f(R)]^{-1} (d^2/dR^2) f(R)$, where $f(R)$ is usually a power $R^n$.

The programs can handle interactions of two structured particles, such as
complex atoms or molecules, but they are not generally applicable to problems
involving more than 2 particles, unless they can be expressed as a set of
coupled equations in a single variable.

The programs are efficient only if the wavefunction at all physically
significant values of $R$ can be expanded compactly in terms of an
$R$-independent basis set of functions in the remaining variables. This is
often not true for strongly bound states of polyatomic molecules, and different
methods are preferable in such cases.

The programs cannot handle interaction potentials that are non-local in the
integration variable, as sometimes occur in nuclear scattering problems.

\MOLSCAT\ cannot apply asymptotic boundary conditions in two different
integration variables, as is common in reactive scattering calculations.

\MOLSCAT\ currently applies boundary conditions that are appropriate only if
the interaction potential decays faster than $R^{-2}$ at long range.
Modifications would be needed to handle scattering boundary conditions for
Coulomb potentials.

\pagebreak
\section{Structure of documentation}\mylabel{parts}

In this manual, coloured vertical bars denote sections that apply to a subset
of the programs, as follows:
\begin{itemize}
\item{\cbcolor{\mcol}\cbstart{\MOLSCAT}\cbend}
\item{\cbcolor{\bcol}\cbstart{\BOUND}\cbend}
\item{\cbcolor{\fcol}\cbstart{\FIELD}\cbend}
\item{\cbcolor{\bfcol}\cbstart{\BOUND\ and \FIELD}\cbend}
\item{\cbcolor{\mfcol}\cbstart{\MOLSCAT\ and \FIELD}\cbend}
\item{\cbcolor{\mbcol}\cbstart{\MOLSCAT\ and \BOUND}\cbend}
\end{itemize}

This documentation is structured as follows:
\begin{itemize}
\item{Chapter \ref{Theory} provides a brief description of the theory
    behind the programs.}
\item{Chapter \ref{use} provides an introduction to use of the programs,
    with some basic examples.}
\item{Chapter \ref{ConstructBasis} describes the arrays used to specify
    basis sets and the input data required for the built-in interaction
    types.}
\item{Chapter \ref{buildVL} describes how to specify interaction
    potentials, including the specification for plug-in potential routines
    \prog{VINIT}/\prog{VSTAR} and \prog{VRTP}.}
\item{Chapter \ref{Energy} describes how to specify energies in \MOLSCAT\
    and \FIELD\ and energy ranges for bound states in \BOUND\ and \FIELD.
    It also describes how to select a reference threshold to be used as a
    (potentially field-dependent) zero of energy.}
\item{Chapter \ref{inputfield} describes how to specify external fields in
    \MOLSCAT\ and \BOUND\ and field ranges for bound states in \FIELD.}
\item{Chapter \ref{propcontrol} describes how to control the propagators
    used to solve the coupled equations.}
\item{\cbcolor{\mcol}\cbstart{}Chapter \ref{processres} describes how to
    specify which properties are calculated from S matrices by
    \MOLSCAT.\cbend}
\item{\cbcolor{\bfcol}\cbstart{}Chapter \ref{processbound} describes how to
    control bound-state calculations in \BOUND\ and \FIELD.\cbend}
\item{Chapter \ref{CommI} provides a complete list of the input variables
    and arrays, in the form of alphabetically organised quick-reference
    lists.}
\item{Chapter \ref{CommII} describes how to control the level of printed
    output.}
\item{Chapter \ref{Auxfiles} describes the auxiliary and scratch files that
    may be produced or used.}
\item{Chapter \ref{testfiles} describes the example input files provided
    with the programs to illustrate their capabilities, the output they
    each produce, and how to interpret it.}
\end{itemize}

Up to this point, the user needs only limited knowledge of internal variables
and subroutines (except for any plug-in routines used for calculating
the interaction potential). The remainder of the documentation is intended for
users who wish to install the program on a new computer system, program more
complicated interaction potentials, implement new plug-in basis-set routines,
use the plug-in basis-set routines provided, or modify other parts of the
program.

\begin{itemize}
\item{Chapter \ref{distribution} describes the files provided with the
    distribution and how to build executables. It includes information
    about potentially machine-dependent features and describes program
    features likely to be needed by users who wish to program their own
    plug-in interaction potential routines or plug-in basis-set suites.}

\item{Chapter \ref{userPotenl} describes the (now seldom used) option to
    supply a complete \prog{POTENL} routine to evaluate the set of
    potential expansion coefficients, rather than the simpler
    \prog{VINIT}/\prog{VSTAR} or \prog{VRTP} routines described in chapter
    \ref{buildVL}.}

\item{Chapter \ref{base9} sets out the specification of a plug-in basis-set
    suite of routines which describe a basis set and calculate coupling
    coefficients different from the built-in interaction types. It is
    intended principally for users who wish to program their own suite.}
		
\item{Chapter \ref{user:gen} describes the two plug-in basis-set suites
    provided as part of the distributed code.}
		
\item{Chapter \ref{supplied-vstar} describes the plug-in potential routines
    provided as part of the distributed code.}

\end{itemize}

\section{Program history}\mylabel{history}

The \MOLSCAT\ program was originally written by Sheldon Green in the 1970s,
incorporating propagators from several different authors and adapting them to
share the same input / output structures and mechanisms for generating matrix
elements of the molecular Hamiltonian and interaction potential. Early versions
of the program handled atom-molecule and molecule-molecule scattering, with a
variety of additional coupling cases added between versions~1 (1973) and 7
(1979) \cite{molscat:NRCC:1980}. Both full space-fixed close-coupling
calculations and decoupling approximations such as coupled states  /
centrifugal-sudden (CS), infinite-order sudden (IOS) and decoupled $L$-dominant
(DLD) approximations were implemented. The program calculated integral elastic,
state-to-state inelastic and line-shape cross sections internally and wrote S
matrices to a file that could be used for post-processors, including DCS
\cite{DCS} for differential cross sections and SBE \cite{SBE} for cross
sections associated with transport and relaxation properties and
Senftleben-Beenakker effects.

Jeremy Hutson became a co-author of \MOLSCAT\ in 1982 and collaborated with
Sheldon Green on its development until Green's death in 1995.  Major changes in
this period included new coupling cases, including the diffractive scattering
of atoms from crystal surfaces, the replacement of some older propagators with
new ones, and code to handle scattering resonances (with post-processor RESFIT
\cite{Hutson:resfit:2007} for fitting resonance parameters). In addition, an
interface was added in version 11 (1992) to allow the inclusion of
plug-in coupling cases with angular momentum algebra and/or interaction
potential expansions that were not already present in the code.

The last version of \MOLSCAT\ produced by Green and Hutson in collaboration was
version 14 in 1994. This was distributed via the CCP6 collaboration in the UK
\cite{molscat:v14} and via the NASA GISS website. Version 14 formed the basis
of a parallel version named {{\scshape PMP}} \MOLSCAT, produced by George
McBane \cite{McBane:PMP:2005}.

\BOUND\ was originally written by Jeremy Hutson in 1984 to calculate bound
states of Van der Waals complexes by coupled-channel methods, using the same
structures as \MOLSCAT\ to generate the coupled equations. Subsequent versions
incorporated basis-set enhancements as they were made in \MOLSCAT. A
fundamental change was made in version 5 (1993) to base the convergence
algorithm on individual eigenvalues of the log-derivative matching matrix
\cite{Hutson:CPC:1994}, rather than its determinant. Versions 4 (1992) and 5
(1993) \cite{Hutson:bound:1993} were distributed via CCP6.

\MOLSCAT\ and \BOUND\ were extended to handle calculations in external electric
and magnetic fields in 2007. \FIELD\ was written by Jeremy Hutson in 2010,
using the same structures as \MOLSCAT\ and \BOUND\ to generate the coupled
equations but designed to locate bound states as a function of external field
at fixed energy, rather than as a function of energy.

There was no fully documented publication of \MOLSCAT\ or \BOUND\ between 1994 and
2019. \FIELD\ was not formally published until 2019. However, there were continuing enhancements to
the capabilities in the intervening years, and updates were provided privately to selected users.
In this documentation we treat all changes from \MOLSCAT\ version 14 and \BOUND\ version 5 as new
in version 2019.0, since that was the first time they were collected together and documented.

\section{Principal changes in version 2019.0}\mylabel{changes-2019-0}

The basis-set plug-in mechanism was extended to allow propagation in basis sets that are not
eigenfuctions of the internal Hamiltonian $H_{\rm intl}$. This made implementing new types of
system much simpler than before, especially where the individual interaction partners have
complicated Hamiltonians.

This functionality was used to add new capabilities to carry out calculations in external fields
(electric, magnetic, and/or photon) and to loop over (sets of) values of the fields.

\cbcolor{\mcol}\cbstart Where necessary, \MOLSCAT\ now tranformed the propagated wavefunction or
log-derivative matrix to a basis set diagonal in $H_{\rm intl}$ before matching to long-range
functions to extract the S matrix.

\MOLSCAT\ was extended to process the S matrix to calculate scattering lengths (or volumes or
hypervolumes) $a_L$ [actually $k$-dependent complex scattering lengths/volumes $a_L(k)$] for any
low-energy scattering channels.

\MOLSCAT\ was extended to extrapolate the real part of $a_0(k)$ to $k=0$ and calculate the
effective range $r_{\rm eff}$.

\MOLSCAT\ was extended to converge on and characterise Feshbach resonances as a function of
external field. This is implemented for both the elastic case, where resonances appear as poles in
the scattering length/volume, and the inelastic case, where resonances are decayed and have more
complicated signatures.

\MOLSCAT\ was extended to calculate a multichannel scattering wavefunction for flux incoming in a
single channel and outgoing in all open channels.\cbend

\cbcolor{\bfcol}\cbstart \BOUND\ and \FIELD\ separated the functions of $R_{\rm mid}$, the distance
where the calculation switches between short-range and long-range propagators, and $R_{\rm match}$,
the distance where the incoming and outgoing wavefunctions are matched.

\BOUND\ and \FIELD\ were modified to calculate the node count from an outwards propagation from
$R_{\rm min}$ to $R_{\rm match}$, an inwards propagation from $R_{\rm max}$ to $R_{\rm match}$, and
the number of negative eigenvalues of the log-derivative matching matrix. This eliminated the need
for a third propagation from $R_{\rm match}$ to $R_{\rm min}$ or $R_{\rm max}$. \cbend

All three programs implemented a more general mechanism for combining propagators for use at short
and long range, which allows any sensible combination.

All the programs allowed more general choices of log-derivative boundary conditions at the starting
points for propagation.

A new propagation approach \cite{MG:symplectic:1995} was included, implemented by George McBane.
This takes advantage of the symplectic nature of the multi-channel radial Schr\"odinger equation
and reformulates it so that symplectic integrators (SIs) may be used to propagate solutions of the
coupled equations. Coefficients for two SIs were included: the five-step fourth-order method of
Calvo and Sanz-Serna (CS4) \cite{CS4} and the six-step fifth-order method of McLachlan and Atela
(MA5) \cite{MA5}. The approach was coded so that other SIs could easily be implemented if desired.

\section{Principal changes in version 2019.1}\mylabel{changes-2019-1}

The fundamental physical constants used by default were updated to the 2018 CODATA recommended
values.

The terminology used in the output from \BOUND\ and \FIELD\ was modified to introduce a \emph{state
number} that is equal to the node count immediately above the state concerned.

The output on the S-matrix save file was modified to allow more general indexing of channel
energies. This is an incompatible change for $\var{ITYP}=3$, 4, 5, 6, so the \iounit{ISAVEU} format
version number \var{IPROGM} (section \ref{Sout}) was increased to 19.

Minor bug fixes as documented on github.

\section{Principal changes in version 2020.0}\mylabel{changes-2020-0}

\MOLSCAT\ can now converge on and characterise a scattering resonance or quasibound state as a
function of energy, as described in ref.\ \cite{Frye:quasibound:2020}; see section
\ref{energyconv}.

\MOLSCAT\ now converges on resonances in the scattering length using the procedures of ref.\
\cite{Frye:resonance:2017} but with the improved algorithm for point selection developed in ref.\
\cite{Frye:quasibound:2020}; this can provide widths that are stabler with respect to variations in
the Hamiltonian.

\MOLSCAT\ now evaluates integral cross sections for collisions of identical molecules using a definition
that gives the same result with and without identical-particle symmetry when the two molecules are in the
same initial or final state; see section \ref{ityp3}.

\BOUND\ and \FIELD\ now allow the log-derivative matching point $R_{\rm match}$ to be at the end of
the propagation range (at $R_{\rm min}$ or $R_{\rm max}$) as well as part-way along it.

\BOUND\ can now calculate the expectation value of the operator for an external field and use it to
evaluate the derivative of the bound-state energy with respect to field, for both the absolute
energy and the energy relative to threshold.

The potential routines and data modules that implement the functional form of the Hannover group
(section \ref{detail:Hannover}) for diatomic potential-energy curves have been generalised to allow the use of potential parameters that do not give exact continuity at the matching points. This is not generally recommended, but it allows the use of potential parameters exactly as published for comparison with other routines.

A new structure has been introduced to allow reinitialisation of the potential
routine for each set of external fields. The default behaviour is unchanged, but the
reinitialisation may be used to interpret a scaling factor \var{SCALAM} (treated as a field) as a
chosen linear or non-linear potential parameter. This allows use of the built-in convergence
capabilities of \MOLSCAT\ or \FIELD\ to adjust the parameter to reproduce precisely a required
value of an observable quantity such as a resonance position or bound-state energy.

The names and structures of the dependency lists in the supplied makefile have been changed
(section \ref{extra:executables}) to simplify the substitution of some routines by special adapted
versions (expert use only).

Minor bug fixes as documented on github.

\section{Principal changes in version 2020.01}\mylabel{changes-2020-01}

\BOUND\ and \FIELD\ can now calculate bound states of the relative motion for atomic and molecular pairs confined in a spherically symmetric harmonic trap; see section \ref{confinement}.

\section{Principal changes in version 2022.0}\mylabel{changes-2022-0}

The Airy propagator can now reuse the step positions at subsequent values of external fields,
to reduce noise due to field-dependent step lengths.

\BOUND\ can now calculate the derivative of the bound-state energy with respect to an applied field, for either the absolute energy or the binding energy with respect to a specified threshold. This can then be used to calculate the closed-channel fraction for a state near threshold.

Harmonic confinement may now be handled by inputting the frequency of relative motion; see section \ref{confinement}.

\begin{shaded} 
\section{Principal changes in version 2025.0}\mylabel{changes-2025-0}

The fundamental constants used in the programs have been updated to the 2022 CODATA values.

The routine \code{FINDRM}, which locates a suitable inner starting point for propagation when $\inpitem{IRMSET} > 0$ on input, has been rewritten. The new version is simpler, with fewer special cases. It returns slightly different results from the old version, which has resulted in small numerical changes in output (within convergence criteria) for several of the examples described in section \ref{testfiles} of this manual.

The code for calculating bound states in the presence of a harmonic trapping potential has been extended to allow input of the trap frequency as an alternative to the harmonic length; see section \ref{confinement}.



All hard-coded unit numbers used for scratch files now lie in the range from 900 to 999. Any additional files used by user-supplied routines should use unit numbers between 800 and 899 to avoid conflicts. As a consequence, values of unit numbers in input data files should not be greater than 799.

Minor bug fixes as documented on github.

The documentation of the basis sets used for nonlinear molecules has been rewritten to integrate the description of symmetric, asymmetric and spherical tops and to clarify their symmetry properties; see section \ref{nonlin}.


Significant changes in the documentation since version \lastversion\ are shaded in the style of this section.

\medskip\noindent
{\bf Changes to supplied plug-in routines}

The routines used to supply parameters for diatomic potential-energy curves with the functional form of the Hannover group have been changed to avoid the use of data modules; see section \ref{detail:Hannover}. The changes are needed to circumvent issues arising from the way that \prog{make} handles module files.

\medskip\noindent
{\bf Changes that do not affect operation}

\MOLSCAT, \BOUND\ and \FIELD\ now use allocatable arrays to manage arrays with dimensions that depend on basis-set size.  The common block \code{MEMORY} and the main storage array \var{X} (of dimension \var{MX}) have been removed. It is no longer necessary to use different main programs, with different dimensions of \var{X}, for different cases.

Arithmetic \code{IF} statements have been replaced with IF blocks of the form \code{IF ... THEN ... ELSEIF ... THEN ... ELSE ... ENDIF}.

Labelled \code{DO} loops have been replaced with \code{DO...ENDDO}, with just a few having construct names.  Where possible, \code{GOTO} statements have been replaced by \code{CYCLE} or \code{EXIT} instructions as appropriate.

Long \code{IF} blocks and \code{DO} loops have comments preceding the beginning and following the end to aid with locating their endpoints.
\end{shaded} 

\chapter{\texorpdfstring{Theory}{\ref{Theory}: Theory}}\mylabel{Theory}

\section{The Hamiltonian}

There are many problems in quantum mechanics in which the total Hamiltonian of
the system may be written
\begin{equation}
H=-\frac{\hbar^2}{2\mu}R^{-1}\frac{\d^2\ }{\d R^2}R
+\frac{\hbar^2 \hat L^2}{2\mu R^2}+H_{\rm intl}(\xi_{\rm intl})+V(R,\xi_{\rm intl}),
\label{eqh}
\end{equation}
where $R$ is a radial coordinate describing the separation of two particles and
$\xi_{\rm intl}$ represents all the other coordinates in the system. $H_{\rm
intl}$ represents the sum of the internal Hamiltonians of the isolated
particles, and depends on $\xi_{\rm intl}$ but not $R$, and $V(R,\xi_{\rm
intl})$ is an interaction potential. The operator $\hbar^2 \hat L^2/2\mu R^2$
is the centrifugal term that describes the end-over-end rotational energy of
the interacting pair.

The internal Hamiltonian $H_{\rm intl}$ is a sum of terms for the two particles
1 and 2,
\begin{equation}
H_{\rm intl}(\xi_{\rm intl}) = H_{\rm intl}^{(1)}(\xi_{\rm intl}^{(1)})
+ H_{\rm intl}^{(2)}(\xi_{\rm intl}^{(2)}),
\end{equation}
with eigenvalues $E_{{\rm intl},i}=E_{{\rm intl},i}^{(1)}+E_{{\rm
intl},i}^{(2)}$, where $E_{{\rm intl},i}^{(1)}$ and $E_{{\rm intl},i}^{(2)}$
are energies of the separated monomers $1$ and $2$. The individual terms can
vary enormously in complexity: each one may represent a structureless atom,
requiring no internal Hamiltonian at all, a vibrating and/or rotating molecule,
or a particle with electron and/or nuclear spins. The problems that arise in
ultracold physics frequently involve pairs of atoms or molecules with electron
and nuclear spins, often in the presence of external electric, magnetic or
photon fields. All these complications can be taken into account in the
structure of $H_{\rm intl}$ and the interaction potential $V(R,\xi_{\rm
intl})$, which may both involve terms dependent on spins and external fields.

\section{Coupled-channel approach}

A standard computational approach for solving the Schr\"o\-ding\-er equation
for the Hamiltonian (\ref{eqh}) is the \emph{coupled-channel} approach, which
handles the radial coordinate $R$ by direct numerical propagation on a grid,
and all the other coordinates using a basis set. In the coupled-channel
approach, the total wavefunction is expanded
\begin{equation} \Psi(R,\xi_{\rm intl})
=R^{-1}\sum_j\Phi_j(\xi_{\rm intl})\psi_{j}(R), \label{eqexp}
\end{equation}
where the functions $\Phi_j(\xi_{\rm intl})$ form a complete orthonormal basis
set for motion in the coordinates $\xi_{\rm intl}$ and the factor $R^{-1}$
serves to simplify the form of the radial kinetic energy operator. The
wavefunction in each {\em channel} $j$ is described by a radial \emph{channel
function} $\psi_{j}(R)$. The expansion (\ref{eqexp}) is substituted into the
total Schr\"odinger equation, and the result is projected onto a basis function
$\Phi_i(\xi_{\rm intl})$. The resulting coupled differential equations for the
channel functions $\psi_{i}(R)$ are
\begin{equation}\frac{\d^2\psi_{i}}{\d R^2}
=\sum_j\left[W_{ij}(R)-{\cal E}\delta_{ij}\right]\psi_{j}(R), \label{eq:se-invlen}
\end{equation}
where $\delta_{ij}$ is the Kronecker delta, ${\cal E}=2\mu E/\hbar^2$, $E$ is
the total energy, and
\begin{equation}
W_{ij}(R)=\frac{2\mu}{\hbar^2}\int\Phi_i^*(\xi_{\rm intl}) [\hbar^2 \hat L^2/2\mu R^2 +
H_{\rm intl}+V(R,\xi_{\rm intl})] \Phi_j(\xi_{\rm intl})\,\d\xi_{\rm intl}. \label{eqWij}
\end{equation}
The different equations are coupled by the off-diagonal terms $W_{ij}(R)$ with $i\ne j$.

The coupled equations may be expressed in matrix notation,
\begin{equation}
\frac{\d^2\boldsymbol{\psi}}{\d R^2}= \left[{\bf W}(R)-{\cal E}{\bf
I}\right]\boldsymbol{\psi}(R). \label{eqcp}
\end{equation}
If there are $N$ basis functions included in the expansion (\ref{eqexp}),
$\boldsymbol{\psi}(R)$ is a column vector of order $N$ with elements
$\psi_{j}(R)$, ${\bf I}$ is the $N\times N$ unit matrix, and ${\bf W}(R)$ is an
$N\times N$ interaction matrix with elements $W_{ij}(R)$.

In general there are $N$ linearly independent solution vectors
$\boldsymbol{\psi}(R)$ that satisfy the Schr\"o\-ding\-er equation subject to
the boundary condition that $\boldsymbol{\psi}(R)\rightarrow0$ in the
classically forbidden region at short range. These $N$ column vectors form a
wavefunction matrix $\boldsymbol{\Psi}(R)$. The various propagators in
\MOLSCAT, \BOUND\ and \FIELD\ work either by propagating $\boldsymbol{\Psi}(R)$
and its radial derivative $\boldsymbol{\Psi}'(R)$ or by propagating the
log-derivative matrix ${\bf
Y}(R)=\boldsymbol{\Psi}'(R)[\boldsymbol{\Psi}(R)]^{-1}$.

The particular choice of the basis functions $\Phi_j(\xi_{\rm intl})$ and the
resulting form of the interaction matrix elements $W_{ij}(R)$ depend on the
physical problem being considered. The complete set of coupled equations often
factorises into blocks determined by the symmetry of the system. In the absence
of external fields, the \emph{total angular momentum} $J_{\rm tot}$ and the
\emph{total parity} are conserved quantities. Different or additional
symmetries arise in different physical situations. The programs are designed to
loop over total angular momentum and parity, constructing a separate set of
coupled equations for each combination and solving them by propagation. These
loops may be repurposed for other symmetries when appropriate.

The programs can also handle interactions that occur in external fields, where
the total angular momentum is no longer a good quantum number.

\section{Convention for quantum numbers}\mylabel{theory:qn}

In bound-state and scattering calculations, it is often necessary to
distinguish between quantum numbers for the individual monomers and for the
pair (supermolecule or interaction complex). It is a widely used convention to
use lower-case letters for the individual monomers and upper-case letters for
the pair. Because of this, monomer quantum numbers that are conventionally
upper-case in single-molecule spectroscopy ($J$, $K$, $M_J$, $F$, etc.) are
often converted to lower-case here ($j$, $k$, $m_j$, $f$, etc.), with the
upper-case letters reserved for the corresponding quantum numbers of the
interacting pair. Where monomer quantum numbers are needed for both species,
they are indicated by a subscript 1 or 2 ($j_1$, $j_2$, etc.).

In keeping with this convention, the end-over-end angular momentum of the
interacting pair is denoted $L$, rather than $l$.

\section{Matrix of the interaction potential}\mylabel{theory:W}

In order to streamline the calculation of matrix elements for the propagation,
\MOLSCAT, \FIELD\ and \BOUND\ express the interaction potential in an expansion
over the internal coordinates,
\begin{equation}
V(R,\xi_{\rm intl})=\sum_\Lambda v_\Lambda(R){\cal V}^\Lambda(\xi_{\rm intl}).
\label{eqvlambda}
\end{equation}
The specific form of the expansion depends on the nature of the interacting
particles. The radial potential coefficients $v_\Lambda(R)$ may either be
supplied explicitly, or generated internally by numerically integrating over
$\xi_{\rm intl}$. The $R$-independent coupling matrices $\bcalV^\Lambda$ with
elements ${\cal V}^\Lambda_{ij}=\langle\Phi_i|{\cal
V}^\Lambda|\Phi_j\rangle_{\rm intl}$ are calculated once and stored for use in
evaluating $W_{ij}(R)$ throughout the course of a propagation.

\section{Matrices of the internal and centrifugal Hamiltonians}\mylabel{theory:Wextra}

Coupled-channel scattering theory is most commonly formulated in a basis set
where $\hat L^2$ and $H_{\rm intl}$ are both diagonal. All the built-in
coupling cases use basis sets of this type. The matrix of $H_{\rm intl}$ is
$\langle\Phi_i|H_{\rm intl}|\Phi_j\rangle_{\rm intl}=E_{{\rm
intl},i}\delta_{ij}$. The diagonal matrix elements of $\hat L^2$ are often of
the form $L_i(L_i+1)$, where the integer quantum number $L_i$ (sometimes called
the partial-wave quantum number) represents the end-over-end angular momentum
of the two particles about one another.

However, the programs also allow the use of basis sets where one or both of
$\hat L^2$ and $H_{\rm intl}$ are non-diagonal. If $H_{\rm intl}$ is
non-diagonal, it is expanded as a sum of terms
\begin{equation}
H_{\rm intl}(\xi_{\rm intl})
=\sum_\Omega h_\Omega {\cal H}^\Omega_{\rm intl}(\xi_{\rm intl}),
\label{eqHomega1}
\end{equation}
where the $h_\Omega$ are scalar quantities, some of which may represent
external fields if desired. The programs generate additional coupling matrices
$\boldsymbol{\cal H}^\Omega$ with elements ${\cal
H}^\Omega_{ij}=\langle\Phi_i|{\cal H}^\Omega_{\rm intl}|\Phi_j\rangle_{\rm
intl}.$ These are also calculated once and stored for use in evaluating
$W_{ij}(R)$ throughout the course of a propagation. A similar
mechanism is used for basis sets where $\hat L^2$ is non-diagonal, with
\begin{equation}
\hat L^2
=\sum_\Upsilon {\cal L}^\Upsilon.
\label{eqL2}
\end{equation}

If $H_{\rm intl}$ is non-diagonal, the allowed energies $E_{{\rm intl},i}$ of
the pair of monomers at infinite separation are the eigenvalues of $H_{\rm
intl}$. The wavefunctions of the separated pair are represented by simultaneous
eigenvectors of $H_{\rm intl}$ and $\hat L^2$.

\section[\texorpdfstring{Results of scattering calculations ({\color{\mcol}\MOLSCAT}
only)}{Results of scattering calculations (MOLSCAT only)}]{Results of
scattering calculations} \mylabel{theory:scatcalcs}

\cbcolor{\mcol}\cbstart The outcome of a collision process is usually described
in quantum mechanics by the scattering matrix (S matrix), which contains
information on the probability amplitudes and phases for the various possible
outcomes. In simple cases (diagonal $H_{\rm intl}$ and $\hat L^2$), each
possible outcome corresponds to one of the channels in the coupled equations.
Alternatively, if $H_{\rm intl}$ is non-diagonal, each outcome corresponds to
an \emph{asymptotic channel} represented by one of the simultaneous
eigenvectors of $H_{\rm intl}$ and $\hat L^2$ with energy eigenvalue $E_{{\rm
intl},i}$. Each asymptotic channel $i$ is {\em open} if it is energetically
accessible as $R\rightarrow\infty$ ($E_{{\rm intl},i}\le E$) or \emph{closed}
if it is energetically forbidden ($E_{{\rm intl},i}>E$).

For each $J_{\rm tot}$ and symmetry block, solutions to the coupled equations
are propagated from deep inside the classically forbidden region at short range
to a distance at long range beyond which the interaction potential may be
neglected. The wavefunction matrix $\boldsymbol{\Psi}(R)$ and its radial
derivative (or the log-derivative matrix ${\bf Y}(R)$) are then matched to the
analytic functions that describe the solutions of the Schr\"odinger equation in
the absence of an interaction potential,
\begin{equation}
\boldsymbol{\Psi}(R)={\bf J}(R)+{\bf N}(R){\bf K}.
\label{K-bc}
\end{equation}
where the matrices ${\bf J}(R)$ and ${\bf N}(R)$ are diagonal and are made up
of Riccati-Bessel functions for the open channels and modified spherical Bessel
functions for the closed channels.\footnote{These boundary conditions are
appropriate only if the interaction potential decays faster than $R^{-2}$ at
long range. Different boundary conditions would be needed to handle long-range
Coulomb potentials.}

For each channel $i$, the Bessel function is of order $L_i$ and its argument is
$k_i R$, where $k_i$ is the asymptotic wavevector such that $\hbar^2
k_i^2/2\mu=|E-E_{{\rm intl},i}|$.

The real symmetric $N\times N$ matrix ${\bf K}$ is then converted to the S
matrix,
\begin{equation}
{\bf S}=({\bf I}+i{\bf K}_{\rm oo})^{-1}({\bf I}-i{\bf K}_{\rm oo}),
\end{equation}
where ${\bf K}_{\rm oo}$ is the open-open portion of ${\bf K}$. ${\bf S}$ is a
complex symmetric unitary matrix of dimension $N_{\rm open}\times N_{\rm
open}$, where $N_{\rm open}$ is the number of open channels.

If $\hat L^2$ and $H_{\rm intl}$ are both diagonal, the asymptotic channels
used for matching to Bessel functions are the same as the channels used to
propagate the wavefunction matrix $\boldsymbol{\Psi}(R)$ or its log-derivative
${\bf Y}(R)$.

If $\hat L^2$ and/or $H_{\rm intl}$ is non-diagonal, \MOLSCAT\ transforms
$\boldsymbol{\Psi}(R)$ or ${\bf Y}(R)$ at $R=R_{\rm max}$ into a basis set that
diagonalises $\hat L^2$ and $H_{\rm intl}$.
\begin{itemize}[leftmargin=13pt]
\item If $\hat L^2$ is diagonal but $H_{\rm intl}$ is not, \MOLSCAT\
    constructs the matrix of $H_{\rm intl}$ for each value of $L$ in turn,
    and diagonalises it. It uses the resulting eigenvectors to transform
    the corresponding block of $\boldsymbol{\Psi}(R_{\rm max})$ or ${\bf
    Y}(R_{\rm max})$ into the asymptotic basis set.
\item If $\hat L^2$ is non-diagonal (whether $H_{\rm intl}$ is diagonal or
    not), \MOLSCAT\ constructs the complete matrix of $H_{\rm intl}$ and
    diagonalises it. If there are degenerate eigenvalues $E_{{\rm intl},i}$
    of $H_{\rm intl}$, it then constructs the matrix of $\hat L^2$ for each
    degenerate subspace and diagonalises it to obtain eigenvalues $L_i$ and
    simultaneous eigenvectors of $\hat L^2$ and $H_{\rm intl}$. It uses the
    simultaneous eigenvectors to transform $\boldsymbol{\Psi}(R_{\rm max})$
    or ${\bf Y}(R_{\rm max})$ into the asymptotic basis set.
\end{itemize}
Finally \MOLSCAT\ uses the transformed $\boldsymbol{\Psi}(R_{\rm max})$ or
${\bf Y}(R_{\rm max})$, together with the eigenvalues $E_{{\rm intl},i}$ and
$L_i$, to extract ${\bf K}$.

Experimental observables that describe completed collisions, such as
differential and integral cross sections, and scattering lengths, can be
written in terms of S-matrix elements. Cross sections typically involve a
\emph{partial-wave sum},\footnote{Note that there is some inconsistency in the
literature in the use of ``partial wave" for multichannel scattering. It
sometimes refers to the total angular momentum, and sometimes to the
end-over-end angular momentum $L$ of the colliding pair.} with contributions
from many values of $J_{\rm tot}$, except at the lowest kinetic energies (in
the ultracold regime). By default \MOLSCAT\ uses its S~matrices to accumulate
degeneracy-averaged state-to-state integral cross sections, which may be
written
\begin{equation}
\sigma_{n_{\rm i}\rightarrow n_{\rm f}} = \frac{\pi}{g_{n_{\rm i}} k_{n_{\rm i}}^2}
\sum_{\substack{J_{\rm tot}\\M}} (2J_{\rm tot}+1)
\sum_{\substack{i\in n_{\rm i}\\f\in n_{\rm f}}}
\left|\delta_{if}-S_{if}^{J_{\rm tot},M}\right|^2.\label{eqsigdef}
\end{equation}
Here $n_{\rm i}$ and $n_{\rm f}$ label initial and final levels (not states) of the colliding
pair,\footnote{For some interaction types, $n_{\rm i}$ and $n_{\rm f}$ each represent several
quantum numbers, not just one.} while $i$ and $f$ indicate the open channels arising from those
levels for total angular momentum $J_{\rm tot}$ and symmetry block $M$. $g_{n_{\rm i}}$ is the
degeneracy of level $n_{\rm i}$. The S matrices may optionally be used to compute line-shape cross
sections (section \ref{pressbroad}), or be written to a file for post-processing in order to obtain
other kinds of collision properties.

\MOLSCAT\ can calculate line-shape cross sections for the broadening, shifting
and mixing of spectroscopic lines for most of the built-in coupling cases.
Line-shape cross sections require scattering calculations for the upper and
lower states of the spectroscopic transition at the same \emph{kinetic} (not
total) energy; if desired, the input energies are interpreted as kinetic
energies and the total energies required are generated internally.

\MOLSCAT\ has features to locate scattering resonances, which may produce sharp features in the
energy-dependence of cross sections and may also be interpreted as predissociating states of Van
der Waals complexes. It can calculate the S-matrix eigenphase sum, which is a generalisation of the
scattering phase shift to multichannel problems. It can converge on resonances in the
eigenphase sum and obtain their positions and widths, and can also calculate partial widths to
individual open channels.

\MOLSCAT\ can output S matrices to auxiliary files for later processing.
Separate programs are available:
\begin{itemize}[nosep]
\item program \prog{DCS} \cite{DCS} to calculate differential cross
    sections;
\item program \prog{SBE} \cite{SBE} to calculate generalised transport,
    relaxation and Senftleben-Beenakker cross sections;
\item program \prog{RESFIT} \cite{Hutson:resfit:2007} to fit to eigenphase
    sums and S-matrix elements to extract resonance positions, widths and
    partial widths as a function of energy or external field.
\end{itemize}

\subsection{Low-energy collision properties}\mylabel{theory:lowE}

\MOLSCAT\ has many features designed to facilitate low-energy scattering
calculations.

\subsubsection{Scattering lengths and volumes}

\MOLSCAT\ can calculate scattering lengths/volumes, which may be complex in the
presence of inelastic channels (section \ref{scatlen}). The diagonal S-matrix
element in an incoming channel 0 may be written in terms of a complex phase
shift $\eta$,
\begin{equation}
S_{00}=\exp(2i\eta).
\end{equation}
For a channel with low kinetic energy, $\tan \eta$ may be expanded in powers of the incoming
wavevector $k$. For a potential that varies as $-C_s / R^{s}$ at long range, the leading term
for $s\geq4$ is proportional to $k$ for $s$-wave scattering and $k^n$ for a partial wave with
$L>0$, with $n=\min(s-2,2l+1)$. By default, \MOLSCAT\ assumes $s=6$ and so calculates scattering
lengths for $L=0$ ($n=1$), volumes for $L=1$ ($n=3$) and hypervolumes for $L>1$ ($n=4$) using the
formula \cite{Hutson:res:2007}
\begin{equation}
a_L(k)=\frac{-\tan\eta}{k^n}=\frac{1}{ik^n}\left(\frac{1-S_{00}}{1+S_{00}}\right).
\label{eq:scatln}\end{equation} It should be emphasised that Eq.~\ref{eq:scatln} is an identity, so
that this is a far more general approach than that used by some other programs that take the limit
of other (often more complicated) functions as $k\rightarrow0$. The scattering length and volume
defined by Eq.~\ref{eq:scatln} become independent of $k$ at sufficiently low $k$.

If desired, the powers $n$ used in Eq.\ \ref{eq:scatln} may be changed by setting a different
value of the power $s$ as variable \var{LRPOW} in subroutine \prog{SPROC}.

\subsubsection{Characterisation of zero-energy Feshbach resonances}

\MOLSCAT\ can converge on and characterise the zero-energy Feshbach resonances
that appear in the scattering length as a function of external fields
\cite{Frye:resonance:2017} (section \ref{fieldconv}). It can do this both for
resonances in elastic scattering, where the scattering length has a simple pole
characterised by its position and width and the background scattering length,
and in inelastic scattering, where the resonant behaviour is more complex and
requires additional parameters \cite{Hutson:res:2007}.

\subsection{Infinite-order sudden approximation}\mylabel{theory:IOS}

\MOLSCAT\ incorporates code for calculating degeneracy-averaged and line-shape
cross sections within the infinite-order sudden (IOS) ansatz. In this
formulation, S matrices are calculated from propagations carried out at fixed
molecular orientations. The cross sections are written in terms of sums of
products of dynamical factors $Q$ and spectroscopic coefficients $F$. The
dynamical factors contain all the information about the collision dynamics;
they are defined as integrals over the fixed-orientation S matrices, which are
evaluated by numerical quadrature. The spectroscopic coefficients contain
information about rotor levels and angular momentum coupling. The IOS code has
not been used much in recent years, but is retained for backwards
compatibility. It does not provide low-energy features such as scattering
lengths. \cbend

\section[\texorpdfstring{Results of bound-state calculations
({\color{\bcol}\BOUND} and {\color{\fcol}\FIELD} only)}{Results of bound-state
calculations (BOUND and FIELD only)}]{Results of bound-state
calculations}\mylabel{theory:boundcalcs}

\cbcolor{\bfcol}\cbstart The quantum-mechanical bound-state problem can be
formulated as a set of coupled differential equations similar to those
encountered in scattering theory. The difference between the two cases is in
the boundary conditions that must be applied. True bound states exist only at
energies where all asymptotic channels are energetically closed, $E<E_{{\rm
intl},i}$ for all $i$. Under these circumstances the bound-state wavefunction
$\boldsymbol{\psi}(R)$ is a column vector of order $N$ that must approach zero
in the classically forbidden regions at both short range, $R\rightarrow 0$, and
long range, $R\rightarrow \infty$.

Continuously differentiable solutions of the coupled equations that satisfy the
boundary conditions at both ends exist only at specific energies $E_n$. These
are the eigenvalues of the total Hamiltonian (\ref{eqh}); we refer to them
(somewhat loosely) as the eigenvalues of the coupled equations, to distinguish
them from eigenvalues of other operators that also enter the discussion below.

Wavefunction matrices $\boldsymbol{\Psi}(R)$ that satisfy the boundary
conditions in \emph{one} of the classically forbidden regions exist at any
energy. We designate these $\boldsymbol{\Psi}^+(R)$ for the solution propagated
outwards from short range and $\boldsymbol{\Psi}^-(R)$ for the solution
propagated inwards from long range. The corresponding log-derivative matrices
are ${\bf Y}^+(R)$ and ${\bf Y}^-(R)$.

It is convenient to choose a matching distance $R_{\rm match}$ where the
outwards and inwards solutions are compared. A solution vector that is
continuous at $R_{\rm match}$ must satisfy
\begin{equation}
\boldsymbol{\psi}(R_{\rm match})=\boldsymbol{\psi}^+(R_{\rm match})=\boldsymbol{\psi}^-(R_{\rm match}).
\end{equation}

Since the derivatives of the outwards and inwards solutions must match too, we
require that
\begin{equation}
\frac{d}{dR}\boldsymbol{\psi}^+(R_{\rm match})=\frac{d}{dR}\boldsymbol{\psi}^-(R_{\rm match})
\end{equation}
so that
\begin{equation}
{\bf Y}^+(R_{\rm match})\boldsymbol{\psi}(R_{\rm match})
= {\bf Y}^-(R_{\rm match})\boldsymbol{\psi}(R_{\rm match}).
\end{equation}
Equivalently,
\begin{equation}
\left[{\bf Y}^+(R_{\rm match}) - {\bf Y}^-(R_{\rm match})\right]
\boldsymbol{\psi}(R_{\rm match}) = 0,
\label{eq:ymatch}
\end{equation}
so that the wavefunction vector $\boldsymbol{\psi}(R_{\rm match})$ is an
eigenvector of the log-derivative matching matrix, $\Delta{\bf Y} = \left[{\bf
Y}^+(R_{\rm match}) - {\bf Y}^-(R_{\rm match})\right]$, with eigenvalue zero
\cite{Hutson:CPC:1994}.

For each $J_{\rm tot}$ and symmetry block, \BOUND\ propagates log-derivative
matrices to a matching point $R_{\rm match}$, both outwards from the
classically forbidden region at short range (or from $R=0$) and inwards from
the classically forbidden region at long range. At each energy $E$, it
calculates the multichannel node count, defined as the number of zeros of
$\boldsymbol{\psi}(R)$ between $R_{\rm min}$ and $R_{\rm max}$. Johnson
\cite{Johnson:1978} showed that this is equal to the number of states that lie
below $E$. It may be calculated as a simple byproduct of the propagations and
the matching matrix. \BOUND\ uses the node count to determine the number of
states in the specified range, and then uses bisection to identify energy
windows that contain exactly one state. In each such window, it uses a
combination of bisection and the Van Wijngaarden-Dekker-Brent algorithm
\cite{VWDB} to converge on the energy where an eigenvalue of the log-derivative
matching matrix $\Delta {\bf Y}$ is zero. This is the energy of a state. The
program extracts the local wavefunction vector $\boldsymbol{\psi}(R_{\rm
match})$, and optionally calculates the complete bound-state wavefunction
$\boldsymbol{\psi}(R)$ using the method of Thornley and Hutson
\cite{THORNLEY:1994}.

\FIELD\ operates in a very similar manner to locate states as a function of external field at fixed
energy (or energy fixed with respect to a field-dependent threshold energy). The one significant
difference is that the multichannel node count is not guaranteed to be a monotonic function of
field, and it is in principle possible to miss pairs of states that cross the chosen energy in
opposite directions as a function of field. In practice this seldom happens. \cbend

\chapter{\texorpdfstring{Using the programs: a basic guide}
{\ref{use}: Using the programs: a basic guide}}\mylabel{use}

\section{Prerequisites}\mylabel{use:prereqs}

The user must always specify:
\begin{itemize}[nosep]
\item the type of system, e.g., atom + linear rigid rotor, diatom + diatom,
    etc.;
\item any dynamical approximations to be applied in setting up the coupled
    equations;
\item the energy levels of the interacting particles, or atomic/molecular
    constants that describe them;
\item the basis set $\{\Phi_j(\xi_{\rm intl})\}$ to be used for the
    internal coordinates $\xi_{\rm intl}$ (everything except the
    interparticle separation $R$).
\item the interaction potential $V(R,\xi_{\rm intl})$;
\item the reduced mass.
\end{itemize}
The programs use this information to set up the required coupled equations.

\medskip\noindent In addition, the user must choose:
\begin{itemize}[nosep]
\item the propagator(s) to be used to solve the coupled equations;
\item parameters to control the energies (and, if appropriate, external
    fields) at which the coupled equations are to be solved;
\item parameters to control the range of interparticle distance over which
    the coupled equations are to be solved, and the propagation step size;
\item parameters to control optional processing, such as calculating
    line-shape cross sections or converging on scattering resonances;
\item parameters to control the level of printed output and optional
    additional output and scratch files.
\end{itemize}

Parameters are input to the program from a plain-text file in namelist format
as described below. The types of system supported are summarised in section
\ref{use:colltype} and the numerical methods available for solving the coupled
differential equations are summarised in section \ref{use:props}.

\section{Input data format}\mylabel{use:input}

The main input file for any of the programs is read on unit 5 (standard input),
and consists of several blocks of namelist data in the following order.

\begin{description}
\item[{\namelist{\&INPUT}}]{is read in subroutine \prog{DRIVER}, and provides overall control
    of the calculation: reduced mass, collision or binding energies, external fields, total
    angular momenta, choice of propagators, propagation ranges and step sizes, specification of
    optional calculations (line-shape cross sections, resonance characterisation, etc.), output
    files and print control. The available input parameters are described in chapters
    \ref{Energy}, \ref{inputfield}, \ref{propcontrol}, \ref{processres}, \ref{processbound},
    \ref{CommII} and \ref{Auxfiles}. Throughout the remainder of this document (with the
    exception of the glossary in chapter \ref{CommI}) items in \namelist{\&INPUT} are coloured
    {\color{red}red}.}

\item [{\namelist{\&BASIS}}]{is read in entry \prog{BASIN} of subroutine
    \prog{BASE}, and describes the type of system, atomic and molecular
    parameters, basis set, and dynamical approximations. The available
    input parameters are described in chapter \ref{ConstructBasis}.
    Throughout the remainder of this document (with the exception of the
    glossary in chapter \ref{CommI}) items in \namelist{\&BASIS} are
    coloured {\color{blue}blue}.}

\item[\namelist{\&POTL}]{is read in the initialisation call of the
    general-purpose version of subroutine \prog{POTENL}, and specifies the
    interaction potential to be used.  The available input parameters are
    described in chapter \ref{buildVL}. Throughout the remainder of this
    document (with the exception of the glossary in chapter \ref{CommI})
    items in \namelist{\&POTL} are coloured {\color{ForestGreen}green}.

    For some cases it may be desirable to substitute a special-purpose
    \prog{POTENL} routine, which does not necessarily read namelist
    \namelist{\&POTL}.  The specification of \prog{POTENL} is given in
    chapter \ref{userPotenl} for use by those who wish to substitute their
    own routine for the general-purpose version.}
\end{description}

All the namelist items are listed in the glossary in chapter \ref{CommI},
together with their default values, a brief description and references to where
the reader can find more information.

In addition, there may be other blocks of input data required by user-supplied
subroutines, which must be inserted in the correct place between or after the
namelist blocks listed above.  In particular, many plug-in basis-set suites
accessed through $\basisitem{ITYPE}=9$ utilise an additional namelist block
named \namelist{\&BASIS9}. These include the suite for interaction of two
alkali-metal atoms, described in chapter \ref{base9}.

Each namelist block in the input file must start with \namelist{\&<name>} or
\namelist{\$<name>}, where \code{<name>} is \code{INPUT}, \code{BASIS} or
\code{POTL} as appropriate, and should be terminated by a slash. Between these
delimiters, the namelist input data consist of entries of the form
$\code{KEYWORD}=$\code{value} or $\code{KEYWORD}=$\code{list of values}, where
\code{KEYWORD} is the variable or array name. Values can be separated by commas,
spaces, tabs, or end-of-line.  A value can be repeated by preceding it with a
multiplier and the \code{*} character. The Fortran 90 standard specifies that
all entries (including the \namelist{\$<name>} and the slash) should begin in
column 2 or later, but most common compilers are more flexible. Most compilers
allow the use of comments; these must be preceded by an exclamation mark and
continue to the end of the line.

Most of the parameters have sensible default values (which are given in this
document); these are used if the parameter is not included in the namelist
block.

\renewcommand\inpitem[1]{{\color{red}\tt #1}\index{{\tt #1} in \namelist{\&INPUT}}}
\renewcommand\basisitem[1]{{\color{blue}\tt #1}\index{{\tt #1} in \namelist{\&BASIS}}}
\renewcommand\potlitem[1]{{\color{ForestGreen}\tt #1}\index{{\tt #1} in \namelist{\&POTL}}}

\section{Interaction types}\mylabel{use:colltype}

The term \emph{interaction type} describes the types of the interacting
monomers and any dynamical approximations to be applied. It was originally
\emph{collision type} in \MOLSCAT, but has been generalised here to include
pairs of monomers that form bound states.

The programs can perform close-coupling calculations (with no dynamical
approximations) for the following interaction types:
\begin{enumerate}[nosep]
\item Atom + linear rigid rotor \cite{Arthurs:1960};
\item Atom + vibrating diatom (rotationally and/or vibrationally inelastic)
    with interaction potentials independent of diatom rotational state
   \cite{Green:1979:vibrational};
\item Linear rigid rotor + linear rigid rotor
    \cite{Green:1975,Green:1977:comment,Heil:1978:coupled};
\item Asymmetric top + linear molecule \cite{Phillips:1995}
\item Atom + symmetric top (also handles near-symmetric tops and linear
    molecules with vibrational angular momentum)
   \cite{Green:1976,Green:1979:IOS};
\item Atom + asymmetric top \cite{Green:1976} (also handles spherical tops
    \cite{Hutson:spher:1994});
\item Atom + vibrating diatom (rotationally and/or vibrationally inelastic)
    with interaction potentials dependent on diatom rotational state
    \cite{Hutson:sbe:1984};
\item Atom + rigid corrugated surface: diffractive (elastic) scattering
    \cite{Wolken:1973:surface,Hutson:1983}. At present, the code is
    restricted to centrosymmetric lattices, for which the potential
    matrices are real;
\item Interaction type specified in a plug-in basis-set suite (which may be
    user-supplied). A substantial number of these plug-in suites exist.
    This release includes two representative examples:
\begin{itemize}[nosep]
\item Structureless atom + $^3\Sigma$ molecule in a magnetic field,
    demonstrated for Mg + NH;
\item Two alkali-metal atoms, including hyperfine coupling and magnetic
    field, demonstrated for $^{85}$Rb$_2$.
\end{itemize}
\end{enumerate}

The quantities that control the quantum states included in the basis set and
the corresponding internal energies of the interacting partners are specified in
namelist \namelist{\&BASIS}.

The computer time required to solve a set of $N$ coupled equations is
approximately proportional to $N^3$. The practical limit on $N$ is from a few
hundred to several thousand, depending on the speed of the computer and the
amount of memory available. The basis sets necessary for converged
close-coupling calculations may easily exceed this limit as scattering energies
increase or rotational constants decrease, particularly for interaction types
other than the very simplest. However, the programs also provide various
approximate (decoupling) methods that reduce the number of coupled equations.
The methods supported at present are:
\begin{itemize}[nosep]
\item{Effective potential approximation \cite{Rabitz:EP} (seldom used
    nowadays);}
\item{Coupled-states (centrifugal sudden) approximation \cite{McG74};}
\item{Decoupled $L$-dominant approximation
    \cite{Green:1976:DLD,DePristo:1976:DLD} (seldom used nowadays);}
\item{\cbcolor{\mcol}\cbstart Infinite-order sudden approximation
    (\MOLSCAT\ only) \cite{Gol77III,Par78,Green:1979:IOS}.\cbend}
\end{itemize}
Not all these approximations are supported for all interaction types, though
the most common ones are. The programs print a warning message and exit if an
unsupported approximation is requested.

\section{Interaction potential}\mylabel{use:interpot}

The programs call a routine (named \prog{POTENL}) to evaluate the radial
potential coefficients $v_\Lambda(R)$ of Eq.~\ref{eqvlambda} that describe the
interaction potential from information provided by the user. The
general-purpose version of \prog{POTENL} obtains information about the
interaction potential either from namelist \namelist{\&POTL}, or from
user-supplied routines that may either provide the radial potential
coefficients directly, or (for some interaction types) may provide values of
the potential at specified points for expansion within \prog{POTENL}. More
information about this given in chapter~\ref{buildVL}.

\section{Propagators}\mylabel{use:props}

The coupled equations may be solved using any one of several methods:
\begin{description}
\item[de Vogelaere propagator (DV) \cite{deVogelaere:1955}:]{\cbcolor{\mcol}\cbstart{}This
    propagates the wavefunction explicitly, but is much slower than more modern methods,
    especially for large reduced masses or high scattering energies. It is not recommended
    except for special purposes.}
\item[R-matrix propagator (RMAT) \cite{Stechel:1978}:]{This is a stable method that works in a
    quasiadiabatic basis. It has relatively poor step-size convergence properties, and has
    largely been superseded by the log-derivative propagators. It is not recommended except for
    special purposes.}\cbend
\item[Log-derivative propagator of Johnson (LDJ) \cite{Johnson:1973,
    Manolopoulos:1993:Johnson}:]{This is a very stable propagator. It has largely been
    superseded by the LDMD propagator, but can be useful in occasional cases where that
    propagator has trouble evaluating node counts.}
\item[Diabatic log-derivative method of Manolopoulos (LDMD) \cite{Manolopoulos:1986}:]{This is
    a very efficient and stable propagator, especially at short and medium range. It is coded
    to detect single-channel cases (including IOS cases) automatically and in that case use a
    more efficient implementation.}
\item[Quasiadiabatic log-derivative propagator of Manolopoulos (LDMA)
    \cite{Manolopoulos:PhD:1988, Hutson:CPC:1994}:]{This\\ is similar to the LDMD propagator,
    but operates in a quasiadiabatic basis. It offers better accuracy than LDMD for very
    strongly coupled problems, but is relatively expensive. It is recommended for production
    runs only for very strongly coupled problems. However, it is also useful when setting up a
    new system, because it can output eigenvalues of the interaction matrix at specific
    distances (adiabats) and nonadiabatic couplings between the adiabatic states.}
\item[Symplectic log-derivative propagators of Manolopoulos and Gray (LDMG)
    \cite{MG:symplectic:1995}:]{\mbox{}\\ This offers a choice of 4th-order or 5th-order
    symplectic propagators. These are 1.5 to 3 times more expensive per step than the LDMD and
    LDJ propagators, but can have smaller errors for a given step size.  They are often the
    most efficient choice when high precision is required.}
\item[AIRY propagator:]{This is the AIRY log-derivative propagator of Alexander
    \cite{Alexander:1984} as reformulated by Alexander and Manolopoulos \cite{Alexander:1987}.
    It uses a quasiadiabatic basis with a linear reference potential (which results in Airy
    functions as reference solutions). This allows the step size to increase rapidly with
    separation, so that this propagator is particularly efficient at long range.}
\item[VIVS propagator \cite{Parker:VIVS}:]{\cbcolor{\mcol}\cbstart{}This is the
    variable-interval variable-step method of Parker \etal\ and is intended for use at long
    range. It is sometimes very efficient, but the interval size is limited when there are
    deeply closed channels, so that it is not efficient at long range in such cases. Control of
    it is considerably more complicated than for other propagators, and it has largely been
    superseded by the AIRY propagator.}
\item[WKB semiclassical integration using Gauss-Mehler quadrature
    \cite{Pac74}:]{This is not a true propagator and can be used only for
    single-channel problems.\cbend}
\end{description}

\cbcolor{\bfcol}\cbstart{In \BOUND\ and \FIELD, only log-derivative propagators are
implemented.}\cbend

All these propagators have options that allow them to use interaction matrices stored at the first
total energy when doing calculations at subsequent energies. For the RMAT, VIVS, and log-derivative
propagators, some of the remaining work is also avoided at subsequent energies, so that in special
circumstances they may cost only 30\% as much CPU time as the first energy.

Gordon's propagator, which was available in early versions of \MOLSCAT, is not implemented in
version \currentversion.

\medskip \noindent \emph{Recommendation:}

For applications that do not need high precision, the LDMD propagator provides a good balance
between stability and efficiency at short and medium range. When high precision is
required, the LDMG propagator may be a better choice.

For problems that do not require long-range propagation, the LDMD or LDMG propagator may be used on
its own. When propagation to very long range is required, it is usually best to combine it with the
AIRY propagator for the long-range part of the propagation.

The remaining propagators should be used only for special purposes by expert users.

\section{Overview of main input file}\mylabel{use:outline}

The main input file specifies the calculation required, and the associated
tasks may be grouped roughly as follows:

\subsection{Scattering and bound-state calculations}\mylabel{outline:scat-and-bound}

\begin{enumerate}[nosep]
\item{Specify the interaction type; see the preamble in chapter
    \ref{ConstructBasis}}
\item{Specify appropriate loops over total angular momentum and/or other
    symmetries; see section \ref{angmom}}
\item{Construct a basis set appropriate to the interacting partners; see the
    rest of chapter \ref{ConstructBasis}}
\item{Construct a potential expansion appropriate to the interacting
    partners; see chapter \ref{buildVL}}
\item{Specify energies and external fields; see chapters \ref{Energy} and
    \ref{inputfield}}
\item{Choose propagator(s) and specify propagation ranges and step sizes;
    see chapter \ref{propcontrol}}
\end{enumerate}

\subsection{Scattering calculations}\mylabel{outline:scats}

\cbcolor{\mcol}\cbstart If the corresponding option is requested:
{\begin{enumerate}[nosep,resume]
\item{Control automated testing of convergence of S-matrix elements with
    respect to propagtion parameters; see section \ref{andconv}}
\item{Specify spectroscopic lines for line-shape cross sections; see section
    \ref{pressbroad}}
\item{Control searches for energy-dependent resonances; see
    section \ref{energyconv}}
\item{Control convergence on field-dependent resonances; see section
    \ref{fieldconv}}
\item{Specify the incoming channel for effective-range calculations; see
    section \ref{effrange}}
\item{Specify the incoming channel for a wavefunction calculation; see
    section \ref{calcwaveM}}
\end{enumerate}}\cbend

\subsection{Bound-state calculations}\mylabel{outline:bound}

If the corresponding option is requested:
\begin{enumerate}[nosep,resume]
\cbcolor{\bcol}\cbstart
\item{Specify expectation values to be calculated (without wavefunctions)
    (\BOUND\ only); see section \ref{calcexp}}
\item{Use automated testing of convergence of bound-state energies and
    expectation values with respect to propagation parameters (\BOUND\
    only); see section \ref{conv}}\cbend
\item{\cbcolor{\bfcol}\cbstart Specify a wavefunction calculation; see
    section \ref{calcwaveBF}.\cbend}

\end{enumerate}
\clearpage
\newgeometry{left=2cm}

\section{Units of mass, length and energy}\mylabel{outline:units}

By default, the programs operate with masses in unified atomic mass units
(Daltons), lengths in \AA\ (1 \AA\ = $10^{-10}$ m) and energies $E$ expressed
as wavenumbers $E/hc$ in cm$^{-1}$. However, all these may be altered using the
variables \inpitem{MUNIT}, \inpitem{RUNIT} and \inpitem{EUNIT}, which give
values for the required units in Daltons, \AA\ and cm$^{-1}$ respectively.

If \inpitem{RUNIT} is not specified, the programs take the unit of length
	from the quantity \var{RM} returned by subroutine \prog{POTENL}. This is
	implemented to provide backward compatibility with input files for program
	versions before 2019.0, and is deprecated in the current version.

Commonly used energy units may be selected with the integer variable
\inpitem{EUNITS} (default 1) in place of \inpitem{EUNIT}. The allowed values
are
\begin{description}[threecol,nosep]
      \item[\inpitem{EUNITS} = 1]{cm$^{-1}$}
      \item[\inpitem{EUNITS} = 2]{Kelvin}
      \item[\inpitem{EUNITS} = 3]{MHz}
      \item[\inpitem{EUNITS} = 4]{GHz}
      \item[\inpitem{EUNITS} = 5]{eV}
      \item[\inpitem{EUNITS} = 6]{erg}
      \item[\inpitem{EUNITS} = 7]{hartree (atomic unit of energy)}
      \item[\inpitem{EUNITS} = 8]{kJ/mol}
      \item[\inpitem{EUNITS} = 9]{kcal/mol}
\end{description}
The value in \inpitem{EUNIT} is used only if \inpitem{EUNITS} is zero.

\section[\texorpdfstring{Examples for {\color{\mcol}\MOLSCAT}}{Examples for MOLSCAT}]{Examples for \MOLSCAT}\mylabel{basic:m:examples}

\subsection{Interpretation of the complete output for a model system}\mylabel{basic:m1}

\tcbset{colback=green!10!white,breakable,skin=enhanced,enlarge left
by=-1cm,enlarge right by=1.5cm,width=\linewidth+2.5cm}

The input file \file{molscat-basic1.input} sets up a small calculation of cross sections for
collisions between an atom and a homonuclear rigid rotor as follows:
\begin{verbatim}
 &INPUT
    LABEL  = 'model system: ITYPE=1',
    URED   =  20.0,
    IPRINT =   1,   ISIGPR =    2,
\end{verbatim}
Here \inpitem{URED} specifies the reduced mass for the collision (in unified atomic mass units).
The print level $\inpitem{IPRINT}=1$ specifies minimal output, and $\inpitem{ISIGPR}=1$ specifies
output of state-to-state cross sections. The program outputs a header with date and time and
reports the input values:
\begin{tcolorbox}[skin=enhancedfirst]
\begin{scriptsize}
\begin{verbatim}
  ---- MOLSCAT ------ MOLSCAT ------ MOLSCAT ------ MOLSCAT ------ MOLSCAT ---
 |                                                                            |
 |                Non-reactive quantum scattering calculations                |
 |                     on atomic and molecular collisions                     |
 |                                                                            |
 |              Copyright (C) 2025 J. M. Hutson & C. R. Le Sueur              |
 |                                                                            |
 |                               Version 2025.0                               |
 |                                                                            |
 |                       Run on xx Xxx 2025   at xx:xx:xx                     |
 |                                                                            |
  ---- MOLSCAT ------ MOLSCAT ------ MOLSCAT ------ MOLSCAT ------ MOLSCAT ---


  This program is free software: you can redistribute it and/or modify it under
  the terms of the GNU General Public License, version 3, as published by
  the Free Software Foundation.

  Publications resulting from the use of this program should cite both
  the version of the program used:

  J. M. Hutson & C. R. Le Sueur, MOLSCAT: a program for non-reactive quantum
  scattering calculation on atomic and molecular collisions,
  Version 2025.0, https//github.com/molscat/molscat

  and the published paper:

  J. M. Hutson & C. R. Le Sueur, Comput. Phys. Commun. 241, 9-18 (2019).


  USING CODATA 2022 RECOMMENDED VALUES OF FUNDAMENTAL PHYSICAL CONSTANTS

  PRINT LEVEL (IPRINT) =  1     OTHER PRINT CONTROLS  ISIGPR = 2

  REDUCED MASS FOR INTERACTION =  20.000000000 ATOMIC MASS UNITS (DALTONS)
\end{verbatim}
\end{scriptsize}
\end{tcolorbox}
The parameters input in namelist \namelist{\&BASIS} specify the interaction
type and energy levels of the colliding partners:
\begin{verbatim}
    ITYPE  = 1,  BE     = 30.0,
    NLEVEL = 4,  JLEVEL =  0, 2, 4, 6,
\end{verbatim}
Here \basisitem{BE} specifies the rotational constant of the rotor, by default in units of
cm$^{-1}$. \basisitem{NLEVEL} specifies that the calculation will include the 4 values of the
rotational quantum number listed in \basisitem{JLEVEL}; these particular values are all even, as
might be appropriate for a homonuclear diatomic molecule. The resulting output is:
\begin{tcolorbox}[skin=enhancedmiddle]
\begin{scriptsize}
\begin{verbatim}
  INTERACTION TYPE IS    LINEAR RIGID ROTOR  -  ATOM.

  MOLECULAR QUANTUM NUMBERS TAKEN FROM JLEVEL INPUT.  NLEVEL =  4

  ENERGY LEVELS OBTAINED FROM B(E) =   30.000000

  QUANTUM NUMBERS FOR INTERACTING PAIR:
  EACH PAIR STATE IS LABELLED BY  1 QUANTUM NUMBER
  EACH CHANNEL FUNCTION IS FORMED BY COMBINING A PAIR STATE WITH A VALUE OF L.
  THE RESULTING BASIS SET IS ASYMPTOTICALLY DIAGONAL.

  PAIR STATE     PAIR STATE QUANTUM NUMBERS    PAIR LEVEL     PAIR ENERGY (CM-1)
                ---------        J ---------
        1                        0                1             0.0000000
        2                        2                2           180.0000000
        3                        4                3           600.0000000
        4                        6                4          1260.0000000
\end{verbatim}
\end{scriptsize}
\end{tcolorbox}
The parameters input in namelist \namelist{\&POTL} items control the interaction potential.  For
$\basisitem{ITYPE}=1$, the potential is expanded in Legendre polynomials $P_\lambda(\cos\theta)$,
and for a homonuclear diatomic molecule only terms with even $\lambda$ exist because of symmetry.
In this calculation each of the $\potlitem{MXLAM}=2$ expansion terms (corresponding to $\lambda=0$
and 2) consists of one or two inverse-power expressions:
\begin{verbatim}
    MXLAM = 2,   LAMBDA =   0,          2,
                 NTERM  =   2,          1,
                 NPOWER = -12,   -6,   -6,
                 A      =   1.0, -2.0, -0.2,

\end{verbatim}
\begin{tcolorbox}[skin=enhancedmiddle]
\begin{scriptsize}
\begin{verbatim}
  GENERAL-PURPOSE POTENL ROUTINE (MAY 18)

  ANGULAR DEPENDENCE OF POTENTIAL EXPANDED IN TERMS OF
  LEGENDRE POLYNOMIALS, P(LAMBDA).

  INTERACTION POTENTIAL FOR EXPANSION TERM NUMBER   1
  WHICH HAS LAMBDA =   0

                 1.00000000E+00 * R **-12
                -2.00000000E+00 * R ** -6

  INTERACTION POTENTIAL FOR EXPANSION TERM NUMBER   2
  WHICH HAS LAMBDA =   2

                -2.00000000E-01 * R ** -6

  POTENL PROCESSING FINISHED.
\end{verbatim}
\end{scriptsize}
\end{tcolorbox}
The units used for length throughout the run  and for interaction energies
returned from the potential routine are specified as \potlitem{RM} and
\potlitem{EPSIL} in namelist \namelist{\&POTL}:
\begin{verbatim}
    RM    = 3.5, EPSIL  =  50.0,
\end{verbatim}
resulting in the following output:
\begin{tcolorbox}[skin=enhancedmiddle]
\begin{scriptsize}
\begin{verbatim}
  POTENTIAL RETURNED IN UNITS OF EPSIL  =  50.000000     CM-1
        CODED WITH R IN UNITS OF RM     =  3.5000000     ANGSTROM

  ALL LENGTHS ARE IN UNITS OF RM (  3.50000000 ANGSTROM ) UNLESS OTHERWISE STATED

  INTERACTION MATRIX USES   2 BLOCKS OF VL ARRAY FOR R-DEPENDENT TERMS IN POTENTIAL
\end{verbatim}
\end{scriptsize}
\end{tcolorbox}

The following items in \namelist{\&INPUT} control the range of the propagation, the step size, and
the propagator used:
\begin{verbatim}
    RMIN   =   0.5, RMAX   =   20.0,
    IPROPS =   6,   DR     =    0.001,
\end{verbatim}
and result in the following output:
\begin{tcolorbox}[skin=enhancedmiddle]
\begin{scriptsize}
\begin{verbatim}
  PROPAGATION METHODS FOR COUPLED EQUATIONS SPECIFIED BY IPROPS =  6

  COUPLED EQUATIONS WILL BE PROPAGATED OUTWARDS IN 1 SEGMENT

  PROPAGATION RANGE IS CONTROLLED BY VARIABLES RMIN AND RMAX, WITH INPUT VALUES
  RMIN =  0.5000     RMAX =   20.00
  ++++++++++++++++++++++++++++++++++++++++++++++++++++++++++++++++++++++++++++++++++++++++++++++++++++++++++++++++++++++
  SEGMENT 1 WILL BE PROPAGATED OUTWARDS

  FROM RMIN CHOSEN USING IRMSET =  9 TO RMAX =     20.00

  COUPLED EQUATIONS SOLVED BY DIABATIC MODIFIED LOG-DERIVATIVE PROPAGATOR OF MANOLOPOULOS

  PROPAGATION STEP SIZE DETERMINED USING DR =  1.000E-03
  STEP SIZE CONSTANT THROUGHOUT RANGE
  STEP SIZE MAY BE ADJUSTED SLIGHTLY SO THAT RANGE IS A WHOLE NUMBER OF STEPS

  LOG-DERIVATIVE MATRIX INITIALISED IN THE LOCAL EIGENBASIS AT RMIN
  LOCALLY CLOSED CHANNELS INITIALISED WITH A WKB BOUNDARY CONDITION
\end{verbatim}
\end{scriptsize}
\end{tcolorbox}
Energies are assumed to be in cm$^{-1}$ by default, and the input file
specifies that only 1 energy is to be used
\begin{verbatim}
    NNRG   =   1,   ENERGY = 1250.0,
\end{verbatim}
\begin{tcolorbox}[skin=enhancedmiddle]
\begin{scriptsize}
\begin{verbatim}
  INPUT ENERGIES ASSUMED TO BE IN UNITS OF CM-1 BY DEFAULT.

  CALCULATIONS WILL BE PERFORMED FOR    1 ENERGY
  ENERGY    1           =   1250.000000     CM-1
\end{verbatim}
\end{scriptsize}
\end{tcolorbox}

The following items in \namelist{\&INPUT} control the values of total angular
momentum for which calculations are executed:
\begin{verbatim}
    JTOTL  =  10,   JTOTU  =   20,     JSTEP = 10,
\end{verbatim}
resulting in
\begin{tcolorbox}[skin=enhancedmiddle]
\begin{scriptsize}
\begin{verbatim}
  TOTAL ANGULAR MOMENTUM JTOT RUNS FROM  10  TO     20  IN STEPS OF  10

  EACH JTOT IS SPLIT INTO A MAXIMUM OF   2 SYMMETRY BLOCKS
\end{verbatim}
\end{scriptsize}
\end{tcolorbox}
This concludes the initialisation procedures.

The program then proceeds to do 4 propagations (2 values of $J_{\rm tot}$, each
of which is factorised into 2 symmetry blocks), which are summarised:
\begin{tcolorbox}[skin=enhancedmiddle]
\begin{scriptsize}
\begin{verbatim}
  =============================================== model system: ITYPE=1 ================================================

  FOR JTOT =   10, SYMMETRY BLOCK =  1, ENERGY(  1) =  1250.000    : MAX DIAG & OFF-DIAG =  9.88E-02 &   1.16E-06

  FOR JTOT =   10, SYMMETRY BLOCK =  2, ENERGY(  1) =  1250.000    : MAX DIAG & OFF-DIAG =  1.75E-01 &   1.29E-03

  FOR JTOT =   20, SYMMETRY BLOCK =  1, ENERGY(  1) =  1250.000    : MAX DIAG & OFF-DIAG =  1.88E-01 &   1.95E-06

  FOR JTOT =   20, SYMMETRY BLOCK =  2, ENERGY(  1) =  1250.000    : MAX DIAG & OFF-DIAG =  1.68E-01 &   2.43E-03
\end{verbatim}
\end{scriptsize}
\end{tcolorbox}

The program calculates an S matrix from each propagation, which is printed if
$\inpitem{IPRINT}\ge 11$ (but not here). It uses the S matrices to calculate
partial cross sections, which are printed if $\inpitem{IPRINT}\ge 5$ (but not
here). At the end of the calculation, the program prints the
degeneracy-averaged state-to-state integral cross sections. These are from
initial levels \code{I} to final levels \code{F}; the level energies are given
here, and the corresponding quantum numbers may be obtained from the list
above.
\begin{tcolorbox}[skin=enhancedmiddle]
\begin{scriptsize}
\begin{verbatim}
  LEVEL   4 WITH ENERGY   1260.000000000000 IS NEVER OPEN

  STATE-TO-STATE INTEGRAL CROSS SECTIONS IN ANGSTROM**2 BETWEEN    3 LEVELS WITH THRESHOLD ENERGIES (IN CM-1):

    1     0.000000000000
    2   180.000000000000
    3   600.000000000000

  *** N.B. CROSS SECTIONS HAVE BEEN MULTIPLIED BY 10.0 TO ACCOUNT FOR JSTEP

   ENERGY (CM-1)    JTOTL  JSTEP  JTOTU      F    I         SIG(F,I)
   1250.000000         10     10     20      1    1          1.81057
   1250.000000         10     10     20      2    1         3.722327E-02
   1250.000000         10     10     20      3    1         6.542671E-07

   1250.000000         10     10     20      1    2         8.697026E-03
   1250.000000         10     10     20      2    2          3.00836
   1250.000000         10     10     20      3    2         8.161625E-05

   1250.000000         10     10     20      1    3         1.398007E-07
   1250.000000         10     10     20      2    3         7.464050E-05
   1250.000000         10     10     20      3    3          5.24476
\end{verbatim}
\end{scriptsize}
\end{tcolorbox}
and the total inelastic integral cross section from each initial level:
\begin{tcolorbox}[skin=enhancedmiddle]
\begin{scriptsize}
\begin{verbatim}
  TOTAL INELASTIC INTEGRAL CROSS SECTIONS IN ANGSTROM**2 FROM LEVEL
   3.72239E-02                                                  1
   8.77864E-03                                                  2
   7.47803E-05                                                  3
\end{verbatim}
\end{scriptsize}
\end{tcolorbox}
The output terminates with a footer message:
\begin{tcolorbox}[skin=enhancedlast]
\begin{scriptsize}
\begin{verbatim}
  ---- MOLSCAT ------ MOLSCAT ------ MOLSCAT ------ MOLSCAT ------ MOLSCAT ---
 |                                                                            |
 |                Non-reactive quantum scattering calculations                |
 |                     on atomic and molecular collisions                     |
 |                                                                            |
 |              Copyright (C) 2025 J. M. Hutson & C. R. Le Sueur              |
 |                                                                            |
 |                               Version 2025.0                               |
 |                                                                            |
 |                      This run used       xxxx cpu secs                     |
 |                                                                            |
  ---- MOLSCAT ------ MOLSCAT ------ MOLSCAT ------ MOLSCAT ------ MOLSCAT ---
\end{verbatim}
\end{scriptsize}
\end{tcolorbox}

\subsection{Higher print level}\mylabel{basic:m2}

The following example (whose output is not listed in full) illustrates the
output at a higher print level, $\inpitem{IPRINT}=11$, for a more sophisticated
case involving collisions of two rigid rotors, specifically para-H$_2$
colliding with ortho-H$_2$. It also uses a realistic potential provided by a
\prog{VRTP} routine.

Only sections of the output file that differ in important ways from section
\ref{basic:m1} are described here.

\file{molscat-basic2.input} contains the input data:
\begin{verbatim}
 &INPUT
    LABEL  = 'p-H2 + o-H2: potential of Zarur and Rabitz supplied by VRTP',
    URED   =  1.00794,
    IPRINT = 11,    ISIGPR =   1,
    RMIN   =  0.43, RMID   =   2.0, RMAX   = 20.0,  IRXSET =  1,
    IPROPS =  6,    IPROPL =   9,   STEPS  = 15.0,
    JTOTL  =  6,    JTOTU  =   6,
    NNRG   =  1,    ENERGY = 700.0,
 /

 &BASIS
    ITYPE  = 3,
    BE     = 2*59.067,
    NLEVEL = 3,   JLEVEL = 0,1,  0,3,  2,1,
 /

 &POTL
    MXLAM  = 4,      NTERM  = 4*-1,
                     LAMBDA = 0,0,0,  2,0,2,  0,2,2,  2,2,4,
    LVRTP  = .TRUE.,
 /
\end{verbatim}
The basis set is described in terms of pair levels and pair states for the
molecule-molecule system, as described in section \ref{basis:lev-state}. The
following output lists the pair state quantum numbers and corresponding pair
levels and pair energies. In this case there are several pair states arising
from one of the pair levels.
\begin{tcolorbox}[skin=enhancedmiddle]
\begin{scriptsize}
\begin{verbatim}
  INTERACTION TYPE IS    LINEAR ROTOR - LINEAR ROTOR.

  PAIR LEVEL QUANTUM NUMBERS TAKEN FROM JLEVEL INPUT.  NLEVEL =  3

  ENERGY LEVELS OF ROTOR 1 OBTAINED FROM B(E) =   59.067000

  ENERGY LEVELS OF ROTOR 2 OBTAINED FROM B(E) =   59.067000

  QUANTUM NUMBERS FOR INTERACTING PAIR:
  EACH PAIR STATE IS LABELLED BY  3 QUANTUM NUMBERS
  EACH CHANNEL FUNCTION IS FORMED BY COMBINING A PAIR STATE WITH A VALUE OF L.
  THE RESULTING BASIS SET IS ASYMPTOTICALLY DIAGONAL.

  PAIR STATE    - PAIR STATE QUANTUM NUMBERS -   PAIR LEVEL     PAIR ENERGY (CM-1)
                       J1        J2       J12
        1               0         1         1       1           118.1340000
        2               0         3         3       2           708.8040000
        3               2         1         1       3           472.5360000
        4               2         1         2       3           472.5360000
        5               2         1         3       3           472.5360000
\end{verbatim}
\end{scriptsize}
\end{tcolorbox}
The general-purpose potential routine produces the following output, which
describes the symmetries of the colliding molecules (both homonuclear), and the
expansion used for the interaction potential. $\potlitem{LVRTP}=\code{.TRUE.}$
specifies that the potential coefficients are to be obtained by quadrature,
with the potential at the quadrature points evaluated by a user-supplied
routine \prog{VRTP}.
\begin{tcolorbox}[skin=enhancedmiddle]
\begin{scriptsize}
\begin{verbatim}
  GENERAL-PURPOSE POTENL ROUTINE (MAY 18)

  UNEXPANDED POTENTIAL IS OBTAINED FROM VRTP ROUTINE.

  A SUITABLE VRTP ROUTINE MUST BE SUPPLIED.

  [OUTPUT FROM INITIALISATION OF SUPPLIED VRTP ROUTINE]

  IHOMO  = 2 SPECIFIES HOMONUCLEAR SYMMETRY FOR ROTOR 1.
  IHOMO2 = 2 SPECIFIES HOMONUCLEAR SYMMETRY FOR ROTOR 2.
  USING   3-POINT QUADRATURE FOR THETA-1
  HOMONUCLEAR SYMMETRY: ONLY HALF OF THE THETA-1 POINTS WILL BE USED
  USING   3-POINT QUADRATURE FOR THETA-2
  HOMONUCLEAR MOLECULE 2: ONLY HALF OF THE THETA-2 POINTS WILL BE USED
  USING   3-POINT QUADRATURE FOR PHI

  ANGULAR DEPENDENCE OF POTENTIAL EXPANDED IN TERMS OF
  CONTRACTED NORMALISED SPHERICAL HARMONICS,
  SUM(M1,M2,M) C(L1,M1,L2,M2,L,M) Y(L1,M1) Y(L2,M2) Y(L,M)
  SEE GREEN, J. CHEM. PHYS. 62, 2271 (1975)

  INTERACTION POTENTIAL FOR EXPANSION TERM NUMBER   1
  WHICH HAS LAM1 =   0,  LAM2 =   0,  LAM =   0

  INTERACTION POTENTIAL FOR EXPANSION TERM NUMBER   2
  WHICH HAS LAM1 =   2,  LAM2 =   0,  LAM =   2

  INTERACTION POTENTIAL FOR EXPANSION TERM NUMBER   3
  WHICH HAS LAM1 =   0,  LAM2 =   2,  LAM =   2

  INTERACTION POTENTIAL FOR EXPANSION TERM NUMBER   4
  WHICH HAS LAM1 =   2,  LAM2 =   2,  LAM =   4


  POTENL PROCESSING FINISHED.

  POTENTIAL RETURNED IN UNITS OF EPSIL  =  24.170000     CM-1
  CODED WITH R IN UNITS OF RM     =  3.4900000     ANGSTROM

  ALL LENGTHS ARE IN UNITS OF RM (  3.49000000 ANGSTROM ) UNLESS OTHERWISE STATED

  INTERACTION MATRIX USES   4 BLOCKS OF VL ARRAY FOR R-DEPENDENT TERMS IN POTENTIAL
\end{verbatim}
\end{scriptsize}
\end{tcolorbox}

The propagation is carried out in 2 parts, with different propagators at short
and long range. This produces the following output:
\begin{tcolorbox}[skin=enhancedmiddle]
\begin{scriptsize}
\begin{verbatim}
  PROPAGATION METHODS FOR COUPLED EQUATIONS SPECIFIED BY IPROPS =  6 AND IPROPL =  9

  COUPLED EQUATIONS WILL BE PROPAGATED OUTWARDS IN 2 SEGMENTS

  PROPAGATION RANGES ARE CONTROLLED BY VARIABLES RMIN, RMID AND RMAX, WITH INPUT VALUES
  RMIN =  0.4300     RMID =   2.000     RMAX =   20.00
  ++++++++++++++++++++++++++++++++++++++++++++++++++++++++++++++++++++++++++++++++++++++++++++++++++++++++++++++++++++++
  SEGMENT 1 WILL BE PROPAGATED OUTWARDS

  FROM RMIN CHOSEN USING IRMSET =  9 TO RMID =      2.00

  COUPLED EQUATIONS SOLVED BY DIABATIC MODIFIED LOG-DERIVATIVE PROPAGATOR OF MANOLOPOULOS

  PROPAGATION STEP SIZE DETERMINED USING STEP =   15.0     (PER HALF WAVELENGTH)
  STEP SIZE CONSTANT THROUGHOUT RANGE
  STEP SIZE MAY BE ADJUSTED SLIGHTLY SO THAT RANGE IS A WHOLE NUMBER OF STEPS

  LOG-DERIVATIVE MATRIX INITIALISED IN THE LOCAL EIGENBASIS AT RMIN
  LOCALLY CLOSED CHANNELS INITIALISED WITH A WKB BOUNDARY CONDITION
  ++++++++++++++++++++++++++++++++++++++++++++++++++++++++++++++++++++++++++++++++++++++++++++++++++++++++++++++++++++++
  SEGMENT 2 WILL BE PROPAGATED OUTWARDS

  FROM RMID =      2.00 TO WHICHEVER IS LARGER OF
  OUTERMOST CENTRIFUGAL TURNING POINT IN OPEN CHANNELS, AND RMAX, WHICH =     20.00

  COUPLED EQUATIONS SOLVED BY VARIABLE-STEP AIRY PROPAGATOR.
  PUBLICATIONS RESULTING FROM THE USE OF THIS PROPAGATOR SHOULD REFERENCE
  M. H. ALEXANDER AND D. E. MANOLOPOULOS,  J. CHEM. PHYS. 86, 2044 (1987).

  PROPAGATION STEP SIZE DETERMINED USING STEP =   15.0     (PER HALF WAVELENGTH)
  STEP SIZES ADJUSTED TO MAINTAIN APPROXIMATE ACCURACY VIA PERTURBATION THEORY
  WITH TOLHI =  1.00E-04 AND POWR =   3.0
\end{verbatim}
\end{scriptsize}
\end{tcolorbox}

The program now enters a loop over total angular momentum \var{JTOT} and
symmetry block \var{IBLOCK}. The first time through the loop is for
$\var{JTOT}=6$ and $\var{IBLOCK}=1$. The output for this begins with a list of
the channels, describing their relationship to the pair states above:

\begin{tcolorbox}[skin=enhancedmiddle]
\begin{scriptsize}
\begin{verbatim}
  *****************************  ANGULAR MOMENTUM JTOT  =   6   AND SYMMETRY BLOCK  =    1  ****************************

  CHANNEL FUNCTION LIST:

  EACH CHANNEL FUNCTION IS FORMED BY COMBINING A PAIR STATE WITH A VALUE OF L.

  CHANNEL  PAIR STATE   - PAIR STATE QUANTUM NUMBERS -           L        PAIR LEVEL   PAIR ENERGY (CM-1)
                              J1        J2       J12
        1           2          0         3         3             4             2       708.8040000
        2           4          2         1         2             4             3       472.5360000
        3           5          2         1         3             4             3       472.5360000
        4           1          0         1         1             6             1       118.1340000
        5           2          0         3         3             6             2       708.8040000
        6           3          2         1         1             6             3       472.5360000
        7           4          2         1         2             6             3       472.5360000
        8           5          2         1         3             6             3       472.5360000
        9           2          0         3         3             8             2       708.8040000
       10           4          2         1         2             8             3       472.5360000
       11           5          2         1         3             8             3       472.5360000
\end{verbatim}
\end{scriptsize}
\end{tcolorbox}
$\inpitem{IRMSET}>0$ (default 9) specifies that the program should locate a
suitable value of $R_{\rm min}$ as described in section \ref{intrange}. This
produces output as follows.
\begin{tcolorbox}[skin=enhancedmiddle]
\begin{scriptsize}
\begin{verbatim}
  INNER CLASSICAL TURNING POINT AT R =  0.734
  RADIAL PROPAGATION WILL START AT R =  0.430
\end{verbatim}
\end{scriptsize}
\end{tcolorbox} The program now enters a loop over energy. For each energy, it
prints the range and number of steps taken for each propagator used:
\begin{tcolorbox}[skin=enhancedmiddle]
\begin{scriptsize}
\begin{verbatim}
  MDPROP. LOG DERIVATIVE MATRIX PROPAGATED FROM       0.4304  TO  2.0000      IN    154  STEPS.
	
  AIRPRP. LOG DERIVATIVE MATRIX PROPAGATED FROM       2.0000  TO  20.000      IN     82  STEPS.
\end{verbatim}
\end{scriptsize}
\end{tcolorbox}
After each propagation, the program outputs the list of open channels and the
resulting S matrix. The open channels are listed in order of increasing
threshold energy (decreasing kinetic energy); the channel index for each open
channel may be used to find its quantum numbers in the full channel list above.
The S-matrix output is lengthy (64 entries in this case), so is abbreviated
here.
\begin{tcolorbox}[skin=enhancedmiddle]
\begin{scriptsize}
\begin{verbatim}
  OPEN CHANNEL   WVEC (1/ANG.)    CHANNEL           L    PAIR LEVEL        PAIR ENERGY (CM-1)  PAIR ENERGY (EPSIL)
           1    5.89835058E+00         4            6         1            118.134000000000     4.887629292511
           2    3.68786851E+00         3            4         3            472.536000000000    19.550517170046
           3    3.68786851E+00         2            4         3            472.536000000000    19.550517170046
           4    3.68786851E+00         6            6         3            472.536000000000    19.550517170046
           5    3.68786851E+00         7            6         3            472.536000000000    19.550517170046
           6    3.68786851E+00         8            6         3            472.536000000000    19.550517170046
           7    3.68786851E+00        10            8         3            472.536000000000    19.550517170046
           8    3.68786851E+00        11            8         3            472.536000000000    19.550517170046

  ROW  COL       S**2                  PHASE/2PI              RE (S)                 IM (S)
    1    1 9.9281505103293E-001  -1.5519889208423E-001   5.5903016764688E-001  -8.2480320240263E-001
    2    1 7.5823637370767E-004   1.4529421384187E-002   2.7421428212003E-002   2.5103084514894E-003

    [ENTRIES FOR REMAINDER OF 8x8 S MATRIX]

    7    8 1.2342642514477E-003   1.6557772306069E-003   3.5130195942515E-002   3.6549211784134E-004
    8    8 9.9351391972436E-001  -2.4320435013539E-001   4.2546698826389E-002  -9.9584320961853E-001
\end{verbatim}
\end{scriptsize}
\end{tcolorbox}
The program outputs the state-to-state partial cross sections. These are the
contributions to the corresponding integral cross sections from the current S
matrix.
\begin{tcolorbox}[skin=enhancedmiddle]
\begin{scriptsize}
\begin{verbatim}
  * * * * * * * * * *  STATE-TO-STATE PARTIAL CROSS SECTIONS (ANGSTROM**2) FROM LEVEL I TO LEVEL F  * * * * * * * * * *
                       FOR JTOT =   6 AND SYMMETRY BLOCK =   1 AT ENERGY(  1) =  700.0000     CM-1

    F  I =      1              3
    1       3.42292E-01    1.43838E-03
    3       2.81148E-03    3.42742E+00

  FOR JTOT =    6, SYMMETRY BLOCK =  1, ENERGY(  1) =  700.0000    : MAX DIAG & OFF-DIAG =  3.43E+00 &   2.81E-03
\end{verbatim}
\end{scriptsize}
\end{tcolorbox}
For subsequent values of \var{JTOT} and/or \var{IBLOCK}, the program also
outputs the values of the integral cross sections accumulated so far; for
$\var{JTOT}=6$ and $\var{IBLOCK}=2$, the corresponding output is
\begin{tcolorbox}[skin=enhancedmiddle]
\begin{scriptsize}
\begin{verbatim}
  * * * * * * * * * *  STATE-TO-STATE PARTIAL CROSS SECTIONS (ANGSTROM**2) FROM LEVEL I TO LEVEL F  * * * * * * * * * *
                       FOR JTOT =   6 AND SYMMETRY BLOCK =   2 AT ENERGY(  1) =  700.0000     CM-1

    F  I =      1              3
    1       1.94114E+00    2.07216E-03
    3       4.05025E-03    2.96884E+00

  FOR JTOT =    6, SYMMETRY BLOCK =  2, ENERGY(  1) =  700.0000    : MAX DIAG & OFF-DIAG =  2.97E+00 &   4.05E-03

  *-*-*-*-*-*-*-*-*-*-STATE-TO-STATE INTEGRAL CROSS SECTIONS: ACCUMULATED FROM JTOT =   6 TO   6 -*-*-*-*-*-*-*-*-*-*

    F  I =      1              3
    1       2.28343E+00    3.51054E-03
    3       6.86173E-03    6.39626E+00
\end{verbatim}
\end{scriptsize}
\end{tcolorbox}
After all propagations are complete, the program summarises the state-to-state
integral cross sections and prints a final footer message as before.

\section[\texorpdfstring{Example for {\color{\bcol}\BOUND}}{Example for BOUND}]{Example for \BOUND}\mylabel{basic:bd+fld}

\file{bound-basic1.input} contains input data for a small calculation of the
bound states for the model system in section \ref{basic:m1}. The only
differences from the corresponding input file for \MOLSCAT\ are that the
collision energies in \inpitem{ENERGY} are replaced with a range specified by
\inpitem{EMIN} and \inpitem{EMAX} and a matching point \inpitem{RMATCH} is
specified. The coupled equations are now propagated in two parts: outwards from
$R_{\rm min}$ to $R_{\rm match}$ and inwards from $R_{\rm max}$ to $R_{\rm
match}$.
\begin{tcolorbox}[skin=enhancedmiddle]
\begin{scriptsize}
\begin{verbatim}
  PROPAGATION METHODS FOR COUPLED EQUATIONS SPECIFIED BY IPROPS =  6 AND IPROPL =  6

  COUPLED EQUATIONS WILL BE PROPAGATED TOWARDS RMATCH IN 2 SEGMENTS

  PROPAGATION RANGES ARE CONTROLLED BY VARIABLES RMIN, RMATCH AND RMAX, WITH INPUT VALUES
  RMIN =  0.5000     RMATCH =   1.000     RMAX =   20.00
  ++++++++++++++++++++++++++++++++++++++++++++++++++++++++++++++++++++++++++++++++++++++++++++++++++++++++++++++++++++++
  SEGMENT 1 WILL BE PROPAGATED OUTWARDS

  FROM RMIN =      0.50 TO RMATCH =      1.00

  COUPLED EQUATIONS SOLVED BY DIABATIC MODIFIED LOG-DERIVATIVE PROPAGATOR OF MANOLOPOULOS

  PROPAGATION STEP SIZE DETERMINED USING DR =  1.000E-03
  STEP SIZE CONSTANT THROUGHOUT RANGE
  STEP SIZE MAY BE ADJUSTED SLIGHTLY SO THAT RANGE IS A WHOLE NUMBER OF STEPS

  LOG-DERIVATIVE MATRIX INITIALISED IN THE LOCAL EIGENBASIS AT RMIN IN THE OUTWARD PROPAGATION PART
  LOCALLY CLOSED CHANNELS INITIALISED WITH A WKB BOUNDARY CONDITION
  ++++++++++++++++++++++++++++++++++++++++++++++++++++++++++++++++++++++++++++++++++++++++++++++++++++++++++++++++++++++
  SEGMENT 2 WILL BE PROPAGATED  INWARDS

  TO RMATCH =      1.00 FROM RMAX =     20.00

  COUPLED EQUATIONS SOLVED BY DIABATIC MODIFIED LOG-DERIVATIVE PROPAGATOR OF MANOLOPOULOS

  INITIAL STEP SIZE TAKEN FROM SIZE OF FINAL STEP OF SHORT-RANGE PROPAGATION
  STEP SIZE CONSTANT THROUGHOUT RANGE
  STEP SIZE MAY BE ADJUSTED SLIGHTLY SO THAT RANGE IS A WHOLE NUMBER OF STEPS

  LOG-DERIVATIVE MATRIX INITIALISED IN THE LOCAL EIGENBASIS AT RMAX IN THE INWARD PROPAGATION PART
  LOCALLY CLOSED CHANNELS INITIALISED WITH A WKB BOUNDARY CONDITION
  LOCALLY  OPEN  CHANNELS INITIALISED WITH THE VALUE  0.000E+00
\end{verbatim}
\end{scriptsize}
\end{tcolorbox}
The matrix of the difference between the two resulting log-derivative matrices
at $R_{\rm match}$ is singular if a bound state exists at that energy. A
singular matrix has at least one eigenvalue that is zero, so \BOUND\ performs a
1D searches for and then converges on zero-valued eigenvalues of the matching
matrix. The input file requests location of bound states with binding energies
between 1~cm$^{-1}$ and 10~cm$^{-1}$ and total angular momentum of 1:
\begin{verbatim}
    JTOTL  =   1,   JTOTU  =  1,   IBFIX  =   2,
    EMIN   = -10.0, EMAX   = -1.0,
\end{verbatim}
In total, this potential supports 5 vibrational states, of which the 4th and
5th are in this energy range:
\begin{tcolorbox}[skin=enhancedmiddle]
\begin{scriptsize}
\begin{verbatim}
  =============================================== model system: ITYPE=1 ================================================

  *****************************  ANGULAR MOMENTUM JTOT  =   1   AND SYMMETRY BLOCK  =    2  ****************************

  CONVERGED ON STATE NUMBER     4 AT                              ENERGY =  -4.992074672     CM-1

  CONVERGED ON STATE NUMBER     5 AT                              ENERGY =  -1.413103892     CM-1
\end{verbatim}
\end{scriptsize}
\end{tcolorbox}

\section{Calculations in external fields}\mylabel{basic:rb2}

The programs can also perform calculations in external (electric, magnetic and
photon) fields. This is particularly important in low-energy atomic and
molecular scattering, where the scattering length may be controlled by varying
external fields in the vicinity of a low-energy Feshbach resonance. The
following set of (related) calculations illustrate this.

\subsection[\texorpdfstring{Using {\color{\mcol}\MOLSCAT} to calculate the
field-dependent scattering length}{Using MOLSCAT to calculate the field-dependent
scattering length}]{Using \MOLSCAT\ to calculate the field-dependent
scattering length}\mylabel{basic:rb2:molscat}

The following \MOLSCAT\ calculation is for collisions of two $^{85}$Rb atoms
(initially in the lowest, $f=2$, $m_f=2$, hyperfine state), with a collision
energy of 100 nK$\times k_{B}$, and in an external magnetic field. Feshbach
resonances appear as features in the scattering length as a function of
magnetic field. This calculation uses a plug-in basis-set suite, described in
section \ref{user:alk-alk}. The potential used is described in chapter
\ref{testfiles}. The input file that specifies this calculation is provided as
\file{molscat-basic\_Rb2.input}.

The basis-set suite used here (\file{base9-alk\_alk\_ucpld.f}) interprets
\inpitem{JTOTL} and \inpitem{JTOTU} as doubled values of $M_{\rm tot}$, which
is the projection of the total angular momentum onto the magnetic field axis;
$M_{\rm tot}$ is the only good angular momentum quantum number in the presence
of a magnetic field. Its value is specified by
\begin{verbatim}
    JTOTL  =   8,   JTOTU  =   8,     IBFIX  =  2,
\end{verbatim}
Setting $\inpitem{IBFIX}=2$ specifies that only the symmetry block with
$j+L+J_{\rm tot}$ even is required; for $\var{JTOT}=8$, this corresponds to
total parity $(-1)^{j+L}=+1$.

The collision energy is often defined with respect to the energy of the
incoming atoms, which is a function of magnetic field. In this run the energy
is specified as a temperature in K ($\inpitem{EUNITS}=2$) and is referred to
the lowest threshold (specified by \inpitem{MONQN}, which in this case contains
doubled values of $f_A$, $m_{fA}$, $f_B$, $m_{fB}$ for the required incoming
channel).
\begin{verbatim}
    EUNITS =   2,   NNRG   =   1,     ENERGY =  1.E-7,
                    DTOL   =   1.E-6, MONQN  =  4, 4, 4, 4,
\end{verbatim}

The scattering length is calculated at 1 G intervals over the range of interest
\begin{verbatim}
    FLDMIN = 800.0, FLDMAX = 900.0,   DFIELD =  1.0,
\end{verbatim}
for any channels that have low enough energy. The output produced at each field is:
\begin{tcolorbox}[skin=enhancedmiddle]
\begin{scriptsize}
\begin{verbatim}
  EFV SET     1:  MAGNETIC Z FIELD =   800.0000000     GAUSS
  REFERENCE ENERGY IS                          -0.1780012287     CM-1    = -0.2561040520     K


  THRESHOLDS CALCULATED FROM ASYMPTOTIC HAMILTONIAN:

  THRESHOLD       L           ENERGY/CM-1                ENERGY/K
          1       0        -0.178001228676            -0.256104052018
  [OTHER THRESHOLDS ALSO LISTED]

  K-DEPENDENT SCATTERING LENGTHS/VOLUMES/HYPERVOLUMES FOR CHANNELS WITH LOW KINETIC ENERGY
  CHAN   L POW     WVEC*BOHR             RE(A)/BOHR            IM(A)/BOHR
    1    0   1 2.2139874716420E-004  -3.9108737692924E+002  -5.5463499106231E-006
    2    2   4 2.2139874578519E-004  -4.0008184039040E+009  -5.0726855300317E+005
\end{verbatim}
\end{scriptsize}
\end{tcolorbox}
The acronym EFV stands for \emph{external field value}. The programs can handle multiple
simultaneous fields (including electric, magnetic and photon fields), so a general field is
specified by a set of values. In this case there is only a magnetic field, along the quantisation
axis, so only one value (here 800~G) is needed to specify it. The reference energy is the energy of
the channel specified by \inpitem{MONQN} at this field.

The output gives the scattering length for the $L=0$ channel and the scattering hypervolume for the
$L=2$ channel. The imaginary part of the s-wave ($L=0$) scattering length is non-zero because there
is some scattering into the d-wave ($L=2$) outgoing channel at the same threshold.

Plotting the (real part of the) s-wave scattering length as a function of
magnetic field yields the following:

\begin{minipage}[b]{0.65\textwidth}

\includegraphics[clip=true, trim=1.5cm 1.5cm 3cm 7.5cm, width=0.98\textwidth]{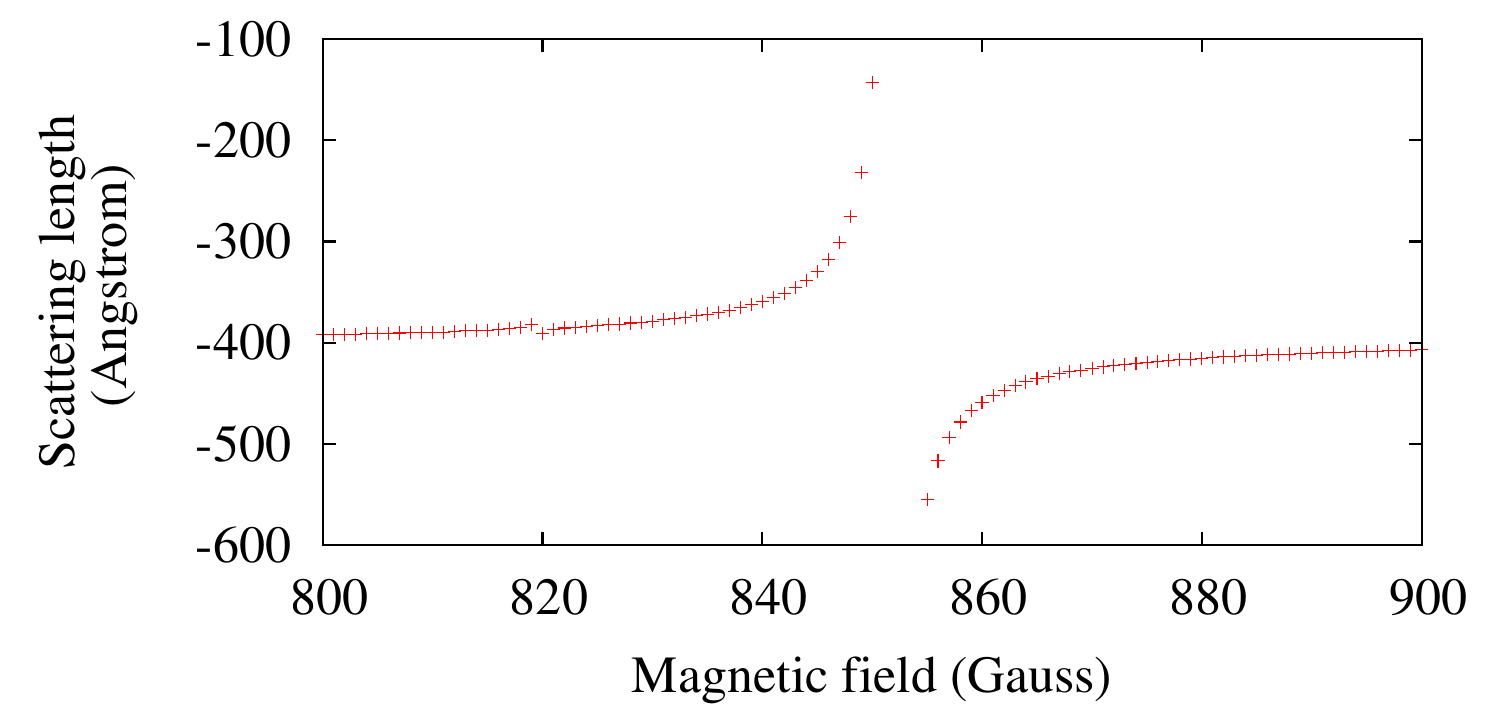}
\end{minipage}
\begin{minipage}[b]{0.35\textwidth}
This plot is dominated by a wide resonance that appears as a pole near 850~G,
and there is also a narrow resonance that is just visible near 820~G. The
narrow resonance would need a finer grid to see well, and in fact there is
also a third resonance that is too narrow to see at all with this grid.
\vspace*{0.5cm}

\end{minipage}

It is not always necessary or desirable to perform calculations on a grid of
fields across every resonance. \MOLSCAT\ offers facilities for converging on
resonances as a function of external field and extracting the parameters that
characterise them (positions, widths, etc.). These are described in section
\ref{fieldconv}.

\subsection[\texorpdfstring{Using {\color{\fcol}\FIELD} to locate threshold crossings}
{Using FIELD to locate threshold crossings}]{Using \FIELD\ to locate threshold
crossings}\mylabel{basic:rb2:field}

Narrow resonances can be hard to locate, and may be missed entirely in
calculations done on a grid which is too coarse. However, resonances occur at
fields where bound states cross the energy of the colliding species. \FIELD\
uses methods similar to \BOUND\ to locate the fields at which bound states have
a specified energy. In the absence of inelastic channels (lower in energy than
the incoming channel), it is therefore possible to use \FIELD\ to locate
\emph{all} the threshold crossings in a given range of fields. To adapt the
\MOLSCAT\ input file described above to do this, it is only necessary to remove
\inpitem{DFIELD} and change \inpitem{ENERGY} from 100 nK.  A collision energy
of 0 (i.e., \emph{at} the threshold) is usually appropriate. It is important to
retain the default boundary condition $\inpitem{BCYOMX}=0.0$ to give continuity
across the threshold.

The input file that specifies this calculation is provided as
\file{field-basic\_Rb2.input}. The resulting output locates the threshold
crossings and contains the lines

\begin{tcolorbox}[skin=enhancedmiddle]
\begin{scriptsize}
\begin{verbatim}
  CONVERGED ON STATE NUMBER  1469 AT                    MAGNETIC Z FIELD =   851.8761789     GAUSS
  .
  .
  CONVERGED ON STATE NUMBER  1470 AT                    MAGNETIC Z FIELD =   819.5625939     GAUSS
  .
  .
  CONVERGED ON STATE NUMBER  1471 AT                    MAGNETIC Z FIELD =   810.9471815     GAUSS
\end{verbatim}
\end{scriptsize}
\end{tcolorbox}
These include both the two resonances visible in the figure above and the third
one near 810~G that is too narrow to see in the figure.

\subsection[\texorpdfstring{Using {\color{\bcol}\BOUND} and {\color{\fcol}\FIELD} to build the bound-state picture}
{Using BOUND and FIELD to build the bound-state picture}]{Using \BOUND\ and \FIELD\ to build the bound-state picture}\mylabel{basic:rb2:bound}

It is possible to build a complete bound-state picture by running \FIELD\ at a
series of energies below threshold, or by running \BOUND\ at a series of values
of the field. In practice a combination of the two approaches is often
beneficial, particularly if some bound states vary relatively fast with field and
others are almost flat.

\restoregeometry

\chapter{\texorpdfstring{Interaction types and basis sets}
{\ref{ConstructBasis}: Interaction types and basis
sets}}\mylabel{ConstructBasis}

\section{Interaction types}

The interaction type is specified by setting \basisitem{ITYPE} in namelist
\namelist{\&BASIS},
\begin{equation*}
\basisitem{ITYPE}=\var{ITYP}+\var{IADD},
\end{equation*}
where
\begin{description}[nosep]
         \item[$\var{ITYP} = 1$:]{Linear rigid rotor + atom}
         \item[$\var{ITYP} = 2$:]{Diatomic vibrotor + atom}
         \item[$\var{ITYP} = 3$:]{Linear rigid rotor + linear rigid rotor}
         \item[$\var{ITYP} = 4$:]{Asymmetric rigid rotor + linear rigid
             rotor}
         \item[$\var{ITYP} = 5$:]{Symmetric top rigid rotor + atom}
         \item[$\var{ITYP} = 6$:]{Asymmetric rigid rotor + atom}
         \item[$\var{ITYP} = 7$:]{Diatomic vibrotor + atom, with different
             interaction potentials for different rotational levels}
         \item[$\var{ITYP} = 8$:]{Atom + rigid corrugated surface}
         \item[$\var{ITYP} = 9$:]{Plug-in code for other interaction types}
\end{description}
and
\begin{description}[nosep]
\item[$\var{IADD} = 0$:]{Full close-coupling (no dynamical approximation)}
\item[$\var{IADD} = 10$:]{Effective potential approximation}
\item[$\var{IADD} = 20$:]{CS (Coupled-states/centrifugal-sudden)
    approximation or helicity decoupling approximation}
\item[$\var{IADD} = 30$:]{Decoupled $L$-dominant approximation}
\item[$\var{IADD} =  100$:]{Infinite-order sudden approximation}
\end{description}	
The allowed combinations are:

\begin{tabular}{|lc|c|c|c|c||c|}
\hline
                             & &CC      &EP      &CS      &DLD     &IOS\\
                             & &(+0)    &(+10)   &(+20)   &(+30)   &(+100)\\
\hline
rigid rotor + atom           &1&\cmark  &\cmark  &\cmark  &\cmark  &\cmark\\
vibrating rotor + atom       &2&\cmark  &\cmark  &\cmark  &\cmark  &\cmark\\
rigid rotor + rigid rotor    &3&\cmark  &\cmark  &\cmark  &\xmark  &\cmark\\
asymmetric top + rigid rotor &4&\cmark  &\xmark  &\cmark  &\xmark  &\xmark\\
symmetric top + atom         &5&\cmark  &\cmark  &\cmark  &\xmark  &\cmark\\
asymmetric top + atom        &6&\cmark  &\cmark  &\cmark  &\xmark  &\cmark\\
vibrating rotor + atom       &7&\cmark  &\cmark  &\cmark  &\cmark  &\xmark\\
atom + corrugated surface    &8&\cmark  &N/A     &N/A     &N/A     &N/A\\
user-defined interaction type  &9&(\cmark)&(\cmark)&(\cmark)&(\cmark)&(\cmark)\\
\hline
\end{tabular}

\smallskip $\var{ITYP}=1$ to 8 are referred to here as \emph{built-in interaction
types}.

\section{Pair levels, pair states and pair basis functions}\mylabel{basis:lev-state}

The programs construct sets of pair basis functions $\Phi_i(\xi_{\rm intl})$
that are used to expand the wavefunction as in Eq.\ \ref{eqexp}. The specific
set of quantum numbers needed to describe each function depends on the
interaction type, as described for the built-in coupling cases in sections
\ref{jlevel}, \ref{couple:monomer}, \ref{couple:cc} and \ref{decouple}.

In constructing basis sets, the programs make a distinction between pair
levels, pair states and pair basis functions. These terms are used somewhat
differently
\begin{enumerate}[nosep]
\item for basis sets that are diagonal in the Hamiltonian $H_{\rm intl}$ of
    the separated monomers and $\hat L^2$;
\item for basis sets that are non-diagonal in $H_{\rm intl}$ and/or $\hat
    L^2$.
\end{enumerate}
The arrays used to select and specify basis sets in these two cases are
described separately in the following subsections.

The programs have outer loops over variables \var{JTOT} and \var{IBLOCK}, as
described in section \ref{angmom}. For the built-in interaction types, these
are (mostly) used for the total angular momentum and a symmetry that is either
total parity or a body-fixed projection quantum number. Before entering these
loops, the programs construct restricted lists of quantum numbers that are
independent of \var{JTOT} and \var{IBLOCK}; inside the loops, they construct
the basis set suitable for the specific set of coupled equations that arises
for that \var{JTOT} and \var{IBLOCK}.

\subsection{\texorpdfstring{Basis sets diagonal in $H_{\rm intl}$ and $\hat
L^2$} {Basis sets diagonal with respect to internal Hamiltonian and centrifugal
operator}}\mylabel{basis:diag}

This class includes all the built-in interaction types.

Each monomer has energy levels identified by a set of quantum numbers. A
separated pair of monomers has levels identified by the product of the two
sets. These are specified by a list of \emph{pair level quantum numbers} held
in the array \basisitem{JLEVEL}. The corresponding set of \emph{pair level
energies} for the separated monomers are held in the array \basisitem{ELEVEL}.

The quantum numbers in \basisitem{JLEVEL} are limited to those needed to label
degeneracy-averaged cross sections. They do {\em not} include quantum numbers
that have no effect on the monomer energies.

The arrays \basisitem{JLEVEL} and \basisitem{ELEVEL} are either input
explicitly or constructed from other input data as described below. If
\basisitem{JLEVEL} is input directly or is constructed in a plug-in basis-set
suite, each set of quantum numbers must be unique. The user needs an
understanding of \basisitem{JLEVEL} and \basisitem{ELEVEL} to construct data
files and interpret degeneracy-averaged cross sections output from \MOLSCAT.

Before entering the loops over \var{JTOT} and \var{IBLOCK}, the pair level
quantum numbers in \basisitem{JLEVEL} are expanded internally into the (often
larger) set of \emph{pair state quantum numbers} in the array \var{JSTATE}.
This includes all the quantum numbers that appear in the basis set for an
interacting pair, including those that do not affect the pair energy, but {\em
excluding} \var{JTOT}, \var{IBLOCK} and the centrifugal quantum number $L$
(the allowed values of which depend on \var{JTOT} and \var{IBLOCK}).

In simple cases, \var{JSTATE} contains the same quantum numbers as
\basisitem{JLEVEL}. However, additional quantum numbers are sometimes needed.
This is described for the built-in coupling cases in section
\ref{couple:monomer}. For example, for diatom + diatom interactions, the two
monomer rotational quantum numbers $j_1$ and $j_2$ may couple to form several
values of a resultant $j$; only pairs ($j_1,j_2$) are in \basisitem{JLEVEL},
but each set ($j_1,j_2,j$) is stored separately in the \var{JSTATE}. The
dimensions of \var{JSTATE} are {\tt(\var{NSTATE},\var{NQN})}, where \var{NSTATE} is the
number of sets and $\var{NQN}-1$ is the number of quantum labels per set.

The programs also need access to the internal energy for each pair state. To
allow this, the last element of \var{JSTATE} for each state is a pointer to the
pair level energy in \basisitem{ELEVEL}.

The user needs an understanding of the array var{JSTATE} to interpret the
program output. The quantum numbers in \var{JSTATE} are printed if
$\inpitem{IPRINT}\ge 1$.

Finally, inside the loops over \var{JTOT} and \var{IBLOCK}, the programs select
the \emph{pair basis functions} that are actually used to solve each set of
coupled equations. Each basis function is specified by an element of the array
\var{JSINDX} (which points to a set of quantum numbers in \var{JSTATE}) and a
corresponding value of $L$ in the array \var{L}. These basis functions are then
used for calculations for all the energies and external fields required for
that \var{JTOT} and \var{IBLOCK}.

For basis sets diagonal in $H_{\rm intl}$ and $\hat L^2$, the pair basis
functions are used to label the channels involved in S-matrix elements.

If $\inpitem{IPRINT}\ge5$, the programs print a list of channels. For each
channel, the list includes \var{JSINDX}, the corresponding quantum numbers in
\var{JSTATE}, the value of $L$, the pair level index and the pair energy. In
cases where $L$ is not a good quantum number ($\basisitem{IBOUND}=1$, for
example in helicity decoupling calculations), it is replaced by the diagonal
matrix element of $\hat{L}^2$.

\subsection{\texorpdfstring{Basis sets non-diagonal in $H_{\rm intl}$ or $\hat
L^2$}{Non-diagonal basis sets}}\mylabel{basis:off-diag}

For the built-in interaction types, the basis functions are eigenfunctions of
the Hamiltonian $H_{\rm intl}$ of the separated monomers. However, for plug-in
basis-set suites this is not essential: there may be off-diagonal (but
$R$-independent) terms in $H_{\rm intl}$ and/or $\hat L^2$ that are programmed
as described in section \ref{base9:calculatecouple}.

Under these circumstances, before entering the loops over \var{JTOT} and
\var{IBLOCK}, a plug-in suite constructs an array \var{JSTATE} that contains
values of all quantum numbers that appear in the basis set for an interacting
pair except \var{JTOT}, \var{IBLOCK} and $L$. Some plug-in suites also
construct the array \var{JLEVEL}, but it is not used outside the plug-in suite
so it has no particular significance.

Inside the loops over \var{JTOT} and \var{IBLOCK}, the plug-in suite constructs
a \emph{primitive basis set} using the arrays \var{JSINDX} and \var{L} in the
same way as described in section \ref{basis:diag}.

If $\inpitem{IPRINT}\ge5$, the programs print a list of the primitive basis
functions. For each basis function, the list gives \var{JSINDX}, the
corresponding quantum numbers in \var{JSTATE} and the value of $L$. In cases
where the basis set is diagonal in $L$ but $L$ does not have an integer value
($\var{NRSQ}=0$ and $\basisitem{IBOUND}=1$), it is replaced by the diagonal
matrix element of $\hat{L}^2$.

The primitive basis set is used to construct the coupled equations. However, a
single basis function does \emph{not} correspond to an energy level of a pair
of separated monomers and cannot be used to label S-matrix elements. Under
these circumstances, the programs diagonalise $H_{\rm intl}$ for each
\var{JTOT} and \var{IBLOCK} (and each external field) to find its eigenvalues
and eigenvectors. The eigenvalues are the channel threshold energies, and are
available for use as reference energies in all three programs. They are printed
if $\inpitem{IPRINT}\ge6$ for \MOLSCAT\ and \BOUND, or 10 for \FIELD\ (because
for \FIELD\ they change as a function of field during the course of locating
each bound state). The corresponding eigenvectors are printed if
$\inpitem{IPRINT}\ge15$.

\BOUND\ and \FIELD\ calculate the log-derivative matching matrix (section
\ref{theory:boundcalcs}) in the primitive basis set. They use the channel
threshold energies only for reference energies. However, \MOLSCAT\ transforms
the wavefunction matrix $\boldsymbol{\Psi}(R_{\rm max})$ (or the log-derivative
matrix ${\bf Y}(R_{\rm max})$) into the \emph{asymptotic basis set} that
diagonalises $H_{\rm intl}$ and $\hat L^2$ (and, optionally, extra operators to
resolve degeneracies in the eigenvalues of $H_{\rm intl}$). Scattering boundary
conditions are then applied in the asymptotic basis set.

\section{Convergence of the basis set}

Calculated scattering and properties and bound-state positions
depend on the size of the basis set. For basis sets off-diagonal in $H_{\rm
intl}$, even the energies of the separated monomers may depend on the size of
the basis set.

The interaction potential and off-diagonal terms in $H_{\rm intl}$ and/or $\hat
L^2$ couple basis functions arising from different monomer levels. It is always
important to establish that the basis set used is large enough to give the
desired accuracy. The size of basis set required may often be estimated by
physical intuition, sometimes assisted by perturbation theory, but there is
usually no substitute for carrying out tests with increasingly large basis sets
and checking the convergence of the properties of interest.

It is \emph{not} usually adequate to carry out scattering calculations that
include only asymptotically open channels. Closed channels can have substantial
effects. Bound-state calculations are often even more sensitive to the
inclusion of high-lying basis functions than scattering calculations.

\section{Units of energy for quantities in \namelist{\&BASIS}}\mylabel{basis:units}

The units of energy for quantities input in \namelist{\&BASIS} are
\emph{independent} of those used for quantities in \namelist{\&INPUT}. They are
specified by \basisitem{EUNITS}, which is an integer that selects a unit of
energy from the list in section \ref{outline:units}. The default for
\basisitem{EUNITS} is 1, indicating energies expressed as wavenumbers in
cm$^{-1}$.

If the energy unit required is not among those listed, \basisitem{EUNITS} may
be set to 0 and the required value (in units of cm$^{-1}$) supplied in
\basisitem{EUNIT}. If this is done, the name of the unit should be supplied in
the character variable \basisitem{EUNAME}.

\basisitem{EUNITS}, \basisitem{EUNIT} and \basisitem{EUNAME} are distinct from
the variables with the same names in namelist \namelist{\&INPUT} (section
\ref{Eunit}), but their allowed values and interpretation are the same.

\section{\texorpdfstring{Built-in interaction types: pair level quantum
numbers and energies} {Built-in interaction types: pair level quantum numbers
and energies}}\mylabel{jlevel}

The user \emph{may} specify a list of pair level quantum numbers in the array
\basisitem{JLEVEL} and a corresponding list of energies in the array
\basisitem{ELEVEL}. Alternatively (and more commonly), \basisitem{JLEVEL}
and/or \basisitem{ELEVEL} are generated from other input quantities:
\begin{description}
\item[\basisitem{NLEVEL}:]{if $> 0$, \basisitem{NLEVEL} indicates that the
    quantum numbers of the levels to be used in constructing the basis set
    are input as \basisitem{NLEVEL} sets of values in the array
    \basisitem{JLEVEL}. If $\basisitem{NLEVEL} = 0$, the quantum numbers
    for the levels are calculated internally as described below.}
\item[\basisitem{JLEVEL}:]{integer array specifying the pair level quantum
    numbers. The array \basisitem{JLEVEL} is structured differently for
    each value of \var{ITYP} as described below. If $\basisitem{NLEVEL} >
    0$, \basisitem{JLEVEL} must be supplied explicitly in the input file.
    If $\basisitem{NLEVEL} = 0$, the pair level quantum numbers are
    calculated internally from input quantities that specify ranges and
    step sizes for them, as described for each value of \var{ITYP} below.
    \basisitem{JLEVEL} is declared as a one-dimensional array (current
    dimension 4000, set in module \module{sizes}), although it is
    conceptually two-dimensional for $\var{ITYP} > 1$.}
\item[\basisitem{ELEVEL}:]{array of \basisitem{NLEVEL} pair level energies,
    corresponding to the pair levels in the array \basisitem{JLEVEL}
    (current dimension 1000, set in module \module{sizes}). If all the
    elements of \basisitem{ELEVEL} are 0.0, or \basisitem{NLEVEL} is 0, the
    energies are calculated from input values of monomer spectroscopic
    constants as described for each value of \var{ITYP} below.}
\end{description}

The methods of specifying the pair levels to be included are independent of any
dynamical approximations employed, so that the information given for each
\var{ITYP} below is applicable to $\basisitem{ITYPE} = \var{ITYP}$,
$\var{ITYP}+10$, $\var{ITYP}+20$ and $\var{ITYP}+30$. For IOS cases
($\basisitem{ITYPE} = \var{ITYP}+100$), the required input is generally the
same, except that rotational energies are not required for IOS calculations.

\subsection{\texorpdfstring{Linear rigid rotor + atom ($\var{ITYP} = 1$)}
{Linear rigid rotor + atom (ITYP = 1)}}\mylabel{ityp1}

The basis set used for a linear rigid rotor is formed from spherical harmonics
$Y^m_j$. These may be functions of either spaced-fixed angles $(\beta,\alpha)$
(for close-coupling calculations) or body-fixed angles $(\theta,\phi)$ (for
coupled-states and helicity-decoupling calculations).

After processing, the array \basisitem{JLEVEL} must contain a list of values of $j$ for monomer
rotational states. It is usually generated from input parameters \basisitem{JMIN}, \basisitem{JMAX}
and \basisitem{JSTEP}, which have the obvious meanings. For special purposes, \basisitem{NLEVEL}
may be set greater than zero and a list of \basisitem{NLEVEL} $j$ values supplied in the array
\basisitem{JLEVEL}.

The energy levels are usually calculated from
\begin{equation}
E(j)= (B_e-\alpha_e/2)j(j+1)-D_e[j(j+1)]^2.
\end{equation}
The input parameters \basisitem{BE}, \basisitem{ALPHAE} and \basisitem{DE} have
the obvious meanings. Since the rotational constant actually used is simply
$\basisitem{BE}-0.5\times\basisitem{ALPHAE}$, this value may be input directly
in \basisitem{BE} with \basisitem{ALPHAE} omitted from the namelist if
preferred.

\subsection{\texorpdfstring{Diatomic vibrotor + atom ($\var{ITYP} = 2\mbox{ and }7$)}
{Diatomic vibrotor + atom (ITYP = 2 and 7)}}\mylabel{ityp2}

The monomer basis set for a vibrating diatom is formed from products of
spherical harmonics and vibrational wavefunctions with quantum number $v$ in
the monomer internuclear separation. Once again the spherical harmonics may be
functions of either spaced-fixed angles $(\beta,\alpha)$ (for close-coupling
calculations) or body-fixed angles $(\theta,\phi)$ (for coupled-states and
helicity-decoupling calculations).

\basisitem{NLEVEL} must be set greater than zero and the pair levels to be
included must be specified as a list of \basisitem{NLEVEL} $(j,v)$ pairs
supplied in the array \basisitem{JLEVEL} (in the order $j_1,v_1,j_2,v_2,$
$\ldots$ $j_\var{NLEVEL},v_\var{NLEVEL}$). There is no option for generating
this list from limits on quantum numbers.

The energy levels may be specified as a list in the array \basisitem{ELEVEL}.
Alternatively, if all \basisitem{NLEVEL} elements of \basisitem{ELEVEL} are
zero (the default), the energy levels are calculated from
\begin{equation}
E(j,v)=
\omega_e v -\omega_ex_e v(v+1)
+\left[B_e-\left(v+\textstyle{\frac{1}{2}}\right)\alpha_e\right]j(j+1)-D_e[j(j+1)]^2.
\end{equation}
The input parameters \basisitem{WE}, \basisitem{WEXE}, \basisitem{BE},
\basisitem{ALPHAE} and \basisitem{DE} have the obvious meanings. Note that the
energies are defined with respect to $E(0,0)$ even if the $(0,0)$ pair is not
included in \basisitem{JLEVEL}.

For IOS calculations the vibrational quantum numbers are selected from the array
\basisitem{JLEVEL}, and any rotational quantum numbers supplied are
ignored, except that the maximum value is used to limit which cross sections
are calculated. If vibrational energies are supplied in \basisitem{ELEVEL}, the
first one encountered for each vibrational manifold is kept. If all
\basisitem{ELEVEL} are zero, energies are generated from \basisitem{WE} and
\basisitem{WEXE}. If the input \basisitem{NLEVEL} is negative, a user-supplied
routine \prog{GET102} must be provided to set \basisitem{NLEVEL},
\basisitem{JLEVEL} and \basisitem{ELEVEL} values.  The distribution includes a
dummy version of this routine.

\subsection{\texorpdfstring{Linear rigid rotor + linear rigid rotor ($\var{ITYP} = 3$)}
{Linear rigid rotor + linear rigid rotor (ITYP = 3)}}\mylabel{ityp3}

The basis set is formed from coupled products of spherical harmonics $Y^{m_1}_{j_1}$ and
$Y^{m_2}_{j_2}$.

After processing, the array \basisitem{JLEVEL} must contain a list of pairs of rotational quantum
numbers $(j_1,j_2)$. It is usually generated from input parameters \basisitem{J1MIN},
\basisitem{J1MAX}, \basisitem{J1STEP} for molecule 1 and \basisitem{J2MIN}, \basisitem{J2MAX},
\basisitem{J2STEP} for molecule 2. For special purposes, \basisitem{NLEVEL} may be set greater than
zero and a list of \basisitem{NLEVEL} $(j_1,j_2)$ pairs supplied in the array \basisitem{JLEVEL}
(in the order $j_{11}, j_{21}$; $j_{12},j_{22};$ $\ldots;$
$j_{1,\var{NLEVEL}},j_{2,\var{NLEVEL}}$).

The energy levels are usually specified as a sum of two terms of the form
\begin{equation}
E(j)= (B_e-\alpha_e/2)j(j+1)-D_e[j(j+1)]^2.
\end{equation}
The input parameters \basisitem{BE}(1,2) \basisitem{ALPHAE}(1,2) and \basisitem{DE}(1,2) specify
the parameters for molecules 1 and 2. If $\basisitem{IDENT}=1$ and the values for molecule 2 are
zero, the program sets them equal to the values for molecule 1. Since the rotational constants
actually used are simply $\basisitem{BE}-0.5\times\basisitem{ALPHAE}$, these values may be input
directly in \basisitem{BE}(1,2) with \basisitem{ALPHAE} omitted from the namelist if preferred.

If the two molecules are identical, the basis functions corresponding to $(j_1,j_2)$ and
$(j_2,j_1)$ are indistinguishable. In this case, the input variable \basisitem{IDENT} (default 0)
should be set to 1 and (if \basisitem{JLEVEL} is supplied as a list) only distinguishable pairs
(i.e., $j_1\ge j_2$) should be included. The programs then carry out separate
calculations for states with odd and even exchange symmetry, in separate symmetry blocks.%
\footnote{In \BOUND\ and \FIELD, a calculation that neglects identical-particle symmetry when
it is present simply combines the coupled equations for odd and even exchange symmetry.
It is less efficient, but gives the same eigenvalues and wavefunctions, provided both sets
of symmetry-related functions are included. In \MOLSCAT, the S-matrix elements obtained in
the symmetrised and unsymmetrised cases are related by simple summations and factors of
$\sqrt{2}$, but some elements that are off-diagonal in the unsymmetrised case are transferred
to the diagonal in the symmetrised case. Because of the different expressions used for elastic
(diagonal) and inelastic (off-diagonal) cross sections (due to the presence of $\delta_{if}$
in Eq.\ \ref{eqsigdef}), there are no simple relationships between cross sections connecting
symmetry-related pairs in the two cases.}

For identical molecules, it is also necessary to specify the statistical weights to be
applied to the different symmetry combinations when calculating cross sections. The statistical
weights for antisymmetric and symmetric combinations of $(j_1,j_2)$ and $(j_2,j_1)$ may be
specified explicitly in the array \basisitem{WT} as \basisitem{WT}(1) and \basisitem{WT}(2)
respectively. If both \basisitem{WT}(1) and \basisitem{WT}(2) are zero (the default), they are
calculated from the single nuclear spin input in \basisitem{SPNUC}: for integer \basisitem{SPNUC}
(bosonic particles), $\basisitem{WT}(1)=\basisitem{SPNUC}/(2*\basisitem{SPNUC}+1)$ and
$\basisitem{WT}(2)=(\basisitem{SPNUC}+1)/(2*\basisitem{SPNUC}+1)$. For half-integer
$\basisitem{SPNUC}$ (fermionic particles), the two values are exchanged.

For identical molecules, all the programs skip any symmetry block for which \basisitem{WT} is zero.

\cbcolor{\mcol}\cbstart There are different definitions in the literature for state-to-state
integral cross sections for two identical particles that are in the same state either before or
after the collision. The issues have been discussed by Huo and Green \cite{Huo:1996}. Versions of
\MOLSCAT\ before 2020.0 evaluated such cross sections for $\var{ITYP}=3$ using the expression of
Takayanagi, Eq.\ (2.18) of ref.\ \cite{Huo:1996}. However, versions from 2020.0 onwards use Eq.\
(2.16) of ref.\ \cite{Huo:1996}; this contains no additional factors of 2 when the initial or final
states are the same, and gives the same values for cross sections $\sigma_{jj\rightarrow j'j'}$
when calculated with or without identical-particle symmetry.\cbend

\begin{shaded} 

\subsection{\texorpdfstring{Symmetries of rotational functions for nonlinear molecules ($\var{ITYP} = 4$, 5 and 6)}
{Symmetries of rotational functions for nonlinear molecules (ITYP = 4, 5 and 6)}}\mylabel{nonlin}

The monomer rotational functions used for rigid nonlinear molecules are based on complex conjugates of Wigner rotation matrices, $D^{j*}_{mk}$, where $k$ is the projection of the angular momentum $j$ onto the $z$ axis of the top (which is usually a symmetry axis if one is present). The rotation matrices may be functions of either spaced-fixed Euler angles $(\alpha,\beta,\gamma)$ (for close-coupling calculations) or body-fixed Euler angles $(\phi,\theta,\chi)$ (for coupled-states and helicity-decoupling calculations).

Wigner rotation matrices transform as 1-dimensional representations of the group $D_2$, whose symmetry elements are the identity $E$ and 2-fold rotations $C_2(z)$, $C_2(y)$ and $C_2(x)$. Symmetrised linear combinations of rotation matrices,
\begin{equation}
D^{j*}_{mk} + \epsilon D^{j*}_{m{}-k},
\label{eq:Dmk-sym}
\end{equation}
with $k\ge0$ and $\epsilon=\pm1$, transform as 1-d representations of a larger group containing a plane of reflection $\sigma_{xz}$, together with the additional operations it generates in combination with the elements of $D_2$: $\sigma_{yz}$, a centre of inversion, and $\sigma_{xy}$. Together, these 8 operations form the group $D_{2\textrm{h}}$.

The interaction potential may conserve some or all of these symmetries. The implementation of $\var{ITYP} = 4$, 5 and 6 requires that the molecular $xz$ plane is a plane of symmetry. It is thus not general enough to handle chiral asymmetric tops, or other molecules without a plane of symmetry. This restriction is required so that the coupling matrices due to the interaction potential between basis functions of the form (\ref{eq:Dmk-sym}) are real and symmetric.

The symmetries that are conserved by the interaction potential depend on the point-group symmetry of the molecule concerned. They are reflected in the terms that are present in the expansion of the atom-molecule interaction potential in spherical harmonics $C_{\lambda\kappa}$, described in section \ref{potl:ityp5or6}. The possible symmetries include:
\begin{enumerate}[nosep]
\item{Axis of rotation $C_2(z)$, which in combination with $\sigma_{xz}$ implies $\sigma_{yz}$. If this exists, $\kappa$ is restricted to even values. The effect of this is to conserve $(-1)^k$ in the coupled equations: functions with $k$ even are not coupled to those with $k$ odd.}
\item{Axis of rotation $C_2(y)$, which in combination with $\sigma_{xz}$ implies a centre of inversion. If this exists, $\lambda$ is restricted to even values. The effect of this is to conserve $(-1)^{j+k} \epsilon$ in the coupled equations: functions with $(-1)^{j+k}\epsilon=+1$ are not coupled to those with $(-1)^{j+k}\epsilon=-1$.}
\item{Axis of rotation $C_2(x)$, which in combination with $\sigma_{xz}$ implies $\sigma_{xy}$. If this exists, $\lambda+\kappa$ is restricted to even values. The effect of this is to conserve the monomer parity $(-1)^j \epsilon$ in the coupled equations: functions with $(-1)^j\epsilon=+1$ are not coupled to those with $(-1)^j\epsilon=-1$.}
\end{enumerate}
If any two of symmetries (1) to (3) are present, the third must be present too. To decide which are present, either consider the point-group symmetry of the molecule or inspect the terms $(\lambda,\kappa)$ that are present in the expansion of the interaction potential.

If any of these symmetries is present, it is most efficient to perform separate calculations for each of the (two or four) independent sets of coupled equations.

\subsection{\texorpdfstring{Symmetric or near-symmetric top + atom ($\var{ITYP} = 5$)}
{Symmetric or near-symmetric top + atom (ITYP = 5)}}\mylabel{ityp5}

The monomer rotational functions used for $\var{ITYP} = 5$ are of the form (\ref{eq:Dmk-sym}), with suitable normalisation factors. For $k=0$, only $\epsilon=+1$ is possible.

After processing, the array \basisitem{JLEVEL} must contain a list of sets of 3 rotational quantum numbers $(j,k,\var{PRTY})$, where $\var{PRTY}=0$ for $\epsilon=+1$ and $\var{PRTY}=1$ for $\epsilon=-1$.

The array \basisitem{JLEVEL} is usually generated from input parameters \basisitem{JMIN} and \basisitem{JMAX}, together with symmetry selectors in \basisitem{JSTEP} and the array \basisitem{ISYM}. The symmetry selectors are described below.

The additional input variable \basisitem{KMAX} (default 0, equivalenced to \basisitem{KSET}) may be used to limit the $k$ values included: if it is negative, only levels with $k=|\basisitem{KSET}|$ are included; if it is zero
or positive, only levels with $k\le\basisitem{KMAX}$ are included. If \emph{all} $k$ levels are required, \basisitem{KMAX} should be set to at least \basisitem{JMAX} (999 recommended).

For special purposes, \basisitem{NLEVEL} may be set greater than zero and a list of \basisitem{NLEVEL} triples $(j,k,\var{PRTY})$ supplied in the array \basisitem{JLEVEL}.

If the energy levels are not given explicitly in \basisitem{ELEVEL}, they are calculated from the standard near-symmetric top equation using rotational constants supplied in the input variables \basisitem{A}, \basisitem{B} and \basisitem{C}. Neglecting centrifugal distortion, $l$-type doubling and tunnelling, the expression is
\begin{align}
E(j,k) =& \frac{(A+B)}{2}\left[j(j+1)-k^2\right] + C k^2 \nonumber\\
&+ \delta_{k1} \epsilon \frac{(A-B)}{4} [j(j+1)-k(k-1)][j(j+1)-(k-1)(k-2)].
\end{align}
Note that \basisitem{A}, \basisitem{B} and \basisitem{C} \emph{must} correspond, respectively, to moments of inertia about the $x$, $y$, and $z$ axes used in the description of the interaction potential; see below.
They are \emph{not} necessarily in descending order of magnitude, as is usual in spectroscopy. For a symmetric top with its axis along $z$, $\basisitem{A}=\basisitem{B}$ and \basisitem{C} is the unique constant.

The symmetry types that are included in the basis set may be restricted using the input variable \basisitem{JSTEP} and the array \basisitem{ISYM}.

If $\basisitem{JSTEP}=2$, only functions with the same value of $(-1)^{j+k}\epsilon$ as $j=\basisitem{JMIN}$, $k=0$ are included.

\basisitem{ISYM}(1) is interpreted bitwise, and a particular basis function is included only if it passes \emph{all} the specified tests. To construct \basisitem{ISYM}(1), start with 0 (which would include all functions) and add $2^n$ for each class of functions to be excluded:

\begin{tabular}{lll}
   If bit 0 is set,& odd  $k$ functions are excluded&add 1 to \basisitem{ISYM}(1)\\
   If bit 1 is set,& even $k$ functions are excluded&add 2 to \basisitem{ISYM}(1)\\
   If bit 2 is set,& functions with $(-1)^j \epsilon = -1$ are excluded&add 4 to \basisitem{ISYM}(1)\\
   If bit 3 is set,& functions with $(-1)^j \epsilon = +1$ are excluded&add 8 to \basisitem{ISYM}(1)\\
\end{tabular}

\begin{tabular}{ll}
   If $\basisitem{ISYM}(2)=0$,& functions with $(-1)^k\epsilon=-1$ are excluded\\
   If $\basisitem{ISYM}(2)=1$,& functions with $(-1)^k\epsilon=+1$ are excluded\\
\end{tabular}

\emph{3-fold symmetry}

A common use of $\var{ITYP}=5$ is for a symmetric top like ammonia, with 3-fold rotational symmetry and inversion doubling.  The code allows the user to set \basisitem{ISYM}(3) to restrict $k$ to values that
satisfy either $3n\pm1$ ($\basisitem{ISYM}(3)=1$) or $3n$ ($\basisitem{ISYM}(3)=3$), and \basisitem{ISYM}(4) to indicate the nature of the 3 identical atoms (0 for bosons and 1 for fermions). In addition, if \basisitem{ROTI}(10) is set non-zero then it is used as the zeroth-order tunnelling splitting $\nu_0$. \basisitem{ROTI}(11) and \basisitem{ROTI}(12) are used for the centrifugal distortion components of the tunnelling splitting, $\nu_a$ and $\nu_b$,
\begin{equation}
\nu=\nu_0-\nu_a\left[j(j+1)-k^2\right]-\nu_bk^2.
\end{equation}

\emph{Higher-order rotational symmetries}

No special code is implemented to handle molecules with axes of rotation $C_n$ with $n>3$. Such molecules can be handled with the current code, but not with optimal efficiency. It would be straightforward to implement additional symmetries using higher elements of the array \var{ISYM}.

\emph{Linear molecules with orbital or vibrational angular momentum}

$\var{ITYP}=5$ can also be used to handle interactions between atoms and linear molecules with either electronic orbital angular momentum $\Lambda$ (electronic state $\Pi$, $\Delta$, etc.) or vibrational angular momentum $l$. The rotational functions for such molecules are formed from rotation matrices $D^{j*}_{m\Lambda}$ or $D^{j*}_{ml}$. In this case the basis set usually contains only a single value of $\Lambda$ or $l$: this may be selected by setting \basisitem{KSET} to a negative value, $\basisitem{KSET}=-|\Lambda|$ or $-|l|$. This option does not handle electron spin.

For molecules with $D_{\infty h}$ symmetry (homonuclear diatomics, or molecules such as CO$_2$ or HCCH), \basisitem{JSTEP} should be set to 2 to include only functions of the same parity as $j=\basisitem{JMIN}$, $k=0$ as above.

\subsection{\texorpdfstring{Asymmetric or spherical top + atom ($\var{ITYP} = 6$)}
{Asymmetric or spherical top + atom (ITYP = 6)}}\mylabel{ityp6}

The monomer rotational functions used for $\var{ITYP} = 6$ are linear combinations of functions of the form (\ref{eq:Dmk-sym}) that diagonalise the Hamiltonian of the isolated asymmetric or spherical top, with suitable normalisation factors.

After processing, the array \basisitem{JLEVEL} must contain a list of pairs $(j,\tau)$, where $\tau$ is an index for the level, $-j\le\tau\le j$.\footnote{An alternative indexing, $1\le\tau\le 2j+1$, is allowed for compatibility with older input files, but is not standard spectroscopic notation and is deprecated.} \basisitem{JLEVEL} is usually generated from input parameters \basisitem{JMIN} and \basisitem{JMAX}, together with symmetry selectors in \basisitem{JSTEP} and the array \basisitem{ISYM}. The symmetry selectors are described below.

The monomer energy levels $E_{j\tau}$ and wavefunction coefficients $a^j_{\tau,k}$ are usually calculated internally from rotational constants. If all three rotational constants \basisitem{A}, \basisitem{B} and \basisitem{C} are specified, the programs do not use \basisitem{NLEVEL} or the input arrays \basisitem{JLEVEL} and \basisitem{ELEVEL}. Instead, they construct and diagonalise the asymmetric top Hamiltonian, which (neglecting centrifugal distortion) is
\begin{equation}
\hat H_{\rm rot} = A \hat{\jmath}_a^2 + B \hat{\jmath}_b^2 + C \hat{\jmath}_c^2.
\end{equation}
Note that (as for $\var{ITYP} = 5$) \basisitem{A}, \basisitem{B} and \basisitem{C} {\em must} correspond respectively to the $x$, $y$ and $z$ axes used in the potential expansion. They are \emph{not} necessarily in descending order of magnitude, as is usual in spectroscopy.

Centrifugal distortion constants $D_J$, $D_{JK}$ and $D_K$ may also be supplied in the input variables \basisitem{DJ}, \basisitem{DJK}, and \basisitem{DK}, respectively, and contribute an energy term
\begin{equation}
-D_J j(j+1) -D_{JK} j(j+1)k^2 -D_K k^4.
\end{equation}
However, these do not correspond to the constants conventionally used for asymmetric tops.

If $\basisitem{IASYMU}<0$, the resulting wavefunctions are written on unit $|\iounit{IASYMU}|$ in the format described below for input from \iounit{IASYMU}. The energies written to \iounit{IASYMU} are in the units
specified by \basisitem{EUNIT} or \basisitem{EUNITS}, so that subsequent calculations must use the same units.

Spherical tops fit into the same framework as asymmetric tops, but with a different rotational Hamiltonian governed by a single rotational constant and a tetrahedral centrifugal distortion constant $d_{\rm t}$ that splits rotational levels with $j>1$ into sets of A$_1$, A$_2$, E, T$_1$ or T$_2$ symmetry (where the last two are commonly called F$_1$ and F$_2$ in the spectroscopic literature). To select this option, set $\basisitem{A} = \basisitem{B} =
\basisitem{C}$ and input a non-zero value $d_{\rm t}$ in \basisitem{DT}. Note that $d_{\rm t}$ must have magnitude greater than about $10^{-7}$ cm$^{-1}$: if it is set to zero, the rotational states are not resolved into their
contributing symmetries.

If rotational constants \basisitem{A}, \basisitem{B} and \basisitem{C} are not input, the programs read the arrays \basisitem{JLEVEL} and \basisitem{ELEVEL} from unit \basisitem{IASYMU}, together with the corresponding wavefunction coefficients $a^j_{\tau,k}$. Each level is described on \iounit{IASYMU} with the triple $j,\tau,E_{j\tau}$ where $E_{j\tau}$ is the rotor energy in \namelist{\&BASIS} energy units. For each level, $2j+1$ wavefunction coefficients $a^j_{\tau,k}$ are required (corresponding to $k=-j,-j+1,..., j$); these must follow the $j,\tau,E_{j\tau}$ line with format \code{(6F12.8)} on $(n_k+5)/6$ subsequent lines. Coefficients need not be normalised, but the programs check that they have valid symmetries. The detailed behaviour depends on the
value of \basisitem{NLEVEL}:
\begin{itemize} [nosep]
\item If $\basisitem{NLEVEL}=0$, the programs read from \iounit{IASYMU} until they reach the end of the file. Each level read is included if \emph{all} the following conditions are met:
\begin{enumerate} [nosep]
\item either $\basisitem{JMAX} \le 0$ or $j$ is in the range \basisitem{JMIN} to \basisitem{JMAX};
\item if \basisitem{JSTEP}=2, the level has the same value of $(-1)^{j+k} \epsilon$ as the level with $j=\basisitem{JMIN}$, $k=0$;
\item $E_{j\tau} \le \basisitem{EMAX}$, or $\basisitem{EMAX} \le 0$.
\end{enumerate}
\item If $\basisitem{NLEVEL}>0$, the programs read the first \basisitem{NLEVEL} levels from \iounit{IASYMU} and include those which meet the same conditions as above.
\item If $\basisitem{NLEVEL}<0$, the programs read from \iounit{IASYMU} until they reach the end of the file but use each level only if its quantum numbers $j$ and $\tau$ match values in $\basisitem{JLEVEL}(2i-1), \basisitem{JLEVEL}(2i), i=1,|\basisitem{NLEVEL}|$.
    If non-zero values are supplied in the array \basisitem{ELEVEL}, they are used in preference to those read from \iounit{IASYMU}.
\end{itemize}

If $\iounit{IASYMU}=5$ (standard input), data records should follow namelist \namelist{\&BASIS} and precede namelist \namelist{\&POTL}; in this case a positive value of \basisitem{NLEVEL} must be set in \namelist{\&BASIS}, to specify the number of rotor functions given in the file.

Note that for IOS calculations ($\basisitem{ITYPE} = 106$) rotor wavefunctions are used only to calculate state-to-state cross sections from the ``generalised IOS" cross sections.  For this case generation of rotational wavefunctions from rotational constants is not implemented; they must be explicitly supplied as data on \iounit{IASYMU} if state-to-state cross sections are required.

Internally, the program assigns one of four symmetry types to each asymmetric top function:

\begin{tabular}{clc}
   \var{PRTY}   &   even/odd $k$   &     $\epsilon$ \\
\hline
     0    &      even      &       $+1$ \\
     1    &      even      &       $-1$ \\
     2    &      odd       &       $+1$ \\
     3    &      odd       &       $-1$ \\
\end{tabular}

Asymmetric top functions input from \iounit{IASYMU} must conform to these symmetries.

The lists of quantum numbers printed for asymmetric and spherical tops include $j$, $\tau$, \var{PRTY} and an index that identifies the location where the corresponding eigenvector is stored internally. The index has no physical
significance.

If the monomer functions are calculated from rotational constants (but not if they are input explicitly from \iounit{IASYMU}), the symmetry types that are actually included in the basis set may be restricted using the input variables \basisitem{JSTEP} and \basisitem{ISYM}(1).

Rotational levels are calculated for $j$ from \basisitem{JMIN} to \basisitem{JMAX}. If $\basisitem{JSTEP}=2$, only functions with the same value of $(-1)^{j+k}\epsilon$ as $j=\basisitem{JMIN}$, $k=0$ are included.

If $\basisitem{EMAX}>0$, a level is kept only if it has energy below \basisitem{EMAX}. A level is also kept only if the corresponding wavefunction coefficients $a^j_{\tau,k}$ meet the symmetry restrictions imposed by \basisitem{ISYM}, which is described below.

\basisitem{ISYM}(1) is interpreted bitwise, and a particular asymmetric rotor function is included if it passes \emph{all} the specified tests. To construct \basisitem{ISYM}(1), start with 0 (which would include all functions) and add $2^n$ for each class of functions to be excluded:

\begin{tabular}{lll}
   If bit 0 is set,& odd  $k$ functions are excluded&add 1 to \basisitem{ISYM}\\
   If bit 1 is set,& even $k$ functions are excluded&add 2 to \basisitem{ISYM}\\
   If bit 2 is set,& functions with $(-1)^j\epsilon = -1$ are excluded&add 4 to \basisitem{ISYM}\\
   If bit 3 is set,& functions with $(-1)^j\epsilon = +1$ are excluded&add 8 to \basisitem{ISYM}\\
   If bit 4 is set,& functions with degeneracy 1 are excluded&add 16 to \basisitem{ISYM}\\
   If bit 5 is set,& functions with degeneracy 2 are excluded&add 32 to \basisitem{ISYM}\\
   If bit 6 is set,& functions with degeneracy 3 are excluded&add 64 to \basisitem{ISYM}\\
   If bit 7 is set,& functions with degeneracy $>$ 3 are excluded&add 128 to \basisitem{ISYM}\\
   \rowcolor{background} If bit 8 is set,& functions that include components $k=3n$ are excluded&add 256 to \basisitem{ISYM}\\
   \rowcolor{background} If bit 9 is set,& functions that include components $k\ne3n$ are excluded&add 512 to \basisitem{ISYM}\\
\end{tabular}

The flags that test degeneracy are intended for use with spherical tops, and allow levels with A, E and F (T) symmetry to be included selectively. Levels of A symmetry may be selected with $\basisitem{ISYM}=224$ (128+64+32). For E levels, both functions in the degenerate pair are needed and may be selected with $\basisitem{ISYM}=208$ (128+64+16). For F (T) levels, only a single function (with even $k$) is needed from each degenerate set, and may be selected with $\basisitem{ISYM}=177$ (128+32+16+1). \label{isym-spher}

The flags that test whether $k$ is a multiple of 3 are intended for cases where a symmetric top with 3-fold symmetry is treated as an asymmetric top with $\basisitem{A}=\basisitem{B}$.

For asymmetric rotors, functions for different values of $\tau$ are non-degenerate. However, near-degeneracies can occur, and the program interprets these as degeneracies if the levels (from diagonalisation) are within a tolerance (currently $10^{-9}$, set as \code{EPS} in \prog{SET6}). To be safe, do not set any bits to select degeneracy for asymmetric rotors.

For the special cases of $J_{\rm tot}=0$ in close-coupling calculations and $K=0$ in coupled-states and helicity-decoupling calculations, the coupled equations separate further into blocks with even and odd values of $(-1)^k\epsilon$. Do not be misled into believing that this symmetry restriction holds in more general cases. If desired, functions with even or odd values of $(-1)^k\epsilon$ may be selected with \basisitem{ISYM}(2):

\begin{tabular}{ll}
   If $\basisitem{ISYM}(2)=0$,& functions with $(-1)^k\epsilon=-1$ are excluded\\
   If $\basisitem{ISYM}(2)=1$,& functions with $(-1)^k\epsilon=+1$ are excluded\\
\end{tabular}

Note that the interpretation of \basisitem{JSTEP} and \basisitem{ISYM}(2) for $\var{ITYP}=6$ have changed slightly in version 2023.0, to provide a more general implementation of the symmetries described in section \ref{nonlin} and for consistency with the extended implementation of symmetry for $\var{ITYP}=5$.

\subsection[\texorpdfstring{Asymmetric rigid rotor + linear rigid rotor ($\var{ITYP} = 4$)}
{Asymmetric rigid rotor + linear rigid rotor (ITYP = 4)}]
{Asymmetric rigid rotor + linear rotor ($\var{ITYP}=4$)\sectionmark
{Asymmetric rigid rotor + linear rotor ($\var{ITYP}=4$)}}\sectionmark
{Asymmetric rigid rotor + linear rotor ($\var{ITYP}=4$)}\mylabel{ityp4}

The basis functions here are combinations of asymmetric top rotor wavefunctions and linear rotor
wavefunctions \cite{Phillips:1995}. After processing, the array \basisitem{JLEVEL} must contain a
list of sets of 3 rotational quantum numbers $(j_1,\tau,j_2)$, where rotor 1 is the asymmetric top.

Input of the asymmetric top functions follows the capabilities for $\var{ITYP} = 6$, as described
above. If all three rotational constants (\basisitem{A}(1), \basisitem{B}(1) and \basisitem{C}(1))
are specified, the programs do not use \basisitem{NLEVEL} or the input arrays
\basisitem{JLEVEL} and \basisitem{ELEVEL}. Instead, they construct asymmetric top wavefunctions
for rotational quantum numbers from \basisitem{JMIN} to \basisitem{JMAX} in steps of
\basisitem{JSTEP} (with synonyms \basisitem{J1MIN}, \basisitem{J1MAX}, and \basisitem{J1STEP}),
\basisitem{EMAX}, and \basisitem{ISYM}.  Otherwise, wavefunctions are read from unit
\iounit{IASYMU}.

The asymmetric top wavefunctions are combined with linear rotor functions
specified by \basisitem{J2MIN}, \basisitem{J2MAX}, \basisitem{J2STEP}; energies
of the linear rotor are obtained from the rotational constant, which must be
supplied as \basisitem{BE}(2). If specified, \basisitem{ALPHAE}(2) and
\basisitem{DE}(2) are also used as for $\var{ITYP} = 1$.  If $\basisitem{EMAX}
> 0$ is specified, it is used to limit the pair levels to those with
energies (asymmetric top plus linear rotor) less than \basisitem{EMAX}.

If rotational constants \basisitem{A}(1), \basisitem{B}(1), and \basisitem{C}(1) are not
supplied and $\basisitem{NLEVEL} < 0$, the programs use $|\basisitem{NLEVEL}|$ sets of quantum
numbers $(j_1,\tau,j_2)$ from the array $\basisitem{JLEVEL}$ and read asymmetric top energy levels
and wavefunction coefficients from unit \iounit{IASYMU} as for $\var{ITYP} = 6$. The asymmetric top
functions for all required $j_1$, $\tau$ must be available on unit \iounit{IASYMU}. In this case it
is also possible to specify the energies (asymmetric top plus linear rotor) in the
array \basisitem{ELEVEL}; otherwise energies are taken from \iounit{IASYMU} for the asymmetric top
and computed from \basisitem{BE}(2) in the same way as for $\var{ITYP}=1$ for the linear rotor. The
format of unit \iounit{IASYMU} is described in section \ref{ityp6}.
\end{shaded} 

\subsection{\texorpdfstring{Atom + rigid corrugated surface ($\var{ITYP} = 8$)}
{Atom + rigid corrugated surface (ITYP = 8)}}\mylabel{ityp8}

\MOLSCAT\ and \BOUND\ can calculate S matrices or bound states respectively for
diffractive scattering of atoms from a rigid corrugated (solid) surface
 \cite{Wolken:1973:surface, Hutson:1983} using $\basisitem{ITYPE}=8$. In this
case the basis functions depend on the energy and angles of incidence, which
are specified in namelist \namelist{\&INPUT}, so both \namelist{\&INPUT} and
\namelist{\&BASIS} are described here.

For $\var{ITYP}=8$ the reduced mass is the same as the atomic mass $m$ and is
input in \inpitem{URED}.

Surface scattering calculations do not require loops over total angular
momentum and symmetry block. The programs therefore use the internal loop over
\var{JTOT} to loop over the polar angle $\theta$ (measured from the surface
normal) and the loop over symmetry block \var{IBLOCK} to loop over azimuthal
angle $\phi$ (measured relative to the surface reciprocal lattice vector
$g_1$). The incident wavevector $\boldsymbol{k}$ may be decomposed into
components $k_\perp=k\cos\theta$, perpendicular to the surface, and a vector
$\boldsymbol{K}$ in the surface plane, with magnitude
$k_\parallel=k\sin\theta$. The loop over $\theta$ is controlled by the input
parameters \inpitem{JTOTL}, \inpitem{JTOTU}, \inpitem{JSTEP}, \inpitem{THETLW}
and \inpitem{THETST}, while that over $\phi$ is controlled by \inpitem{MXPHI},
\inpitem{PHILW} and \inpitem{PHIST}. The logic used is equivalent to:

\begin{verbatim}
DO JTOT = JTOTL, JTOTU, JSTEP
  THETA = THETLW + THETST*JTOT
  DO M = 1, MXPHI
    PHI = PHILW + PHIST*(M-1)
          ..
          ..
\end{verbatim}
\begin{itemize}
\item[]{Scattering calculation for parallel momentum corresponding to
    angles \var{THETA}, \var{PHI} and incident energy \inpitem{ENERGY}(1),
    or}
\item[]{bound-state calculations for parallel momentum corresponding to angles
\var{THETA}, \var{PHI} and wavevector 100 \AA$^{-1}$.}
\end{itemize}
\begin{verbatim}
          ..
          ..
  ENDDO
ENDDO
\end{verbatim}

The basis functions for $\basisitem{ITYPE}=8$ are formed from surface
reciprocal-lattice vectors $\boldsymbol{G}=(g_1,g_2)$, and are proportional to
\begin{equation}
\exp[{\rm i}(\boldsymbol{K}+\boldsymbol{G})\cdot \boldsymbol{R}],
\end{equation}
where $\boldsymbol{R}$ is the position of the atom within the surface unit
cell. The dimensions and symmetry of the surface unit cell are specified in the
input array \basisitem{ROTI}. \basisitem{ROTI}(1) and \basisitem{ROTI}(2) are
the lengths of the real-space lattice vectors (in \AA, irrespective of the
units of length used elsewhere). \basisitem{ROTI}(3) is the lattice angle in
degrees.

Note that, for surface scattering, subsequent energies have a rather
non-intuitive meaning, because the ``energy" that appears in the coupled
equations is $\hbar^2 k_\perp^2/2m$, while the parallel component
$\boldsymbol{K}$ of the momentum enters in the threshold energies
$\hbar^2|\boldsymbol{K}+\boldsymbol{G}|^2/2m$. The programs interpret
subsequent energies as having the same parallel momentum as the first energy,
but a different perpendicular momentum. In \BOUND, this allows calculations of
energies as a function of $\boldsymbol{K}$ (a band-structure diagram). In
\MOLSCAT\, however, it corresponds to a change in the polar angle as well as
the scattering energy, and the program calculates and prints the new polar
angle.

\cbcolor{\bcol}\cbstart The same loops are used by \BOUND\ to specify the
parallel component of the momentum, so the bound states located are for the
same $k_\perp$ as at \inpitem{EMIN}, rather than at the same polar angle. The
same is in principle true for \FIELD, but it is hard to think of a use for
$\basisitem{ITYPE}=8$ with \FIELD.\cbend

Basis sets for surface scattering are generated in two steps. At the time that
namelist \namelist{\&BASIS} is read, the program sets up a master list of basis
functions that \emph{may} be included in subsequent scattering calculations.
This takes place before entering the loops over incident angles and energies,
so must be angle- and energy-independent. If $\basisitem{NLEVEL} = 0$, the list
of basis functions is generated from \basisitem{J1MAX}, \basisitem{J2MAX};
$g_1$ loops from $-\basisitem{J1MAX}$ to $+\basisitem{J1MAX}$, and $g_2$ loops
from $-\basisitem{J2MAX}$ to $+\basisitem{J2MAX}$. However, each basis function
$\boldsymbol{G} = (g_1,g_2)$ is included only if the parallel kinetic energy
$\hbar^2|\boldsymbol{G}|^2/2m < \basisitem{EMAX}$.

Alternatively, if $\basisitem{NLEVEL}>0$, the array \basisitem{JLEVEL} must
contain a list of $(g_1,g_2)$ pairs. The test involving \basisitem{EMAX} is
bypassed in this case.

Subsequently, for each value of incident $\theta$, $\phi$ and energy, the
program calculates the parallel component $\boldsymbol{K}$ of the incident
momentum, and selects from the master list those basis functions for which
$\hbar^2|\boldsymbol{K}+\boldsymbol{G}|^2/2m < \basisitem{EMAXK}$. This occurs
even if the array \basisitem{JLEVEL} is specified explicitly in namelist
\namelist{\&BASIS}.

Thus, only those basis functions that satisfy both the \basisitem{EMAX} and
\basisitem{EMAXK} criteria are ultimately included in the basis set. Since the
\basisitem{EMAXK} criterion is usually the most sensible physically,
\basisitem{EMAX} serves principally to keep the automatically generated master
list within manageable bounds; the master list must not contain more than
\var{MXELVL} functions (current value 1000, set in module \module{sizes}).

For the common case of surface scattering with the incident beam approaching
along a symmetry direction, the programs automatically construct the
appropriate symmetrised linear combinations of basis functions. This takes
place as part of the second step described above, since it is not until that
stage that the program knows the incident angle.

\section[{\texorpdfstring{Additional quantum numbers in \code{JSTATE} but not \code{JLEVEL}}
{Additional quantum numbers in JSTATE but not JLEVEL}}]{Additional quantum numbers in \var{JSTATE} but not \basisitem{JLEVEL}}\mylabel{couple:monomer}

The array of pair state quantum numbers \var{JSTATE} always includes all the
pair level quantum numbers in \basisitem{JLEVEL} and an entry (the last one for
each pair state) that identifies the state-to-state cross sections to which it
contributes for basis sets diagonal in $H_{\rm intl}$ and $\hat L^2$. This
entry is a pointer to the arrays \basisitem{JLEVEL} and \basisitem{ELEVEL}. For
asymptotically non-diagonal basis sets, this entry is not used, and instead the
asymptotic energies are stored in the array \var{ELEVEL} as they are
calculated, and an array \var{INDLEV} is used to index them.

For $\var{ITYP}=1$, 2, 5 and 7, the quantum numbers in \var{JSTATE} are the
same as those in \basisitem{JLEVEL}.

For $\var{ITYP}=3$, both species have angular momentum, $j_1$ and $j_2$. The
basis sets used couple these together to form a resultant $j$, which can take
values from $|j_1-j_2|$ to $j_1+j_2$. \var{JSTATE} has an additional entry for
$j$ and lists each resulting pair state separately, so $\var{NQN}=4$. The
exception to this is the effective potential approximation
($\basisitem{ITYPE}=13$), where there is no $j$ quantum number and
$\var{NQN}=3$.

For $\var{ITYP}=6$, \var{JSTATE} includes an entry for \var{PRTY}, which
describes the symmetry of the rotor function as discussed in section
\ref{ityp6}. In addition, \var{JSTATE} contains $i_{j\tau}$ and
$n_{j\tau}=2j+1$, resulting in $\var{NQN}=6$; $i_{j\tau}$ is a pointer to the
first of $n_{j\tau}$ wavefunction coefficients $a^j_{\tau,k}$ in the array
\var{ATAU}. 

For $\var{ITYP}=4$, \var{JSTATE} includes all the additional entries described
for $\var{ITYP}=6$ and one for the resultant $j$ of $j_1$ and $j_2$, so
$\var{NQN}=8$.

\section{Close-coupling calculations}\mylabel{couple:cc}

The programs implement close-coupling calculations (calculations without
dynamical approximations) in a space-fixed basis set for $\basisitem{ITYPE}=1$
to 7. The full space-fixed basis functions are formed by coupling $j$ to the
end-over-end quantum number $L$ to form the total angular momentum $J_{\rm
tot}$
\begin{equation}
|(j,L)J_{\rm tot}M_{\rm tot}\rangle=\sum_{m_j,M_L}
\langle jm_j,LM_L|J_{\rm tot}M_{\rm tot}\rangle|jm_j\rangle|LM_L\rangle.
\end{equation}
The resulting coupled equations are independent of $M_{\rm tot}$. The
calculations are carried out for one value of $J_{\rm tot}$ and symmetry block
\var{IBLOCK} at at time. For close-coupling calculations, \var{IBLOCK} encodes
the total parity $(-1)^{j+L}$, which is $(-1)^{J_{\rm tot}+\var{IBLOCK}}$. For
$\basisitem{ITYPE}=3$, \var{IBLOCK} also encodes identical particle symmetry if
present, with the 2 blocks for odd exchange symmetry followed by the 2 blocks
for even exchange symmetry.

The full basis set includes all combinations of $L$ with functions specified by
\var{JSTATE} that have the required $J_{\rm tot}$ and total parity. Inside the
loops over \var{JTOT} and \var{IBLOCK}, the programs construct arrays
\var{JSINDX} and \var{L}, of dimension $N$. For each basis function $i$,
$\var{JSINDX}(i)$ is a pointer to quantum numbers in the array \var{JSTATE} and
$\var{L}(i)$ is the corresponding value of $L$. For specific values of $J_{\rm
tot}$ and parity, some functions in \var{JSTATE} may not appear at all, and
others may appear multiple times in combination with different values of $L$.

For special purposes, basis functions for high $L$ may be excluded by setting
\basisitem{ISYM2}(1) to the upper limit required for $\var{L}(i)$.

\section{Decoupling approximations}\mylabel{decouple}

\subsection{Effective potential}\mylabel{basis:EF}

The effective potential method of Rabitz \cite{Rabitz:EP} is supplied and coded
as $\var{IADD}=10$, but it has not been used for many years and is no longer
supported.  It uses `effective rotational states', which are nondegenerate and
do not couple to the orbital angular momentum $L$, so that $L$ and $J_{\rm
tot}$ are identical.

\subsection{CS (Coupled-states / centrifugal sudden) and helicity decoupling}\mylabel{basis:CS}

The monomer basis functions can be expressed relative to a rotating, body-fixed
coordinate system, where now the projection quantum numbers $m$ (designated $K$
in this case) refer to projection on the interparticle axis rather than the
space-fixed $Z$ axis.  The resulting body-fixed basis set is related to the
space-fixed set by a unitary transformation, but now the interaction matrix is
diagonal in $K$. However, the operator $\hat{L}^2$ is non-diagonal in $K$.

There are several different approximations that neglect the matrix elements of
$\hat{L}^2$ which are off-diagonal in $K$. All these are implemented with
$\var{IADD}=20$, with symmetry block \var{IBLOCK} used for
coupled equations with $K=\var{IBLOCK}-1$. Since $-K$ is equivalent to $+K$,
only $K\ge0$ is implemented.

The helicity decoupling approximation neglects matrix elements off-diagonal in
$K$ but evaluates the diagonal matrix elements exactly as
\begin{equation}
\langle jKJ_{\rm tot} | \hat{L}^2 | jKJ_{\rm tot} \rangle
= J_{\rm tot}(J_{\rm tot}+1) + j(j+1) - 2K^2
\end{equation}
This is invoked with $\basisitem{IBOUND}=1$. In this approach $\var{JTOT}$ is
interpreted as $J_{\rm tot}$, so $K\le J_{\rm tot}$ and only basis functions
with $j\ge K$ are included in the basis set.

The helicity decoupling approximation is very useful for bound states of some
Van der Waals complexes. However, the diagonal matrix elements of $\hat L^2$ do
not correspond to values that can be expressed as $L(L+1)$ with integer $L$.
\MOLSCAT\ applies boundary conditions using Riccati-Bessel functions of
non-integer order that properly take account of the diagonal centrifugal
potentials.\footnote{In versions of \MOLSCAT\ before 2014, the integer values
contained in \var{L} were used in the scattering boundary conditions.}

The $L$-labelled coupled-states approximation of McGuire and Kouri \cite{McG74}
is also implemented with $\var{IADD}=20$, but requires $\basisitem{IBOUND}=0$.
This makes the \emph{centrifugal sudden} approximation in which $L$ is set to
the same value for all basis functions in each set of coupled equations. The
sums over $J_{\rm tot}$ that appear in cross sections (e.g., Eq.\
\ref{eqsigdef}) are replaced by sums over $L$, and $\var{JTOT}$ is interpreted
as $L$ not $J_{\rm tot}$. In this case, although the basis set is still
restricted to functions $j\ge K$, $K$ is \emph{not} restricted by $K\le
\var{JTOT}$.

By default, CS and DLD calculations (described below) are executed for all values of
the body-fixed projection number $K$ up to the largest value of $j$ in the
basis set. However, if cross sections are required between only the lowest few
levels, this is unnecessary. If \basisitem{JZCSMX} is set $> -1$ on input, $K$
is limited by \basisitem{JZCSMX} instead of \basisitem{JMAX}. Cross sections
involving levels with $j > \basisitem{JZCSMX}$ are then not valid.

In CS calculations, \basisitem{JZCSFL} (default 0, allowed values $-1, 0, 1$)
sets the orbital quantum number in each channel \code{I} to $\var{L}(i) =
\code{IABS( JTOT + JZCSFL*}j(i))$. This is a historical remnant and values
other than the default are not recommended.

\subsection{\texorpdfstring{Decoupled $L$-Dominant}{Decoupled L-Dominant}}\mylabel{DLD}

This decoupling approximation \cite{DePristo:1976:DLD} is supplied and coded as
$\var{IADD}=30$, but it has not been used for many years and is no longer
supported.

\basisitem{JZCSMX} may be used with DLD calculations, as described for CS
calculations above.

\section{Loops over \var{JTOT} and \var{IBLOCK}}\mylabel{angmom}

\renewcommand{\thefootnote}{\fnsymbol{footnote}}

The loop over \var{JTOT} is controlled by the three input variables
\inpitem{JTOTL}, \inpitem{JTOTU} and \inpitem{JSTEP}; the program loops from
\inpitem{JTOTL} to \inpitem{JTOTU} in steps of \inpitem{JSTEP}.

The loop over symmetry blocks \var{IBLOCK} is used for different purposes for
different interaction types. By default, the programs loop over all possible
values of \var{IBLOCK} for the interaction type concerned. However, if the
input variable \inpitem{IBFIX} is non-zero and $\inpitem{IBHI}<\inpitem{IBFIX}$,
the programs perform calculations only for $\var{IBLOCK}=\inpitem{IBFIX}$. If
$\inpitem{IBHI}\ge\inpitem{IBFIX}$, calculations are performed for the range of
\var{IBLOCK} values from \inpitem{IBFIX} to \inpitem{IBHI}.

It should be noted that bound-state energies can shift quite substantially between different
\var{JTOT} and \var{IBLOCK} values, so that an energy range that is appropriate for one case may
not be appropriate for another. In such cases \BOUND\ and \FIELD\ are usually used for one
\var{JTOT} and \var{IBLOCK} at a time, with $\inpitem{JTOTU}=\inpitem{JTOTL}$ (and often
$\inpitem{IBFIX}>0$, $\inpitem{IBHI}=0$). These choices are also often suitable in \MOLSCAT\ when
searching for and characterising energy-dependent resonances in the eigenphase sum (section
\ref{energyconv}) or field-dependent resonances in the scattering length (section \ref{fieldconv})
and convergence testing (section \ref{andconv}).

\section[\texorpdfstring{Contracting the basis set ({\color{\bcol}\BOUND} only)}
{Contracting the basis set (BOUND only)}] {Contracting the basis
set\sectionmark{Contracting the basis set}}
\sectionmark{Contracting the basis set}\mylabel{contract}

\cbcolor{\bcol}\cbstart

If $\inpitem{RCTRCT}>0$, \BOUND\ calculates and diagonalises the Hamiltonian
matrix at the fixed distance \inpitem{RCTRCT}, and then discard eigenvectors of
this matrix whose eigenvalues are greater than \inpitem{ECTRCT}.  This
contracted basis set is used in the remainder of the calculation.

This has not proved a particularly useful approach.\cbend

\section[\texorpdfstring{IOS calculations ({\color{\mcol}\MOLSCAT} only, $\var{IADD}=100$)}
{IOS calculations (MOLSCAT only, IADD = 100)}]{IOS calculations\sectionmark{IOS calculations}}
\sectionmark{IOS calculations}
\mylabel{angmom:IOS}

\cbcolor{\mcol}\cbstart

IOS calculations do not use basis sets constructed from monomer rotational
functions. Instead calculations are done for fixed intermolecular angles and
the resulting fixed-orientation terms \cite{Gol78, Green:1979:IOS} are
integrated with appropriate angular factors by quadrature to obtain the
collision dynamics factors and hence the state-to-state cross sections. If set
greater than 0, \inpitem{LMAX} and \inpitem{MMAX} specify the highest $L$ and
$M$ values for which generalised IOS cross sections, $Q(L,M,M')$ are
accumulated. For $\basisitem{ITYPE}=103$ (section \ref{ityp3}), \inpitem{LMAX}
and \inpitem{MMAX} identify the maximum $L$ for rotors 1 and 2 respectively.

\begin{description}
\item[\basisitem{IOSNGP}]{is an integer array of dimension 3. It specifies
    the number of (Gauss) integration points to use for quadrature over
    fixed-orientation cross sections in an IOS calculation.
    \basisitem{IOSNGP}(1) is the number of points for $\theta$. In
    \basisitem{ITYPE}=105 and 106, \basisitem{IOSNGP}(2) is the number of
    points for $\chi$; in \basisitem{ITYPE}=103, the 3 values are for
    $\theta_1$, $\theta_2$ and $\phi_1-\phi_2$.

If values are not given, the program tries to choose the
minimum number needed for requested values of \inpitem{LMAX},
\inpitem{MMAX} and/or input basis set rotor levels.}

\item[\basisitem{IPHIFL}]{controls the type of numerical
    quadrature on $\phi$ for $\basisitem{ITYPE} = 103$, 105 and 106. The
    default of 0 requests equally spaced points (recommended; algebraically
    equivalent to Gauss-Chebyshev quadrature on $\cos\phi$) while
    $\basisitem{IPHIFL} \ne 0$ requests Gauss-Legendre.}

\end{description} \cbend

\section{The BCT Hamiltonian}\mylabel{BCT}

The Bohn-Cavagnero-Ticknor (BCT) Hamiltonian  \cite{Bohn:BCT:2009} provides a simple model for the
collision of two species that interact through a dipole-dipole potential. It assumes that the
dipoles have a fixed direction in space, at an angle $\theta$ to the interparticle vector. The
resulting interaction is anisotropic, coupling different partial waves $L$.

The coupled equations for the BCT Hamiltonian are closely analogous to those for linear rotor
+ atom in the coupled-states approximation, $\basisitem{ITYPE}=21$; they differ only in that the
centrifugal Hamiltonian is different and that the ``linear molecule" rotational quantum number $j$
in \MOLSCAT\ and \BOUND\ is interpreted as the BCT quantum number $L$. Use of this Hamiltonian is
invoked by choosing $\basisitem{ITYPE}=21$ and setting the logical variable \basisitem{BCT}
to \code{.TRUE.}, with \basisitem{BE} set to a tiny value (such as \code{1.D-99}).
\inpitem{JTOTL}, \inpitem{JTOTU} and \inpitem{IBOUND} must be zero.

The coupled-states quantum number $K$ is interpreted as the BCT quantum number $M_L$. It is taken
from $\var{IBLOCK}-1$; by default, \var{IBLOCK} loops from 1 to $\basisitem{JMAX}+1$, so $M_L$
loops from 0 to \basisitem{JMAX}, unless limited by \basisitem{JZCSMX} as described below. A single
value of $M_L$ may be selected by setting \inpitem{IBFIX} to $M_L+1$.

The interaction potential for the BCT Hamiltonian contains terms for $\lambda=2$ only. A complete
Hamiltonian suitable for coupled-channel calculations needs either a short-range boundary
condition, set with \inpitem{BCYOMN}, or an additional isotropic potential that is repulsive at
short range.

The BCT Hamiltonian does not couple functions with even $L$ to those with odd $L$. The even
block may be selected with $\basisitem{JMIN}=0$ and $\basisitem{JSTEP}=2$, while the odd block may
be selected with $\basisitem{JMIN}=1$ and $\basisitem{JSTEP}=2$. For distinguishable particles,
blocks with even $L$ and odd $L$ both exist, but are independent. For identical bosons only the
even block exists, whereas for identical fermions only the odd block exists.

Examples of bound-state calculations with \BOUND, using Hamiltonians of this form, are given by
Karman \etal\ \cite{Karman:dipole:2018}.

For the BCT Hamiltonian, all channels are asymptotically degenerate, so all cross sections are
``elastic" in the sense that the collisions do not release kinetic energy. \MOLSCAT\ nevertheless
gives integral cross sections $\sigma_{LL'}$ that are summed over $M_L$ but not over $L$ and $L'$.
The actual elastic cross section must be calculated externally from $\sum_{LL'} \sigma_{LL'}$.

For distinguishable particles, cross sections for even $L$ and odd $L$ should be calculated
separately and added together. For this, \inpitem{JSTEP} in \namelist{\&INPUT} must be 1 (the
default), even though \basisitem{JSTEP} in \namelist{\&BASIS} is 2. For identical particles (either
bosons or fermions), \inpitem{JSTEP} should be set to 2 to achieve the doubling required in the
cross sections.

By default, BCT calculations are executed for all values of $M_L$ up to the largest value of
$L$ in the basis set. However, if cross-section contributions are required between only the lowest
few values of $L$, this is unnecessary. If \basisitem{JZCSMX} is set $> -1$ on input, $M_L$ is
limited by \basisitem{JZCSMX} instead of \basisitem{JMAX}. Naturally, doing this means that
contributions to cross sections from $L$ or $L' > \basisitem{JZCSMX}$ are incomplete.

It should be noted that scattering calculations with the BCT Hamiltonian often need extremely
large values of \inpitem{RMAX} for convergence, particularly at very low energy.

\section[{\texorpdfstring{Plug-in basis-set suites ($\code{ITYPE}=9$)}
{Plug-in basis-set suites (ITYPE = 9)}}]{Plug-in basis-set suites ($\basisitem{ITYPE}=9$)}\mylabel{ityp9}

The programs include an interface for users to specify interaction types
different from the built-in ones. A considerable number of such routines have
been written, though the interface has developed and become more sophisticated
over the years and some older routines would need work to update and test.

In this distribution we provide plug-in basis-suites for two interaction types
of current interest:
\begin{description}
\item[\file{base9-1S\_3Sigma\_cpld.f}]{handles interactions of a
    structureless ($^1$S) atom and a molecule in
    a $^3\Sigma$ state in a magnetic field;}
\item[\file{base9-alk\_alk\_ucpld.f}]{handles interactions of two alkali-metal
    atoms in $^2$S states in a magnetic field, including hyperfine
    coupling.}
\end{description}
These basis-set suites are described in chapter \ref{user:gen}; the description
of each of them is divided approximately into sections for \emph{users} of the
suites and additional sections for people who wish to understand their internal
working, perhaps in order to program their own basis-set suite for a different
case.

Chapter \ref{base9} gives a more formal description of the components of a
plug-in basis-set suite, and specifies the calling sequence of each routine
involved. It is intended primarily for people who wish to program their own
basis-set suite.

Plug-in basis set suites usually have access to many variables input in
namelist \namelist{\&BASIS}, which are placed in module \module{basis\_data}
and are listed in section \ref{module:basis-data}. These can be used as desired
by the programmer.

\chapter{\texorpdfstring{Constructing the interaction potential}
{\ref{buildVL}: Constructing the interaction potential}}\mylabel{buildVL}

\section{The potential expansion}\mylabel{buildVL:expand}

The interaction potential $V(R,\xi_{\rm intl})$ is formally a function of all
the coordinates. However, it has the additional property that it is invariant
to rotations of the whole system. In practice interaction potentials
are usually written as a function of relative coordinates, with angles
expressed with respect to the interparticle vector $\boldsymbol{R}$.

The programs internally require an expansion of the interaction potential in a
set of orthogonal functions of the internal coordinates\footnote{This is not
true for the IOS code in \MOLSCAT, which can use unexpanded potentials
directly; see section \ref{potl:IOS}.},
\begin{equation}
V(R,\xi_{\rm intl})=\sum_\Lambda v_\Lambda(R){\cal V}^\Lambda(\xi_{\rm intl}),
\end{equation}
where the labels that comprise $\Lambda$ depend on \var{ITYP} and are held in
an array \potlitem{LAMBDA}.

{\def\mytabwidth{1.75in}
\parindent 0pt
\vspace*{\baselineskip}
\label{lambda-table}
\begin{tabular}{llll}
\multispan{2}{\hfill Interaction type\hfill} & $\Lambda$ & notes\\
\hline
linear rigid rotor + atom & 1 & $\lambda$ \\
linear vibrotor + atom & 2 & $\lambda\, v\, v'$ & \parbox[t]{\mytabwidth}
{\raggedright$\lambda\, v\, v'\equiv\lambda\, v'\, v$; only one should be supplied}\\
linear rigid rotor + linear rigid rotor & 3 & $\lambda_1\,\lambda_2\,\lambda$ &
\parbox[t]{\mytabwidth}{\raggedright$\lambda_1\,\lambda_2\,\lambda\equiv\lambda_2\,\lambda_1\,\lambda$;
both \emph{must} be supplied}\\
non-linear rigid rotor + linear rigid rotor & 4 & $\lambda_1\,\kappa_1\,\lambda_2\,\lambda$ \\
non-linear rigid rotor + atom & 5 \& 6 & $\lambda\,\kappa$ \\
linear vibrotor + atom & 7 & $\lambda\, v\, j\, v'\, j'$ &
\parbox[t]{\mytabwidth}{\raggedright$\lambda\, v\, j\, v'\, j'\equiv\lambda\, v'\, j'\, v\, j$;
only one should be supplied}\\
atom + surface & 8 & $g_1\,g_2$\\
\hline
\end{tabular}
\vspace*{\baselineskip}}

The terms included in the potential expansion, and values for the radial
potential coefficients $v_\Lambda(R)$, are provided by subroutine
\prog{POTENL}. Versions of the programs before 2019.0 were documented in the
expectation that \prog{POTENL} would usually be a plug-in routine. This is
still possible, as described in chapter \ref{userPotenl} below. However, the
programs are now supplied with a general-purpose version of \prog{POTENL} that
is adequate for most purposes. This section documents the capabilities of the
general-purpose version of \prog{POTENL}.

The radial potential coefficients may be generated in two different ways.
\begin{enumerate}
\item Coefficients supplied explicitly. Very simple coefficients (sums of
    exponential terms and inverse-power terms) can be set up in
    \prog{POTENL} itself. More complicated coefficients must be provided by
    a user-supplied routine \prog{VINIT} (with entry points \prog{VSTAR},
    \prog{VSTAR1} and \prog{VSTAR2}) as described in section
    \ref{vrtp:false} below.
\item For most built-in interaction types, a user-supplied routine
    \prog{VRTP} may be provided to evaluate the interaction potential at
    values of the coordinates specified by \prog{POTENL}, as described in
    section \ref{vrtp:true} below. \prog{POTENL} then evaluates the radial
    potential coefficients itself using appropriate quadratures.
\end{enumerate}
Both \prog{VINIT}/\prog{VSTAR} and \prog{VRTP} have much simpler specifications
than \prog{POTENL} itself.

On initialisation, \prog{POTENL} reads namelist \namelist{\&POTL} and returns
integer labels for the potential expansion terms in the array
\potlitem{LAMBDA}. These labels are usually generated from namelist
\namelist{\&POTL} items \potlitem{LMAX}, \potlitem{MMAX}, \potlitem{L1MAX} and
\potlitem{L2MAX} (as appropriate for the value of \var{ITYP}), but in special
cases the array \potlitem{LAMBDA} may be specified as an explicit list. In this
case \potlitem{MXLAM} must be the number of expansion functions.

The type of functions used for expanding the interaction potential depends on
\var{ITYP}:

\subsection{\texorpdfstring{Rigid rotor + atom ($\var{ITYP}=1$)}
{Rigid rotor + atom (ITYP = 1)}}\mylabel{potl:ityp1}
The interaction potential is expanded in Legendre polynomials
\begin{equation}
V(R,\theta)=\sum_{\lambda} v_{\lambda}(R) P_\lambda(\cos\theta).
\end{equation}
Note that the Legendre polynomials $P_\lambda$ are normalised with
$P_\lambda(1)=1$, so that they are orthogonal but not orthonormal.

If $\potlitem{MXLAM}=0$ (the default), \potlitem{LMAX} sets the maximum value
for $\lambda$, and \prog{POTENL} creates an array \potlitem{LAMBDA} containing
entries for all non-zero terms from $\lambda=0$ upwards. Setting
$\potlitem{IHOMO}=2$ creates an array \potlitem{LAMBDA} that contains only even
values of $\lambda$; this is appropriate if the molecule has symmetry
$D_{\infty h}$ (e.g., if it is a homonuclear diatomic molecule).

If $\potlitem{MXLAM}>0$, the array \potlitem{LAMBDA} is read as an explicit
list of \potlitem{MXLAM} values.

\subsection{\texorpdfstring{Diatomic vibrotor + atom ($\var{ITYP}=2$ or 7)}
{Diatomic vibrotor + atom (ITYP = 2 or 7)}}\mylabel{potl:ityp2or7}

The potential matrix element between each pair of molecular vibration-rotation
functions is expanded in Legendre polynomials
\begin{equation}
\langle v'j' | V(R,r,\theta)| vj \rangle
=\sum_{\lambda} v_{\lambda v j v' j'}(R) P_\lambda(\cos\theta).
\end{equation}
The Legendre polynomials $P_\lambda$ are normalised as for $\var{ITYP}=1$. For
$\var{ITYP}=7$ each term in the expansion is represented by 5 consecutive
elements of the array \potlitem{LAMBDA}: $\lambda, v, j, v', j'$. For
$\var{ITYP}=2$ the effect of monomer centrifugal distortion on the potential
matrix elements is neglected; the $j$ labels are omitted, and each term is
represented by 3 indices: $\lambda, v, v'$.

If $\var{ITYP}=2$ and $\potlitem{MXLAM}=0$ (the default), \potlitem{LMAX} sets
the maximum value for $\lambda$, and $\potlitem{IHOMO}=2$ may be used, as for
$\var{ITYP}=1$. The vibrational quantum numbers $v$ and $v'$ loop from
\potlitem{IVMIN} to \potlitem{IVMAX}.

For $\var{ITYP}=7$, $\potlitem{MXLAM}$ must be set greater than zero and the
array \potlitem{LAMBDA} must be provided as an explicit list.

If $\potlitem{MXLAM}>0$, the array \potlitem{LAMBDA} is read as an explicit
list of \potlitem{MXLAM} triples $\lambda, v, v'$ (for $\var{ITYP}=2$) or
quintuples $\lambda, v, j, v', j'$ (for $\var{ITYP}=7$). The interaction
potential is invariant to exchange of the labels $v$ and $v'$ (or $v,j$ and
$v',j'$), and only one should be supplied.

\subsection{\texorpdfstring{Rigid rotor + rigid rotor ($\var{ITYP}=3$)}
{Rigid rotor + rigid rotor (ITYP = 3)}}\mylabel{potl:ityp3}

The interaction potential is expanded in coupled products of spherical
harmonics for the two molecules \cite{Green:1975},
\begin{IEEEeqnarray}{rCcl}
V(R,\theta_1,\phi_1,\theta_2,\phi_2)
&=&\sum_{\lambda_1,\lambda_2,\lambda} v_{\lambda_1,\lambda_2,\lambda}(R)
\sum_{\mu}&\langle \lambda_1\mu,\lambda_2,-\mu|\lambda,0\rangle
\left(\frac{2\lambda+1}{4\pi}\right)^{1/2}\nonumber\\
&&&\times Y^{\mu}_{\lambda_1}(\theta_1,\phi_1)
Y^{-\mu}_{\lambda_2}(\theta_2,\phi_2).
\end{IEEEeqnarray}
The interaction potential actually depends only on the relative angle
$\phi_1-\phi_2$. The spherical harmonics $Y^\mu_\lambda$ are normalised by
integrating over angles, so the set of functions is orthonormal.\footnote{This
normalisation differs from that for the Legendre polynomials used for
$\var{ITYP}=1$, 2 and 7, and has the important consequence that $v_{000}(R)$ is
\emph{not} the spherical average of the potential.}

The coupled expansion is not the only one commonly in the literature. An
alternative is an uncoupled expansion,
\begin{equation}
V(R,\theta_1,\phi_1,\theta_2,\phi_2)=\sum_{\lambda_1,\lambda_2,\mu}
v'_{\lambda_1,\lambda_2,\mu}(R)
\left(\frac{[(2\lambda_1+1)(2\lambda_2+1)]^{1/2}}{4\pi}\right)
Y^{\mu}_{\lambda_1}(\theta_1,\phi_1) Y^{-\mu}_{\lambda_2}(\theta_2,\phi_2).
\end{equation}
The uncoupled expansion is not directly implemented in these programs, but
coefficients may readily be converted between the two forms. The coupled
expansion has the advantage that electrostatic interactions between multipole
moments on the two monomers each result in a single long-range term.

Each term in the expansion is represented by 3 consecutive elements of the
array \potlitem{LAMBDA}: $\lambda_1$, $\lambda_2$ and their vector sum
$\lambda$.

If $\potlitem{MXLAM}=0$ (the default), \potlitem{L1MAX} and \potlitem{L2MAX}
set the maximum values for $\lambda_1$ and $\lambda_2$, and $\lambda$ takes all
integer values from $|\lambda_1-\lambda_2|$ to $\lambda_1+\lambda_2$.
\potlitem{IHOMO} and/or \potlitem{IHOMO2} can be set to 2 to indicate that the
corresponding molecule has symmetry $D_{\infty h}$.\footnote{in this particular
instance \potlitem{ICNSYM} is a deprecated synonym for \potlitem{IHOMO2},
retained for backwards compatibility.}

If $\potlitem{MXLAM}>0$, the array \potlitem{LAMBDA} is read as an explicit
list of \potlitem{MXLAM} triples $\lambda_1, \lambda_2, \lambda$.

If the two rotors are identical, potential terms $v_{\lambda_1,\lambda_2,\lambda}(R)$
and $v_{\lambda_2,\lambda_1,\lambda}(R)$ must be identical and both must be included.

\subsection{\texorpdfstring{Non-linear molecule + atom ($\var{ITYP}=5$ or 6)}
{Non-linear molecule + atom (ITYP = 5 or 6)}}\mylabel{potl:ityp5or6}
The interaction potential is expanded in spherical harmonics
\cite{Green:1976},\footnote{This normalisation differs from that for the
Legendre polynomials used for $\var{ITYP}=1$, and has the important consequence
that $v_{00}(R)$ is \emph{not} the spherical average of the potential.}
\begin{equation}
V(R,\theta,\chi)=\sum_{\lambda,\kappa} v_{\lambda,\kappa}(R)Y^{\kappa}_{\lambda}(\theta,\chi).
\end{equation}
$\var{ITYP}=5$ or 6 are coded only for molecules for which the molecule-fixed
$xz$ plane is a plane of symmetry, and in this case the expansion becomes
\begin{equation}
V(R,\theta,\chi)=\sum_{\lambda,\kappa\ge0} v_{\lambda,\kappa}(R) (1+\delta_{\kappa,0})^{-1}
\left[Y^{\kappa}_{\lambda}(\theta,\chi)+(-1)^\kappa Y^{-\kappa}_{\lambda}(\theta,\chi)\right].
\end{equation}

Each term in the expansion is represented by 2 consecutive elements of the
array \potlitem{LAMBDA}: $\lambda$ and $|\kappa|$.

If $\potlitem{MXLAM}=0$ (the default), \potlitem{LMAX} and \potlitem{MMAX} set
the maximum values for $\lambda$ and $|\kappa|$. If $\potlitem{ICNSYM}>0$, only
terms where $|\kappa|$ is a multiple of \potlitem{ICNSYM} are included; this is
appropriate for molecules where the $z$ axis is a proper axis of rotation of
order \potlitem{ICNSYM}.

If $\potlitem{MXLAM}>0$, the array \potlitem{LAMBDA} is read as an explicit
list of \potlitem{MXLAM} pairs $\lambda$, $|\kappa|$.

The interaction potential may also be expanded in terms of renormalised
(Racah-normalised) spherical harmonics
$C_{\lambda\kappa}(\theta,\chi)=[4\pi/(2\lambda+1)]^{1/2}Y_\lambda^\kappa(\theta,\chi)$
in place of the normalised spherical harmonics $Y_\lambda^\kappa(\theta,\chi)$.
This form of expansion is indicated by setting \potlitem{CFLAG} to 1.

\subsection{\texorpdfstring{Asymmetric rotor + diatom ($\var{ITYP}=4$)}
{Asymmetric rotor + diatom (ITYP = 4)}}\mylabel{potl:ityp4} The interaction
potential is expanded in coupled products of rotation matrices for the
asymmetric rotor and spherical harmonics for the diatom \cite{Phillips:1995}.
This is a generalisation of the expansion for $\var{ITYP}=3$, with an extra
angle $\chi$ representing the rotation of the asymmetric rotor about its $z$
axis. The spherical harmonics for the asymmetric rotor are replaced by
normalised rotation matrices, with an index $\kappa_1$ corresponding to the
angle $\chi$.

Each term in the expansion is represented by 4 consecutive elements of the
array \potlitem{LAMBDA}: $\lambda_1$ (the tensor order of the expansion for the
non-linear molecule), $\kappa_1$ (the projection of $\lambda_1$ on the $z$ axis
of the non-linear molecule), $\lambda_2$ (the tensor order of the expansion for
the diatom) and $\lambda$ (the vector sum of $\lambda_1$ and $\lambda_2$).

There is no option to generate the array \potlitem{LAMBDA} from limits supplied
in other namelist items. \potlitem{MXLAM} must be set greater than 0, and the
array \potlitem{LAMBDA} is read as an explicit list of \potlitem{MXLAM} sets of
4 consecutive values $\lambda_1$, $\kappa_1$, $\lambda_2$ and $\lambda$.

\subsection{\texorpdfstring{Atom + corrugated solid surface ($\var{ITYP}=8$)}
{Atom + corrugated solid surface (ITYP = 8)}}\mylabel{potl:ityp8}

The interaction potential is expanded in surface reciprocal lattice vectors,
\begin{equation}
V(\boldsymbol{r})=V(\boldsymbol{R},z)
=\sum_{g_1,g_2} v_{\boldsymbol{G}}(z)\exp({\rm i}\boldsymbol{G}\cdot\boldsymbol{R}).
\end{equation}

\potlitem{LAMBDA} has 2 entries per term in the expansion: $g_1$ and $g_2$,
which are the two components of the reciprocal lattice surface vector
$\boldsymbol{G}$.  The general-purpose version of \prog{POTENL} does not
generate the array \potlitem{LAMBDA} from limits supplied in other namelist
items, and it must be input as an explicit list of \potlitem{MXLAM} pairs
$(g_1,g_2)$.

The interaction potential may sometimes be described in terms of a sum of
pairwise interactions between the colliding atom and the particles that make up
the surface:
\begin{equation}
V(\boldsymbol{r})=\sum_j U(|\boldsymbol{r}-\boldsymbol{r}_j|),
\end{equation}
where $\boldsymbol{r}_j$ is the $j$th lattice site.  This gives for the Fourier
components of the atom-surface potential
\begin{equation}
v_{\boldsymbol{G}}(z)=\frac{1}{a_c}\int\sum_j U(|\boldsymbol{r}-\boldsymbol{r}_j|)
\exp(-{\rm i}\boldsymbol{G}\cdot\boldsymbol{R})\,\d^2\boldsymbol{R},
\end{equation}
where $a_c$ is the area of the direct-space unit cell.

\section{Units of length and energy for the interaction potential}\mylabel{buildVL:epsil}

The quantities \potlitem{RM} and \potlitem{EPSIL} specify the units of length
and energy used by the potential routine. They are specified in namelist
\namelist{\&POTL} in units of {\AA}ngstrom and cm$^{-1}$ respectively.

Any values of \potlitem{RM} and \potlitem{EPSIL} set in namelist
\namelist{\&POTL} are passed to the user-supplied routine \prog{VINIT} (if
\potlitem{LVRTP} is set \code{.FALSE.}) or \prog{VRTP} (if \potlitem{LVRTP} is
set \code{.TRUE.}) and are overwritten if those routines change their values.

If \potlitem{RM} is not specified, it defaults to the value of \inpitem{RUNIT}
in namelist \namelist{\&INPUT}. If \inpitem{RUNIT} is not specified,
\potlitem{RM} is used as the unit of length throughout the programs. If neither
is specified, both are set to 1.0 (indicating length units of \AA).

\section{Evaluating the radial potential coefficients}\mylabel{buildVL:EvalCoeffs}

\subsection{\texorpdfstring{Potential pre-expanded in internal coordinates}{Potential
pre-expanded in internal coordinates (supplied as data in \namelist{\&POTL} or
by \prog{VINIT}/\prog{VSTAR})}}\mylabel{vrtp:false}

If the radial potential coefficients are to be supplied explicitly (as opposed
to generated by quadrature in \prog{POTENL}), the array \potlitem{NTERM} sets
out how they are to be calculated:
\begin{itemize}[nosep]
\item{If $\potlitem{NTERM}(i)\ge 0$, the $i$th coefficient is evaluated as
    a sum of $\potlitem{NTERM}(i)$ separate terms, each of which is either
    an exponential in $R$ or an inverse power of $R$.  The function for
    each of these terms is specified by the arrays \potlitem{NPOWER},
    \potlitem{E} and \potlitem{A}, as described below.}
\item{If $\potlitem{NTERM}(i)<0$, \prog{POTENL} obtains the coefficient by
    a call to \prog{VSTAR}, as described below.}
\end{itemize}

\subsubsection{Expansion coefficients supplied as data in \namelist{\&POTL}}

The arrays \potlitem{NPOWER}, \potlitem{E} and \potlitem{A} are interpreted as
in the code fragment:\footnote{In versions before 2019.0, any positive values
of \potlitem{NPOWER} were converted to negative values. Positive values are now
allowed and left unchanged.}
\begin{verbatim}
IEXP  = 0
ITERM = 0
DO I = 1, MXLAM
  P(I) = 0.D0
  IF (NTERM(I).LT.0) CALL VSTAR(R, I, P(I))
  DO JTERM = 1, NTERM(I)
    ITERM = ITERM + 1
    IF (NPOWER(ITERM).NE.0) THEN
      P(I) = P(I) + A(ITERM) * R**NPOWER(ITERM)
    ELSE
      IEXP = IEXP + 1
      P(I) = P(I) + A(ITERM) * EXP(R * E(IEXP))
    ENDIF
  ENDDO
ENDDO
\end{verbatim}

For example, consider a potential with 3 expansion terms:
\begin{itemize}[nosep]
\item the first is provided by a user-supplied \prog{VINIT}/\prog{VSTAR}
    subroutine;
\item the second is a Lennard-Jones potential $C_{12}R^{-12}-C_6R^{-6}$;
\item the third is an exponential-6 potential $a\exp(-\beta R)-bR^{-6}$.
\end{itemize}
In this case, the input file would contain: $\potlitem{NTERM}=-1$, 2, 2,
$\potlitem{NPOWER}=-12$, $-6$, 0, $-6$, $\potlitem{A}=C_{12}$, $-C_6$, $a$,
$-b$ and $\potlitem{E}=-\beta$.

\subsubsection{Expansion coefficients supplied by \prog{VINIT}/\prog{VSTAR}}

If any of the elements of \potlitem{NTERM} are negative, user-supplied
subroutines \prog{VINIT}, \prog{VSTAR}, \prog{VSTAR1} and \prog{VSTAR2} must be
provided.  It is simplest to provide the latter 3 as entry points to
\prog{VINIT}. \prog{VINIT} is called once for each term in the potential
expansion for which $\potlitem{NTERM}(i)<0$. Its purpose is to read any data
required and to carry out any $R$-independent processing desired, to save time
on future calls. Its specification is
\begin{verbatim}
SUBROUTINE VINIT(I, RM, EPSIL)

DOUBLE PRECISION, INTENT(INOUT) :: RM, EPSIL
INTEGER,          INTENT(IN)    :: I
\end{verbatim}
The index of the potential expansion term is passed in \code{I}. The current
values of the length units factor \potlitem{RM} and the energy units factor
\potlitem{EPSIL} are passed in as arguments, and may be overwritten if desired.
Calls to \prog{VINIT} for different values of \code{I} should not return
different values of \potlitem{RM} or \potlitem{EPSIL}.

If desired, \prog{VINIT} may supply names for the units \potlitem{RM} and
\potlitem{EPSIL} in the character variables \var{RMNAME} and \var{EPNAME} in
module \module{potential}. These variables are pre-populated with \code{'RM'}
and \code{'EPSIL'} respectively, but if $\inpitem{RUNIT}=1.0$ after the call to
\prog{POTENL}, the programs set \var{RUNAME} to \code{'ANGSTROM'}; similarly,
if $\potlitem{EPSIL}=1.0$, they set \var{EPNAME} to \code{'CM-1'}.

The entry points (or subroutines) \prog{VSTAR}, \prog{VSTAR1} and \prog{VSTAR2}
are subsequently called many times to evaluate the radial potential
coefficients for either the interaction potential or its first or second radial
derivatives respectively. The specification of these routines is
\begin{verbatim}
SUBROUTINE VSTAR(I, R, V)

DOUBLE PRECISION, INTENT(IN)  :: R
DOUBLE PRECISION, INTENT(OUT) :: V
INTEGER,          INTENT(IN)  :: I
\end{verbatim}
and similarly for \prog{VSTAR1} and \prog{VSTAR2} (but see the next paragraph).
The argument \code{R} contains the distance $R$ (in units of \potlitem{RM}) and
the routine must return the value of the radial potential coefficient \code{I}
(or its first or second radial derivative) in the argument \code{V}, in units
of \potlitem{EPSIL}.

If required, \prog{VINIT} and \prog{VSTAR} can access the array \potlitem{LAMBDA}
by using the module \module{potential}, described in section
\ref{module:potential}. This facility is particularly useful if the ordering of
the potential terms (or which ones are needed) depends on the basis set.

The only functions of the current programs that use radial potential
derivatives are the VIVS propagator and calculations of nonadiabatic matrix
elements at high print level in the LDMA propagator. Even for these,
derivatives may be calculated numerically by specifying
\inpitem{NUMDER}=\code{.TRUE.} in namelist \namelist{\&INPUT}. For all other
purposes it is adequate to supply \prog{VSTAR1} and \prog{VSTAR2} routines that
prints an error message and stop if they are called.

\subsection{\texorpdfstring{Potential supplied by \prog{VRTP} as function of internal
coordinates} {Potential supplied by \prog{VRTP} as function of internal
coordinates}}\mylabel{vrtp:true}

It is often more convenient to specify the interaction potential $V(R,\xi_{\rm
intl})$ explicitly as a function of internal coordinates $\xi_{\rm intl}$,
rather than as an expansion over internal functions. In such cases
\potlitem{LVRTP} must be set to \code{.TRUE.} and the user-supplied routine
\prog{VRTP} must supply values for the interaction potential at a given set of
internal coordinates.

The expansion of the interaction potential in terms of sets of orthogonal
functions can be inverted to give
\begin{equation}
v_\Lambda(R)
=N_\lambda \int \d \xi_{\rm intl}\,V(R,\xi_{\rm intl})\bcalV^\Lambda(\xi_{\rm intl}),
\end{equation}
where $N_\lambda$ depends on the normalisation of the expansion functions (and
is 1 if they are orthonormal). The integral can be approximated by numerical
quadrature
\begin{equation}
v_\Lambda(R)=N_\lambda \sum_{i=1}^{n}w_{i} \bcalV^\Lambda({\xi}_{i})V(R,\xi_i),
\end{equation}
where $\xi_i$ and $w_{i}$ are the $n$ points and weights required for the
quadrature scheme over the functions $\bcalV^\Lambda(\xi)$.

The quadrature scheme is a combination of one or more of the following types of
quadrature over each individual internal coordinate.

\vspace*{\baselineskip}
\begin{tabular}{lll}
\hline
type of quadrature & $x_i$ & $w_i$ \\
\hline
Gauss-Legendre & $i$th root of $P_n(x)$ & $2(P_n(1))^2/[(1-x_i^2)(P_n'(x_i))^2]$\\
Gauss-Hermite & $i$th root of $H_n(x)$ & $2^{n-1}n!\sqrt{\pi}/[n^2(H_{n-1}(x_i))^2]$\\
Equally spaced & $[2i-1]\pi/2n$ & $\pi/n$ \\
\hline
\end{tabular}
\vspace*{\baselineskip}

\prog{POTENL} calculates the points and weights required for each type of
Gaussian quadrature needed for the interaction types 1, 2, 3, 5 and 6.

\vspace*{\baselineskip}
\begin{tabular}{llll}
\hline
\var{ITYP} & \potlitem{NPTS}(1) (or \potlitem{NPT}) & \potlitem{NPTS}(2)
(or \potlitem{NPS}) & \potlitem{NPTS}(3) \\
\hline
& quadrature scheme & quadrature scheme & quadrature scheme \\
\hline
1 & Gauss-Legendre over $\cos\theta$\\
2 & Gauss-Legendre over $\cos\theta$ & Gauss-Hermite over $r$\\
3 & Gauss-Legendre over $\cos\theta_1$ & Gauss-Legendre over $\cos\theta_2$ & equally spaced \\
& & & over $\phi_1-\phi_2$ \\
5 \& 6 & Gauss-Legendre over $\cos\theta$ & equally spaced over $\chi$ \\
\hline
\end{tabular}

\vspace*{\baselineskip} Note that Gauss-Legendre quadrature is not optimum for
the integral over $\cos\theta$ for symmetric top functions with $k>0$, but it
is still used.

The number of points required for these quadrature schemes may be given in
\potlitem{NPTS}. It should be at least $\lambda+1$ (Gauss-Legendre quadrature),
$\mu+1$ or $\kappa+1$ (equally spaced) or $v+v'+1$ (Gauss-Hermite). If the
value supplied for each quadrature is lower than this, \prog{POTENL} prints a
warning and increases it to the minimum valid value.

If $\potlitem{IHOMO}=2$, only points for $\theta\ge 90^\circ$ are used. If
$\potlitem{ICNSYM}>1$, \potlitem{NPTS}(2) points are used in the range
$0<\phi_1-\phi_2<\pi/\potlitem{ICNSYM}$.

Subroutine \prog{VRTP} must be supplied by the user. Its specification is:
\begin{verbatim}
SUBROUTINE VRTP(IDERIV, R, V)

USE angles

DOUBLE PRECISION, INTENT(INOUT) :: R
DOUBLE PRECISION, INTENT(OUT)   :: V
INTEGER,          INTENT(IN)    :: IDERIV
\end{verbatim}
On an initialisation call, $\var{IDERIV}=-1$. The routine may read any data
required and carry out any coordinate-independent processing desired, to save
time on future calls. It may also specify the units of length and energy to be
used in subsequent calls by overwriting the variables \potlitem{RM}
($=\var{R}$) and \potlitem{EPSIL} ($=\var{V}$).  It may also set the variables
\potlitem{IHOMO}, \potlitem{IHOMO2}, \potlitem{ICNSYM}, and \potlitem{ICNSY2}
in module \module{angles}; see below for
specification.\footnote{Before version 2019.0, the quantities included in module
\module{angles} were in a common block named \common{ANGLES}. Earlier versions of
\prog{VRTP} must be modified to use module \module{angles}.}

If desired, an initialisation call to \prog{VRTP} may supply names for the
units \potlitem{RM} and \potlitem{EPSIL} in the character variables
\var{RMNAME} and \var{EPNAME} in module \module{potential}.

Subsequent calls specify $\var{IDERIV} = 0$, 1, 2 for the interaction potential
and its first and second derivatives respectively. The distance (in units of
\potlitem{RM}) is specified in argument \var{R} whilst the internal coordinates
are supplied in the array \var{COSANG} in module \module{angles}, as described below.
\prog{VRTP} must return the corresponding interaction potential (in units of
\var{EPSIL}) in argument \var{V}.

The only functions of the current programs that use radial potential
derivatives are the VIVS propagator and calculations of nonadiabatic matrix
elements at high print level in the LDMA propagator. Even for these,
derivatives may be calculated numerically by specifying
\inpitem{NUMDER}=\code{.TRUE.}\ in namelist \namelist{\&INPUT}. For all other
purposes it is adequate to supply a \prog{VRTP} routine that traps calls with
$\var{IDERIV}> 0$, prints an error message, and stops.

For $\var{ITYP} = 1$, 2, 3, 5, and 6, $\var{COSANG}(1)$ is $\cos\theta$.

For $\var{ITYP} = 3$, $\var{COSANG}(2)$ is $\cos\theta_2$ and $\var{COSANG}(3)$ is
$\phi_1-\phi_2$.

For $\var{ITYP} = 5$ and 6, $\var{COSANG}(2)$ is $\chi$.

For non-IOS $\var{ITYP} = 2$ cases, \code{\var{COSANG}(2)} is $q$, the reduced
harmonic oscillator coordinate, such that the ground-state vibrational
wavefunction of the diatom is $\exp(-q^2/2)$; the quadrature code in
\prog{POTENL} uses \emph{harmonic} vibrational wavefunctions for the vibrating
diatomic molecule. If a more sophisticated integration is required, it must be
programmed separately (e.g., in \prog{VINIT}).

\subsubsection{\texorpdfstring{Module \module{angles}}{Module
angles}}\mylabel{angles:module}

The specification of module \module{angles} is
\begin{verbatim}
USE sizes, ONLY: MXANG
INTEGER          IHOMO, ICNSYM, IHOMO2, ICNSY2
DOUBLE PRECISION COSANG(MXANG), FACTOR
\end{verbatim}

The array dimension \var{MXANG} is set in module \module{sizes} and is
currently 7.

\begin{shaded} 
\section{Auxiliary files in \prog{VINIT}/\prog{VSTAR} or \prog{VRTP}}\mylabel{potential_files}

User-supplied potential routines may if desired use auxiliary input, output or scratch files.
If these have hard-coded unit numbers, they should be in the range 800 to 899 to avoid
conflicts with unit numbers supplied in data (1 to 799) or used internally by the main
parts of the programs (900 to 999).
\end{shaded} 

\cbcolor{\mcol}\cbstart
\section[\texorpdfstring{IOS calculations ({\color{\mcol}\MOLSCAT} only, $\var{IADD}=100$)}
{IOS calculations (MOLSCAT only, IADD = 100)}]
{IOS calculations\sectionmark{IOS calculations}}\sectionmark{IOS calculations}\mylabel{potl:IOS}

IOS calculations perform propagations at fixed orientations. This is most
naturally done by setting $\potlitem{LVRTP}=\code{.TRUE.}$ to use the
\prog{VRTP} method described above. There is no need to obtain radial potential
coefficients by numerical quadrature, so \potlitem{MXLAM} is reset internally
to 1, and the quantum labels in \potlitem{LAMBDA} are ignored.\footnote{In
versions before 2019.0, when $\potlitem{LVRTP}=\code{.TRUE.}$ and
$\potlitem{MXLAM}>0$, \MOLSCAT\ projected out the potential expansion terms
using quadrature, and then re-summed them to obtain the interaction potential
at each orientation. This was inefficient and unnecessarily approximate.}

For $\basisitem{ITYPE} = 102$, \prog{VRTP} does not use $\var{COSANG}(2)$, and
must return in argument \var{V} a vector of matrix elements $\langle
v'|V(R,\theta)|v\rangle$ in the order expected by \prog{POTENL}.
\potlitem{MXLAM} must be set to a negative value, with $v$ and $v'$ values
supplied as $\potlitem{LAMBDA}(3i-1)$ and $\potlitem{LAMBDA}(3i)$ for
$i=1,|\potlitem{MXLAM}|$. $\potlitem{LAMBDA}(3i-2)$, which would normally be
the order of the Legendre polynomial, is not used.

\potlitem{IHOMO} and \potlitem{ICNSYM} would normally be determined
automatically by examining the array \potlitem{LAMBDA}, but this can be done
only if the interaction potential is expanded in angular functions. They may be
set explicitly in the input file if desired.

If $\potlitem{LVRTP}=\code{.FALSE.}$, the terms of the potential expansion
should be indicated as for non-IOS calculations and the
\prog{VINIT}/\prog{VSTAR} mechanism may be used if desired.\cbend

\chapter{\texorpdfstring{Specifying input energies}
{\ref{Energy}: Specifying input energies}}\mylabel{Energy}

For each total angular momentum \var{JTOT} and symmetry block \var{IBLOCK}, the
programs loop over energies and external fields. The coupling matrices for
external fields must be implemented in plug-in basis-set suites and are
described in chapter \ref{inputfield} below. This chapter describes control of
the energy. For \MOLSCAT\ this is the energy for the scattering calculation.
For \BOUND\ or \FIELD\ it is the proposed energy for a bound state.

The energy may optionally be specified with respect to the energy of separated
monomers in specified states (a scattering threshold). We therefore include
here a description of how to interpret the indices used to identify scattering
channels in the output, to assist both in choosing the right index and in
identifying S-matrix elements and quantities obtained from them.

\section{Units of energy}\mylabel{Eunit}

The units of energy for quantities input in \namelist{\&INPUT} and for most
output energies are \emph{independent} of those used for quantities in
\namelist{\&BASIS}. They are specified by \inpitem{EUNITS}, which is an integer
that selects a unit of energy from the list in section \ref{outline:units}. The
default for \inpitem{EUNITS} is 1, indicating energies expressed as wavenumbers
in cm$^{-1}$.

If the energy unit required is not among those listed, \inpitem{EUNITS} may be
set to 0 and the required value (in units of cm$^{-1}$) supplied in
\inpitem{EUNIT}. If this is done, the name of the unit should be supplied in
the character variable \inpitem{EUNAME}.

\inpitem{EUNITS}, \inpitem{EUNIT} and \inpitem{EUNAME} are distinct from the
variables with the same names in namelist \namelist{\&BASIS} (section
\ref{basis:units}), but their allowed values and interpretation are the same.

The input quantities affected by \inpitem{EUNITS} or \inpitem{EUNIT} are
\inpitem{ENERGY}, \inpitem{DNRG}, \inpitem{EMIN} and \inpitem{EMAX},
\inpitem{EREF}, the (energy) convergence criteria \inpitem{DTOL} (for
\BOUND\ only), and \inpitem{DEGTOL}.

The input energies are converted immediately into cm$^{-1}$ using the
appropriate conversion factor. The programs often give energies in both
cm$^{-1}$ and in the units specified by \inpitem{EUNITS}. Any energies in the
output that are not given an explicit unit are in cm$^{-1}$.

\section[\texorpdfstring{Specifying energies for {\color{\mcol}\MOLSCAT} and
{\color{\fcol}\FIELD} calculations}{Specifying energies for MOLSCAT and FIELD calculations}]
{Specifying energies for \MOLSCAT\ and \FIELD\ calculations\sectionmark
{Specifying energies for \MOLSCAT\ and \FIELD\ calculations}}
\sectionmark{Specifying energies for \MOLSCAT\ and \FIELD\ calculations}\mylabel{EMF}

\cbcolor{\mfcol}\cbstart The total energies at which calculations are performed
are controlled by the array \inpitem{ENERGY} and the variables \inpitem{NNRG},
\inpitem{DNRG} and \inpitem{LOGNRG}. Calculations are performed for
\inpitem{NNRG} energies. If $\inpitem{DNRG}\ne0.0$, calculations are performed
at \inpitem{NNRG} equally spaced energies, \inpitem{DNRG} apart, starting at
\inpitem{ENERGY}(1).  If $\inpitem{DNRG}=0.0$ and $\inpitem{LOGNRG} =
\code{.FALSE.}$, the \inpitem{NNRG} energies must be supplied in the
array \inpitem{ENERGY}. If $\inpitem{DNRG}=0.0$ and $\inpitem{LOGNRG} =
\code{.TRUE.}$ the \inpitem{NNRG} energies are geometrically spaced between
\inpitem{ENERGY}(1) and \inpitem{ENERGY}(2).   The maximum allowed value of
\inpitem{NNRG} is set in the variable \var{MXNRG}, which is 2000 in \MOLSCAT\
and 100 in \FIELD. \cbend

\section[\texorpdfstring{Internally generated energies in {\color{\mcol}\MOLSCAT}}
{Internally generated energies in MOLSCAT}]{Internally generated energies in \MOLSCAT\sectionmark
{Internally generated energies in \MOLSCAT}}
\sectionmark{Additional energy specifications for \MOLSCAT}
\mylabel{scatE}

\cbcolor{\mcol}\cbstart There are special uses of the \inpitem{NNRG} and
\inpitem{ENERGY} that cause \MOLSCAT\ to generate energy lists internally for:
\begin{enumerate}[nosep]
\item{Searching for energy-dependent resonances in the S-matrix eigenphase sum;}
\item{Calculating line-broadening cross sections.}
\end{enumerate}
These are described separately in sections \ref{energyconv} and \ref{pressbroad}
respectively.\cbend

\section[\texorpdfstring{Temperatures for thermal averaging ({\color{\mcol}\MOLSCAT} only)}
{Temperatures for thermal averaging (MOLSCAT only)}]{Temperatures for thermal averaging
\sectionmark{Temperatures for thermal averaging}}
\sectionmark{Temperatures for thermal averaging}\mylabel{thermal}

\cbcolor{\mcol}\cbstart An alternative form of input is available to facilitate
thermal averaging of cross sections using Gaussian quadrature. This is
controlled by the array \inpitem{TEMP} and the variables \inpitem{NTEMP} and
\inpitem{NGAUSS}. If $\inpitem{NTEMP}>0$, \MOLSCAT\ calculates appropriate
energies and weighting factors which correspond to \inpitem{NGAUSS}-point
Gaussian quadratures for each of the \inpitem{NTEMP} different temperatures (in
Kelvin) in the array \inpitem{TEMP}.  Scattering calculations are then
performed at each of the $\inpitem{NGAUSS}\times\inpitem{NTEMP}$ energies. The
maximum allowed values are $\inpitem{NTEMP} = 5$ and $\inpitem{NGAUSS} = 6$.

The thermal averaging itself must be done outside \MOLSCAT. Note that Gaussian
quadrature is not a reliable way of thermally averaging some types of cross
sections, particularly if there are resonances present, and use of this option
is generally not recommended for precise work.\cbend

\section[\texorpdfstring{Specifying energies for {\color{\bcol}\BOUND} calculations}
{Specifying energies for BOUND calculations}]{Specifying energies for \BOUND\ calculations
\sectionmark{Specifying energies for \BOUND\ calculations}}
\sectionmark{Specifying energies for \BOUND\ calculations}\mylabel{specener}

\cbcolor{\bcol}\cbstart The energies at which calculations are performed in
\BOUND\ are governed by the namelist items \inpitem{EMAX} and \inpitem{EMIN}.
\inpitem{EMAX} must be greater than \inpitem{EMIN}. \BOUND\ calculates the node
counts at \inpitem{EMAX} and \inpitem{EMIN}. The difference between them is the
number of states in the energy interval. By default, \BOUND\ attempts to locate
all these states. However, if \inpitem{NODMAX} and/or \inpitem{NODMIN} are also
set non-zero, only states between \inpitem{EMIN} and \inpitem{EMAX} with node
counts in the range \inpitem{NODMIN} to \inpitem{NODMAX} are located.

It is permissible for \inpitem{EMAX} to refer to an energy above the lowest threshold, where states are no longer bound by the interaction potential. Above this energy, quantisation is produced by the boundary conditions at \inpitem{RMAX}, or by an external confining potential as described in section \ref{confinement}.  If calculations on these states are not desired, \inpitem{EMAXBD} may be set to \code{.TRUE.} to reduce the maximum energy for the search to the energy of the lowest threshold for the current basis set and combination of EFVs.
\cbend

\section{The reference energy}\mylabel{EREF}

By default, the zero of energy used for total energies is the one used for monomer energies, as
described in chapter \ref{ConstructBasis}, or defined by the monomer Hamiltonians programmed in a
plug-in basis-set suite. However, it is sometimes desirable to use a different zero of energy
(reference energy), such as the energy of a particular scattering threshold (which may depend on
external fields). This energy can be specified in several different ways:
\begin{itemize}
\item{If $\inpitem{EREF}\ne 0$, it is used as the reference energy for all
    scattering and bound-state energies.}
\item{If $H_{\rm intl}$ is diagonal (which includes all the built-in coupling cases and plug-in
    basis-set suites with $\var{NCONST}=0$):
    \begin{itemize}[nosep]
       \item{If $\inpitem{MONQN}(1)=-99999$ (the default) and $\inpitem{IREF}>0$, the
           programs use the energy of the pair level with index \inpitem{IREF}.}
       \item{If $\inpitem{MONQN}(1)\ne-99999$, the values supplied in the array
           \inpitem{MONQN} are quantum labels for the reference threshold, which must match
           those in a row of the array \var{JLEVEL}.}
    \end{itemize}}
\item{If $H_{\rm intl}$ is non-diagonal (which includes plug-in basis-set suites with
    $\var{NCONST}>0$):
    \begin{itemize}[nosep]
       \item{If $\inpitem{MONQN}(1)=-99999$ (the default) and $\inpitem{IREF}>0$, the
           programs use the threshold energy of the channel with index \inpitem{IREF}. The
           user must identify the index of the required channel before the full calculation
           (which often requires a pilot calculation; see section \ref{threshchoice}
           below).}
       \item{If $\inpitem{MONQN}(1)\ne-99999$, the code requires a routine \prog{THRSH9} as
           part of a plug-in basis-set suite to calculate the reference energy from the
           values in the array \inpitem{MONQN}. See section \ref{base9:asympthresh} for
           further details.}
    \end{itemize}}
\end{itemize}

\subsection{Specifying the index of the reference threshold}\mylabel{threshchoice}

For the built-in interaction types ($\var{ITYP} \ne 9$), or for plug-in
basis-set suites that use a basis set in which $H_{\rm intl}$ is diagonal, it
is almost always easiest to specify the reference threshold via the array
\inpitem{MONQN}.

Some older plug-in basis-set suites for basis sets in which $H_{\rm intl}$ is
non-diagonal do not implement the routine \prog{THRSH9}, so it is not possible
to specify a reference energy via the array \inpitem{MONQN}. In this case it
may be desired to specify the reference energy via a threshold index
\inpitem{IREF}.

It is usually necessary to carry out a pilot calculation to identify the index
required. If $H_{\rm intl}$ is non-diagonal, the programs print a list of
threshold energies calculated from the internal Hamiltonian if
$\inpitem{IPRINT}\ge10$. Any threshold index with the required threshold energy
may be specified; the value of $L$ is immaterial.

The \inpitem{IREF} mechanism is intended for calculations where there is no
loop over \var{JTOT} or \var{IBLOCK}, and does \emph{not} work in cases where
the threshold required does not exist for the first \var{JTOT}/\var{IBLOCK}
combination. In complicated cases it may be quite involved to identify the
index required, and under these circumstances it is often better to use (or
implement) a \prog{THRSH9} routine in the basis-set suite, so that the
reference energy can be specified via \inpitem{MONQN} instead.

\chapter{\texorpdfstring{External fields and scaling the interaction potential}
{\ref{inputfield}: External fields and scaling the interaction potential}}\mylabel{inputfield}

\MOLSCAT\ and \BOUND\ were extended in 2007 to incorporate the effects of
external magnetic and/or electric fields \cite{Gonzalez-Martinez:2007}. Version
2019.0 introduced a more general structure that allows multiple external fields
(which may be static, such as electric or magnetic fields, or oscillatory, such
as photon fields). None of the built-in interaction types include any external
fields, so they must be implemented within a plug-in basis-set suite for
$\var{ITYP}=9$. The current release includes two examples of plug-in basis-set
suites that include external magnetic fields. If the user wishes to take
advantage of this feature, they should also refer to the specification of the
module \module{efvs} in section \ref{module:efvs} in order to work out exactly
what they need to program.

Each calculation is carried out with external fields specified by \var{NEFV}
real numbers \var{EFV}, referred to here as external field variables (EFVs).
These can include field strengths, frequencies, relative angles, or other
variables. The programmer of the plug-in basis-set suite specifies \var{NEFV}
and may use the elements of \var{EFV} in any way desired. The programs normally
consider all but one of the EFVs as fixed, but allow one of them (identified by
the index \inpitem{IFVARY}) to be varied. In the simplest case the variation is
specified as a grid, but \MOLSCAT\ and \FIELD\ can locate resonances or bound
states, respectively, as a function of the varying EFV.

The interaction potentials that are used for coupled-channel calculations are
seldom known exactly. Exploring the sensitivity of calculated properties to
parameters of the interaction potentials is a complicated task, but some
estimate may be obtained by scaling the potential. Ultracold scattering
properties often show extreme sensitivity to such variations. The programs
allow such a scaling to be incorporated by treating it as an artificial EFV;
\MOLSCAT\ can converge on resonances as a function of potential scaling, and
\FIELD\ can find bound states as a function of potential scaling. By default
the entire potential is scaled by the same factor, but the subroutine that
performs the scaling (\prog{SCAPOT}) can be replaced with a bespoke version
that can apply a different scaling to each term in the potential expansion.

\section{Using external fields}\mylabel{ExtVar}

The items in namelist \namelist{\&INPUT} that are used to control the values
and types of external field variables included, are:
\begin{description}
\item[\inpitem{IFVARY}]{(deprecated synonym \inpitem{MAGEL}) specifies the
    index of the EFV to be varied in the array \var{EFV}. If it is 0, the
    potential scaling factor is varied.}
\item[\inpitem{NFVARY}]{specifies the number of varying EFVs included in
    the array \inpitem{FIELD}.\hfil\break \inpitem{NFVARY} is usually 0
    or 1. $\inpitem{NFVARY}>1$ is implemented \emph{only} for varying EFVs
    supplied explicitly as a list of values in the array \inpitem{FIELD}.
    In this case \inpitem{IFVARY} is a list of indices of the variable
    EFVs.}
\item[\inpitem{FIXFLD}]{is an array of dimension \var{NEFV} (limited by
    $\var{MXEFV}=10$, set in module \module{efvs}). It specifies values of
    all the EFVs in the array \var{EFV}; the value for element
    \inpitem{IFVARY} is ignored, so \inpitem{FIXFLD} is not needed if
    \var{NEFV} is 1.}
\item[\inpitem{FLDMIN}]{is the lower end of the range of the (single)
    varying EFV.}
\item[\inpitem{FLDMAX}]{is the upper end of the range of the (single)
    varying EFV.}
\item[\inpitem{DFIELD}]{is the step size between values of the (single)
    varying EFV. \cbcolor{\mcol}\cbstart In \MOLSCAT, \inpitem{DFIELD} is
    ignored if a resonance is to be characterised ($\inpitem{IFCONV}>0$,
    section \ref{fieldconv}) or a specific value of the scattering length
    is to be located (section \ref{fieldval}). \cbend \cbcolor{\fcol}\cbstart
    In \FIELD, \inpitem{DFIELD} (default $10^{30}$) must be greater than
    $\inpitem{FLDMAX}-\inpitem{FLDMIN}$ if bound states are to be
    located.\cbend}
\item[\inpitem{DTOL}]{\cbcolor{\mfcol}\cbstart is the convergence criterion
    used when converging on quantities as a function of the (single)
    varying EFV: on resonances or values of the scattering length/volume in
    \MOLSCAT; or on the position of bound states in \FIELD. \cbend
    (\cbcolor{\bcol}\cbstart In \BOUND, \inpitem{DTOL} is the convergence
    criterion for the energy of the bound state).\cbend}
\item[\inpitem{NFIELD}]{If both \inpitem{FLDMIN} and \inpitem{FLDMAX} are
    0.0 (the default), \inpitem{NFIELD} sets of values of \var{EFV} must be
    supplied in the array \inpitem{FIELD} (but only for the EFVs whose
    indices were given in \inpitem{IFVARY}).}
\item[\inpitem{FIELD}]{is an array of dimension \var{MXFLD} ($=10000$)
    containing values of \var{EFV} for the varying EFVs.}
\item[\inpitem{IFIELD}]{(obsolete, but retained for backwards
    compatibility) specifies the index of the first EFV in the \var{VCONST}
    array in certain plug-in basis-set suites.  This quantity is now
    unnecessary and any value in \namelist{\&INPUT} namelist is ignored.}
\end{description}

\section{Potential scaling}\mylabel{potscale}

The programs implement a potential scaling factor (\inpitem{SCALAM}) that
scales the whole interaction potential.  The scaling factor is handled in the
same way as an EFV.

If the scaling factor is to be held fixed while an EFV is varied, it must be
specified in \inpitem{SCALAM}.

Alternatively, setting $\inpitem{IFVARY}=0$ instructs the programs to treat the
scaling factor as the varying quantity. The scaling factor is handled in the
same way as a varying EFV, so that:
\begin{description}
\item[\inpitem{FLDMIN}]{is the lower bound value of the scaling factor.}
\item[\inpitem{FLDMAX}]{is the upper bound of the scaling factor.}
\item[\inpitem{DFIELD}]{is the step size between values of the scaling
    factor. If a resonance is to be characterised as a function of
    \inpitem{SCALAM} (section \ref{inputres}), \inpitem{DFIELD} is not
    used.}
\item[\inpitem{DTOL}]{is the convergence criterion used when converging on
    quantities as a function of the scaling factor: on resonances or values
    of the scattering length/volume in \MOLSCAT; or on the position of
    bound states in \FIELD.}
\end{description}

Note that, if $\inpitem{NFVARY}>1$, including 0 among the indices of varying
EFVs in \inpitem{IFVARY} means that the grid of values held in the
array \inpitem{FIELD} include the scaling factor.  If, however, the scaling
factor is to be held constant at a value other than 1, it must be set in
\inpitem{SCALAM}, not in the array \inpitem{FIXFLD}.

\chapter{\texorpdfstring{Controlling the propagators}
{\ref{propcontrol}: Controlling the propagators}}\mylabel{propcontrol}

\section{Propagator choice}\mylabel{propchoice}

Versions of the programs before 2019.0 implemented a variety of propagators to
solve the coupled equations, and allowed some specific combinations of
them as ``hybrid" propagators that combine a propagator suitable at short range
with a different one suitable at long range. From version 2019.0, this
mechanism was generalised to allow \emph{any} sensible combination of a
short-range propagator with a long-range propagator.

The methods used to propagate solutions to the coupled-channel equations are
controlled by \inpitem{IPROPS} to specify the short-range propagator and
\inpitem{IPROPL} to specify the long-range propagator. If only \inpitem{IPROPS}
is specified, \inpitem{IPROPL} is set the same as \inpitem{IPROPS}. If neither
is specified, the default combinations are
$\inpitem{IPROPS}=\inpitem{IPROPL}=6$ for \BOUND\ and \FIELD, and
$\inpitem{IPROPS}=6$ and $\inpitem{IPROPL}=9$ for \MOLSCAT.

The propagator codes are:
\begin{itemize}[nosep,twocol]
\item[{\bf 2}]{de Vogelaere (DV)}
\item[{\bf 3}]{R-matrix (RMAT)}
\item[{\bf 4}]{Variable-interval variable-step (VIVS)}
\item[{\bf 5}]{Johnson log-derivative (LDJ)}
\item[{\bf 6}]{Manolopoulos diabatic modified log-derivative (LDMD)}
\item[{\bf 7}]{Manolopoulos quasiadiabatic modified log-derivative (LDMA)}
\item[{\bf 8}]{Manolopoulos-Gray symplectic log-derivative (LDMG)}
\item[{\bf 9}]{Alexander-Manolopoulos Airy (AIRY)}
\item[$\boldsymbol{-1}$]{WKB phase integrals by quadrature}
\end{itemize}

If $\inpitem{IPROPL} = 0$, it is set the same as \inpitem{IPROPS}.

For backwards compatibility, \inpitem{INTFLG} is a deprecated synonym for
\inpitem{IPROPS}, except that there are two values of \inpitem{INTFLG} that
have special meanings:
\begin{description}[nosep]
\item[$\inpitem{INTFLG}=4$]{sets $\inpitem{IPROPS}=5$ and
    $\inpitem{IPROPL}=4$, corresponding to the VIVAS hybrid propagator of
    Parker \etal\ \cite{Parker:LOGD-VIVS}.}
\item[$\inpitem{INTFLG}=8$]{sets $\inpitem{IPROPS}=6$ and
    $\inpitem{IPROPL}=9$, corresponding to the hybrid LDMD/AIRY propagator
    of Alexander and Manolopoulos \cite{Alexander:1984, Alexander:1987}.}
\end{description}
Further information on the individual propagators is given in section
\ref{propdetail} below.

\section{Units of length}\mylabel{lengthunit}

The programs operate in units of lengths specified by \inpitem{RUNIT} in
namelist \namelist{\&INPUT}. If \inpitem{RUNIT} is unset, it taken from the
value returned by the initialisation call to \prog{POTENL}). In the
general-purpose version of \prog{POTENL}, the value may be input as
\potlitem{RM}, though this may be overwritten by code in user-supplied
\prog{VINIT} or \prog{VRTP} routines. If neither \inpitem{RUNIT} nor
\potlitem{RM} is set, the programs operate in length units of \AA.

Input length variables that control the propagation (\inpitem{RMIN}, \inpitem{RMAX},
\inpitem{RMID}, \inpitem{RMATCH}, \inpitem{DRS}, \inpitem{DRL}) are in units of \inpitem{RUNIT}.
\inpitem{RUNIT} itself is specified in units of \AA. For example $\inpitem{RUNIT}=1.0$ (the
default) indicates that all distances are in \AA, while $\inpitem{RUNIT}=0.529177210903$ (2018
value) indicates that they are in units of the bohr radius (atomic units). Most output quantities
with dimensions of length (including scattering lengths, but not cross sections) are output in
units of \inpitem{RUNIT}.

\section{Units of reduced mass}\mylabel{massunit}

The units of the reduced mass \inpitem{URED} are specified by \inpitem{MUNIT}.
\inpitem{MUNIT} itself is specified in units of unified atomic mass units
$m_{\rm u}$ (Daltons). For example $\inpitem{MUNIT}=1.0$ (the default)
indicates that \inpitem{URED} is in Daltons, while
$\inpitem{MUNIT}=5.48579909065$D-4 (2018 value) indicates that \inpitem{URED} is
in units of the electron mass $m_e$ (atomic units).

\section{Internal units of energy }\mylabel{intunit}

Quantities with dimensions of energy (total energy $E$, interaction matrix
${\bf W}(R)$, etc.)\ are processed internally as reduced energies $2\mu
E/\hbar^2$, with dimensions of [length]$^{-2}$. This reduces the coupled
equations to the form (\ref{eq:se-invlen}).

Energies expressed as wavenumbers in cm$^{-1}$ may be converted into reduced
energies in \AA$^{-2}$ by multiplying by
$\inpitem{MUNIT}\times\inpitem{URED}/\var{BFCT}$, where \var{BFCT} = [$\hbar$/(J s)] /
$(4 \pi [c/$(m/s)]  [$m_{\rm u}$/kg]) [m/cm] [m/\AA]$^2 = 16.85762919164$ (2018
value).

The programs use two conversion factors: dividing by $\var{CM2RU} =
(\inpitem{RUNIT})^2\times\inpitem{MUNIT}\times\inpitem{URED}/\var{BFCT}$ converts reduced energies
in units of $(\inpitem{RUNIT})^{-2}$ into wavenumbers in cm$^{-1}$, while dividing by $\var{EP2RU}
= \var{CM2RU} \times \potlitem{EPSIL}$ converts reduced energies into the units \potlitem{EPSIL}
used for the interaction potential.

\section{Ranges of propagation}\mylabel{intrange}

\cbcolor{\mcol}\cbstart
\MOLSCAT\ propagates the coupled equations outwards from $R_{\rm min}$ to
$R_{\rm max}$, with the option to switch propagation method at $R_{\rm mid}$.\cbend

\cbcolor{\bfcol}\cbstart
\BOUND\ and \FIELD\ propagate the coupled equations outwards from $R_{\rm min}$
to $R_{\rm match}$ and inwards from $R_{\rm max}$ to $R_{\rm match}$. The
propagation method may be switched at $R_{\rm mid}$, which {\em may} be the
same as $R_{\rm match}$ but can be different if desired.\cbend

$R_{\rm min}$, $R_{\rm max}$ and $R_{\rm mid}$ are \emph{based on} the input
variables \inpitem{RMIN}, \inpitem{RMAX}, and \inpitem{RMID}, but are not
necessarily {\em equal to} them.

\subsection{\texorpdfstring{Inner limit $R_{\rm min}$}{Inner limit Rmin}}

If the origin is energetically accessible at the energy of the
calculation, $R_{\rm min}$ should be zero. If there is an infinite hard wall at
short range, it should be placed at the hard wall. Otherwise, it should be far
enough into the classically forbidden region at short range that that the
wavefunction at $R<R_{\rm min}$ does not contribute significantly to the
calculated quantities. If a WKB boundary condition is used at short range, as
described in section \ref{Y-bc}, it is sufficient to place $R_{\rm min}$
slightly further out, but still well inside the inner turning point.

If $\inpitem{IRMSET}=0$ (default 9), $R_{\rm min}$ is set to \inpitem{RMIN}.

If $\inpitem{IRMSET} > 0$, the programs obtain $R_{\rm min}$ from a
semiclassical estimate of a distance such that the wavefunction amplitude in
all channels is less than $10^{-\inpitem{IRMSET}}$ at $R_{\rm min}$. This
estimate is calculated separately for each \var{JTOT}, \var{IBLOCK} at the
highest energy value in \inpitem{ENERGY} (for \MOLSCAT\ and \FIELD) or at
\inpitem{EMAX} (for \BOUND).

There are some cases where $R$ is not a radial coordinate, such as
scattering from a solid surface ($\basisitem{ITYPE}=8$). To accommodate these,
$R_{\rm min}$ is allowed to be negative. However, there are certain program
features that are clearly inappropriate and should not be used when $R$ can
pass through zero, such as step sizes proportional to a non-zero power of $R$
(section \ref{dr-power}).

\subsection{\texorpdfstring{Outer limit $R_{\rm max}$}{Outer limit Rmax}}

\cbcolor{\bfcol}\cbstart

\BOUND\ and \FIELD\ are designed to operate with a classically
forbidden region at long range. Under these circumstances, $R_{\rm max}$ should
be far enough into the classically forbidden region that the wavefunction at
$R>R_{\rm max}$ does not contribute significantly to the calculated quantities.
If a WKB boundary condition is used at long range, as described in section
\ref{Y-bc}, it is sufficient to place $R_{\rm max}$ further in, but still well
outside the outer turning point.\footnote{For calculations with \BOUND\ and
\FIELD\ that are designed to locate quasibound states above the lowest
threshold, $R_{\rm max}$ should be placed well into the classically forbidden
for the channels that support the quasibound states of interest. Under these
circumstances there are additional artificial levels that arise from
quantisation of the open-channel continua by the boundary conditions at $R_{\rm
max}$; the energies of these artificial levels depend on $R_{\rm max}$, and
they perturb the quasibound states unphysically when they come close to them.}
This may require very large values of $R_{\rm max}$ for near-threshold states.

\BOUND\ and \FIELD\ always set $R_{\rm max}$ to \inpitem{RMAX}.\cbend

\cbcolor{\mcol}\cbstart For \MOLSCAT, $R_{\rm max}$ should be
large enough that both
\begin{enumerate}[nosep]
\item the interaction potential makes no significant contribution to the
    wavefunction in the open channels outside $R_{\rm max}$, so that the
    open-channel wavefunctions are well represented by the Riccati-Bessel
    boundary conditions (\ref{K-bc});
\item $R_{\rm max}$ is well outside the outer turning point in any closed
    channels that support scattering resonances of interest, so that the
    wavefunction in those channels is well represented by the
    closed-channel boundary condition (which may be a WKB boundary
    condition as described in section \ref{Y-bc}).
\end{enumerate}
Ultracold scattering calculations may require values of $R_{\rm max}$
comparable to the scattering length, which may be very large, particularly near
a Feshbach resonance.

An additional consideration is that open-channel matching does not work well if
$R_{\rm max}$ is so far inside a centrifugal barrier in an open channel that
the wavefunction has decayed far below its asymptotic amplitude. This is not
usually an issue in ultracold scattering, but can be important in cross-section
calculations that require high $L$ for convergence.

If $\inpitem{IRXSET}=0$ (the default), \MOLSCAT\ also sets $R_{\rm max}$ to
\inpitem{RMAX}.

If $\inpitem{IRXSET}=1$, \MOLSCAT\ calculates the turning point, at every
energy $E_j$ in the \inpitem{ENERGY} list, for the pure centrifugal potential
of every open channel $i$ with $L_i>0$. This is \begin{equation} R_{\rm
cent}^{ij}=\left[\frac{\hbar^2 L_i(L_i+1)}{2\mu(E_j-E_{{\rm
intl},i})}\right]^\frac{1}{2}. \end{equation} The value used for $R_{\rm max}$
is the largest of all these values and \inpitem{RMAX}.

This approach generally works well for cross-section calculations at energies
well above thresholds, but it is \emph{not} always adequate and convergence
tests should always be carried out. However, it can be drastically inefficient
(and is unnecessary) for ultracold scattering including channels with $L>0$, so
the default value of \inpitem{IRXSET} was changed from 1 to 0 from version
2019.0. \cbend

\subsection{\texorpdfstring{Propagator switch point $R_{\rm mid}$}{Propagator switch point Rmid}}

The LDJ, LDMD, LDMA and LDMG propagators have good step-size
convergence in regions where the interaction potential is strong, but cannot
take very long steps even when it is weak. The RMAT, VIVS and AIRY
propagators have poorer step-size convergence when the interaction potential is
strong, but can take much longer steps when it is weak. If different
propagators or step-size algorithms are used at short and long range, $R_{\rm
mid}$ should be chosen for optimum efficiency. It should usually be placed well
outside the potential minimum, and it is often effective to place it between
75\% and 99\% of the way up the attractive limb of the potential.

\cbcolor{\bfcol}\cbstart For \BOUND\ and \FIELD, $R_{\rm mid}$ is set to
\inpitem{RMID}. If no value for \inpitem{RMID} is provided, $R_{\rm mid}$ is
set to \inpitem{RMATCH}.\cbend

\cbcolor{\mcol}\cbstart For \MOLSCAT, if $\inpitem{RVFAC}=0.0$ (the default),
$R_{\rm mid}$ is set to \inpitem{RMID}.

If $\inpitem{RVFAC}>0.0$, $R_{\rm mid}$ is set to $\inpitem{RVFAC}\times R_{\rm
turn}$, where $R_{\rm turn}$ is an estimate of the position of the classical
turning point in the lowest channel, calculated for each \var{JTOT} and
symmetry block at the highest energy in \inpitem{ENERGY}. Values of
$\inpitem{RVFAC}$ from 1.3 to 2.0 are often satisfactory for cross-section
calculations. If $\inpitem{IRMSET}=0$ and $\inpitem{RVFAC}>0$, $R_{\rm mid}$ is
set to $\inpitem{RVFAC}\times R_{\rm min}$.

\inpitem{RVIVAS} is a deprecated synonym for \inpitem{RMID}.\cbend

If $R_{\rm mid} < R_{\rm min}$, \inpitem{IPROPS} is not used. If $R_{\rm mid} >
R_{\rm max}$, \inpitem{IPROPL} is not used.

\subsection{\texorpdfstring{Bound-state matching point $R_{\rm
match}$}{Bound-state matching point Rmatch}}\mylabel{sec:match}

\cbcolor{\bfcol}\cbstart The value of $R_{\rm match}$ does not affect the energies or fields at
which the matching condition (\ref{eq:ymatch}) is satisfied, so it does not affect converged
bound-state energies or fields. However, it does affect the log-derivative matching matrix at other
energies or fields, so it can affect the rate of convergence on states (and sometimes the success
of convergence).

A value of $R_{\rm match}$ somewhat inside the outer turning point (near the maximum in the
outermost lobe of the wavefunction) is usually optimal for near-threshold bound states. For deeply
bound states a value slightly outside the inner classical turning point usually gives rapid
convergence.

It is usually inappropriate to place $R_{\rm match}$ far into a classically forbidden region, or at
a distance where the log-derivative matrix has very large eigenvalues (such as very close to a hard
wall in the interaction potential).

$R_{\rm match}$ is set to \inpitem{RMATCH}. If no value for \inpitem{RMATCH} is
provided, $R_{\rm match}$ is set to \inpitem{RMID}. At least one of
\inpitem{RMID} and \inpitem{RMATCH} must be provided. \cbend

\section{Step size}\mylabel{stepsize}

The programs offer three different approaches for choosing the propagation step
size (length of propagation step). These are
\begin{itemize}[nosep]
\item equally spaced steps;
\item step size proportional to a power of $R$;
\item adaptive step size based on error estimates.
\end{itemize}
Not all propagators implement all these approaches.

The step size(s) for the short-range propagator (\inpitem{IPROPS}) are
controlled by variables \inpitem{DRS} or \inpitem{STEPS} and \inpitem{EPS}.
Variable-step propagators use the additional variables \inpitem{TOLHIS} and
\inpitem{POWRS}. The step size(s) for the long-range propagator
(\inpitem{IPROPL}) are controlled by corresponding variables \inpitem{DRL},
\inpitem{STEPL}, \inpitem{EPL}, \inpitem{TOLHIL} and \inpitem{POWRL}.

\inpitem{DRL} defaults to \inpitem{DRS}.

\inpitem{STEPL} defaults to \inpitem{STEPS}.

\inpitem{EPS} and \inpitem{EPL} each default to 0.0, and setting one has no
effect on the other.

\inpitem{TOLHIL} defaults to \inpitem{TOLHIS}.

\inpitem{POWRL} defaults to $1.33\dot{3}$ if $\inpitem{TOLHIL}=0$ or \inpitem{POWRX} which itself defaults to 3.0 if
$\inpitem{TOLHIL}>0$. \inpitem{POWRS} defaults to $0.0$ (equally spaced steps)
except for the AIRY propagator ($\inpitem{IPROPS}=9$) with
$\inpitem{TOLHIS}>0$, when it defaults to  3.0.

\inpitem{DR} and \inpitem{TOLHI} are deprecated synonyms for \inpitem{DRS} and
\inpitem{TOLHIS}. \inpitem{POWRX} is a deprecated synonym for \inpitem{POWRL}.

In the remainder of Chapter \ref{propcontrol}, these variables are referred to
without suffices as \var{DR}, \var{STEP}, \var{EP}, \var{TOLHI} and \var{POWR}.

The VIVS propagator has additional control variables that are not distinguished
by suffices \code{S} and \code{L}, as described in section \ref{iprop4} below.

\subsection{Equally spaced steps}\mylabel{dr-fixed}

All propagators except VIVS offer the option of equally spaced steps. For
propagators where equally spaced steps are not the only option, they are
selected by setting $\var{POWR}=0.0$.

\subsection[\texorpdfstring{Step size proportional to $R^\var{POWR}$}
{Step size proportional to R**\var{POWR}}] {Step size proportional to
$R^\var{POWR}$}\mylabel{dr-power}

The RMAT and AIRY propagators allow a step size proportional to
$R^{\var{POWR}}$. $\var{POWR}=0.0$ generates equally spaced steps. For the
RMAT propagator, this mechanism is always used. For the AIRY propagator, it
is used only if $\var{TOLHI}=0.0$.

For $\var{POWR}\ne 1.0$, this option is implemented by choosing steps whose
boundaries are equally spaced in the transformed variable $R^{1-\var{POWR}}$.
For the special case $\var{POWR}=1.0$, it is implemented with the size of each
step proportional to $R$ at the \emph{inner} end of the step.

Most systems of interest have interaction matrices that include centrifugal
terms that are asymptotically diagonal in the adiabatic representation and
decay as $R^{-2}$ at long range, with first and second derivatives that decay
as $W^{(1)}\propto R^{-3}$ and $W^{(2)}\propto R^{-4}$. For the RMAT
propagator, the error in a single step is proportional to $W^{(1)} \delta R^3$
so the step size at long range should be proportional to $R$, which is achieved
with $\var{POWR}=1.0$.  For the AIRY propagator the error is proportional to
$W^{(2)} \delta R^3$, so $\var{POWR}=1.33\dot{3}$ is appropriate at long range.

Different values of \var{POWR} may be appropriate if a different inverse power
is dominant. For the AIRY propagator, the error is proportional to $W^{(2)}
\delta R^3$ for terms that are diagonal in the quasiadiabatic representation,
but $W^{(1)} \delta R^3$ for terms that are off-diagonal. If the interaction
potential decays as $R^{-n}$, it may be appropriate to use a step size
proportional to  $R^{(n+2)/3}$ if the $R^{-n}$ potential terms are purely
diagonal, or $R^{(n+1)/3}$ if they are off-diagonal. For example, for s-wave
collisions between two atoms in S states, with diagonal potential terms that
decay as $R^{-6}$, $\var{POWR}=8/3=2.66\dot{6}$ would be expected to offer
optimum efficiency for the AIRY propagator at long range.

\subsection{Adaptive step size}

The VIVS and AIRY propagators can use adaptive step-size algorithms, with the
size of each step based on an estimate of the errors in the previous step. The
algorithms used are described in sections \ref{iprop4} and \ref{iprop9}.

\subsection{Initial step size}

\subsubsection{Setting initial step size with \inpitem{DRS} or \inpitem{DRL}}

The default behaviour is to take the initial step size $\delta R$ from
\var{DR}. The input step size is modified slightly if necessary to give an
integer number of steps over the propagation range.

If the step size is proportional to a power of $R$, \var{DR} is interpreted as
the step size at the \emph{inner} end of the range, even for inwards
propagations. This allows the same value of \var{DR} to be used in \BOUND,
\FIELD\ and \MOLSCAT.

\subsubsection{Setting initial step size with \inpitem{STEPS} and \inpitem{EPS}
or \inpitem{STEPL} and \inpitem{EPL}}

If $\var{STEP}>0.0$, the step size is calculated from \var{STEP} and \var{EP}.
This is interpreted as the number of steps per half-wavelength for the channel
with the highest asymptotic kinetic energy $E_{\rm kin}=E-E_{{\rm intl},i}$.
The step size is calculated from $\pi/(k\times \var{STEP})$, where
$k^2=2\mu(E_{\rm kin}+\var{EP})/\hbar^2$. Thus \var{EP} may be input to
estimate the depth of the interaction potential in the relevant region, to take
account of the fact that the wavefunction varies faster over a potential well
than asymptotically. \var{EP} is usually needed only if using the \var{STEP}
mechanism at asymptotic kinetic energies smaller than the potential well depth.
A value of \var{STEP} between 10 and 20 is usually adequate.

If $\inpitem{ISCRU}>0$, $\delta R$ is calculated at the highest energy in
\inpitem{ENERGY} and this value is used for the propagations at all energies.

The value of $\delta R$ obtained from \var{STEP} and \var{EP} is modified
slightly if necessary to give an integer number of steps over the propagation
range.

If the step size is proportional to a power or $R$, the step size obtained from
\var{STEP} and \var{EP} is used at the \emph{inner} end of the propagation
range, even for inwards propagations. This allows the same values of \var{STEP}
and \var{EP} to be used in \BOUND, \FIELD\ and \MOLSCAT.

\subsection[\texorpdfstring{3-segment propagation in {\color{\bcol}\BOUND} and {\color{\fcol}\FIELD}}
{3-segment propagation in BOUND and FIELD}] {3-segment propagation in
\BOUND\ and \FIELD}\mylabel{3partprop}

\cbcolor{\bfcol}\cbstart
The wavefunction matching distance \inpitem{RMATCH}
may be different from the propagator switching point \inpitem{RMID}, as
described in section \ref{sec:match}. If \inpitem{RMID} and \inpitem{RMATCH}
are different, the programs perform a 3-segment propagation: outwards from
\inpitem{RMIN} to the smaller of \inpitem{RMID} and \inpitem{RMATCH}; inwards
from \inpitem{RMAX} to the larger of \inpitem{RMID} and \inpitem{RMATCH}; and
from \inpitem{RMID} to \inpitem{RMATCH} (in whichever direction is
needed).

If the propagation has three segments, there are two possibilities. The
directions of propagation and the initial step sizes used are:
\begin{itemize}
\item[1: ]{$R_{\rm mid}<R_{\rm match}$}
\end{itemize}
\begin{tikzpicture}
\node[above] at (2,9) {short range}; \node[above] at (8,9) {long range}; \node
(rmin) at (0,10) {$R_{\rm min}$}; \node (rmax) at (12,10) {$R_{\rm max}$};
\node (rmid) at (4,10) {$R_{\rm mid}$}; \node (rmatch) at (8,10) {$R_{\rm
match}$}; \node[left, text width=1in] at (0,9) {propagator ranges}; \draw
[ultra thick, |-|,shorten >=0.3em] (0,9)--(4,9); \draw [ultra thick,
|-|,shorten >=0.3em] (12,9)--(4,9); \node[above] at (2,8) {\tt DRS};
\node[above] at (6,8) {$\delta R$ at $R_{\rm mid}$}; \node[above] at (10,8) {\tt DRL}; \node[left,
text width=1in] at (0,8) {\color{\bfcol} direction of propagation}; \draw
[ultra thick, ->, >=latex, color=\bfcol] (0,8)--(4,8); \draw [ultra thick, ->,
>=latex, color=\bfcol] (4,8)--(8,8); \draw [ultra thick, ->, >=latex,
color=\bfcol] (12,8)--(8,8);
\end{tikzpicture}
\begin{itemize}
\item[2: ]{$R_{\rm match}<R_{\rm mid}$}
\end{itemize}
\begin{tikzpicture}
\node[above] at (4,9) {short range}; \node[above] at (10,9) {long range}; \node
(rmin) at (0,10) {$R_{\rm min}$}; \node (rmax) at (12,10) {$R_{\rm max}$};
\node (rmid) at (8,10) {$R_{\rm mid}$}; \node (rmatch) at (4,10) {$R_{\rm
match}$}; \node[left, text width=1in] at (0,9) {propagator ranges}; \draw
[ultra thick, |-|,shorten >=0.3em] (0,9)--(8,9); \draw [ultra thick,
|-|,shorten >=0.3em] (12,9)--(8,9); \node[above] at (2,8) {\tt DRS};
\node[above] at (6,8) {\tt DRS}; \node[above] at (10,8) {\tt DRL}; \node[left,
text width=1in] at (0,8) {\color{\bfcol} direction of propagation}; \draw
[ultra thick, ->, >=latex, color=\bfcol] (0,8)--(4,8); \draw [ultra thick, <-,
>=latex, color=\bfcol] (4,8)--(8,8); \draw [ultra thick, ->, >=latex,
color=\bfcol] (12,8)--(8,8);
\end{tikzpicture}

\cbend

\section{Convergence of propagations}\mylabel{propconv}

It is very important to test convergence of coupled-channel calculations with respect to
propagation range and step size. Lack of convergence can give very poor results, whereas
unnecessarily conservative settings can waste large amounts of computer time. It is always
advisable to conduct careful convergence tests (which can usually be done with a small basis set)
before embarking on a major set of calculations. The programs provide automated mechanisms to
facilitate this, described in section \ref{andconv} for \MOLSCAT\ and section \ref{conv} for
\BOUND.

\section{Specific propagators}\mylabel{propdetail}

Most propagators are constructed to be exact for some \emph{reference
potential}, and incorporate deviations from the reference potential by
approximation (or for some propagators not at all). Propagators that use a
trivial reference potential (such as ${\bf W}(R)={\cal E}{\bf I}$ in Eq.\
\ref{eqcp}) are termed \emph{solution-following} methods, whereas those that
use a reference potential closer to the real potential, often without
corrections, are termed \emph{potential-following} methods.

The DV, LDJ and LDMG propagators are solution-following methods, while the RMAT
and AIRY propagators are potential-following methods. The VIVS, LDMD and LDMA
propagators are use intermediate schemes, incorporating both a reference
potential and corrections to it. Solution-following methods must always take
steps that are (much) smaller than the local wavelength. Potential-following
and intermediate methods can sometimes be accurate even for step sizes larger
than the wavelength, though the methods used to calculate node counts in
\BOUND\ and \FIELD\ may fail if there is more than one node in any channel in a
single step.

Propagators may operate either in the primitive basis set or in a
quasiadiabatic representation, defined by diagonalising the interaction matrix
${\bf W}(R)$ at some point within the step. A quasiadiabatic representation is
unchanged throughout the step, so should be distinguished from a true adiabatic
representation, which changes continuously with $R$; in a true adiabatic
representation the coupled equations contain first-derivative coupling terms
(involving $d/dR$ rather than just $d^2/dR^2$), and such terms are not handled
by the propagators in \MOLSCAT, \BOUND\ and \FIELD. The DV, LDJ, LDMD
and LDMG propagators operate in the primitive basis set, while the RMAT,
VIVS, LDMA and AIRY propagators operate in a quasiadiabatic representation.

\subsection[\texorpdfstring{DV propagator (propagator 2) ({\color{\mcol}\MOLSCAT} only)}
{DV propagator (propagator 2) (MOLSCAT only)}]{DV
propagator (propagator 2)}\mylabel{iprop2}

\cbcolor{\mcol}\cbstart The DV propagator
uses the solution-following method of de Vogelaere \cite{deVogelaere:1955} to propagate the wavefunction matrix
$\boldsymbol{\Psi}(R)$ and its radial derivative $\boldsymbol{\Psi}'(R)$ in the
primitive basis set. It allows only equally spaced steps.

The error in each step is proportional to $\delta R^5$, where $\delta R$ is the
step size. The total error after propagating across a range is proportional to
$\delta R^4$.

The de Vogelaere method is potentially unstable for channels that are locally
closed (i.e., in a classically forbidden region): the exponential growth of
closed-channel wavefunctions can lead to a loss of linear independence of the
solutions. To avoid this, the DV propagator re-imposes linear
independence every \inpitem{NSTAB} steps. The default is usually adequate.
\cbend

\subsection[\texorpdfstring{RMAT propagator (propagator 3) ({\color{\mcol}\MOLSCAT} only)}
{RMAT propagator (propagator 3) (MOLSCAT only)}]{RMAT propagator
(propagator 3)}\mylabel{iprop3}

\cbcolor{\mcol}\cbstart The RMAT propagator uses the potential-following
method of Stechel \etal\ \cite{Stechel:1978} to propagate the R matrix, which
is the inverse of the log-derivative matrix.

The propagation is done in a quasiadiabatic basis set obtained by diagonalising
the interaction matrix ${\bf W}(R)$ at the centre of each step. In each step,
it uses a constant reference potential that is the interaction potential
evaluated at the centre of the step in the quasiadiabatic basis.

The error in each step is proportional to $\delta R^3$, where $\delta R$ is the
step size. The total error after propagating across a range is proportional to
$\delta R^2$. This is poorer than for the log-derivative methods and the
RMAT propagator is recommended only for special purposes.

The RMAT propagator is implemented with equally spaced or power-law steps
as described in sections \ref{dr-fixed} and \ref{dr-power}. It is usually
recommended to use it with equally spaced steps at short range and power-law
steps at long range. It does not implement the adaptive step-size algorithm
described in ref.\ \cite{Stechel:1978}. \cbend

\subsection[\texorpdfstring{VIVS propagator (propagator 4) ({\color{\mcol}\MOLSCAT} only)}
{VIVS propagator (propagator 4) (MOLSCAT only)}]{VIVS propagator (propagator
4)}\mylabel{iprop4}

\cbcolor{\mcol}\cbstart The VIVS propagator is an intermediate method that
propagates the R matrix by the variable-interval variable-step method of Parker
\etal\ \cite{Parker:VIVS}. It operates in a quasiadiabatic representation, with
both a variable interval length and a variable step length. A single
diagonalising transformation is used over the whole of each interval, which may
consist of several steps. The error in each step or interval is proportional to
$\delta R^4$, where $\delta R$ is the step or interval size.

The VIVS propagator is designed for use at long range, but it usually offers
poorer performance and stability than the AIRY propagator, and is recommended
only for special purposes.

The step size and interval size algorithms used by VIVS attempt to take the longest step that gives
the required accuracy at each point in the propagation. However, the optimum step size at one
scattering energy is not necessarily safe at another, and VIVS can sometimes give inaccurate
results at subsequent energies if the groups of energies in a particular run are not chosen with
care. In particular, one should avoid:
\begin{enumerate}[nosep]
\item{A first energy that is close to (or above) a channel threshold and a
    subsequent energy that is far below it.}
\item{A first energy that is far above a channel threshold and a subsequent
    energy that is close to it.}
\end{enumerate}

\subsubsection{Interval sizes}\mylabel{intsize4}

VIVS accumulates perturbation corrections to the wavefunction as it propagates,
and uses these to obtain a suitable length for the next interval. The input
parameter \var{DR} described above is used as the size of the first interval,
and subsequent interval lengths are obtained using the input tolerance
\var{TOLHI}; the criterion is that some functional $t$ of the perturbation
corrections should be not greater than \var{TOLHI} over any interval. Within an
interval, $t$ is tested against \var{TOLHI} at each step, and a new interval is
started (with a new diagonalising transformation) if it appears likely to
exceed \var{TOLHI} over the next step.

Even when the perturbation corrections are small, the algorithm used to obtain
interval sizes limits the factor by which the interval size may be increased
each time. The limit is more conservative in the presence of closed channels.
In addition, when closed channels are present, the interval size is limited to
$4/|k_n|$, where $k_n$ is the local wavevector for the most deeply closed
channel. Because of this, the VIVS propagator is not very efficient at long
range in the presence of closed channels, and the AIRY propagator is more
suitable in such cases.

\subsubsection{Step sizes}\mylabel{stepsize4}

Each interval is divided into \inpitem{IALPHA} steps. Within an interval, the
step sizes increase geometrically, with each step being a factor $\alpha$
larger than the previous one, subject to a maximum step size set by
\inpitem{DRMAX}. The quantity $\alpha$ may be specified in either of 2 ways:
\begin{enumerate}[nosep]
\item{If the logical input parameter \inpitem{IALFP} is \code{.FALSE.},
    $\alpha$ increases linearly from \inpitem{ALPHA1} at the starting point
    for VIVS to \inpitem{ALPHA2} at \inpitem{RMAX}.}
\item{If \inpitem{IALFP} is \code{.TRUE.} on input, the program starts with
    an initial value of \inpitem{ALPHA1}, and adjusts this as it
    propagates. \inpitem{ALPHA2} is then not used.}
\end{enumerate}

\subsubsection{Automatic step and interval lengths}\mylabel{auto4}

A useful special option is obtained by setting $\inpitem{IALPHA} = 0$ on input.
Intervals then consist of a variable number of steps, and the decision to start
a new interval is based solely on the magnitude of the perturbation
corrections; a new interval is started (and a new diagonalising transformation
obtained) whenever the quantity $t$ approaches \var{TOLHI}. The initial step
size is taken from \var{DR}, and subsequent steps use a criterion based on
\var{TOLHI}.

The $\inpitem{IALPHA} = 0$ option can be very efficient, and often requires
remarkably few intervals/steps to produce converged results.

\subsubsection{Perturbation corrections}\mylabel{pert4}

There are several logical input variables that control the extent to which VIVS
calculates and uses perturbation corrections to the wavefunction. The three
variables \inpitem{IV}, \inpitem{IVP} and \inpitem{IVPP} control the
calculation of perturbation corrections due to the potential itself
(\inpitem{IV}) and to its first (\inpitem{IVP}) and second (\inpitem{IVPP})
derivatives. The perturbation corrections thus calculated are used in
calculating interval sizes, but are included in the wavefunction only if
\inpitem{IPERT} is \code{.TRUE.}. If \inpitem{ISHIFT} is \code{.TRUE.}, the
second derivative is used to shift the reference potential to give the best fit
to the true potential.

For production runs, \inpitem{IV}, \inpitem{IVP}, \inpitem{IVPP},
\inpitem{IPERT} and \inpitem{ISHIFT} should usually all be \code{.TRUE.} This
may be forced by setting $\inpitem{IDIAG} = \code{.TRUE.}$, which overrides any
\code{.FALSE.} values for the individual variables.

If \inpitem{IVP}, \inpitem{IVPP} or \inpitem{ISHIFT} are \code{.TRUE.}, VIVS
requires radial derivatives of the interaction potential. These are supplied
properly for simple potentials by the general-purpose version of \prog{POTENL}
described below, but for some potentials they can be difficult to evaluate. In
this case, the input variable \inpitem{NUMDER} may be set \code{.TRUE.}, in
which case the necessary derivatives are calculated numerically, and
\prog{POTENL} is never called with $\var{IC} > 0$; see section \ref{eval}.

\subsubsection{Other variables}\mylabel{other4}

\begin{description}
\item[\inpitem{ISYM}]{If \inpitem{ISYM} is \code{.TRUE.}, the R matrix is
    forced to be symmetric at the end of each interval. This is usually
    advisable for production runs.}

\item[\inpitem{XSQMAX}]{controls the application of perturbation
    corrections to deeply closed channels. If a channel is locally closed
    by more than \inpitem{XSQMAX} reduced units, perturbation corrections
    for it are not calculated. The default should be adequate.}
\end{description}
\cbend

\subsection{LDJ, LDMD, LDMA and LDMG propagators (propagators 5 to 8)}\mylabel{iprop5678}

The LDJ \cite{Johnson:1973, Manolopoulos:1993:Johnson}, LDMD
\cite{Manolopoulos:1986}, LDMA \cite{Manolopoulos:PhD:1988, Hutson:CPC:1994}
and LDMG \cite{MG:symplectic:1995} methods all propagate the log-derivative
matrix ${\bf Y}(R) = \boldsymbol{\Psi}'(R) [\boldsymbol{\Psi}(R)]^{-1}$.
Although ${\bf Y}(R)$ is discontinuous, the invariant-imbedding methods used to
derive the propagators actually expand $\boldsymbol{\Psi}(R)$ and
$\boldsymbol{\Psi}'(R)$, which are continuous functions.

The LDJ and LDMD propagators both operate in the primitive basis set. The LDJ
propagator is a solution-following method, with the entire interaction matrix
incorporated through quadrature. The LDMD propagator is an intermediate method
that uses a reference potential that is the diagonal part of the interaction
matrix ${\bf W}(R)$ at the centre of each step, and treats only deviations from
it by quadrature. The LDJ and LDMD propagators avoid the need to diagonalise
${\bf W}(R)$ at each step, but can be inefficient at long range if $H_{\rm
intl}$ and/or $\hat L^2$ is non-diagonal.

The LDMA propagator is an intermediate method that uses a quasiadiabatic basis
set defined by diagonalising ${\bf W}(R)$ at the centre of each step.
Deviations from the reference potential are included by quadrature.

The LDJ, LDMD and LDMA propagators use a quadrature based on Simpson's rule to
obtain an error in each step proportional to $\delta R^5$, where $\delta R$ is
the step size. This advantage applies only with an even number of steps, so
internally the routines all propagate using half-steps of size $\delta R/2$.
The total error after propagating across a range is proportional to $\delta
R^4$.

The LDMG propagators are solution-following methods that take advantage of the
symplectic nature of the multi-channel Schr\"odinger equation and reformulate
it so that symplectic integrators (SIs) may be used to propagate solutions of
the coupled equations. The variable \inpitem{IMGSEL} specifies whether to use
the five-step 4th-order method of Calvo and Sans-Serna ($\inpitem{IMGSEL}=4$)
or the six-step 5th-order method of McLachlin and Atela ($\inpitem{IMGSEL}=5$).
Both of these operate in the primitive basis set.

All these propagators operate only with equally spaced steps, and power-law
steps are not currently implemented. For the LDJ, LDMD and LDMA propagators, a
variable step size would lose the advantage of Simpson's rule and introduce
errors proportional to a lower power of $\delta R$.

\subsection{AIRY propagator (propagator 9)}\mylabel{iprop9}

The AIRY propagator uses the solution-following method of Alexander
\cite{Alexander:1984}, as reformulated by Alexander and Manolopoulos
\cite{Alexander:1987}. It operates in a quasiadiabatic basis set defined by
diagonalising ${\bf W}(R)$ at the central point $R_i$ of each step $i$. It
propagates the log-derivative matrix across the step using a linear reference
potential $W_i(R) = W_i^{(0)} + W_i^{(1)} (R-R_i)$ in the quasiadiabatic
representation.\footnote{The independent solutions of this piecewise linear
potential are Airy functions of argument $W_i^{(0)} (W_i^{(1)})^{-2/3} +
(W_i^{(1)})^{1/3} (R-R_i)$.  For small arguments, the propagators are evaluated
using the routines of Alexander and Manolopoulos \cite{Alexander:1987}.
However, as $W_i^{(1)}$ tends to zero at long range, the arguments of the Airy
functions become asymptotically large; this can cause a loss of precision and
may lead to numerical noise in derived quantities such as the eigenphase sum.
Version 2019.0 and later implement a modified algorithm for evaluating the
propagators at long range, using the expansions given in appendix~A of ref.\
\cite{Karman:2014}; these are accurate as $W_i^{(1)} \rightarrow 0$. The
asymptotic expansions are used if $\log( | W_i^{(0)} (W_i^{(1)})^{-2/3} | ) >
\var{AC}$; the value of \var{AC} is hard-coded as 3.0 in subroutine
\prog{SPROPN} but can be changed for special purposes.} Matrix elements
off-diagonal in the quasiadiabatic basis set are neglected. The step size can
increase rapidly with separation, so that this propagator is particularly
efficient at long range.

\var{DR} is the size of the innermost step, except in the case of inwards
propagation with $\var{TOLHI}>0$, when it sets the size of the \emph{first}
step.

\subsubsection{Adaptive step size}

By default the AIRY propagator uses an adaptive step-size algorithm based on the variables
\var{TOLHI} and \var{POWR}. \var{TOLHI} is a tolerance that is used to adjust the step size to try
to maintain the same accuracy throughout the propagation. Values of \var{TOLHI} between $10^{-4}$
and $10^{-10}$ are generally useful.

The algorithm used to obtain step sizes for outwards propagations is that
described by Alexander \cite{Alexander:1984}. It calculates quantities that
estimate two different sources of error:
\begin{description}[nosep]
\item[\var{CDIAG}]{is proportional to the error due to neglected
    second-derivative terms in the potential that are diagonal in the local
    adiabatic basis set;}
\item[\var{COFF}]{is proportional to the error due to neglected
    first-derivative terms in the potential that are off-diagonal in the
    local adiabatic basis set.}
\end{description}
The error from each source is proportional to the cube of the current step
size, so the basic algorithm is for the step size in the next step to change
from that in the current step by a factor
$[{\max\{\var{CDIAG},\var{COFF}\}/\var{TOLHI}]^{1/\var{POWR}}}$;
\var{POWR} defaults to \inpitem{POWRX} (which defaults to 3.0) when $\var{TOLHI}>0$, but larger values may be input
to limit the rate of increase/decrease. Any increase is limited to a factor of
2 in each step for stability.

The adaptive step-size algorithm is based on the expectation that the neglected
derivatives of the potential are similar at the next step to those in the
current step. For outwards propagation, the resulting step size generally
increases with $R$ at long range in a way that is sufficient to take account of
the decreasing values of the neglected derivatives. However, for inwards
propagation the derivatives are often larger at the next step than at the
current one: for a potential $R^{-n}$, the first and second derivatives are
proportional to $R^{-n-1}$ and $R^{-n-2}$, and the basic algorithm might
generate an over-long step. To maintain the error closer to a constant value,
the predicted step size for inwards propagations is multiplied by an additional
factor $(1-|\delta R|/R)^3$, which is sufficiently conservative to take account
of potentials with inverse powers $n\le 7$.

The AIRY propagator prints a warning if $\inpitem{IPRINT}\ge10$ and the value of
\var{CDIAG} or \var{COFF} in any step is greater than $5\times\var{TOLHI}$. This can
occur if the step midpoint is close to a narrow avoided crossing between
eigenvalues of ${\bf W}(R)$, since such avoided crossings produce narrow spikes
in first-derivative terms that are off-diagonal in the local adiabatic basis
set. However, such warnings can also occur if there is an unphysical
discontinuity in the potential or its derivative, and this should be checked.

\subsubsection{Noise due to adaptive step-size algorithm}

The adaptive step-size algorithm can produce numerical noise in the values of
$R$ where the potential is evaluated, particularly when the range of
eigenvalues of ${\bf W}(R)$ is large and part of the Hamiltonian (such as an
external field) varies between calculations. This results in noise in S
matrices and log-derivative matching matrices. The noise is usually small in
absolute terms, but can impede convergence on resonances and bound states.
\phantomsection\label{ss-noise}

Step-size noise can be eliminated by setting $\inpitem{ISCRU}>0$, as described in section
\ref{CommIII:iscru}, so that subsequent propagations for the same $J+{\rm tot}$ and
symmetry block use exactly the same steps as the first.

Step-size noise does not occur when $\var{TOLHI}=0$, specifying equally spaced or power-law
step sizes, as described in sections \ref{dr-fixed} and \ref{dr-power}, in place of
the adaptive step-size algorithm with $\var{TOLHI}>0$ (but see warnings below).

\subsubsection{Choice of adaptive or power-law step sizes}

The adaptive step-size algorithm is easy to control and robust for outwards
propagation. \var{POWR} should almost always be set to 3.0, and the accuracy is controlled by the
single variable \var{TOLHI}. If the step size is initially too small, the algorithm quickly
increases it without much inefficiency. For outwards propagations, it allows the step size to
increase very fast in regions where the error is dominated by potential terms proportional to
$R^{-n}$ with $n>2$, and then moderates the rate of increase when centrifugal terms become
dominant. It usually takes very long steps at long range, so that using very large values of
\inpitem{RMAX} is inexpensive.

\cbcolor{\bfcol}\cbstart The adaptive step-size algorithm is less robust for inwards
propagation, as often chosen for \BOUND\ and \FIELD\ at long range. When some channels have $L>0$,
the curvature of the centrifugal potential usually prevents excessive step sizes and the algorithm
works well. However, the case where all channels have $L=0$ is problematic (and occurs, for
example, for atomic collisions with $L_{\rm max}=0$). In this case the step size can grow very
large, and not decrease fast enough to provide sufficient points at shorter range. For inwards
propagation in cases where $L=0$ for all channels, step sizes proportional to $R^\var{POWR}$ are
recommended; for potentials that decay as $1/R^6$, as for neutral atoms, $\var{POWR}=2.66\dot{6}$
is often appropriate. \cbend

Step sizes proportional to $R^\var{POWR}$ eliminate step-size noise, but take
more expertise to specify and are often less efficient for a given accuracy. In
this case, \var{DR} must be chosen with care, since the step size is
proportional to it throughout the range; it is very inefficient to use too
small a value. A single value of \var{POWR} must be used throughout the range,
and it must be chosen conservatively, usually as $\var{POWR}=1.33\dot{3}$ to
accommodate centrifugal terms as described in section \ref{dr-power}; this may
mean forgoing fast increases in step size achieved by the adaptive algorithm at
shorter range. Finally, the values of \var{DR} and \var{POWR} needed at shorter
range may result in much smaller steps at long range than are achieved by the
adaptive algorithm, and this in turn may require greater care in choosing a
value of \inpitem{RMAX} that is not unnecessarily large.

For these reasons, we recommend using the AIRY propagator with the adaptive
step-size algorithm ($\var{TOLHI}>0$) in the first instance, and switching to
power-law steps ($\var{TOLHI}=0$) only if convergence difficulties due to
step-size noise are encountered, with careful evaluation of the different
values that may be needed for \var{DR}, \var{POWR} and \inpitem{RMAX}.

\subsection[\texorpdfstring{WKB integration (propagator -1) ({\color{\mcol}\MOLSCAT} only)}
{WKB integration (MOLSCAT only)}]{WKB integration (propagator
-1)}\mylabel{intflgwkb}

\cbcolor{\mcol}\cbstart WKB integration is not strictly a propagator and is
suitable only for single-channel cases (particularly in IOS calculations). It
evaluates the WKB integral for the phase shift by Gauss-Mehler quadrature.  It
may not be combined with any other propagator. It is controlled by input
variables \inpitem{NGMP} and \inpitem{TOLHI}.

\begin{description}
\item[\inpitem{NGMP}]{Dimension 3.  $N$-point Gaussian integration is
    performed starting with $N = \inpitem{NGMP}(1)$, incrementing by
    \inpitem{NGMP}(2), until \inpitem{NGMP}(3). Starting with the 2nd pass,
    the phase shift is compared with the previously calculated value until
    it has converged to within a tolerance specified by \inpitem{TOLHI}.}

\item[\inpitem{TOLHI}]{is the convergence tolerance for the WKB phase
    shift.}
\end{description}\cbend

\section{Adiabats and nonadiabatic matrix elements}

The LDMA propagator includes features to print adiabats and
nonadiabatic matrix elements between the adiabatic functions. These are often
useful in interpreting bound states and scattering.

The adiabats $U_i(R)$ are the eigenvalues of
\begin{equation}
H_{\rm intl}+V(R,\xi_{\rm intl})+\frac{\hbar^2 \hat L^2}{2\mu R^2}
\end{equation}
at fixed values of $R$. They are simply the eigenvalues of the interaction
matrix ${\bf W}(R)$, rescaled into energy units. The adiabatic functions
$\Phi_i^{\rm ad}(\xi_{\rm intl};R)$ are obtained from the corresponding
eigenvectors; the semicolon indicates that the functions depend only
parametrically on $R$.

The nonadiabatic matrix elements $B_{ij}(R)$ are
\begin{equation}
B_{ij}(R)=\int\Phi_i^{{\rm ad}*}(\xi_{\rm intl};R) \frac{d}{dR}
\Phi_j^{\rm ad}(\xi_{\rm intl};R)\,\d\xi_{\rm intl}. \label{eqBij-def}
\end{equation}
The diagonal elements $B_{ii}(R)$ are zero and the off-diagonal elements are
evaluated using the Hellmann-Feynman theorem \cite{Hutson:CBO:1980},
\begin{equation}
B_{ij}(R)=\frac{\displaystyle\int\Phi_i^{{\rm ad}*}(\xi_{\rm intl};R) \left[\frac{d}{dR} \left(
V(R,\xi_{\rm intl})+\frac{\hbar^2 \hat L^2}{2\mu R^2} \right) \right]
\Phi_j^{\rm ad}(\xi_{\rm intl};R)\,\d\xi_{\rm intl}}{U_j(R)-U_i(R)}. \label{eqDij-eval}
\end{equation}

At the midpoint of each step, the program evaluates the matrix $d{\bf W}/dR$,
transforms it into the adiabatic basis set using the eigenvectors of ${\bf
W}(R)$, and then divides by the appropriate denominators to obtain $B_{ij}(R)$.
It should be noted that the nonadiabatic matrix elements peak sharply at values
of $R$ that correspond to narrow avoided crossings between adiabats, where
$U_j(R)-U_i(R)$ can be very small. The absolute signs of the eigenvectors of
${\bf W}(R)$ are numerically arbitrary and may change between propagation
steps; as a result, the signs of the matrix elements produced from Eq.\
\ref{eqDij-eval} may also change sign arbitrarily.

There are also both diagonal and off-diagonal matrix elements of the operator
$d^2/dR^2$ as described in ref.\ \cite{Hutson:CBO:1980}. These are not
currently evaluated but would be straightforward to add.

The print levels needed for these features are described in section
\ref{output:props}.

\section{Boundary conditions}\mylabel{Y-bc}

All the programs carry out outwards propagations starting at $R_{\rm min}$, and
\BOUND\ and \FIELD\ also carry out inward propagations starting at $R_{\rm
max}$. Boundary conditions are needed to initialise these propagations.
Versions of the programs before 2019.0 used a variety of boundary conditions,
which were adequate for most cases, but better control is needed for special
purposes. From version 2019.0, the choice of boundary conditions for
log-derivative propagators has been unified, and variables \inpitem{ADIAMN},
\inpitem{ADIAMX}, \inpitem{BCYCMN}, \inpitem{BCYCMX}, \inpitem{BCYOMN},
\inpitem{BCYOMX}, \inpitem{WKBMN} and \inpitem{WKBMX} have been introduced to
control them. The variables \code{...MX} are not used in \MOLSCAT.

The boundary condition is always expressed as a diagonal log-derivative matrix,
but it may be diagonal either in the primitive basis set or in the locally
adiabatic representation that diagonalises the coupling matrix ${\boldsymbol
W}(R)$. In versions of the programs before 2019.0, the boundary conditions were
applied in the primitive basis set for diabatic propagators and in the
adiabatic basis set for adiabatic propagators. From version 2019.0 the choice
is controlled by the namelist items \inpitem{ADIAMN} and \inpitem{ADIAMX}:
these are logical variables (default \code{.TRUE.}) that specify whether the
boundary conditions at $R_{\rm min}$ and $R_{\rm max}$, respectively, are
applied in the adiabatic basis. If necessary, the resulting diagonal
log-derivative matrix is transformed into the basis set used by the propagator
chosen.

For a single channel, the log-derivative is defined as $Y=\psi'(R)/\psi(R)$. At
the wall of a box, $\psi(R_{\rm wall})=0$, so $Y=\infty$. However, it is
sometimes useful to define a boundary condition corresponding to $\psi'(R)=0$
and $\psi(R)$ finite, so that $Y=0$. Lastly, if the wavefunction follows the
WKB approximation in the classically forbidden region, then
\begin{eqnarray}
\psi(R)&=&[k(R)]^{-\frac{1}{2}} \exp\left(\pm\int_{R_{\rm turn}}^R k(R')\,\d R'\right),\\
\psi'(R)&=&[k(R)]^{-\frac{1}{2}}\left[\pm k(R)-\frac{1}{2}\frac{k'(R)}{k(R)} \right]
\exp\left(\pm\int_{R_{\rm turn}}^R k(R')\,\d R'\right),\\
Y(R)&=&\pm k(R)-\frac{1}{2}\frac{k'(R)}{k(R)},\label{eq:bcwkb}
\end{eqnarray}
where $k(R) = [2\mu(V(R)-E)/\hbar^2]^{1/2}$ and $V(R)$ is an effective
potential energy for the channel concerned. The + sign applies inside the inner
turning point (where the phase integral is itself negative) and the $-$ sign
applies outside the outer turning point. The first term in Eq.\ \ref{eq:bcwkb}
dominates either when $k(R)$ is large (in a strongly classically forbidden
region) or when the interaction potential is nearly constant (at very long range).
The term involving $k'(R)$ is therefore neglected in the implementation of WKB
boundary conditions.

\emph{Locally closed channels}

For locally closed channels, the default is to use WKB boundary conditions at
both $R_{\rm min}$ and $R_{\rm max}$ ($\inpitem{WKBMN} = \inpitem{WKBMX} =
\code{.TRUE.}$). WKB boundary conditions usually give the fastest convergence
with respect to $R_{\rm min}$ and $R_{\rm max}$ for problems where the
interaction potential remains finite in the classically forbidden region. For
special purposes, \inpitem{WKBMN} and/or \inpitem{WKBMX} may be set
\code{.FALSE.} and explicit values provided in the variables \inpitem{BCYCMN}
and/or \inpitem{BCYCMX}. These values are used for all locally closed channels
at $R_{\rm min}$ and $R_{\rm max}$, respectively.

\emph{Locally open channels at $R_{\rm min}$}

It is relatively rare to start a propagation in the presence of locally open
channels at $R_{\rm min}$. By default, \inpitem{BCYOMN} is unset and the
programs stop if they detect locally open channels at $R_{\rm min}$. If locally
open channels actually exist, \inpitem{BCYOMN} must be set explicitly.

For systems that have no classically forbidden region at short range, it may be
appropriate to start the propagation at $R=0$. When $R$ is a true radial
coordinate with volume element $R^k$ and $k>0$, $Y(0)$ is usually $\infty$ and
\inpitem{BCYOMN} should be set to a large positive value (e.g., $10^8$) to
represent this as described below. For 1-dimensional systems where $R$ does not
represent a radial coordinate, states that are symmetric about $R=0$ require
$Y(0)=0$ and those that are antisymmetric require $Y(0)=\infty$.

\cbcolor{\bfcol}\cbstart \emph{Locally open channels at $R_{\rm max}$}

It is more common to start an inwards propagation in the presence of locally
open channels at $R_{\rm max}$. It often occurs, for example, when running
\FIELD\ at the energy of a scattering threshold to locate the fields at which
bound states cross the threshold. It is also sometimes desired to use \BOUND\
or \FIELD\ to estimate the positions of quasibound states that lie above
threshold, applying a boundary condition with \inpitem{BCYOMX} at long range to
quantise the open channels. By default, $\inpitem{BCYOMX}=0.0$. The reason for
this choice is that the WKB boundary condition (\ref{eq:bcwkb}) implies that
$Y(R_{\rm max})\rightarrow 0$ as the energy $E$ approaches $V(R_{\rm max})$
from below. The choice $\inpitem{BCYOMX}=0.0$ provides continuity across this
energy when WKB boundary conditions are used for locally closed channels.
Discontinuities in boundary conditions may cause discontinuities in the node
count and disrupt convergence on bound-state positions.\cbend

\emph{Finite values representing infinity}

There are several situations where the physical boundary condition required is
$Y(R)=\infty$. These include an infinite hard wall and the behaviour at the
origin in polar or spherical polar coordinates. Such boundary conditions should
be specified with large positive values (e.g., $10^8$) for $\inpitem{BCYCMN}$
or $\inpitem{BCYOMN}$ and with large negative values (e.g., $-10^8$) for
$\inpitem{BCYCMX}$ or $\inpitem{BCYOMX}$. Using large values with signs
different from these has little effect on scattering properties or bound-state
positions, but in \BOUND\ and \FIELD\ it may affect the node count, as it
causes an extra node to appear very close to the end-point in each channel.

Values larger than $10^8$ may be used if required, but can sometimes cause
problems if the initialisation is done in one representation (adiabatic or
diabatic) and then transformed to the other. This is particularly true if
different boundary conditions are used for different channels (e.g., WKB
boundary conditions for locally closed channels and large values for locally
open channels). In such circumstances, initialisation should be done in the
same representation as the propagation.

\section{Propagator scratch file}
\mylabel{CommIII:iscru}

All the propagators have options to save some information between propagations on a scratch file.
If $\inpitem{ISCRU} > 0$, this file is created on unit \iounit{ISCRU}. It can be large.

The propagator scratch file has two purposes. The first is to store energy-independent information
from the first propagation for each $J_{\rm tot}$ and symmetry block to use at subsequent energies.
The information includes the values of $R$ at which interaction matrices are evaluated, the
interaction matrices themselves, and (for adiabatic propagators) their eigenvectors. This option
saves CPU time at the expense of disc I/O. It is often advantageous for the LDMD, LDMA, AIRY, RMAT
and VIVS propagators if \inpitem{NNRG} is not 1, but for the DV and LDJ propagators it does not
usually save resources overall unless the interaction potential itself is very expensive to
evaluate. This is particularly true on machines where the scratch file is accessed over a network.
Note, however, that $\inpitem{ISCRU} > 0$ may save considerable time for single-channel IOS cases
with the LDMD propagator, since the computer time for these is often dominated by potential
evaluations.

For the special case of the AIRY propagator, the option $\inpitem{ISCRU} > 0$ also ensures that the
steps taken by the adaptive step-size algorithm are identical for subsequent propagations to those
at the first energy and EFV set (which is typically the one corresponding to \inpitem{FLDMAX} when
locating bound states as a function of field, and \inpitem{FLDMIN} otherwise). This may be useful
in reducing step-size noise, as described on p.\ \pageref{ss-noise}.  From version 2022.0 onwards,
if there are multiple EFV sets and multiple energies, a second scratch file is created on unit
$\iounit{ISCRU}+1$ to save just the values of $R$. If there are multiple EFV sets but only a single
energy (or, for \FIELD, also for multiple energies), the values of $R$ alone are saved on unit
\iounit{ISCRU}.

\cbcolor{\mcol}\cbstart For the special case of \MOLSCAT\ calculations for a
single value of \var{JTOT} and \var{IBLOCK} and a single set of EFVs, the
\iounit{ISCRU} file from one run may be used as input for the next run at a
different set of energies. The routine \file{mol.driver.f} must first be
recompiled with a different \code{OPEN} statement for unit \iounit{ISCRU},
which must omit the specification \code{STATUS='SCRATCH'}. The relevant lines
of code are currently commented out in \file{mol.driver.f}. The first run then
produces a file on unit \iounit{ISCRU} that is preserved. In subsequent runs,
\inpitem{ISCRU} should be set negative; the program then expects to find the
file from the first run on unit $|\inpitem{ISCRU}|$. It reads the header on
this file to check that it contains valid information, and then proceeds with
``subsequent energy" calculations for all the energies requested. \cbend

\chapter[\texorpdfstring{Controlling scattering calculations ({\color{\mcol}\MOLSCAT} only)}
{\ref{processres}: Controlling scattering calculations (MOLSCAT only)}]
{Controlling scattering calculations\chaptermark{Controlling scattering
calculations}}\chaptermark{Controlling scattering
calculations}\mylabel{processres} \cbcolor{\mcol}\cbstart

\section{Asymptotic basis sets}\mylabel{prop:YtoS}

If the operators $H_{\rm intl}$ and $\hat L^2$ are diagonal in the basis set
used for the propagation, the propagated wavefunction matrix $\Psi(R_{\rm
max})$ and its derivative $\Psi'(R_{\rm max})$, or the log-derivative matrix
${\bf Y}(R_{\rm max})$, are matched directly to analytic radial functions that
describe the solutions of the Schr\"odinger equation in the absence of an
interaction potential to obtain the K matrix. This is then converted into the
scattering S matrix, as described in section \ref{theory:scatcalcs}. This
procedure is used for all the built-in interaction types and for some plug-in
basis-set suites.

If one or both of $H_{\rm intl}$ and $\hat L^2$ are not diagonal in the basis
set used for the propagation, the propagated functions are transformed into an
asymptotic basis set that diagonalises them. In the simplest (and most common)
case, \MOLSCAT\ constructs the matrices of $H_{\rm intl}$ and/or $\hat L^2$ as
necessary, diagonalises them, and transforms the wavefunction or log-derivative
matrices before matching to analytic radial functions to obtain the K and S
matrices. The rows and columns of the K and S matrices are labelled by the
asymptotic basis functions, but these functions are not generally described in
any simple way by the quantum numbers in the array \var{JSTATE}.

\section{Resolving degeneracies with extra operators}\mylabel{prop:add-op}

In some cases, the requirement that the asymptotic basis functions are
eigenfunctions of $H_{\rm intl}$ and $\hat L^2$ is not enough to define them
uniquely. This occurs if two or more channels with the same value of $L$ are
degenerate in energy (or very nearly degenerate). Numerical diagonalisation of
the matrix $\boldsymbol{H}_{\rm intl}$ may then produce eigenvectors that are
linear combinations of the physically relevant vectors. Plug-in basis-set
suites may be programmed to specify extra operators $\hat P_i$ to aid in
resolving such \mbox{(near-)degeneracies}. These extra operators, which need
not necessarily contribute to $H_{\rm intl}$ or $\hat L^2$, must nevertheless
commute with them. In any \mbox{(near-)degenerate} subspace of the eigenvectors
of $H_{\rm intl}$ and $\hat L^2$, \MOLSCAT\ constructs the matrix of the first
such operator ($\hat P_1$) and finds linear combinations of the degenerate
functions that are eigenfunctions of it. If the eigenvalues of $\hat P_1$ are
sufficiently non-degenerate, the process ends; if not, it is repeated with
operator $\hat P_2$, and so on.

The namelist item \inpitem{DEGTOL} is used as a threshold for degeneracy for
both $H_{\rm intl}$ and the extra operators. It is treated as an energy and so
is scaled according to \inpitem{EUNITS} or \inpitem{EUNIT}.

\section{Channel indices for open channels}\mylabel{scatchan}

The S matrix has dimension $N_{\rm open}\times N_{\rm open}$, where $N_{\rm
open}$ is the number of open channels. The open channels for each propagation
are sorted in order of increasing threshold energy, and \MOLSCAT\ prints a list
of them after each propagation if $\inpitem{IPRINT}\ge10$.

The list of open channels gives the open-channel index and the index of the
corresponding channel in the complete channel list (described in sections
\ref{basis:diag} and \ref{basis:off-diag}). It also repeats the associated
value of $L$, the index of the pair level and the corresponding pair energy
(threshold energy).  If $\basisitem{IBOUND}=1$, the diagonal matrix element
$\langle \hat L^2 \rangle$ stored in the array \var{CENT} is printed in place
of the integer $L$.

\section{S matrix}\mylabel{printSmatrix}

If $\inpitem{IPRINT}\ge11$, \MOLSCAT\ prints the S matrix in the main output
file. Each element is labelled by the initial and final open-channel indices.
The output gives the square modulus, the phase, and the real and imaginary
parts. Only elements with square modulus greater than $10^{-20}$ are printed.

If the S matrix is to be read by an external program, it is usually more
convenient to obtain it from an auxiliary output file written on channel
\inpitem{ISAVEU}, as described in section \ref{CommII:SandK}, rather than from
the main output file.

\section{Automated testing of propagator convergence}
\mylabel{andconv}

If $\inpitem{NCONV}>0$, \MOLSCAT\ performs \inpitem{NCONV} extra calculations
of the S matrix, with different values of \inpitem{RMIN}, \inpitem{RMAX} or the
step size, and compares the results. Which one of these lengths is varied is
governed by \inpitem{ICON} as follows:
\begin{description}[nosep]
          \item[$\inpitem{ICON} = 1$]{doubles the initial step size each
              time}
          \item[$\inpitem{ICON} = 2$]{decreases \inpitem{RMAX} by
              \inpitem{DRCON} each time}
	  \item[$\inpitem{ICON} = 3$]{increases \inpitem{RMIN} by
\inpitem{DRCON} each time}
\end{description}
\MOLSCAT\ calculates an S matrix and prints the root-mean-square change in
S-matrix elements and transition probabilities each time. This provides an
automated means of choosing propagation parameters capable of providing the
required accuracy. Only a single S matrix is stored between calculations, so
the program must not loop over angular momenta, symmetry blocks, energies, or
sets of EFVs; thus \inpitem{JTOTU} must be equal to
\inpitem{JTOTL}, \inpitem{IBFIX} must be set (and \inpitem{IBHI} must be
unset), \inpitem{NNRG} must be 1 and \inpitem{NFIELD} must be 1. In addition,
\inpitem{ISCRU} must be 0.

If $\inpitem{ICONVU}=0$ (the default), the results from each propagation are
compared with the first. If $\inpitem{ICONVU} > 0$ on input, the first S matrix
is written (unformatted) to unit \iounit{ICONVU}; if $\inpitem{ICONVU} < 0$, a
previously saved S matrix is read from unit $|\inpitem{ICONVU}|$, and used as
the reference S matrix in calculating root-mean-square (rms) errors.

Remember that, in some modes, \MOLSCAT\ determines \inpitem{RMIN},
\inpitem{RMID} and \inpitem{RMAX} internally, and it is safest to test
convergence with these options switched off: $\inpitem{IRMSET}=0$,
$\inpitem{RVFAC}=0.0$, $\inpitem{IRXSET}=0$.

\section{Elastic and inelastic cross sections}\mylabel{CS}
\subsection{Basis sets diagonal in $H_{\rm intl}$ and $\hat L^2$}

For the built-in interaction types, \MOLSCAT\ calculates degeneracy-averaged elastic and inelastic
cross sections between the levels in the array \basisitem{JLEVEL}; see Eq.~\ref{eqsigdef}. Cross
sections are labelled by the indices of initial and final levels. These are given as {\tt PAIR
LEVEL} in the list of pair states that is printed if $\inpitem{IPRINT}\ge1$.

The same structures are used for plug-in basis-set routines that implement basis sets in which
$H_{\rm intl}$ and $\hat L^2$ are diagonal.

\MOLSCAT\ calculates degeneracy-averaged cross sections as defined by Eq.\ \ref{eqsigdef}.
However, in some cases there is an ambiguity in the appropriate degeneracy factor to use. The
factors used are coded in entry \prog{DEGENF} in subroutine \code{BASE} for built-in coupling
cases, or in \prog{DEGEN9} for plug-in basis-set suites. The values coded are suitable for most
cases, but there are a few where care is needed:
\begin{itemize}
\item For identical pairs of molecules, there are two options for the degeneracy factor as
    described by Huo and Green \cite{Huo:1996}. From version 2020.0, \MOLSCAT\ implements the
    preferred Eq.\ 2.16 of ref.\ \cite{Huo:1996}.
\item For symmetric tops, basis functions exist with both positive and negative values of $k$.
    The programs use symmetrised basis functions that are even and odd linear combinations of
    these. The even and odd functions are degenerate unless split by tunnelling or similar
    splittings (such as $l$-type doubling for a linear molecule with vibrational angular
    momentum).  The programs always treat the two levels as separate, and the cross sections
    involving them are typically different. If cross sections averaged over the pairs are
    required, they must be constructed by hand as described below.
\item For asymmetric tops, states with even and odd symmetry with respect to $k\leftrightarrow
    -k$ are physically distinct and usually non-degenerate. From version 2022.0, they are
    always treated as distinct, with separate cross sections. However, in versions 2019 and
    2020, if the two states were very close together (typically within $10^{-8}$ cm$^{-1}$, set
    in subroutine \prog{SET6C}), they were treated as associated with a single level; when this
    occurred, \MOLSCAT\ produced degeneracy-averaged cross sections that were summed over the
    states but \emph{not} divided by a factor of 2 to reflect the extra degeneracy. Such cross
    sections were thus too large by a factor of 2.
\item For spherical tops, there are levels of A, E and F (T) symmetry that do not interconvert
    in collisions, as described on p.~\pageref{isym-spher}. \MOLSCAT\ gives correct
    degeneracy-averaged cross sections for A levels (with $\basisitem{ISYM}=224$) because there
    is no additional degeneracy, and for F (T) levels (with $\basisitem{ISYM}=177$) because
    only one of each set of 3 degenerate levels is included. However, for E levels (with
    $\basisitem{ISYM}=208$), the degeneracy-averaged cross sections obtained with the default
    $\inpitem{IBFIX}=0$ are too large by a factor of 2 (even in version \currentversion)
    because they are summed over both states of the degenerate pair but not divided by the
    additional factor of 2 needed to account for the degeneracy. In close-coupling
    calculations, the sets of coupled equations for E symmetry are identical for even and odd
    parity, so a workaround that produces correct degeneracy-averaged cross sections is (for E symmetry, but not for A or F (T)
    symmetry) to set
    \inpitem{IBFIX} to 1 or 2 instead of the default 0.
\end{itemize}

\subsubsection{Constructing degeneracy-averaged cross sections by hand}

Consider the case where \MOLSCAT\ identifies two or more initial levels $n_{\rm i}$ and $n_{\rm j}$
with degeneracies $g_{n_{\rm i}}$ and $g_{n_{\rm j}}$, and two final levels $n_{\rm e}$ and $n_{\rm
f}$ with degeneracies $g_{n_{\rm e}}$ and $g_{n_{\rm f}}$. The corresponding cross sections between
the distinct levels are $\sigma_{n_{\rm i}\rightarrow n_{\rm f}}$, etc. When $I=\{n_{\rm i},n_{\rm
j}\}$ and $F=\{n_{\rm e},n_{\rm f}\}$ are treated as degenerate, the appropriate
degeneracy-averaged cross section is
\begin{equation}
\sigma_{I\rightarrow F} = \left(\sum_{n_{\rm i}\in I} g_{n_{\rm i}}\right)^{-1}
\sum_{\substack{n_{\rm i}\in I\\n_{\rm f}\in F}} g_{n_{\rm i}} \sigma_{n_{\rm i}\rightarrow n_{\rm f}}
.\label{eqdegsig}
\end{equation}

\subsection{Basis sets off-diagonal in $H_{\rm intl}$ and/or $\hat L^2$}

If one or both of $H_{\rm intl}$ and $\hat L^2$ is non-diagonal, the asymptotic energy levels for a
particular symmetry block are not known until the basis set for that symmetry block is constructed
and cannot be described by \var{JLEVEL}. In this case, if $\inpitem{MXSIG}>0$, \MOLSCAT\ analyses
the array of eigenvalues of $H_{\rm intl}$ from successive values of \var{JTOT} and \var{IBLOCK}
and constructs a master array of pair level energies in the array \var{ELEVEL}. Channels that have
energies that are degenerate to within a convergence criterion \inpitem{DEGTOL} are assumed to
originate from the same pair level. \MOLSCAT\ assigns a level index to each open channel, stored in
the array \var{INDLEV}, and accumulates cross sections between the stored levels. If
$\inpitem{IPRINT}\ge10$, \MOLSCAT\ prints the assignment of channels to levels for each
\var{JTOT}/\var{IBLOCK} combination. After the loops over \var{JTOT} and/or \var{IBLOCK}, \MOLSCAT\
has a complete list of levels that are open at one or more collision energies. It prints this list,
with the corresponding energies, immediately before the final output of elastic and inelastic
cross sections.

Pair levels are stored only if they will be energetically accessible (open) at one or more
energies in the run. If collision energies are specified with respect to a non-zero reference
energy \inpitem{EREF}, it is important that it is the same for all values of \var{JTOT} and all
symmetry blocks. This may be achieved by specifying \inpitem{EREF} either explicitly or via the
array \inpitem{MONQN}. If the reference energy is obtained from \inpitem{IREF}, cross sections are
calculated only in runs that are limited to a single value of \var{JTOT} and a single symmetry
block.

The value of \inpitem{MXSIG} limits the number of pair levels that are stored and used to calculate
cross sections. Cross sections are calculated between the \emph{first} \inpitem{MXSIG} levels
encountered, not the \emph{lowest} \inpitem{MXSIG} levels, so it is important for \inpitem{MXSIG}
to be large enough to reserve storage for all levels that are open at any energy in the run.

This scheme for identifying levels is compatible with calculations at multiple collision energies,
but \emph{not} with calculations at multiple values of external fields, since threshold energies
depend on external fields. \MOLSCAT\ calculates cross sections only if there is just one set of
EFVs in a run\footnote{\MOLSCAT\ can, however, calculate cross sections for scans across the
potential scaling factor}.

For basis sets in which $H_{\rm intl}$ is non-diagonal, \MOLSCAT\ has no knowledge of any monomer
quantum numbers associated with individual levels. The levels must be identified on physical
grounds, based on their energies. If necessary, the eigenvectors that connect the asymptotic
channels to the primitive basis functions (for each \var{JTOT} and \var{IBLOCK}) may be printed as
described in section \ref{basis:off-diag}.

For basis sets in which $H_{\rm intl}$ is non-diagonal, \MOLSCAT\ has no way of knowing the
degeneracy of the levels. The cross sections are \emph{summed} over all levels that are found to be
degenerate (according to tolerance \inpitem{DEGTOL}), but they are not divided by the degeneracy of
the incoming level as in Eq.\ \ref{eqsigdef}. If degeneracy-averaged cross sections are required,
they must be divided by the appropriate degeneracy factor for the incoming level by hand.

\subsection{Partial cross sections}\mylabel{pCS}

The contributions to state-to-state cross sections from individual values of \var{JTOT} and
\var{IBLOCK} (partial cross sections) are calculated after each propagation and printed if
$\inpitem{IPRINT}\ge5$.  There was previously an option to print them out to unit \iounit{IPARTU}
if $\inpitem{IPARTU}>0$; this option is currently disabled, but would not be difficult to revive if
needed.

For basis sets off-diagonal in $H_{\rm intl}$ and/or $\hat L^2$, the levels between which partial
cross sections are calculated are limited by \inpitem{MXSIG} as described above.

\subsection{Integral cross sections}\mylabel{tCS}

The state-to-state partial cross sections are accumulated to form the state-to-state integral cross
sections for each collision energy. By default, \MOLSCAT\ reserves storage for, and accumulates,
integral cross sections between every pair of levels included in the calculation. This may use a
substantial amount of storage, and is often not required, particularly if some of the levels are
energetically inaccessible at all collision energies. If $\inpitem{MXSIG}>0$, only cross sections
between the first \inpitem{MXSIG} levels are calculated.

If $\inpitem{ISIGU}>0$, a direct access-file is opened on unit \iounit{ISIGU} and is used to store
the accumulated state-to-state cross sections. It is updated after each propagation, so contains
useful information on the calculation so far, even if the program terminates abnormally.

\subsubsection{Automated convergence with respect to \var{JTOT}}

Integral cross sections are calculated by accumulating cross sections for all
possible values of \var{JTOT} from $\inpitem{JTOTL}$ to $\inpitem{JTOTU}$ in
steps of $\inpitem{JSTEP}$. However, if $\inpitem{JTOTU} \ge 99999$ (or
$\inpitem{JTOTU} < \inpitem{JTOTL}$) (the default)), the loop terminates when
contributions from successive values of \var{JTOT} are negligible. Termination
of the loop is controlled by the variables \inpitem{DTOL}, \inpitem{OTOL} and
\inpitem{NCAC}: \var{JTOT} starts at \inpitem{JTOTL} and is incremented by
\inpitem{JSTEP} until \inpitem{NCAC} successive values of \var{JTOT} each
contribute less that \inpitem{DTOL} to any diagonal cross section and less than
\inpitem{OTOL} to any off-diagonal cross section.

Contributions to cross sections sometimes oscillate as a function of
\var{JTOT}, and may sometimes remain low for several consecutive values of
\var{JTOT}. The program may mistakenly detect convergence and terminate too
early if \inpitem{NCAC} is too small.

The final cross sections are multiplied by \inpitem{JSTEP} to account
(approximately) for incomplete sampling of \var{JTOT} values, unless
$\var{JHALF}=0$, indicating that \var{JTOT} is not used for $J_{\rm tot}$.

This option should be used with care to ensure that the calculation converges
before the job runs out of time. Note that elastic cross sections usually
converge very much more slowly than inelastic ones.

It is sometimes desired to test the convergence of cross sections with respect
to \var{JTOT} for a group of energies together, rather than one energy at a
time. This option is controlled by the namelist item \inpitem{NNRGPG}, which
specifies the number of (successive) energies to be considered together.

\subsection{Restarting a run to calculate cross sections}\mylabel{restart}

If a particular calculation terminates prematurely (e.g., crashes or runs out of time), but the S
matrices have been written to unit \iounit{ISAVEU}, it is possible to restart the run from midway
through the sequence of propagations. If $\inpitem{IRSTRT}>0$, \MOLSCAT\ checks that the results
stored on \iounit{ISAVEU} are for the same calculation and recalculates cross sections from S
matrices included in the file. It then continues with subsequent propagations and S matrices.
\begin{itemize}[nosep]
\item If $\inpitem{IRSTRT}=3$, the program restarts after the last complete propagation;
\item If $\inpitem{IRSTRT}=2$, the program restarts after the last complete symmetry block;
\item If $\inpitem{IRSTRT}=1$, the program restarts after the last completed value of
    \var{JTOT}. \item If $\inpitem{IRSTRT}=-1$, the program extends the set of values for
    \var{JTOT} upwards to a new (and larger) value of \inpitem{JTOTU}.
\end{itemize}


\section{Line-shape cross sections}\mylabel{pressbroad}

For many of the built-in interaction types, \MOLSCAT\ can calculate the cross
sections that characterise pressure broadening, pressure shifting and
pressure-induced mixing of spectroscopic lines. Each cross section is labelled
by a pair of spectroscopic lines, $n_{\rm a}\rightarrow n_{\rm b}$ and $n'_{\rm
a}\rightarrow n'_{\rm b}$. The real and imaginary parts of diagonal cross
sections ($n_{\rm a}= n'_{\rm a}$ and $n_{\rm b}= n'_{\rm b}$) describe
pressure broadening and shifting (respectively) of an isolated line, while the
off-diagonal matrix elements describe line mixing.

As an example, the line-shape cross sections for close-coupling calculations on
atom + vibrating diatom collisions, where $n$ represents vibrational and
rotational quantum numbers $v$ and $j$, are \cite{Green:1979:vibrational}
\begin{IEEEeqnarray}{rCll}\label{eqnPRBR}
\sigma(v_{\rm a}j_{\rm a},v_{\rm b}j_{\rm b}|v_{\rm a}'j_{\rm a}',v_{\rm
b}'j_{\rm b}';E_{\rm kin})& =&\frac{\pi}{k_j^2} \sum_{\substack{LL'\\J_{\rm
a}J_{\rm b}}} &\sixj{j_{\rm a}}{q}{j_{\rm b}}{J_{\rm b}}{L}{J_{\rm
a}}\sixj{j_{\rm a}'}{q}{j_{\rm b}'}{J_{\rm b}}{L'}{J_{\rm a}}\\\nonumber
&&\hfill\times&\left[I -S_{i_{\rm b}f_{\rm b}}^{J_{\rm b}}(E_{\rm b}^{\rm
kin})^* S_{i_{\rm a}f_{\rm a}}^{J_{\rm a}}(E_{\rm a}^{\rm kin})\right],
\end{IEEEeqnarray}
where $E_{\rm x}^{\rm kin}=E_{\rm kin}+E_{\rm x}$ and $I$ is shorthand for
$\delta_{v_{\rm a}v_{\rm a}'}\delta_{v_{\rm b}v_{\rm b}'}\delta_{j_{\rm
a}j_{\rm a}'}\delta_{j_{\rm b}j_{\rm b}'}\delta_{L,L'}$. $q$ is the tensor
order of the spectroscopic transition, and unprimed and primed quantities refer
to values before and after a collision. It should be noted that the two S
matrices involved here are evaluated at the same {\em kinetic} energy $E_{\rm
kin}$ but different {\em total} energies.

Calculations of line-shape cross sections are implemented for the
\basisitem{ITYPE}s indicated with reference numbers or black tick marks as
follows.

\definecolor{light-gray}{gray}{1.0}
\newcommand\cgmark{{\color{light-gray}\cmark}}
\newcommand\xgmark{{\color{light-gray}\xmark}}
\newcommand\gna{{\color{light-gray}N/A}}
\newcommand\na{{N/A}}
\begin{tabular}{|lc|c|c|c|c||c|}
\hline
                             & &CC       &EP       &CS       &DLD      &IOS\\
                             & &(+0)     &(+10)    &(+20)    &(+30)    &(+100)\\
\hline
rigid rotor + atom           &1&\cite{Shafer:1973}   &\cite{Fisanick-Englot:1975} &\cite{Goldflam:pb:1977}   &\cite{Green:pb:1977}   &\cite{Goldflam:IOS:1977}\\
vibrating rotor + atom       &2&\cite{Shafer:1973, Blackmore:1988}   &\cmark   &\cmark   &\cgmark  &\cgmark\\
rigid rotor + rigid rotor    &3&\cite{Green:HCl-H2:1977}   &\cgmark  &\cgmark  &\xgmark  &\cgmark\\
asymmetric top + rigid rotor &4&\cgmark   &\xgmark  &\cgmark  &\xgmark  &\xgmark\\
symmetric top + atom         &5&\cite{Green:1976}   &\cmark   &\cite{Green:1976}   &\xgmark  &\cite{Green:sym-top-IOS:1979}\\
asymmetric top + atom        &6&\cite{Green:1976}   &\cmark   &\cite{Green:1976}   &\xgmark  &\cgmark\\
vibrating rotor + atom       &7&\cite{Shafer:1973, Hutson:sbe:1984}   &\cmark   &\cmark   &\cgmark  &\xgmark\\
atom + corrugated surface    &8&\na  &\na      &\na      &\na      &\na\\
\hline
\end{tabular}

Line-shape cross sections are calculated if $\inpitem{NLPRBR} > 0$ on input;
the value of \inpitem{NLPRBR} specifies the number of (pairs of) spectroscopic
lines for which line-shape calculations are required. The lines themselves are
specified by the array \inpitem{LINE}, of $4\times\inpitem{NLPRBR}$ elements.
Each successive quartet of elements in \inpitem{LINE} specifies the two pairs
of levels involved in the two transitions ($n_{\rm a}$, $n_{\rm b}$, $n_{\rm
a}'$, $n_{\rm b}'$) as pointers to the arrays \var{JLEVEL} and \var{ELEVEL}; see
section \ref{ConstructBasis}) for details.

Line-shape cross sections require S-matrix elements involving the initial and
final (spectroscopic) levels at the same kinetic (not total) energy. If
$\inpitem{IFEGEN} > 0$, the program treats the input \inpitem{ENERGY} values as
kinetic energies ($E_{\rm kin}$ in Eq.~\ref{eqnPRBR}), and generates the
necessary total energies for the lines requested (i.e., $E^{\rm kin}_{\rm a}$
and $E^{\rm kin}_{\rm b}$ in Eq.~\ref{eqnPRBR}). If $\inpitem{IFEGEN} = 0$,
only those requested lines for which cross sections can be constructed from the
total energies actually specified by \inpitem{NNRG} and \inpitem{ENERGY} are
calculated. Only total energies needed for the requested line-shape
calculations are retained. In addition, specifying $\inpitem{IFEGEN} > 1$
suppresses calculations for individual combinations of \var{JTOT}, \var{IBLOCK}
and energy that do not contribute S matrices needed for the requested
line-shape cross sections. \emph{Warning:} some of the state-to-state integral
cross sections may be incomplete (missing contributions from some values of
\var{JTOT} and \var{IBLOCK}) when $\inpitem{IFEGEN} > 1$.

The tensor order of the spectroscopic transition (i.e., 1 for dipole
transitions and 0 or 2 for isotropic or anisotropic Raman scattering
respectively) is specified in the input array \inpitem{LTYPE}. If the default
value is found, \inpitem{LTYPE} is calculated as the difference between the
rotational quantum numbers of the levels specified (taken from the \var{JLEVEL}
array). The default is usually adequate \emph{except} for Q-branch lines in
$\var{ITYP}=5$ or 6 or anisotropic Raman spectra with $\Delta j\ne2$.

\section{Line-shape cross sections in the IOS approximation}\mylabel{IOScs}

This code is at present unsupported, but is left in case someone wants to
develop it.

\section{Locating scattering resonances as a function of energy}\mylabel{energyconv}

This section is for resonances that appear in the energy dependence of S matrices. Resonances that
appear as a function of external field at constant kinetic energy, such as magnetically tunable
Feshbach resonances in low-energy collisions, should be located as described in section
\ref{fieldconv}.

These options are not supported for IOS calculations ($\basisitem{ITYPE} > 100$).

Scattering resonances and predissociating states of Van der Waals molecules appear as
characteristic features in the energy dependence of S matrices. The eigenphase sum ${\cal S}$,
which is the sum of phases of the eigenvalues of the S matrix \cite{Ashton:1983}, follows a
Breit-Wigner form in the vicinity of a resonance,
\begin{equation}
{\cal S}(E) = {\cal S}_{\rm bg}(E) + \arctan\left(\frac{\Gamma}{2(E_{\rm res}-E)}\right),
\end{equation}
where $E_{\rm res}$ is the energy of the resonance, $\Gamma$ is its width, and
${\cal S}_{\rm bg}(E)$ is a slowly varying background phase. If $\inpitem{IPHSUM}>0$, the
eigenphase sum is calculated and a summary of the eigenphases is output on unit \iounit{IPHSUM}.

The product state distribution from decay of a quasibound state is characterized by a set of
partial widths $\Gamma_i$ for each open channel $i$. For an isolated narrow resonance, the partial
widths sum to the total width $\Gamma$. Across a resonance, each S-matrix element describes a
circle in the complex plane \cite{Brenig:1959, Taylor:1972},
\begin{equation}
S_{ii'}(E)=S_{\textrm{bg},ii'}(E)-\frac{\textrm{i}g_ig_{i'}}{E-E_\textrm{res}+\textrm{i}\Gamma/2}.
\label{eq:S_circle}
\end{equation}
The partial widths are defined as real quantities, $\Gamma_i=|g_i|^2$, and the circles in the
complex plane have radii $\sqrt{\Gamma_i\Gamma_{i'}}/\Gamma$.

In cases where the background phase does not vary significantly across the width of the resonance,
and the location of the resonance is approximately known in advance, \MOLSCAT\ can converge on the
resonance and obtain its position, width and background phase using the algorithm of Frye and
Hutson \cite{Frye:quasibound:2020}. If the energy dependence of the background phase can be
estimated independently, a user-supplied routine \prog{BCKGRD} may be provided to subtract it from
the calculated phase.

If $\inpitem{IECONV} = 4$, \MOLSCAT\ performs scattering calculations at the first 3 energies
specified as described in Chapter \ref{Energy}. It then uses the algorithm of Frye and Hutson
\cite{Frye:quasibound:2020} to attempt to converge on and characterise the resonance. The variables
\inpitem{TLO}, \inpitem{THI} and \inpitem{XI} set values for the parameters $t_{\rm lo}$, $t_{\rm
hi}$ and $\xi$ of ref.\ \cite{Frye:quasibound:2020}. The characterisation terminates when one of
the three points is within \inpitem{DTOL} of the estimated value of $E_\textrm{res}$ and the other
two are within the ranges prescribed by \inpitem{TLO}, \inpitem{THI} and \inpitem{XI}. It also ends
if \var{MXLOC} (set in module \module{sizes}) propagations are performed without convergence. The
current estimates of the resonance parameters are printed at each step.

If $\inpitem{IECONV} = 5$, \MOLSCAT\ uses an algorithm based on the fully complex procedure of
ref.\ \cite{Frye:resonance:2017} to attempt to converge on a resonance in a single diagonal
S-matrix element and extract parameters $E_{\rm res}$, $\Gamma$, $g_i^2$ and $S_{\textrm{bg},ii}$.
The channel concerned is identified by its open-channel index \inpitem{ICHAN} (section
\ref{scatchan}). The sequence of points is controlled by \inpitem{TLO}, \inpitem{THI} and
\inpitem{XI} in the same way as for $\inpitem{IECONV} = 4$.

The default values $\inpitem{TLO}=-0.1$ and $\inpitem{THI}=1.0$ are usually appropriate for narrow
resonances. However, larger values of $t_\textrm{lo}$ and $t_\textrm{hi}$ are sometimes needed for
extremely narrow resonances, to reduce the effects of numerical noise, and smaller values may be
needed for very wide resonances, to reduce variation in the background across the range. The
default value $\inpitem{XI}=0.25$ is usually appropriate for general resonance characterisation,
but much smaller values may be needed for special purposes (such as least-squares fitting to
determine interaction potentials) to avoid small discontinuities in calculated resonance properties
as a function of potential parameters.

If convergence succeeds when \inpitem{IECONV} is either 4 or 5, \MOLSCAT\ also calculates and
prints partial widths based on the two final points closest to and furthest from $E_{\rm res}$.

To handle cases where the automated algorithm is unsatisfactory, \MOLSCAT\ outputs K matrices
instead of S matrices on channel \inpitem{ISAVEU} when $\inpitem{IPHSUM}>0$. If the run is for a
single value of \var{JTOT} and a single symmetry block, the K-matrix file is suitable for input to
the separate program \prog{SAVER}, which can accumulate results from several different runs. The
accumulated K matrices may then be processed by external program \prog{RESFIT}
\cite{Hutson:resfit:2007} to obtain resonance positions, widths and partial widths.

Broad resonances may often be identified from features in cross sections
(usually peaks, but sometimes troughs or more complicated features). However,
very narrow resonances (corresponding to long-lived quasibound states) can be
hard to find. This is compounded by the fact that, although the eigenphase sum
${\cal S}$ increases smoothly by $\pi$ across the width of a resonance, it is
numerically defined only modulo $\pi$ when evaluated from a K or S matrix.
\MOLSCAT\ chooses the integer part of ${\cal S}/\pi$ so that it is around at 10
at the first energy, and then chooses the integer part at each subsequent
energy to be within 0.5 of the value at the previous energy. Thus, if
successive energies fall more than about $\Gamma/5$ below and above a
resonance, the resonance may appear as an irregularity in ${\cal S}/\pi$,
rather than a smooth increase through 1. If the spacing is much larger than
$\Gamma$, the presence of the resonance may be hard to spot.

\MOLSCAT\ has a capability to identify a resonance from calculations in its wings, and
to step towards it when the \emph{curvature} of the eigenphase sum is dominated by the resonant
contribution. This may succeed in locating a resonance from further away than the automated
algorithm of ref.\ \cite{Frye:quasibound:2020}, particularly in the presence of an unknown
background slope. If $\inpitem{IECONV}=-5$ or $\inpitem{NNRG}<0$, \MOLSCAT\ generates 5 initial
energies from \inpitem{ENERGY}(1) and \inpitem{DNRG} and performs $n$ groups of 5 equally spaced
calculations, where $n$ is the integer part of $|\inpitem{NNRG}|/5$. After each group, the program
tries to interpret the 5 eigenphase sums as the ``tail" of a resonance, and estimate the width and
the position of the resonance centre. These estimates are then used to choose the next group of 5
energies. This option can be useful if a reasonably good estimate of the resonance energy is
already available, and in favourable cases may succeed in converging towards a narrow resonance
from as far as $10^5$ widths away. However, convergence is \emph{not} guaranteed, and it is not
usually useful to do more than 3 sets of 5 energies in a single run. Furthermore, once the
eigenphase sums span the resonance (or come very close to it), the automated algorithm fails to
provide further improvement. At this point the resonance should be characterised using
$\inpitem{IECONV} = 4$ or 5, or in difficult cases by performing additional calculations on an
appropriate equally spaced grid of energies.

Searches for resonances in the eigenphase sum as a function of energy should
usually be done with log-derivative propagators, since they have good stability
in the presence of closed channels. Since many energies are needed to
characterise a resonance, it is usually most efficient to use the \inpitem{ISCRU}
option to save energy-independent matrices on a scratch file. For calculations
with $\inpitem{JTOTU} = \inpitem{JTOTL}$ and $\inpitem{IBFIX} > 0$, the
\iounit{ISCRU} file from one run may be used as input for the next run at a
different set of energies, as described in section \ref{CommIII:iscru}.

\section{Low-energy collision properties}\mylabel{inputres}

\subsection{Scattering lengths/volumes}\mylabel{scatlen}

\MOLSCAT\ calculates energy-dependent complex scattering lengths (or volumes or
hypervolumes for $L>0$) using Eq.~\ref{eq:scatln} as described in section
\ref{theory:lowE}. These are output in units $(\inpitem{RUNIT})^n$, where $n=1$
for lengths ($L=0$ channels), $n=3$ for volumes ($L=1$ channels) and $n=4$ for
hypervolumes ($L=2$ channels). They are calculated only for `low-energy'
channels where the wavevector $k$ is less than 0.01 $(\inpitem{RUNIT})^{-1}$.
This rather arbitrary threshold value may be changed by altering the value of
the parameter \var{AWVMAX} in subroutine \prog{DRIVER} (in the file
\file{mol.driver.f}).

\subsection{Scanning the scattering length/volume over an EFV}\mylabel{fieldscan}

If $\inpitem{IFCONV} = 0$ (the default), \MOLSCAT\ performs a scan from
\inpitem{FLDMIN} to \inpitem{FLDMAX} in steps of \inpitem{DFIELD}. This option
may be used with any value of \inpitem{NNRG}.

\subsection{Characterising a resonance in the scattering length as a function of EFV}
\mylabel{fieldconv}

At low collision energy, scattering resonances appear as characteristic features in
the scattering length as a function of external field. In the absence of inelastic scattering, the
scattering length shows a simple pole, but in the presence of inelasticity the behaviour is more
complicated \cite{Hutson:res:2007}.

If $\inpitem{IFCONV} = 1$, 2 or 3, \MOLSCAT\ attempts to converge on and characterise a resonance
in the scattering length/volume, as a function of the varying EFV, for the incoming channel with
open-channel index \inpitem{ICHAN} (section \ref{scatchan}). It calculates scattering
lengths/volumes at 3 different values (\inpitem{FLDMIN}, \inpitem{FLDMAX}, and [\inpitem{FLDMIN} +
\inpitem{FLDMAX}]/2), and uses the algorithms of Frye and Hutson \cite{Frye:quasibound:2020,
Frye:resonance:2017} to converge on and characterise the resonance.

This option requires $\inpitem{NNRG}=1$.

\inpitem{IFCONV} specifies the type of resonance, and which algorithm is used to characterise it:
\begin{description}
\item[$\inpitem{IFCONV}=1$]{indicates that the resonance is elastic, with a
    pole in scattering length,
\begin{equation}
a(B)=a_{\rm bg}\left(1-\frac{\Delta}{B-B_{\rm res}}\right),
\label{eq:pole}
\end{equation}
    where $B$ represents the varying EFV. In this case the program uses the
    elastic procedure of ref.\ \cite{Frye:resonance:2017}. The parameters obtained are the
    pole position $B_{\rm res}$, the resonance width $\Delta$ and the real background
    scattering length $a_{\rm bg}$.}
\item[$\inpitem{IFCONV}=2$]{indicates that the resonance is weakly decayed, meaning that the
    pole is suppressed and there is a peak in inelastic scattering at resonance, but that there
    is no significant background inelasticity away from resonance. In this case
    the program uses the weakly inelastic procedure of ref.\ \cite{Frye:resonance:2017}. the
    parameters obtained are as for $\inpitem{IFCONV}=1$ but with the addition of the resonant
    scattering length $a_{\rm res}$, which is real and may be processed to obtain an inelastic
    width $\Gamma_{\rm inel}$.}
\item[$\inpitem{IFCONV}=3$]{indicates that the resonance is strongly decayed, meaning that
    there is significant background inelasticity away from resonance. The resonant peak in
    inelastic scattering is then asymmetric. In this case the program uses the
    fully complex procedure of ref. \cite{Frye:resonance:2017}. The parameters obtained are
    $B_{\rm res}$, $\Delta$ and the real and imaginary parts of $a_{\rm res}$ and $a_{\rm
    bg}$.}
\end{description}

The values $\inpitem{IFCONV}=4$ and 5 are also implemented, to converge on resonances in the
eigenphase sum and in a single S-matrix element as a function of EFV, by analogy with
$\inpitem{IECONV}=4$ and 5 described in section \ref{energyconv}. It should be noted that the value
of $B_{\rm res}$ obtained in this way differs from that obtained from the scattering length, except
at zero kinetic energy.

For any positive value of $\inpitem{IFCONV}$, the variables \inpitem{TLO}, \inpitem{THI} and
\inpitem{XI} set values for the parameters $t_{\rm lo}$, $t_{\rm hi}$ and $\xi$ of ref.\
\cite{Frye:quasibound:2020}.\footnote{The implementation from version 2020.0 onwards differs
slightly from that in version 2019.0 and 2019.1, with \inpitem{TLO}, \inpitem{THI} and \inpitem{XI}
replacing \code{TOLMIN} and \code{TOLMAX}.} The considerations that apply to the choice of
\inpitem{TLO}, \inpitem{THI} and \inpitem{XI} are the same as described in section
\ref{energyconv}; the defaults are usually adequate, except for wide resonances. The
characterisation terminates when one of the three points is within \inpitem{DTOL} of the estimated
value of $B_\textrm{res}$ and the other two are within the ranges prescribed by \inpitem{TLO},
\inpitem{THI} and \inpitem{XI}. It also ends if \var{MXLOC} (set in module \module{sizes})
propagations are performed without convergence. The current estimates of the resonance parameters
are printed at each step.

By default the algorithms neglect variation in $a_{\rm bg}(B)$. If the $B$-dependence can be estimated
independently, a user-supplied routine \prog{BCKGRD} may be provided to subtract it from the calculated
scattering length.

In earlier work, a resonance width $\Delta_0$ was sometimes defined such that
$a(B_{\rm res}+\Delta_0)=0$, so that the scattering length crosses zero a distance $\Delta_0$ from
the pole. This is equivalent to the definition based on Eq.\ \ref{eq:pole} when $a_{\rm bg}$ does
not vary across the width of the resonance, as is approximately true for most narrow resonances.
However, $\Delta_0$ does not properly capture the behaviour near the pole when there is significant
variation in $a_{\rm bg}(B)$. It may in principle be obtained from the present algorithm by setting
\inpitem{XI} to a small value and $\inpitem{THI}=+1.0$ for a resonance with positive $\Delta$ or
$\inpitem{THI}=-1.0$ for a resonance with negative $\Delta$. However, such a large value of
\inpitem{THI} may give poor convergence for a wide resonance. If this occurs, and the value
$\Delta_0$ based on the zero crossing is truly required, it may be obtained by locating the
position of the pole with a smaller value of $|\inpitem{THI}|$ and converging separately on the
position of the zero crossing as described in section \ref{fieldval}.

\subsection{Converging on a specific value of the scattering length/volume}\mylabel{fieldval}

If $\inpitem{IFCONV} = -1$, \MOLSCAT\ attempts to converge on a value of the
EFV where the scattering length in open channel \inpitem{ICHAN} satisfies
$a-\inpitem{AZERO}=0$. This can be used to converge on a zero crossing if
$\inpitem{AZERO}=0.0$. Convergence is attempted only if the value of
$a-\inpitem{AZERO}$ changes sign between \inpitem{FLDMIN} and \inpitem{FLDMAX}.
Convergence uses the Van Wijngaarden-Dekker-Brent method \cite{VWDB} and
terminates when the predicted step is less than \inpitem{DTOL}.

This option requires $\inpitem{NNRG}=1$.

\subsection{Effective range}\mylabel{effrange}

In the absence of inelastic scattering, the $s$-wave scattering length $a_0$ is
real. The near-threshold dependence of the $s$-wave scattering phase shift
$\eta$ on kinetic energy $E_{\rm kin}$ or wavevector $k$ is often characterised
by an effective-range expansion at small collision momentum
\begin{equation}
k\cot\eta(k)=-\frac{1}{a_0(0)}+\textstyle{\frac{1}{2}}r_{\rm eff}k^2+\cdots
\end{equation}
Re-expressing this using the definition $a_0(k)=-\tan\eta(k)/k$ gives two
different expressions for the scattering length in terms of the effective range
$r_{\rm eff}$:
\begin{equation}
[a_0(k)]^{-1}=[a_0(0)]^{-1}-\textstyle{\frac{1}{2}}r_{\rm eff}k^2+\cdots, \label{eq:eff-range1}
\end{equation}
or
\begin{equation}
a_0(k)=a_0(0)+\textstyle{\frac{1}{2}}r_{\rm eff}[a_0(0)]^2k^2+\cdots.\label{eq:eff-range2}
\end{equation}
At external fields far from a resonance or zero crossing in $a_0$, either of these relationships
may be used to evaluate $r_{\rm eff}$, using finite differences between scattering lengths
evaluated at different kinetic energies. However, Eq.~\ref{eq:eff-range1} is numerically unstable
near a zero crossing and Eq.~\ref{eq:eff-range2} is numerically unstable near a pole. For this
reason, for each energy after the first, \MOLSCAT\ calculates the effective range for open channel
\inpitem{ICHAN} from both
\begin{itemize}[nosep]
\item{a quadratic expansion of $1/[a_0(k)]$ at $\inpitem{ENERGY}(i)$
    ($i>1$) and \inpitem{ENERGY}(1) and}
\item{a quadratic expansion of $a_0(k)$.}
\end{itemize}
\inpitem{ENERGY}(1) should be small enough for $a_0(k)$ to be very close to
$a_0(0)$ but not so small that numerical noise dominates the evaluation of
$1-S_{00}$.

This option requires \inpitem{NNRG} to be greater than 1 and so must be done
separately from characterisation of resonances or convergence on specific
values. The effective range is calculated for the second and subsequent
energies, when the energy is low enough that the scattering length itself is
calculated, as described in section \ref{scatlen}.

\section{Scattering wavefunctions}\mylabel{calcwaveM}

If $\inpitem{IWAVE}>0$ and $\inpitem{IPROPS}=6$ ($\inpitem{IPROPL}=0$ or 6), \MOLSCAT\ calculates
the energy-normalised multichannel scattering wavefunction that is incoming only in channel
\inpitem{ICHAN}. The wavefunction is written on unit \iounit{IWAVE} in the format described in
section \ref{CommII:bwave}. \cbend

\chapter[\texorpdfstring{Controlling bound-state calculations
({\color{\bcol}\BOUND} and {\color{\fcol}\FIELD} only)} {\ref{processbound}:
Controlling bound-state calculations (BOUND and FIELD only)}] {Controlling
bound-state calculations\chaptermark{Controlling bound-state calculations}}
\chaptermark{Controlling bound-state calculations}\mylabel{processbound}

\cbcolor{\bfcol}\cbstart
\section{State numbers and the node count}\mylabel{nodecount}

The multichannel node count used by \BOUND\ and \FIELD\ was introduced by
Johnson \cite{Johnson:1978}. It is defined as a function of energy and is equal
to the number of bound-state solutions of a set of coupled equations that lie
below that energy.  It is evaluated during propagation of the coupled
equations.

In versions of \BOUND\ and \FIELD\ before version 2019.0, the algorithm used
required an additional propagation from either $R_{\rm min}$ or $R_{\rm max}$
to $R_{\rm mid}$. From version 2019.0, this additional propagation is no longer
performed; instead the node count is evaluated by summing the node counts from
the outward and inward propagations and adding the number of negative
eigenvalues of the log-derivative matching matrix.

The node count algorithm is quite reliable, but has been known to generate
additional nodes in the classically forbidden region. For this reason,
convergence algorithms based on the node count very occasionally fail.

\BOUND\ and \FIELD\ attempt to label each eigenvalue with a state number
related to the node count. However, the node count as defined above
\emph{changes} at an eigenvalue. The programs define the state number as the
node count at an energy just above the state.

The procedure used to assign a state number is not completely reliable, because
rounding errors may cause the node count to change slightly above or below the
eigenvalue. For this reason, the programs print warnings if the node count is
not as expected, but this does not necessarily indicate that the bound-state
position is in error and the programs do not terminate when it happens. \cbend

\section[\texorpdfstring{Locating bound states with {\color{\bcol}\BOUND}}
{Locating bound states with BOUND}] {Locating bound states with
\BOUND\sectionmark {Locating bound states with \BOUND}}\sectionmark {Locating
bound states with \BOUND}\mylabel{eigenB}

\cbcolor{\bcol}\cbstart \BOUND\ begins by propagating the log-derivative matrix
at \inpitem{EMIN} and \inpitem{EMAX} and using the node counts to ascertain how
many bound states to locate. The program searches for bound states with node
counts between \inpitem{NODMIN} and \inpitem{NODMAX} that lie between
\inpitem{EMIN} and \inpitem{EMAX}.  It proceeds initially by bisection, until
it has identified a range of energies within which the node count changes by
exactly 1.  It continues to use bisection until this range is smaller than
100$\times$\inpitem{DTOL}, or until the minimum eigenvalue of the matching
matrix changes from negative at the low-energy bound to positive at the
high-energy bound. At this point it switches to using the Van
Wijngaarden-Dekker-Brent algorithm \cite{VWDB} to converge on a bound state.
Convergence terminates when the predicted step size is less than \inpitem{DTOL}
(in \namelist{\&INPUT} energy units). The number of
energy values allowed in converging on each bound state is limited by the
internal variable \var{NITER}, which is currently set to 20. This value should
be sufficient if \inpitem{RMATCH} is chosen appropriately, unless exceptionally
stringent convergence is required.

When searching for closely spaced states, it is important that 100$\times$\inpitem{DTOL} is significantly smaller than the spacing between the states, as otherwise the program may decide that bisection has failed before it switches to the Van Wijngaarden-Dekker-Brent algorithm.

In special circumstances it may be desirable to carry out a scan of the node
count and the smallest eigenvalue of the matching matrix as a function of
energy. This is done by setting the absolute value of \inpitem{DNRG} to less
than the absolute value of $\inpitem{EMAX}-\inpitem{EMIN}$. \BOUND\ then scans
over energies in the range \inpitem{EMIN} to \inpitem{EMAX} using a step size
of \inpitem{DNRG}, without attempting to converge on eigenstates.

The namelist item \inpitem{MXCALC} limits the number of energy values in a
complete run. \cbend

\section[\texorpdfstring{Locating bound states with {\color{\fcol}\FIELD}}
{Locating bound states with FIELD}] {Locating bound states with \FIELD
\sectionmark{Locating bound states with \FIELD}}
\sectionmark{Locating bound states with \FIELD}\mylabel{eigenF}

\cbcolor{\fcol}\cbstart
\FIELD\ reverses the order of the loops over energy and
EFV value, and locates values of the EFV at which bound states with a specified
energy exist. \FIELD\ is particularly useful for estimating resonance positions
at the lowest threshold of a particular symmetry, for use in subsequent
scattering calculations, and also for mapping bound states whose energies vary
very fast with the EFV.

An additional complication for \FIELD\ is that the node count is not guaranteed
to be a monotonic function of the EFV (though it often is, at least locally).
This arises because states may pass through the energy of the calculation from
either higher or lower values of the EFV. If the absolute value of
\inpitem{DFIELD} is less than the absolute value of
$\inpitem{FLDMAX}-\inpitem{FLDMIN}$, \FIELD\ scans over EFV values in the range
\inpitem{FLDMIN} to \inpitem{FLDMAX} using a step size of \inpitem{DFIELD},
without attempting to converge on eigenstates. This enables the user to
identify EFV ranges within which the node count increases or decreases
monotonically.

The algorithm used to converge on bound states is essentially the same as
described above for \BOUND: \FIELD\ calculates the node counts at
\inpitem{FLDMAX} and \inpitem{FLDMIN}. It then searches for and attempts to
converge on bound states with node counts between \inpitem{NODMIN} and
\inpitem{NODMAX} that lie between \inpitem{FLDMIN} and \inpitem{FLDMAX}.
Convergence for each bound state terminates when the predicted step size is
less than \inpitem{DTOL} (which is interpreted as having the same units as the
EFV concerned).\cbend

\cbcolor{\bfcol}\cbstart
\section{Harmonic confinement}\mylabel{confinement}

Cold atoms and molecules are sometimes confined in a trapping potential created with external fields,
as for example in an optical lattice or tweezer. If the confining potential is harmonic, the motion may be factorized at least approximately into terms involving the relative and centre-of-mass coordinates of the pair. If the atomic masses are $m_1$ and $m_2$ and the corresponding single-atom harmonic frequencies are $\omega_1$ and $\omega_2$, the frequencies for relative and centre-of-mass motion are \cite{Deuretzbacher:2008}
\begin{eqnarray}
\omega_\textrm{rel} &= \sqrt{\left({m_2\omega_1^2+m_1\omega_2^2}\right)\big/\left(m_1+m_2\right)},\\
\omega_\textrm{com} &= \sqrt{\left(m_1\omega_1^2+m_2\omega_2^2\right)/\left(m_1+m_2\right)},
\label{eq:conf-freq}
\end{eqnarray}
together with a coupling term between the two that is proportional to
\begin{equation}
\Delta\omega_ = \sqrt{\omega_1^2-\omega_2^2}.
\end{equation}
The relative and centre-of-mass motions are thus uncoupled if the atoms are identically trapped and the trapping is harmonic. If the confinement is non-spherical, these equations apply along each principal axis $x$, $y$, $z$ of the trap.

If $\inpitem{CONFRQ} > 0.0$ or $\inpitem{CONLEN} > 0.0$, \BOUND\ and \FIELD\ add a spherically symmetric harmonic potential of the form
\begin{equation}
V_{\rm confine}(R) = \frac{1}{2}\mu \omega_\textrm{rel}^2 R^2 = \frac{\hbar^2}{2\mu} \frac{R^2}{\beta_\textrm{rel}^4},
\end{equation}
where $\omega_\textrm{rel}$ is the harmonic frequency and $\beta_\textrm{rel} = \sqrt{\hbar/(\mu\omega_\textrm{rel})}$ is the harmonic length for \emph{relative} motion. \inpitem{CONFRQ} specifies $\hbar\omega_\textrm{rel}$ in \namelist{\&INPUT} energy units. If \inpitem{CONFRQ} is zero, \inpitem{CONLEN} specifies $\beta_\textrm{rel}$. For identical atoms, $\omega_\textrm{rel}=\omega_1=\omega_2$ but, because $\mu=m_1/2$, $\beta_\textrm{rel}=\sqrt{2}\beta_1$.

It should be noted that the separation between successive confined levels produced by \BOUND, in the absence of centrifugal or other potential terms, is $2\hbar\omega_\textrm{rel}$ rather than $\hbar\omega_\textrm{rel}$. This arises because \BOUND\ solves the coupled equations on the range $0 \le R <\infty$ instead of $-\infty < R < \infty$, and effectively places a node in the wavefunction at the origin.

Anisotropic or anharmonic confining potentials can in principle be handled with a plug-in basis-set
suite, but this has not yet been implemented.

\section{Bound-state wavefunctions}\mylabel{calcwaveBF}

If $\inpitem{IWAVE}\ne 0$, \BOUND\ and \FIELD\ calculate wavefunctions and output them on channel
\iounit{IWAVE}. This is currently implemented only for the LDMD and AIRY propagators, so both
\inpitem{IPROPS} and \inpitem{IPROPL} must be either 6 or 9: if any other propagator is requested,
the request for wavefunctions is cancelled but the rest of the calculation proceeds.

To do this, once a bound state has been located, \BOUND\ or \FIELD\ repeats the propagation at the
converged energy or EFV, saving the log-derivative matrices to a temporary file, and then
back-substitutes from $R_{\rm match}$ to $R_{\rm min}$ and $R_{\rm max}$ to obtain the wavefunction,
as described in ref.~\cite{THORNLEY:1994}. The wavefunction is normalised on the range
from $R_{\rm min}$ to $R_{\rm max}$ using the alternative extended Simpson's rule \cite{AESimpson},
neglecting any part of the wavefunction that lies outside the range.

The wavefunction is written on unit \iounit{IWAVE} in the format described in section
\ref{CommII:bwave}.\cbend

\section[\texorpdfstring{Expectation values ({\color{\bcol}\BOUND} only)}
{Expectation values (BOUND only)}]{Expectation values\sectionmark {Expectation
values}}\sectionmark{Expectation values}\mylabel{calcexp}

\cbcolor{\bcol}\cbstart In addition to calculating bound-state energies, \BOUND\ implements the
calculation of expectation values using the finite-difference approach \cite{Hutson:expect:88}.
After a bound state is located at energy $E^{(0)}$, \BOUND\ repeats the calculation with a small
perturbation $a \hat A(R)$ added to the Hamiltonian to obtain a modified energy $E(a)$. From
perturbation theory,
\begin{equation} E_n(a) = E_n^{(0)} + a
\langle\hat A\rangle_n + {\cal O}(a^2),
\end{equation}
where ${\cal O}(a^2)$ are second-order terms. The finite-difference approximation to the
expectation value $\langle\hat A\rangle_n$ is
\begin{equation}
\langle\hat A\rangle_n = \frac{E_n(a) - E_n^{(0)}}{a},
\end{equation}
and is accurate to order $a$.

For built-in interaction types, \BOUND\ can calculate expectation values of an operator $\hat A$
that is made up of a product of one of the angular functions in the potential expansion and a power
of $R$; see the documentation on potential expansions for each interaction type (section
\ref{listpotl}). More complicated functions of $R$ can be handled by modifying subroutine
\prog{PERTRB}. For coupling cases implemented in plug-in basis-set suites, any required operator
can be implemented in the array \var{VL}.

The expectation values to be calculated are specified by a variable \inpitem{NPERT} and arrays
\inpitem{IPPERT}, \inpitem{NPOW}, \inpitem{DELTA} and \inpitem{FACTOR} in namelist
\namelist{\&INPUT}. For each bound state located, the program attempts to calculate \inpitem{NPERT}
(up to 20) expectation values; if the operator whose coupling matrix (in the array \var{VL}) is
\code{A(IPPERT(IP))}, the \var{IP}th value calculated is the expectation value of
\begin{equation*}
\code{FACTOR(IP) \times A(IPPERT(IP))} / R^\code{NPOW(IP)}.
\end{equation*}
The program applies a perturbation of $\inpitem{DELTA}(\var{IP}) \times
\code{A(}\inpitem{IPPERT}(\var{IP}))$ in order to use the finite-difference approximation. Small
values of \inpitem{DELTA} give smaller higher-order contributions to the result, but poorer
numerical stability; $\inpitem{DELTA}=0.001$ cm$^{-1}$ usually gives good results for operators
with matrix elements of order of unity.

\section[\texorpdfstring{Electric dipoles and magnetic moments ({\color{\bcol}\BOUND} only)}
{Electric dipoles and magnetic moments (BOUND only)}]{Electric dipoles and magnetic
moments\sectionmark {Electric dipoles and magnetic moments}}\sectionmark{Electric dipoles and
magnetic moments}\mylabel{moments}

The capability to calculate expectation values described in section \ref{calcexp} may be used to
evaluate electric dipoles and magnetic moments. The space-fixed electric dipole or magnetic moment
$\mu_n$ of state $n$ is related to the derivative of its energy $E_n$ with respect to an external
electric or magnetic field $B$,
\begin{equation}
\mu_n=-\frac{dE_n}{dB}.
\end{equation}

If \inpitem{IPPERT}(\var{IP}) indexes an operator for an EFV and $\inpitem{NPOW}(\var{IP})=0$,
\BOUND\ calculates the derivative of the bound-state energy with respect to the EFV, for both the
absolute energy and the energy relative to threshold. \inpitem{DELTA}(\var{IP}) is interpreted in
units of the EFV.

If the EFV is handled by setting $\var{NDGVL}>0$ in a plug-in basis set suite,
\inpitem{IPPERT}(\var{IP}) must be set negative and the program then calculates the derivative of
the bound-state energy with respect to \var{EFV}(-\inpitem{IPPERT}(\var{IP})).

\section[\texorpdfstring{Closed-channel fraction ({\color{\bcol}\BOUND} only)}
{Closed-channel fraction (BOUND only)}]{Closed-channel fraction\sectionmark {Closed-channel
fraction}}\sectionmark{Closed-channel fraction}\mylabel{closed-fraction}

An important application of the derivative with respect to a threshold energy $E_i$ is in the
calculation of the closed-channel fraction $Z(B)$ for a near-threshold bound state near a Feshbach
resonance in incoming channel $i$. This is given by \cite{Chin:RMP:2010}
\begin{equation}
Z(B)= \frac{d(E_n-E_i)/dB}{d(E_{n,\textrm{bare}}-E_i)/dB}.
\end{equation}

Here $E_{n,\textrm{bare}}$ is the energy of the bare bound state, which may usually be obtained
from bound-state calculations far enough below threshold to be unaffected by the resonance. If the
threshold energy itself varies non-linearly with energy, it is best to evaluate $dE_i/dB$ in the
denominator at $B$, rather than far from resonance.

For the special case of a derivative with respect to an EFV, \inpitem{FACTOR}(\var{IP}) should contain $d(E_{n,\textrm{bare}}-E_i)/dB$: the derivative of the binding energy with respect to threshold is \emph{divided} by this to obtain the bound-state fraction, rather than multiplied by it as for other expectation values.

Note that some care is needed with sign conventions. \BOUND\ outputs energy derivatives, which have
opposite sign to magnetic moments, but ref.\ \cite{Chin:RMP:2010} defines the quantity $\delta\mu$,
commonly used in the theory of Feshbach resonances, as $\delta\mu =
\mu_\textrm{thresh}-\mu_{n,\textrm{bare}} = d(E_{n,\textrm{bare}}-E_i)/dB$; similarly, it defines
$\delta\mu_n = \mu_\textrm{thresh}-\mu_n = d(E_n-E_i)/dB$.
\cbend

\section[\texorpdfstring{Automated convergence testing ({\color{\bcol}\BOUND} only)}
{Automated convergence testing (BOUND only)}]{Automated convergence testing
\sectionmark{Automated convergence testing}}\sectionmark{Automated convergence testing}\mylabel{conv}

\cbcolor{\bcol}\cbstart
If $\inpitem{NCONV}>0$, \BOUND\ performs
\inpitem{NCONV} extra calculations of each eigenvalue or expectation value,
with different values of \inpitem{RMIN}, \inpitem{RMAX} or \var{DR}. This is
useful in testing the convergence with respect to these parameters, or in
estimating the error due to the use of a finite step size. Which one of these
lengths is varied is governed by \inpitem{ICON} as follows:
\begin{description}[nosep]
          \item[$\inpitem{ICON} = 1$]{doubles the initial
              step size each time}
          \item[$\inpitem{ICON} = 2$]{decreases \inpitem{RMAX} by
              \inpitem{DRCON} each time}
	  \item[$\inpitem{ICON} = 3$]{increases \inpitem{RMIN} by
\inpitem{DRCON} each time}
\end{description}

Note that, if \BOUND\ does not succeed in converging on an eigenvalue for one
value of the parameter concerned, the parameter is NOT changed before the next
calculation. Make sure that \BOUND\ can find the eigenvalue before attempting
to test convergence.\cbend

\section[\texorpdfstring{Richardson extrapolation to zero step size ({\color{\bcol}\BOUND} only)}
{Richardson extrapolation to zero step size (BOUND only)}]
{Richardson extrapolation to zero step size
\sectionmark{Richardson extrapolation to zero step size}}
\sectionmark{Richardson extrapolation to zero step size}

\cbcolor{\bcol}\cbstart
For propagators that use equally spaced or power-law steps, the error in
bound-state energies due to a finite step size is proportional to a power of
the step size (in the limit of small steps). This power is 4 for the LDJ, LDMD,
LDMA and LDMG(CS4) propagators and 5 for the LDMG(MA5) propagator. It is thus
possible to obtain an improved estimate of the bound-state energy by performing
calculations with two different step sizes and extrapolating to zero step size.
\BOUND\ does this automatically and gives the extrapolated results as well as
the ones using the step sizes specified when \var{DR} is varied using
$\inpitem{ICON}=1$ as described in section \ref{conv}.

If the short-range and long-range propagators are different, the power used for
Richardson extrapolation is that appropriate for the short-range propagator. It
is thus important to ensure that
\begin{itemize}[nosep]
\item \emph{either} the short-range and long-range propagators have the
    same step-size convergence properties, and both are used with either
    equally spaced steps or step sizes proportional to a power of $R$;
\item \emph{or} the short-range propagator is used throughout the well
    region, so that any errors due to the long-range propagator are
    negligible.
\end{itemize}
\cbend

{
\renewcommand\inpitem[1]{{\tt #1}\index{{\tt #1} in \namelist{\&INPUT}}}
\renewcommand\basisitem[1]{{\tt #1}\index{{\tt #1} in \namelist{\&BASIS}}}
\renewcommand\potlitem[1]{{\tt #1}\index{{\tt #1} in \namelist{\&POTL}}}
\chapter{\texorpdfstring{Complete list of input parameters}
{\ref{CommI}: Complete list of input parameters}}\mylabel{CommI}

The \namelist{\&INPUT} item \inpitem{LASTIN} controls whether the current set
of namelist blocks is the last set to be used in the current run. If set to 0,
the \prog{DRIVER} routine loops back to the start, resetting all namelist items
to their default values, and start a new set of calculations by reading in the
next set of namelist blocks in the input file.

\section{\texorpdfstring{Items in namelist \namelist{\&INPUT}}
{Items in namelist \&INPUT}}\mylabel{listinput}

The namelist items are here listed alphabetically, followed by their default
values in square brackets, then a short description of what they are and
finally the section in which a more complete description may be found.  If more
than one default value is listed, they refer to different default values in the
programs \MOLSCAT, \BOUND\ and \FIELD\ using the colour coding described
immediately below. If a namelist item is coloured, it is used by only a subset
of the programs. The key is as follows:
\begin{description}
\item[if a keyword or default is red,]{it is used only by {\color{\mcol}{\MOLSCAT}},}
\item[if blue]{then it is used only by {\color{\bcol}{\BOUND}},}
\item[if green]{then it is used only by {\color{\fcol}{\FIELD}},}
\item[if cyan]{then it is used only by {\color{\bfcol}{\BOUND\ and
    \FIELD}},}
\item[if brown]{it is used only by {\color{\mfcol}{\MOLSCAT\ and \FIELD}},
    and}
\item[if purple]{it is used only by {\color{\mbcol}{\MOLSCAT\ and
    \BOUND}}.}
\end{description}

\begin{description}
\item[\inpitem{ADIAMN} {[T]}:]{Controls whether adiabatic basis or
    primitive basis is used for boundary conditions at $R_{\rm min}$
    (section \ref{Y-bc}).}
\item[{\color{\bfcol}\inpitem{ADIAMX}} {[T]}:]{Controls whether adiabatic
    basis or primitive basis is used for boundary conditions at $R_{\rm
    max}$ (section \ref{Y-bc}).}
\item[{\color{\mcol}\inpitem{ALPHA1}} {[1.0]}:]{Start value of step size
    for VIVS propagator (section \ref{iprop4}).}
\item[{\color{\mcol}\inpitem{ALPHA2}} {[1.5]}:]{End value of step size for
    VIVS propagator (section \ref{stepsize4}).}
\item[{\color{\mcol}\inpitem{AZERO}} {[0.0]}:]{Value of the (real part of
    the) scattering length/volume to converge upon (section
    \ref{fieldval}).}
\item[\inpitem{BCYCMN} {[$\boldsymbol{-1.0}$]}:]{Controls value used in
    construction of log-derivative matrix for channels that are locally
    closed at $R_{\rm min}$ (section \ref{Y-bc}).}
\item[{\color{\bfcol}\inpitem{BCYCMX}} {[$\boldsymbol{-1.0}$]}:]{Controls
    value used in construction of log-derivative matrix for channels that
    are locally closed at $R_{\rm max}$ (section \ref{Y-bc}).}
\item[\inpitem{BCYOMN} {[unset]}:]{Controls value used in construction of
    log-derivative matrix at channels that are locally open at $R_{\rm min}$
    (section \ref{Y-bc}).}
\item[{\color{\bfcol}\inpitem{BCYOMX}} {[0.0]}:]{Controls value used in
    construction of log-derivative matrix at channels that are locally open at
    $R_{\rm max}$ (section \ref{Y-bc}).}
\item[{\color{\bfcol}\inpitem{CONFRQ}} {[0.0]}:]{If greater than 0.0, sets the confinement frequency (in the coordinate for relative motion) for pairs confined in a spherically symmetric harmonic potential (section \ref{confinement}).}
\item[{\color{\bfcol}\inpitem{CONLEN}} {[0.0]}:]{If greater than 0.0 (and $\inpitem{CONFRQ}\le0.0$), sets the confinement
    length for pairs confined in a spherically symmetric harmonic potential (section \ref{confinement}).}
\item[\inpitem{DEGTOL} {[$\boldsymbol{10^{-10}}$]}:]{Criterion used to test
    degeneracy, given in \namelist{\&INPUT} energy units; see sections
    \ref{prop:add-op} and \ref{CS}.}
\item[{\color{\bcol}\inpitem{DELTA}} {[0.001]}:]{Array of step sizes used in the calculation of
    expectation values by finite differences (section \ref{calcexp}).}
\item[\inpitem{DFIELD} {[{\color{\mbcol}1.0}/{\color{\fcol}$\bf
    10^{30}$}]}:]{Step size for varying EFV (sections \ref{ExtVar},
    \ref{potscale}, \ref{fieldscan} and \ref{eigenF}).}
\item[\inpitem{DNRG} {[{\color{\mfcol} 0.0}/{\color{\bcol} $\mathbf 10^{30}$}]}:]{Step size for
    energies (sections \ref{EMF}, \ref{energyconv} \ref{eigenB}).}
\item[\inpitem{DR}:]{Deprecated synonym for \inpitem{DRS}.}
\item[\inpitem{DRAIRY}:]{Deprecated synonym for \inpitem{DRL}.}
\item[\inpitem{DRCON} {[0.1]}:]{Size of change to
    \inpitem{RMID} or \inpitem{RMAX} used to test convergence in \MOLSCAT\
    (section \ref{andconv}) and \BOUND\ (section \ref{conv}).}
\item[\inpitem{DRL} {[unset]}:]{
    Initial value of $\delta R$ used for the long-range part of the
    propagation (if appropriate) (sections \ref{stepsize} and
    \ref{3partprop}).}
\item[{\color{\mcol}\inpitem{DRMAX}} {[5.0]}:]{Maximum allowed step size
	for VIVS propagator (section \ref{iprop4}).}
\item[{\color{\mcol}\inpitem{DRNOW}}:]{Deprecated synonym for \inpitem{DR}.}
\item[\inpitem{DRS} {[unset]}:]{
    Initial value of $\delta R$ used for the short-range part of the
    propagation (if appropriate) (sections \ref{stepsize} and
    \ref{3partprop}).}
\item[\inpitem{DTOL}
    {[{\color{\mcol}0.3}/{\color{\bfcol}$\boldsymbol{10^{-7}}$}]}:]
    {Convergence criterion used for convergence on bound-state energies in
    \BOUND\ (sections \ref{eigenB}), bound-state EFVs in \FIELD\ (section
    \ref{eigenF}), and the positions of EFV-dependent Feshbach resonances
    (section \ref{fieldconv}) and specific values of scattering lengths
    (section \ref{fieldval}) in \MOLSCAT. Also applied to the convergence
    of diagonal (elastic) cross sections with respect to $J_{\rm tot}$
    (section \ref{tCS}).}
\item[{\color{\bcol}\inpitem{ECTRCT}} {[$\boldsymbol{-10^{30}}$]}:]{Energy
    cutoff for contraction of basis set in RMAT propagator (section
    \ref{contract}).}
\item[{\color{\bcol}\inpitem{EMAX}} {[$\bf10^{30}$]}:]{Upper end of energy
    range for calculations in \BOUND\ (sections \ref{specener},
    \ref{eigenB}).}
\item[{\color{\bcol}\inpitem{EMAXBD}} {[F]}:]{If \code{.TRUE.}, restricts $E_{\rm max}$ for the current basis set and set of EFVs to the lowest threshold energy (section \ref{specener}).}
\item[{\color{\bcol}\inpitem{EMIN}} {[$\boldsymbol{-10^{30}}$]}:]{Lower end
    of energy range for calculations in \BOUND\ (sections \ref{specener},
    \ref{eigenB}).}
\item[{\color{\mfcol}\inpitem{ENERGY}} {[0.0]}:]{Array of energies for which calculations are
    requested (sections \ref{EMF}, \ref{pressbroad}, \ref{energyconv} and \ref{effrange}).}
\item[\inpitem{EPL} {[0.0]}:]{Estimated maximum depth of the interaction
    potential in the long-range region; used only when the initial step
    size of the long-range propagator is determined from \inpitem{STEPL}
    (section \ref{stepsize}). Note that \inpitem{EPL} is in
    \namelist{\&INPUT} energy units, \emph{not} \namelist{\&POTL} energy
    units.}
\item[\inpitem{EPS} {[0.0]}:]{Estimated maximum depth of the interaction
    potential in the short-range region; used only when the initial step
    size of the short-range propagator is determined from \inpitem{STEPS}
    (section \ref{stepsize}). Note that \inpitem{EPS} is in
    \namelist{\&INPUT} energy units, \emph{not} \namelist{\&POTL} energy
    units.}
\item[\inpitem{EREF} {[0.0]}:]{Reference energy used for input and output
    of energies (scattering energy, depth of bound states) (section
    \ref{EREF}).}
\item[\inpitem{EUNAME} {[\code{'E UNITS'}]}:]{Character string output to
    describe energy unit specified as \inpitem{EUNIT}.}
\item[\inpitem{EUNITS} {[1]}:]{Integer to choose unit of energy for
    quantities input in namelist \namelist{\&INPUT} and most output, chosen
    from the list in section \ref{outline:units}. Note that there
    are independent values of \inpitem{EUNITS}, \inpitem{EUNIT} and
    \inpitem{EUNAME} in namelist blocks \namelist{\&INPUT} and
    \namelist{\&BASIS}.}
\item[\inpitem{EUNIT} {[1.0]}:]{Unit of energy (in cm$^{-1}$), used only
    if \inpitem{EUNITS} is set to 0.}
\item[{\color{\bcol}\inpitem{FACTOR}} {[1.0]}:]{Array of factors that calculated
    expectation values are multiplied by (section \ref{calcexp}).}
\item[\inpitem{FIELD} {[0.0]}:]{Array of values of varying EFVs for which
    calculations are requested (section \ref{ExtVar}).}
\item[\inpitem{FIXFLD}(\var{MXEFV}) {[0.0]}:]{Array of values of fixed
    EFVs (section \ref{ExtVar}).}
\item[\inpitem{FLDMAX} {[0.0]}:]{Upper end of range of varying EFV
    (sections \ref{ExtVar}, \ref{potscale}, \ref{fieldscan},
    \ref{fieldconv}, \ref{fieldval} \ref{eigenF}).}
\item[\inpitem{FLDMIN} {[0.0]}:]{Lower end of range of varying EFV
    (sections \ref{ExtVar}, \ref{potscale}, \ref{fieldscan},
    \ref{fieldconv}, \ref{fieldval}, \ref{eigenF}).}
\item[{\color{\mcol}\inpitem{IALFP}} {[F]}:]{Controls whether
    \inpitem{ALPHA2} is used as end value for step size in VIVS propagator
    (section \ref{iprop4}).}
\item[{\color{\mcol}\inpitem{IALPHA}} {[6]}:]{Controls whether step size
    can be variable in VIVS propagator (section \ref{iprop4}).}
\item[{\color{\bfcol}\inpitem{IBDSUM}} {[0]}:]{Unit number for summary of
    bound-state locations (sections \ref{ksave:bf} and
    \ref{output:boundstates}).}
\item[\inpitem{IBFIX} {[0]}:]{If positive, lower bound for symmetry block
    index.  If greater than \inpitem{IBHI}, restricts symmetry block index
    to this value alone (section \ref{angmom}).}
\item[\inpitem{IBHI} {[0]}:]{Highest value of symmetry block index
    for which calculations are required (section \ref{angmom}).}
\item[{\color{\mcol}\inpitem{ICHAN}} {[1]}:]{Index of open channel to use in characterising an
    EFV-dependent Feshbach resonance (section \ref{fieldconv}), converging on an EFV with a
    specified value of the scattering length (section \ref{fieldval}), calculating the
    effective range (section \ref{effrange}) or calculating a scattering wavefunction (section
    \ref{calcwaveM}).}
\item[\inpitem{ICON} {[1]}:]{Controls which variable is to
    be varied in convergence tests in \MOLSCAT\ (section \ref{andconv}) and
    \BOUND\ (section \ref{conv}).}
\item[{\color{\mcol}\inpitem{ICONVU}} {[0]}:]{Unit number for storing or
    retrieving an S matrix for use in convergence tests in \MOLSCAT\
    (section \ref{andconv}).}
\item[{\color{\mcol}\inpitem{IDIAG}} {[F]}:]{Master control for
    perturbation correction type in VIVS propagator (section
    \ref{iprop4}).}
\item[{\color{\mcol}\inpitem{IECONV}} {[0]}:]{Specifies a scan of
    energies or a method for converging on a scattering resonance
    as a function of energy (section \ref{energyconv}).}
\item[{\color{\mcol}\inpitem{IFCONV}} {[0]}:]{Specifies a scan of
    scattering length as a function of EFV (section \ref{fieldscan}), the
    procedure to be used for convergence on a low-energy Feshbach resonance
    (section \ref{fieldconv}), or convergence on an EFV with a specified value of
    the scattering length (section \ref{fieldval}).}
\item[{\color{\mcol}\inpitem{IFEGEN}} {[0]}:]{Control of generation of
    total energies for calculation of line-shape cross sections (section
    \ref{pressbroad}).}
\item[{\color{\mcol}\inpitem{IFLS}}:]{Deprecated synonym for
    \inpitem{NLPRBR}.}
\item[\inpitem{IFVARY}(\var{MXEFV}) {[$\boldsymbol{\min\{1,\var{NEFV}\}}$]}:]{index (or
	indices) of varying EFVs (section \ref{ExtVar}).}
\item[{\color{\mcol}\inpitem{ILDSVU}} {[0]}:]{Unit number for output of
    log-derivative matrix (section \ref{CommII:LD}).}
\item[\inpitem{IMGSEL} {[4]}:]{ Selects the method used for the symplectic
    (LDMG) propagator (section \ref{iprop5678}).}
\item[\inpitem{INTFLG}:]{Deprecated variable to control propagator
    selection.}
\item[{\color{\mcol}\inpitem{IPARTU}} {[0]}:]{Unit number for partial cross sections (not
    currently implemented; see sections \ref{pCS} and \ref{CommII:ipartu}).}
\item[{\color{\mcol}\inpitem{IPERT}} {[T]}:]{Master control for inclusion
    of perturbation corrections in VIVS propagator (section \ref{iprop4}).}
\item[{\color{\mcol}\inpitem{IPHSUM}} {[0]}:]{Unit number for summary of
    eigenphase sums and scattering lengths (section \ref{ksave:mol}).}
\item[{\color{\bcol}\inpitem{IPPERT}} {[1,2,3,\ldots]}:]{Array of indices of operators for
    which expectation values are to be calculated (section \ref{calcexp}).}
\item[\inpitem{IPRINT} {[2]}:]{Controls extent of printed output; see
    chapter \ref{CommII}.}
\item[\inpitem{IPROPL} {[{\color{\mcol}9}/{\color{\bfcol}6}]}:]{Propagator
    code for propagation between $R_{\rm mid}$ and $R_{\rm max}$ (section
    \ref{propchoice}).}
\item[\inpitem{IPROPS} {[6]}:]{Propagator code for propagation between
    $R_{\rm min}$ and $R_{\rm mid}$ (section \ref{propchoice}).}
\item[\inpitem{IREF} {[0]}:]{Index of the threshold whose energy is used as
    the reference energy (section \ref{EREF}).}
\item[\inpitem{IRMSET} {[9]}:]{If positive, is used in choice of starting
    point for propagations (section \ref{intrange}). }
\item[{\color{\mcol}\inpitem{IRSTRT}} {[0]}:]{Indicates whether the
    calculation is a continuation of previous scattering calculations
    (section \ref{restart}).}
\item[{\color{\mcol}\inpitem{IRXSET}} {[0]}:]{Indicates whether propagation
    is to be extended beyond \inpitem{RMAX} to the outermost turning point
    of the centrifugal potential (section \ref{intrange}).}
\item[{\color{\mcol}\inpitem{ISAVEU}} {[0]}:]{Unit number for output of S or K matrices
    (sections \ref{restart}, \ref{energyconv} and \ref{CommII:SandK}).}
\item[\inpitem{ISCRU}  {[0]}:]{Scratch unit number for energy-independent
    matrices (section \ref{CommIII:iscru}).}
\item[{\color{\mcol}\inpitem{ISHIFT}} {[F]}:]{Controls whether 2nd
    derivative is used to shift reference potential for VIVS propagator
    (section \ref{iprop4}).}
\item[{\color{\mcol}\inpitem{ISIGPR}} {[0]}:]{Controls printing of cross
    sections (section \ref{CommII:isigpr}).}
\item[{\color{\mcol}\inpitem{ISIGU}} {[0]}:]{Unit number for accumulated
    cross sections (section \ref{CommII:isigu}).}
\item[{\color{\mcol}\inpitem{ISYM}} {[T]}:]{Controls whether R matrix is
    forced to be symmetric for VIVS propagator (section \ref{iprop4}).}
\item[{\color{\mcol}\inpitem{IV}} {[T]}:]{Controls calculation of
    perturbation corrections due to potential for VIVS propagator  (section
    \ref{iprop4}).}
\item[{\color{\mcol}\inpitem{IVP}} {[F]}:]{Controls calculation of
    perturbation corrections due to 1st derivative of interaction potential
    for VIVS propagator (section \ref{iprop4}).}
\item[{\color{\mcol}\inpitem{IVPP}} {[F]}:]{Controls calculation of
    perturbation corrections due to 2nd derivative of interaction potential
    for VIVS propagator (section \ref{iprop4}).}
\item[\inpitem{IWAVE} {[0]}:]{Unit number for output of wavefunction; see section
    \ref{calcwaveBF} for bound-state wavefunctions, section \ref{calcwaveM} for scattering
    wavefunctions and section \ref{CommII:bwave} for description of content.}
\item[\inpitem{IWAVEF} {[\code{.TRUE.}]}:]{Flag to specify whether the auxiliary output file
    for wavefunctions is formatted or unformatted (section \ref{CommII:bwave}).}
\item[\inpitem{IWVSTP} {[1]}:]{Controls the values of $R$ at which wavefunctions are written
    (section \ref{CommII:bwave}).}
\item[\inpitem{JSTEP} {[1]}:]{Step length for loop over \var{JTOT} (section
    \ref{angmom}).}
\item[\inpitem{JTOTL} {[0]}:]{Lowest value of \var{JTOT} (section
    \ref{angmom}).}
\item[\inpitem{JTOTU} {[0]}:]{Highest value of \var{JTOT} (section
    \ref{angmom}).}
\item[{\color{\mcol}\inpitem{KSAVE}}:]{Deprecated synonym for
    \inpitem{IPHSUM}.}
\item[{\color{\bfcol}\inpitem{KSAVE}}:]{Deprecated synonym for
    \inpitem{IBDSUM} in \BOUND\ and \FIELD.}
\item[\inpitem{LABEL} {[\code{' '}]}:]{\inpitem{LABEL} is a title for the
    run, up to 80 characters.}
\item[\inpitem{LASTIN} {[1]}:]{Controls whether programs process another
    complete input data set after the current one (starting with another
    \namelist{\&INPUT} block), or terminate; see page \pageref{CommI}.}
\item[{\color{\mcol}\inpitem{LINE}} {[0]}:]{Array of quartets of indices
    for levels, specifying spectroscopic lines for calculations of
    line-shape cross sections (section \ref{pressbroad}).}
\item[{\color{\mcol}\inpitem{LMAX}} {[0]}:]{Highest $L$ value for which IOS
    cross sections are accumulated (section \ref{angmom:IOS}).}
\item[{\color{\mcol}\inpitem{LOGNRG}} {[F]}:]{If \code{.TRUE.}, indicates
    that energies are to be generated in a geometric series (section
    \ref{EMF}).}
\item[{\color{\mcol}\inpitem{LTYPE}} {[$\boldsymbol{-1}$]}:]{Tensor order
    of spectroscopic transition involved in calculations of line-shape
    cross sections (section \ref{pressbroad}).}
\item[\inpitem{MAGEL} {[1]}:]{Deprecated. Indicated which of the EFVs was
    referred to by \inpitem{FLDMIN}, \inpitem{FLDMAX} and \inpitem{DFIELD}.
    This has now been replaced by \inpitem{IFVARY} (with a different
    specification).}
\item[\inpitem{MHI}:]{Deprecated synonym for \inpitem{IBHI}.}
\item[{\color{\mcol}\inpitem{MMAX}} {[0]}:]{Highest $M$ value for which IOS
    cross sections are accumulated (section \ref{angmom:IOS}).}
\item[\inpitem{MONQN} {[$\boldsymbol{\inpitem{MONQN}(1)=-99999}$]}:]{Array
    of values (typically quantum labels) used to specify the reference
    energy (section \ref{EREF}).}
\item[\inpitem{MSET}:]{Deprecated synonym for \inpitem{IBFIX}.}
\item[\inpitem{MUNIT} {[1.0]}:]{Unit of mass, in Daltons.}
\item[{\color{\bfcol}\inpitem{MXCALC}} {[1000]}:]{Maximum number of
    propagations (energies or EFV sets) in a run (sections \ref{eigenB} and
    \ref{eigenF}).}
\item[{\color{\mcol}\inpitem{MXPHI}} {[1]}:]{Number of values of the
    azimuthal angle for surface scattering calculations
    ($\basisitem{ITYPE}=8$, section \ref{ityp8}).}
\item[{\color{\mcol}\inpitem{MXSIG}} {[0]}:]{If positive, only cross
    sections between the first \inpitem{MXSIG} open channels are calculated
    (section \ref{CS}).}
\item[{\color{\mcol}\inpitem{NCAC}} {[14]}:]{Convergence criterion for
    accumulation of cross sections (section \ref{tCS}).}
\item[{\color{\bcol}\inpitem{NCONV}} {[0]}:]{Number of extra sets of
    calculations performed in convergence testing by \MOLSCAT\ (section
    \ref{andconv}) or \BOUND\ (section \ref{conv}).}
\item[{\color{\mbcol}\inpitem{NFIELD}} {[1]}:]{Number of sets of EFVs for
    which calculations are requested (section \ref{ExtVar}).}
\item[\inpitem{NFVARY} {[-1]}:]{Number of varying EFVs (section
    \ref{ExtVar}).}
\item[{\color{\mcol}\inpitem{NGAUSS}} {[3]}:]{Number of Gaussian quadrature
    points generated for use in thermal averaging of cross sections
    (section \ref{scatE}).}
\item[{\color{\mcol}\inpitem{NGMP}} {[(8,1,16)]}:]{Specifies numbers of
    quadrature points used in WKB integration (section \ref{intflgwkb}).}
\item[{\color{\mcol}\inpitem{NLPRBR}} {[0]}:]{Number of line-shape cross
    sections to be calculated, each specified by a pair of spectroscopic
    lines (section \ref{pressbroad}).}
\item[{\color{\mfcol}\inpitem{NNRG}} {[1]}:]{Number of values in
    array \inpitem{ENERGY} (section \ref{EMF}).}
\item[{\color{\mcol}\inpitem{NNRGPG}} {[1]}:]{Size of group of energies to
    be considered together in testing convergence of cross sections with
    respect to \var{JTOT} (section \ref{tCS}).}
\item[{\color{\bfcol}\inpitem{NODMAX}} {[99999]}:]{Upper bound of node
    count for bound states to be located (sections \ref{specener},
    \ref{eigenB}, \ref{eigenF}).}
\item[{\color{\bfcol}\inpitem{NODMIN}} {[0]}:]{Lower bound of node count
    for bound states to be located (sections \ref{specener},
    \ref{nodecount}, \ref{eigenB}, \ref{eigenF}).}
\item[{\color{\bcol}\inpitem{NPERT}} {[0]}:]{Number of operators for which
    expectation values are to be calculated (section \ref{calcexp}).}
\item[{\color{\bcol}\inpitem{NPOW}} {[0]}:]{Array of powers of $R$ to be used in calculating
    expectation values (section \ref{calcexp}).}
\item[{\color{\mcol}\inpitem{NSTAB}} {[5]}:]{Controls how often linear
    independence is re-established for DV propagator (section
    \ref{iprop2}).}
\item[{\color{\mcol}\inpitem{NTEMP}} {[0]}:]{Number of temperatures for
    which energies are to be generated for use in (external) thermal
    averaging (section \ref{scatE}).}
\item[{\inpitem{NUMDER}} {[F]}:]{Controls whether derivatives of potential
    coefficients are calculated numerically when required; needed only for
    the VIVS propagator (section \ref{iprop4}) and for calculating
    nonadiabatic matrix elements with the LDMA propagator (section
    \ref{iprop5678}).}
\item[\inpitem{NWVCOL} {[8]}:]{The number of columns used to write radial channel functions on
    each line of the formatted wavefunction file (section \ref{CommII:bwave}). Note that the
    number of channel functions written per line is \inpitem{NWVCOL} for \BOUND\ and \FIELD,
    but $\inpitem{NWVCOL}/2$ for \MOLSCAT\ because scattering wavefunctions are complex.}
\item[{\color{\mcol}\inpitem{OTOL}} {[0.005]}:]{Convergence criterion
    applied to convergence of off-diagonal (usually inelastic) cross
    sections with respect to \var{JTOT} (section \ref{tCS}).}
\item[\inpitem{PHILW} {[0.0]}:]{First value for azimuthal angle in surface
    scattering calculations ($\basisitem{ITYPE}=8$, section \ref{ityp8}).}
\item[\inpitem{PHIST} {[0.0]}:]{Step size for azimuthal angle in surface
    scattering calculations ($\basisitem{ITYPE}=8$, section \ref{ityp8}).}
\item[\inpitem{POWRL} {[$\bf 1.33\dot{3}$ or 3.0]}:]{Power used in step-size
    algorithm for RMAT and AIRY propagators at long range (sections
    \ref{dr-power}, \ref{iprop3} and \ref{iprop9}).}
\item[\inpitem{POWRS} {[0.0 or 3.0]}:]{Power used in step-size algorithm
    for RMAT and AIRY propagators at short range (sections
    \ref{dr-power}, \ref{iprop3} and \ref{iprop9}).}
\item[\inpitem{POWRX} {[3.0]}:]{Deprecated synonym for \inpitem{POWRL}.}
\item[\inpitem{PRNTLV}:]{Deprecated synonym for \inpitem{IPRINT}.}
\item[{\color{\bcol}\inpitem{RCTRCT}} {[0.0]}:]{If greater than 0.0, value
    of $R$ at which basis set is contracted in RMAT propagator (section
    \ref{contract}).}
\item[{\color{\bfcol}\inpitem{RMATCH}} {[unset]}:]{Value of $R$ at
    which log-derivative matrices for outwards and inwards propagation
    parts are matched; see sections \ref{intrange} and \ref{3partprop}. If
    left unset, set internally to \inpitem{RMID} (if set).}
\item[\inpitem{RMAX} {[10.0]}:]{Maximum value of $R$ for propagation
    (sections \ref{intrange}, \ref{3partprop}, \ref{iprop3}, \ref{iprop4}
    and \ref{iprop9}).}
\item[\inpitem{RMID}
    {[{\color{\mcol}$\boldsymbol{10^{30}}$}/{\color{\bfcol}\inpitem{RMATCH}}]}:]{Value
    of $R$ at which propagation method switches for propagations which
    comprise more than one part (sections \ref{intrange}, \ref{3partprop},
    \ref{iprop4} and \ref{iprop9}). Also used as the value where the step
    size starts to increase in the RMAT propagator (section
    \ref{iprop3}). If left unset in \BOUND\ or \FIELD, set internally to
    \inpitem{RMATCH} (if set).}
\item[\inpitem{RMIN} {[0.8]}:]{Minimum value of $R$ for propagation
    (sections \ref{intrange}, \ref{3partprop}).}
\item[\inpitem{RUNAME} {[\code{'RUNIT'}]}:]{Character string output to
    describe unit of length specified as \inpitem{RUNIT} (section \ref{lengthunit})}
\item[\inpitem{RUNIT} {[\code{unset}]}:]{Unit of length, in \AA, for
    quantities in namelist \namelist{\&INPUT} and for most output.
    If unset, unit of length is taken from variable \var{RM} returned by
    subroutine \prog{POTENL}.}
\item[{\color{\mcol}\inpitem{RVFAC}} {[0.0]}:]{If set greater than 0.0,
    controls how $R_{\rm mid}$ is chosen.}
\item[{\color{\mcol}\inpitem{RVIVAS}}:]{Deprecated synonym for
    \inpitem{RMID}.}
\item[\inpitem{SCALAM} {[1.0]}:]{Value of the interaction potential scaling
    factor used in the current calculation (section \ref{potscale}).}
\item[\inpitem{STEPL} {[$\boldsymbol{-10.0}$]}:]{If positive, number of steps
    per half wavelength for
    long-range propagator (section \ref{stepsize}). Otherwise \inpitem{DRL}
    is used to set the step length.}
\item[\inpitem{STEPS} {[$\boldsymbol{-10.0}$]}:]{If positive, number of steps
    per half wavelength for
    short-range propagator (section \ref{stepsize}). Otherwise
    \inpitem{DRS} is used to set the step length.}
\item[{\color{\mcol}\inpitem{TEMP}} {[0.0]}:]{Array of temperatures for
    which energies are to be generated for use in (external) thermal
    averaging (section \ref{scatE}).}
\item[\inpitem{THETLW} {[0.0]}:]{First value for polar angle in surface
    scattering calculations ($\basisitem{ITYPE}=8$, section \ref{ityp8}).}
\item[\inpitem{THETST} {[0.0]}:]{Step size for polar angle in surface scattering calculations
    ($\basisitem{ITYPE}=8$, section \ref{ityp8}).}
\item[{\color{\mcol}\inpitem{THI}} {[1.0]}:]{controls positioning of the outermost of the 3
    target points used in characterising a scattering resonance as a function of either energy
    (section \ref{energyconv}) or EFV (section \ref{fieldconv}).}
\item[{\color{\mcol}\inpitem{TLO}} {[$\boldsymbol{-0.1}$]}:]{controls positioning of the second
    nearest of the 3 target points used in characterising a scattering resonance as a function
    of either energy (section \ref{energyconv}) or EFV (section \ref{fieldconv}).}
\item[\inpitem{TOLHIL} {[0.0001]}:]{Controls step size for variable-step propagators when used
    at long range (sections \ref{iprop3}, \ref{iprop4}, \ref{iprop9}).}
\item[\inpitem{TOLHIS} {[0.0001]}:]{Controls step size for variable-step
    propagators when used at short range (sections \ref{iprop3},
    \ref{iprop4}, \ref{iprop9}). Also used as the convergence criterion for
    the WKB phase shift (section \ref{intflgwkb}).}
\item[\inpitem{TOLHI}:]{Deprecated synonym for \inpitem{TOLHIS}.}
\item[\inpitem{URED} {[no default]}:]{Reduced mass for calculation (chapter
    \ref{ConstructBasis}).}
\item[\inpitem{WKBMN} {[\code{.TRUE.}]}:]{Flag to use WKB boundary
    conditions in locally closed channels at $R_{\rm min}$ (section \ref{Y-bc}).}
\item[{\color{\bfcol}\inpitem{WKBMX}} {[\code{.TRUE.}]}:]{Flag to use WKB boundary conditions
    in locally closed channels at $R_{\rm max}$ (section \ref{Y-bc}).}
\item[{\color{\mcol}\inpitem{XI}} {[0.25]}:]{controls the tolerance for the positioning of the
    two outer target points used in characterising a scattering resonance as a function of
    either energy (section \ref{energyconv}) or EFV (section \ref{fieldconv}).}
\item[{\color{\mcol}\inpitem{XSQMAX}} {[10000]}:]{Controls application of
    perturbation correction for VIVS propagator (section \ref{iprop4}).}
\end{description}

\section{\texorpdfstring{Items in namelist \namelist{\&BASIS}}{Items in namelist \&BASIS}}\mylabel{listbasis}

The parameters input in namelist \namelist{\&BASIS} specify the interaction
type, the quantum numbers and energies of the levels to be used in the basis
set, and the dynamical approximations (if any) to be used in constructing the
coupled equations.
\begin{description}
\item[\basisitem{A} {[0.0]}:]{Array of dimension 2.  The rotational
    constant about the $x$ axis for a symmetric or asymmetric top; see
    sections \ref{ityp4}, \ref{ityp5} and \ref{ityp6}. \basisitem{A}(2) is
    solely for use in plug-in basis-set routines.}
\item[\basisitem{ALPHAE} {[0.0]}:]{Array of dimension 2.  The vibrational
    dependence of the rotational constant for a vibrotor; see sections
    \ref{ityp1}, \ref{ityp2}, \ref{ityp3} and \ref{ityp4}.}
\item[\basisitem{B} {[0.0]}:]{Array of dimension 2.  The rotational
    constant about the $y$ axis for a symmetric or asymmetric top; see
    sections \ref{ityp4}, \ref{ityp5} and \ref{ityp6}. \basisitem{B}(2) is
    solely for use in plug-in basis-set routines. Note that \basisitem{B}
    is \emph{not} the correct namelist item for the rotational constant for
    a linear rotor or vibrotor.}
\item[\basisitem{BCT} {[F]}:]{Logical variable used to indicate that the
    centrifugal potential for the BCT Hamiltonian is to be used (section
    \ref{BCT}).}
\item[\basisitem{BE} {[0.0]}:]{Array of dimension 2.  The rotational
    constant for a linear rotor or vibrotor.  See sections \ref{ityp1},
    \ref{ityp2}, \ref{ityp3} and \ref{ityp4}.}
\item[\basisitem{C} {[0.0]}:]{Array of dimension 2.  The rotational
    constant about the $z$ axis for a symmetric or asymmetric top; see
    sections \ref{ityp4}, \ref{ityp5} and \ref{ityp6}. \basisitem{C}(2) is
    solely for use in plug-in basis-set routines.}
\item[\basisitem{DE} {[0.0]}:]{Array of dimension 2.  The centrifugal
    distortion constant for a linear vibrotor; see sections \ref{ityp1},
    \ref{ityp2}, \ref{ityp3} and \ref{ityp4}.}
\item[\basisitem{DJ} {[0.0]}:]{The centrifugal distortion constant $D_J$
    for a symmetric or asymmetric top; see sections \ref{ityp5} and
    \ref{ityp6}.}
\item[\basisitem{DJK} {[0.0]}:]{The centrifugal distortion constant
    $D_{JK}$; see sections \ref{ityp5} and \ref{ityp6}.}
\item[\basisitem{DK} {[0.0]}:]{The centrifugal distortion constant $D_K$;
    see sections \ref{ityp5} and \ref{ityp6}.}
\item[\basisitem{DT} {[0.0]}:]{The spherical top tetrahedral centrifugal
    distortion constant $d_{\rm t}$; see section \ref{ityp6}.}
\item[\basisitem{ELEVEL} {[0.0]}:]{Array of dimension \var{MXELVL} (which
    is set in module \module{sizes} to be 1000).  The energy levels for
    basis functions specified in \basisitem{JLEVEL}.  See section
    \ref{jlevel}}
\item[\basisitem{EMAX} {[0.0]}:]{If $\basisitem{EMAX}> 0.0$, it is used to
    limit the selection of pair levels for $\var{ITYP}=4$, $\var{ITYP}=6$
    and $\var{ITYP}=8$; see sections \ref{ityp4}, \ref{ityp6} and
    \ref{ityp8}.}
\item[\basisitem{EMAXK} {[0.0]}:]{If $\basisitem{EMAXK}> 0.0$, it is used
    to limit the selection of basis functions basis functions for
    $\var{ITYP}=8$; see section \ref{ityp8}.}
\item[\basisitem{EUNAME} {[\code{'EN UNITS'}]}:]{Character string output to
    describe energy unit specified as \basisitem{EUNIT}.}
\item[\basisitem{EUNITS} {[1]}:]{Integer to choose units of energy for
    quantities in namelist \namelist{\&BASIS} from the list in section
    \ref{outline:units}; see section \ref{basis:units}. Note that there
    are independent values of \basisitem{EUNITS}, \basisitem{EUNIT} and
    \basisitem{EUNAME} in namelist blocks \namelist{\&BASIS} and
    \namelist{\&INPUT}.}
\item[\basisitem{EUNIT} {[1.0]}:]{Units of energy (in cm$^{-1}$) if
    \basisitem{EUNITS} is set to 0.}
\item[\basisitem{IASYMU} {[0]}:]{Unit number for input and output of
    asymmetric top functions; see sections \ref{ityp4} and \ref{ityp6}.}
\item[\basisitem{IBOUND} {[0]}:]{If \basisitem{IBOUND} is set non-zero, it
    indicates that the basis set and centrifugal energy in calculations for
    $\basisitem{ITYPE}=21$ to 27 should be for helicity decoupling instead
    of $L$-labelled coupled states; see section \ref{decouple}. Also used
    by plug-in basis-set suites; see section \ref{bas9in}.}
\item[\basisitem{IDENT} {[0]}:]{Indicates whether interaction partners are
    identical ($0=$non-identical, $1=$identical); see section \ref{ityp3}.}
\item[\basisitem{IOSNGP} {[0]}:]{Array of dimension 3: numbers of
    quadrature points over angles in IOS calculations; see section
    \ref{angmom:IOS}.}
\item[\basisitem{IPHIFL} {[0]}:]{Controls the type of quadrature over
    $\phi_1-\phi_2$ or $\chi$ in IOS calculations; see section
    \ref{angmom:IOS}.}
\item[\basisitem{ISYM} {[$\boldsymbol{-1}$]}:]{Array of dimension
    \var{MXSYMS} (set to 10 in module \module{sizes}), used for control of
    symmetry types included for symmetric and asymmetric tops; see sections
    \ref{ityp6}, \ref{ityp4} and \ref{ityp5}.}
\item[\basisitem{ISYM2} {[$\boldsymbol{-1}$]}:]{Array of dimension
    \var{MXSYMS} (set to 10 in module \module{sizes}), available for future
    expansion in analogy with \basisitem{ISYM}. \basisitem{ISYM2}(1) is
    also used to limit $L_{\rm max}$ (section \ref{couple:cc}).}
\item[\basisitem{ITYPE} {[no default]}:]{see chapter \ref{ConstructBasis}.}
\item[\basisitem{IVLU} {[0]}:]{if non-zero, indicates that the coupling
    matrices are stored on unit \iounit{IVLU} to save memory.}
\item[\basisitem{J1MAX} {[0]}:]{Maximum value of $J$ used in creating lists
    of pair levels for first structured particle.  See section \ref{ityp3}.
    It is equivalenced to \basisitem{JMAX} and so can be specified using
    this name also.}
\item[\basisitem{J1MIN} {[0]}:]{Minimum value of $J$ used in creating lists
    of pair levels for first structured particle.  See section \ref{ityp3}.
    It is equivalenced to \basisitem{JMIN} and so can be specified using
    this name also.}
\item[\basisitem{J1STEP} {[1]}:]{Step used in creating lists of pair levels
    for first structured particle.  See section \ref{ityp3}.  It is
    equivalenced to \basisitem{JSTEP} and so can be specified using this
    name also.}
\item[\basisitem{J2MAX} {[0]}:]{Maximum value of $J$ used in creating lists
    of pair levels for second structured particle.  See sections
    \ref{ityp3} and \ref{ityp4}.  It is equivalenced to \basisitem{KSET}
    and to \basisitem{KMAX}, and so may be specified using either of these
    names also.}
\item[\basisitem{J2MIN} {[0]}:]{Minimum value of $J$ used in creating lists
    of pair levels for second structured particle.  See sections
    \ref{ityp3} and \ref{ityp4}.}
\item[\basisitem{J2STEP} {[1]}:]{Step used in creating lists of pair levels
    for second structured particle.  See sections \ref{ityp3} and
    \ref{ityp4}.}
\item[\basisitem{JHALF} {[1]}:]{See section \ref{base9:loops}.  Use in
    namelist deprecated. Should be set (if required) in \prog{BAS9IN}
    (section \ref{bas9in}).}
\item[\basisitem{JLEVEL} {[0]}:]{Array of dimension \var{MXJLVL} (which is
    set to 4000 in module \module{sizes}). Contains labels for the
    \basisitem{NLEVEL} monomer levels used in constructing the basis set.
    See section \ref{jlevel}.}
\item[\basisitem{JMAX} {[0]}:]{Maximum value of $J$ used in creating lists
    of pair levels for first structured particle.  See sections
    \ref{ityp1}, \ref{ityp2}, \ref{ityp5}, \ref{ityp6} and \ref{ityp8}.  It
    is equivalenced to \basisitem{J1MAX} and so may be specified using this
    name also.}
\item[\basisitem{JMIN} {[0]}:]{Minimum value of $J$ used in creating lists
    of pair levels for first structured particle.  See sections
    \ref{ityp1}, \ref{ityp2}, \ref{ityp5}, \ref{ityp6} and \ref{ityp8}.  It
    is equivalenced to \basisitem{J1MIN} and so may be specified using this
    name also.}
\item[\basisitem{JSTEP} {[1]}:]{Step used in creating lists of pair levels
    for first structured particle.  See sections \ref{ityp1}, \ref{ityp2},
    \ref{ityp5} ,\ref{ityp6} and \ref{ityp8}.  It is equivalenced to
    \basisitem{J1STEP} and so may be specified using this name also.}
\item[\basisitem{JZCSFL} {[0]}:]{See section \ref{decouple}}
\item[\basisitem{JZCSMX} {[$\boldsymbol{-1}$]}:]{See section
    \ref{decouple}}
\item[\basisitem{KMAX} {[0]}:]{If set greater than or equal to 0, limits
    the values of $k$ for pair levels for $\var{ITYP}=5$ to be less than or
    equal to \basisitem{KMAX}; see section \ref{ityp5}.  It is equivalenced
    to \basisitem{J2MAX} and \basisitem{KSET} and so may be specified using
    these names also.}
\item[\basisitem{KSET} {[0]}:]{If set less than 0, limits the values of $k$
    for pair levels for $\var{ITYP}=5$ to be equal to $|\basisitem{KSET}|$;
    see section \ref{ityp5}.  It is equivalenced to \basisitem{J2MAX} and
    \basisitem{KMAX} and so may be specified using these names also.}
\item[\basisitem{NLEVEL} {[0]}:]{Number of sets of pair level quantum
    numbers to be read in. See chapter \ref{ConstructBasis}.}
\item[\basisitem{ROTI} {[0.0]}:]{Array of dimension \var{MXROTS} (which is
    set to 12 in module \module{sizes}), containing rotational and
    vibrational constants for interaction partners.  Most elements are
    equivalenced to other variables, as described below, and so may be specified
    using other, more specific, names:
		
\begin{tabular}{lll|lll}
\hline
\basisitem{ROTI}(1)$\equiv$ & \basisitem{A}(1), &\basisitem{BE}(1)&
\basisitem{ROTI}(2)$\equiv$ & \basisitem{A}(2), &\basisitem{BE}(2)\\
\basisitem{ROTI}(3)$\equiv$ & \basisitem{B}(1), &\basisitem{ALPHAE}(1) \qquad\qquad\qquad&
\basisitem{ROTI}(4)$\equiv$ & \basisitem{B}(2), &\basisitem{ALPHAE}(2) \\
\basisitem{ROTI}(5)$\equiv$ & \basisitem{C}(1), &\basisitem{DE}(1) &
\basisitem{ROTI}(6)$\equiv$ & \basisitem{C}(2), &\basisitem{DE}(2) \\
\basisitem{ROTI}(7)$\equiv$ & \basisitem{DJ},   &\basisitem{WE}(1) &
\basisitem{ROTI}(8)$\equiv$ & \basisitem{DJK},  &\basisitem{WE}(2) \\
\basisitem{ROTI}(9)$\equiv$ & \basisitem{DK},   &\basisitem{WEXE}(1) &
\basisitem{ROTI}(10)$\equiv$ & \basisitem{DT},  &\basisitem{WEXE}(2) \\
\hline 
\end{tabular}
		}
\item[\basisitem{SPNUC} {[0]}:]{If non-zero, used to calculate the
    statistical weights for interactions involving identical partners; see
    section \ref{ityp3}.}
\item[\basisitem{WE} {[0.0]}:]{Array of dimension 2.  The equilibrium
    vibrational frequency of vibrating linear rotors; see sections
    \ref{ityp2} and \ref{ityp4}.}
\item[\basisitem{WEXE} {[0.0]}:]{Array of dimension 2.  The anharmonicity
    constant of vibrating linear rotors; see sections \ref{ityp2} and
    \ref{ityp4}.}
\item[\basisitem{WT} {[0.0]}:]{Array of dimension 2 used to supply nuclear
    spin statistics weighting for identical partners; see section
    \ref{ityp5}.}
\end{description}

\section{\texorpdfstring{Items in namelist \namelist{\&POTL}}{Items in namelist \&POTL}}\mylabel{listpotl}

The general-purpose version of subroutine \prog{POTENL} supplied in this
distribution reads a namelist block named \namelist{\&POTL}. The parameters
that may be input in \namelist{\&POTL} are as follows; see chapter~\ref{buildVL}
for a full description:
\begin{description}
\item[\potlitem{CFLAG} {[0]}:]{If 1, indicates that the interaction
    potential is expanded in modified spherical harmonics
    $C_{\lambda\kappa}$ rather than spherical harmonics $Y_\lambda^\kappa$;
    see section \ref{potl:ityp5or6}.}
\item[\potlitem{EPSIL} {[1.0]}:]{Specifies the energy units of quantities
    input in \namelist{\&POTL} and of the potential coefficients returned
    by \prog{POTENL}; see section \ref{buildVL:epsil}. When a routine
    \prog{VINIT}/\prog{VSTAR} or \prog{VRTP} is supplied, \potlitem{EPSIL}
    is usually coded there instead of being supplied in namelist
    \namelist{\&POTL}. \potlitem{EPSIL} must be specified in cm$^{-1}$. The
    value given to \potlitem{EPSIL} does not affect the interpretation of
    energies input in \namelist{\&INPUT} and \namelist{\&BASIS} or output
    by the programs.}
\item[\potlitem{ICNSYM} {[1]}:]{For nonlinear molecules, denotes symmetry
    about $z$-axis; see sections \ref{potl:ityp5or6} and \ref{vrtp:true}
		(but note the caveat below in the description of \potlitem{IHOMO2}).
		For example, for NH$_3$,
    $\potlitem{ICNSYM} = 3$ indicates three-fold symmetry. For interaction
    potentials that are not expanded in angular functions this value should
    be set here or internally in \prog{VRTP}; for interaction potentials
    that are expanded in angular functions, it is determined from the
    array \potlitem{LAMBDA}.}
\item[\potlitem{IHOMO} {[1]}:]{If 2, indicates homonuclear symmetry
    (reflection about $\theta=\pi/2$); see sections \ref{potl:ityp1},
    \ref{potl:ityp2or7} and \ref{potl:ityp3}.  The default value of 1
    indicates heteronuclear symmetry. For interaction potentials that are
    not expanded in angular functions, this value should be set here or
    internally in \prog{VRTP}; see section \ref{vrtp:true}). For
    interaction potentials that are expanded in angular functions, it is
    determined from the array \potlitem{LAMBDA}.}
\setlength{\myitembox}{\widthof{\bf{\tt ICNYSY2} [1]:}}
\setlength{\myitemlength}{\hsize-\myitembox-0.7em}
\setlength{\mythirdlength}{\myitembox+\itemindent+0.65em}
\setlength{\myfourthlength}{\hsize+\itemindent-0.65em}
\item[\raisebox{-0.6em}{\parindent
    0pt\hskip-0.5em\begin{tabular}{l}\potlitem{ICNSY2}
    [1]:\\\potlitem{IHOMO2}
    [1]:\end{tabular}}]{\parbox[t]{\myitemlength}{\parshape=5 -0.5em
    \myitemlength -0.5em \myitemlength -\mythirdlength \myfourthlength
    -\mythirdlength \myfourthlength -\mythirdlength \myfourthlength As
    \potlitem{IHOMO} and \potlitem{ICNSYM}, but for molecule 2. For
    historical reasons, \potlitem{ICNSYM} may also be used to describe
    homonuclear symmetry of the second rotor for rigid rotor + rigid rotor
    interactions ($\var{ITYP} = 3$).}}
    \setlength{\myitembox}{\widthof{\bf{\tt IVMIN} [$\boldsymbol{-1}$]:}}
    \setlength{\myitemlength}{\hsize-\myitembox-0.7em}
    \setlength{\mythirdlength}{\myitembox+\itemindent+0.65em}
    \setlength{\myfourthlength}{\hsize+\itemindent-0.65em}
\item[\raisebox{-0.6em}{\parindent
    0pt\hskip-0.5em\begin{tabular}{l}\potlitem{IVMIN}
    [$\boldsymbol{-1}$]:\\\potlitem{IVMAX}
    [$\boldsymbol{-1}$]:\end{tabular}}]{\parbox[t]{\myitemlength}{\parshape=4
    -0.5em \myitemlength -0.5em \myitemlength -\mythirdlength
    \myfourthlength -\mythirdlength \myfourthlength For $\var{ITYP}=2$, if
    \potlitem{LMAX} is used with $\potlitem{LVRTP}=\code{.TRUE.}$ to
    generate the array \potlitem{LAMBDA} internally, \potlitem{IVMIN} must
    be given a non-negative value  to indicate the lowest vibrational level
    in the basis set. \potlitem{IVMAX}, if greater than \potlitem{IVMIN},
    indicates the highest vibrational level; otherwise \potlitem{IVMAX} is
    set equal to \potlitem{IVMIN}; see section \ref{potl:ityp2or7}.}}
\item[\potlitem{LAMBDA} {[0]}:]{An array of labels for the \potlitem{MXLAM}
    different terms included in the interaction potential; see section
    \ref{buildVL:expand}.} \setlength{\myitembox}{\widthof{\bf{\tt L2MAX}
    [$\boldsymbol{-1}$]:}}
    \setlength{\myitemlength}{\hsize-\myitembox-0.7em}
    \setlength{\mythirdlength}{\myitembox+\itemindent+0.65em}
    \setlength{\myfourthlength}{\hsize+\itemindent-0.65em}
\item[\raisebox{-1.6em}{\parindent
    0pt\hskip-0.5em\begin{tabular}{l}\potlitem{LMAX}
    [$\boldsymbol{-1}$]:\\\potlitem{MMAX}
    [$\boldsymbol{-1}$]:\\\potlitem{L1MAX}:\\\potlitem{L2MAX}
    [$\boldsymbol{-1}$]:\end{tabular}}]{\parbox[t]{\myitemlength}{\parshape=8
    -0.5em \myitemlength -0.5em \myitemlength -0.5em \myitemlength -0.5em
    \myitemlength -\mythirdlength \myfourthlength -\mythirdlength
    \myfourthlength -\mythirdlength \myfourthlength -\mythirdlength
    \myfourthlength When given non-negative values, the appropriate subset
    of these quantities specify the highest term(s) to include in the
    potential expansion.  See sections \ref{potl:ityp1},
    \ref{potl:ityp2or7}, \ref{potl:ityp5or6} and \ref{potl:ityp3} for more
    details. \potlitem{IHOMO} and \potlitem{ICNSYM} (described above) may
    be used to exclude values not allowed by symmetry.}}

\item[\potlitem{LVRTP} {[\code{.FALSE.}]}:]{If $\potlitem{LVRTP} =
    \code{.FALSE.}$, the interaction potential is specified in terms of its
    expansion in angular functions; see section \ref{vrtp:false}. If
    $\potlitem{LVRTP} = \code{.TRUE.}$, subroutine \prog{VRTP} is called as
    described in section \ref{vrtp:true} to evaluate the interaction
    potential at specified interaction coordinates.  Note that
		\potlitem{LVRTP} is forced to be \code{.TRUE.} if $\potlitem{MXLAM} \le 0$.}
\item[\potlitem{MXLAM} {[0]}:]{If positive, the number of terms in the
    expansion of the interaction potential.}
\item[\potlitem{MXSYM}:]{Deprecated synonym for \potlitem{MXLAM}.}
\item[\potlitem{NPTS} {[0]}:]{Array specifying the numbers of Gauss
    integration points used in projecting the angular components of the
    interaction potential; see section \ref{vrtp:true}.}
\item[\potlitem{NPT}:]{Deprecated synonym for \potlitem{NPTS}(1).}
\item[\potlitem{NPS}:]{Deprecated synonym for \potlitem{NPTS}(2).}

\item[\potlitem{NTERM} {[$\boldsymbol{-1}$]}:]{An array of \potlitem{MXLAM}
    integers, describing how to evaluate the interaction potential if
    $\potlitem{LVRTP} = \code{.FALSE.}$; see section \ref{vrtp:false}. Each
    element of the array \potlitem{NTERM} corresponds to one element in the
    expansion of the interaction potential.

If $\potlitem{NTERM}(i)<0$, this element of the potential array is
evaluated using the \prog{VINIT}/\prog{VSTAR} mechanism.

If $\potlitem{NTERM}(i)$ is positive, this element of the potential array
is evaluated as a sum of $\potlitem{NTERM}(i)$ exponential or inverse-power
terms, specified by \potlitem{A}, \potlitem{E} and \potlitem{NPOWER}.
\begin{description}
\item[\potlitem{NPOWER} {[0]}:]{Array of integers.  If zero, the
    corresponding term has the form
    $\potlitem{A}\exp(-\potlitem{E}\times R)$, and if positive it has
    the form $\potlitem{A}/R^{\potlitem{NPOWER}}$.}
\item[\potlitem{A} {[0.0]}:]{Array of prefactors for interaction
    potential terms.}
\item[\potlitem{E} {[0.0]}:]{Array of inverse lengths that specify the
    range parameter in exponential potential terms.}\end{description}}
\item[\potlitem{RM} {[1.0]}:]{specifies the unit of length that is used
    by the potential routine, and throughout the programs if
    \inpitem{RUNIT} is not specified in namelist \namelist{\&INPUT}.
    \potlitem{RM} is specified in units of \AA. When a routine
    \prog{VINIT}/\prog{VSTAR} or \prog{VRTP} is supplied, \potlitem{RM} is
    often coded there instead of being supplied in namelist
    \namelist{\&POTL}. Subsequent calls to \prog{POTENL} handle distances
    in these units.}
\end{description}
} 

\chapter{\texorpdfstring{Controlling the print level}
{\ref{CommII}: Controlling the print level}}\mylabel{CommII}

The level of output written to the standard output channel (Fortran unit 6) is
controlled by the integer variable \inpitem{IPRINT}; sensible values of
\inpitem{IPRINT} vary from 1 (when only integral cross sections are required
from \MOLSCAT, or only details of converged energies or fields are required for
bound-state calculations) to 40 (when debugging).

\cbcolor{\mcol}\cbstart For \MOLSCAT, state-to-state integral cross sections
are printed if $\inpitem{ISIGPR}>0$ (regardless of the value of
\inpitem{IPRINT}), scattering lengths/volumes are printed if
$\inpitem{IPRINT}\ge 6$,
and complete S matrices are printed if $\inpitem{IPRINT}\ge 11$. Voluminous
debugging output starts appearing at $\inpitem{IPRINT} = 15$.\cbend

\cbcolor{\bfcol}\cbstart For bound-state calculations (\BOUND\ and \FIELD),
located bound states are printed if $\inpitem{IPRINT}=1$ or greater,
$\inpitem{IPRINT} = 5$ prints the larger components of the bound-state
wavefunction at the matching point and $\inpitem{IPRINT} = 6$ prints brief
information on the progress of locating bound states. Voluminous debugging
output starts appearing at $\inpitem{IPRINT} = 12$.\cbend

A certain amount is always printed, regardless of the value of
\inpitem{IPRINT}, including:
\begin{itemize}[nosep]
\item{Any error conditions that cause the program to stop prematurely}
\item{Warnings about namelist values that have had to be changed because of
    incompatibilities between them}
\item{Some, but not all, other warnings}
\item{Some output from older sections of infrequently used code, such as
    for IOS calculations and line-shape cross sections}
\end{itemize}
Note that this is not a complete list.

\section{Headers and loops}

If $\inpitem{IPRINT}\ge1$, all the programs print:
\begin{itemize}[nosep]
\item a header;
\item the run label;
\item information about propagator choice, ranges and step size control;
\item information about the generation of basis sets;
\item information about energies and sets of EFVs;
\item information about the interaction potential.
\end{itemize}

The programs then loop over \var{JTOT}, symmetry block \var{IBLOCK}, sets of
EFVs and energies as required.

For each \var{JTOT} and symmetry block \var{IBLOCK}, the programs print

\begin{itemize}[nosep]
\item{details of the specific basis set used for that \var{JTOT} and
    \var{IBLOCK} (which for non-diagonal Hamiltonians is the primitive set)
    if $\inpitem{IPRINT}\ge5$.}
\end{itemize}

For each set of EFVs, the programs print

\begin{itemize}[nosep]
\item{for basis sets non-diagonal in $H_{\rm intl}$ or $\hat L^2$,
    information on the asymptotic basis set and how it relates to the
    primitive basis set, and the resulting thresholds if
    $\inpitem{IPRINT}\ge6$ (10 for \FIELD).}
\end{itemize}

\cbcolor{\mbcol}\cbstart
In \MOLSCAT\ and \BOUND, the loop over energies is inside the loop over field.
At each energy, the programs print
\begin{itemize}[nosep]
\item{the energy relative to the reference energy \var{EREF} if $\inpitem{IPRINT}\ge7$;}
\item{the absolute energy if $\inpitem{IPRINT}\ge8$ and $\var{EREF}\ne 0.0$.}
\end{itemize}
\cbend

\cbcolor{\fcol}\cbstart
In \FIELD, the loop over EFVs is inside the loop over energy.
At each EFV, \FIELD\ prints
\begin{itemize}[nosep]
\item{the reference energy \var{EREF} if $\inpitem{IPRINT}\ge10$;}
\item{the absolute energy if $\inpitem{IPRINT}\ge9$ and $\var{EREF}\ne 0.0$.}
\end{itemize}
\cbend

The behaviour of the programs differs within the innermost loop, and so does
the output, but the description of individual propagators is common to all of
them.

\section{Basis sets and quantum numbers}\mylabel{output:basis}

The list of pair state quantum numbers stored in the array \var{JSTATE},
described in sections \ref{basis:diag} and \ref{basis:off-diag}, is printed if
$\inpitem{IPRINT}\ge1$. If $H_{\rm intl}$ and $\hat L^2$ are diagonal, the pair
level index and energy that correspond to each set of quantum numbers are also
printed. Once the basis set has been chosen for a particular set of coupled
equations (i.e., for a particular \var{JTOT} and symmetry block), the basis
functions included are printed if $\inpitem{IPRINT}\ge5$.  For basis sets
diagonal in $H_{\rm intl}$ and $\hat L^2$, each basis function corresponds to a
scattering channel, and its energy is also given.

\section{Coupling matrices}\mylabel{output:potl}

If $\inpitem{IPRINT}\ge26$, the coupling matrices $\boldsymbol{{\cal
V}}^\Lambda$ for the current \var{JTOT} and symmetry block are printed. The
programs print a warning if all the elements of a particular coupling matrix
are zero and $\inpitem{IPRINT}\ge14$. Once all the coupling matrices have been
calculated, the programs print the number of them that are all zero if
$\inpitem{IPRINT}\ge10$.

\section{Reference energy, threshold energies and channel indices}\mylabel{output:Elist}

If energies are specified relative to a non-zero reference energy as described
in section \ref{EREF}, then the method of choosing the reference energy
is printed if $\inpitem{IPRINT}\ge1$. If the value of the reference energy is
independent of the symmetry block and of the values of EFVS, it is also given.
If it depends on the symmetry block or the values of EFVS, it is printed each
time it is calculated if $\inpitem{IPRINT}\ge3$ (for \BOUND), 4 (for \MOLSCAT)
or 6 (for \FIELD).

A list of the channel (or threshold) energies is printed if
$\inpitem{IPRINT}\ge6$ (for \MOLSCAT\ and \BOUND) or 10 (for \FIELD).

\cbcolor{\mcol}\cbstart A list of the open channels is printed if
$\inpitem{IPRINT}\ge10$.\cbend

\subsection{Threshold energies for asymptotically non-diagonal basis sets}\mylabel{output:extra}

For asymptotically non-diagonal basis sets, the internal Hamiltonian is
constructed and diagonalised before propagation in order to obtain the
threshold energies. If $\hat L^2$ is diagonal, the basis set is first separated
into sets corresponding to different values of $L$ (or, if
$\basisitem{IBOUND}=1$, values of \var{CENT} that are the same to within
\inpitem{DEGTOL}). The constant coupling coefficients, $h_{\Omega}$ and (later)
the eigenvalues of the internal Hamiltonian are printed if
$\inpitem{IPRINT}\ge10$. For each set, the matrix of the internal Hamiltonian
is printed if $\inpitem{IPRINT}\ge25$. The eigenvalues are printed if
$\inpitem{IPRINT}\ge 15$. The eigenvectors are printed if $\inpitem{IPRINT}\ge
25$.

If $\hat L^2$ is non-diagonal, it is treated as an extra operator as described
in the following section.

\subsection{Resolving degeneracies amongst threshold energies}\mylabel{output:degens}
\cbcolor{\mcol}\cbstart

If extra operators are used to resolve (near-)degeneracies between threshold
energies, \MOLSCAT\ works through $H_{\rm intl}$ and the extra operators in
turn. The constant coupling coefficients $h_{\Omega}$ and the coefficients for
the extra operators are printed if $\inpitem{IPRINT}\ge6$. Any sets of
eigenvalues of the current operator that are degenerate to within
\inpitem{DEGTOL} are printed if $\inpitem{IPRINT}\ge10$. The print levels for
matrices, eigenvalues and eigenvectors of the submatrices are as in section
\ref{output:extra}. Transformed submatrices are printed if $\inpitem{IPRINT}\ge
30$.

If the eigenvectors need to be reordered to match their ordering in the
internal Hamiltonian, the complete set of reordered eigenvalues is printed if
$\inpitem{IPRINT}\ge15$ and the reordered eigenvectors are printed if
$\inpitem{IPRINT}\ge25$.

Once \MOLSCAT\ has worked through all the operators, the eigenvalues for all
operators are printed for each channel if $\inpitem{IPRINT}\ge6$. The channel
numbering in this list is the one used for the channel threshold energies
described in section \ref{basis:off-diag}. A warning is printed if any channels
are degenerate in all operators.\cbend

\section{Propagations}\mylabel{output:props}

If $R_{\rm min}$ and/or $R_{\rm max}$  have been altered from the input values
as described in section \ref{intrange}, a message is printed to this effect if
$\inpitem{IPRINT}\ge9$. More detailed information about the search for a
suitable value of $R_{\rm min}$ is printed if $\inpitem{IPRINT}\ge13$ and
$\inpitem{IRMSET}>0$.

All propagators (except the WKB integrator) print the values of $R$ at which
the propagation starts and ends, and the number of steps taken, if
$\inpitem{IPRINT}\ge8$. In addition:
\begin{itemize}
\item{The WKB integrator prints the WKB phase shift and how many quadrature
    points were used to obtain it if $\inpitem{IPRINT}\ge4$.  It also gives
    progress information about the searches for turning points if
    $\inpitem{IPRINT}\ge13$; see section \ref{intflgwkb}.}

\item{The DV propagator prints prints the values of $R$ at which
    stabilisation is done if $\inpitem{IPRINT}\ge13$; see section
    \ref{iprop2}.}

\item{The RMAT propagator prints the largest and smallest eigenvalues
    of the matching matrix at each step if $\inpitem{IPRINT}\ge20$; see
    section \ref{iprop3}.}

\item{The VIVS propagator prints the value of $R$ and an estimate of the
    size of the derivative of the irregular solution $f_2^{i'}(R)$ every
    time a new interval is started if $\inpitem{IPRINT}\ge13$, and all the
    control data if $\inpitem{IPRINT}\ge20$; see section \ref{iprop4}.}

\item{The LDMA propagator prints (at each step of the propagation) the
    lowest 30 adiabats if $\inpitem{IPRINT}\ge19$ and the first 30 diagonal
    elements of $\hbar^2{\bf W}/2\mu$ if $\inpitem{IPRINT}\ge20$, the first 9
    radial potential coefficients $v_\Lambda(R)/\potlitem{EPSIL}$ if
    $\inpitem{IPRINT}\ge21$, the interaction matrix ${\bf W}(R)$ (in
    reduced units) if $\inpitem{IPRINT}\ge22$, the matrix of the
    nonadiabatic couplings $\d /\d R$ between the adiabatic states if
    $\inpitem{IPRINT}\ge23$ and the eigenvectors that define the adiabatic
    states if $\inpitem{IPRINT}\ge24$.}

\item{The AIRY propagator prints the maximum values of \var{CDIAG} and
    \var{COFF} and the number of steps in which they exceeded 5 times the
    accuracy tolerance \var{TOLHI} if $\inpitem{IPRINT}\ge12$. It prints
    the individual steps that exceeded this criterion and the sizes of the
    smallest and largest steps if $\inpitem{IPRINT}\ge13$. It prints the
    midpoints, sizes, and diagonal and off-diagonal correction terms for
    each step if $\inpitem{IPRINT}\ge20$; see section \ref{iprop9}.}
\end{itemize}
All the values of \inpitem{IPRINT} in this subsection are increased by 10 for
IOS calculations.

\section[\texorpdfstring{Scattering calculations ({\color{\mcol}\MOLSCAT} only)}
{Scattering calculations (MOLSCAT only)}] {Scattering
calculations\sectionmark{Scattering calculations}}
\sectionmark{Scattering calculations}\mylabel{output:molscat}

\cbcolor{\mcol}\cbstart
\subsection{S-matrix elements}\mylabel{output:Smat}

The open-channel basis functions and wavevectors are printed if
$\inpitem{IPRINT}\ge10$, and S-matrix elements with square modulus greater than
$10^{-20}$ are printed if $\inpitem{IPRINT}\ge11$.

For IOS calculations, the S-matrix elements are printed if
$\inpitem{IPRINT}\ge15$.

\subsection{Eigenphase sums}\mylabel{output:eigsums}

If $\inpitem{IPHSUM}>0$ and $\inpitem{IPRINT}\ge6$, \MOLSCAT\ prints the S-matrix eigenphase sum.

\subsection{Scattering lengths/volumes}\mylabel{output:scatlens}

If $\inpitem{IPRINT}\ge6$, \MOLSCAT\ prints the scattering length/volume for
each channel that has a low kinetic energy. For comparison, various approximate
forms are additionally printed if $\inpitem{IPRINT}\ge26$. These are (where $n$
is 1 if $L=0$, 3 if $L=1$ and 4 otherwise):
\begin{equation}
a_L\approx-\frac{\arg S}{2k^n};\qquad{\rm and}\qquad a_L\approx\frac{(1-S)}{{\rm i}k^n}.
\end{equation}

\subsection{Convergence of S-matrix elements}\mylabel{output:Sconv}

If automatic convergence testing of S-matrix elements is requested, as
described in section \ref{andconv}, convergence information is printed without
additional print controls.

\subsection{Cross sections (non-IOS calculations)}\mylabel{output:CS99}

If $\inpitem{ISIGPR}>0$, some cross section information is printed in the main
output if relevant: if $\inpitem{IPRINT}\ge3$, partial cross sections are
printed after every propagation. If $\inpitem{IPRINT}\ge11$ the state-to-state
integral cross sections accumulated thus far are also printed.

If $\inpitem{ISIGU}>0$ the state-to-state integral cross sections accumulated
thus far are updated on unit \iounit{ISIGU}.

\subsection{Cross sections (IOS only)}\mylabel{output:CS100}

IOS total cross sections are printed regardless of the value of
\inpitem{IPRINT}, as is their average over orientations, which is equal to the
total scattering $Q^t(0,0,0)-Q^s(0,0,0)$. IOS total elastic and inelastic
scattering are also printed regardless of the value of \inpitem{IPRINT}, as are
the contributing dynamical factors $Q^{s{\rm \ or\ }t}(L,M_a,M_b)$.

The collision dynamics factors are also written to unit \iounit{ISAVEU} for
every collision energy and value of \var{JTOT} if $\inpitem{ISAVEU}>0$.

Setting \inpitem{IPRINT} to a positive value gives the following:
\begin{description}
\item[$\inpitem{IPRINT}\ge2$]{prints the collision energy and current value
    of \var{JTOT}.}
\item[$\inpitem{IPRINT}\ge10$]{prints the current contributions to the
    collision dynamics factors, together with the totals accumulated thus
    far.}
\item[$\inpitem{IPRINT}\ge13$]{prints the current contributions to the
    total cross sections, together with the totals accumulated thus far.}
\item[$\inpitem{IPRINT}\ge20$]{prints the current contributions to the T
    matrix with the totals accumulated thus far.}
\end{description}

\subsection{Line-shape cross sections}\mylabel{output:PRCS}

When calculating line-shape cross sections (section \ref{pressbroad}) without
using the IOS approximation, the total accumulated line-shape cross sections
are always printed at the end of the calculation. If $\inpitem{IPRINT}\ge1$,
the running totals are printed after each \var{JTOT} and \var{IBLOCK}. If
$\inpitem{IPRINT}\ge4$, the contributions from individual S-matrices are
printed as they are calculated.

For calculations using the IOS approximation, no additional information is
printed depending on the value of \inpitem{IPRINT}.

\subsection{Characterisation of a resonance or quasibound state as a function of energy or EFV}\mylabel{output:ephase}

For characterisation of a resonance with $\inpitem{IECONV}=4$ or 5 (section \ref{energyconv}) or $\inpitem{IFCONV}=1$ to 5
(section \ref{fieldconv}):

If $\inpitem{IPRINT}\ge2$, the converged resonance location is printed.\hfil\break
If $\inpitem{IPRINT}\ge3$, the resonance parameters at the final step are printed. \hfil\break
If $\inpitem{IPRINT}\ge5$ and \inpitem{IECONV} or \inpitem{IFCONV} is 4 or 5, the partial widths are printed. \hfil\break
If $\inpitem{IPRINT}\ge6$, the current estimates of the resonance parameters are printed after each step.\hfil\break
If $\inpitem{IPRINT}\ge7$, information is printed about the logic used to choose the next energy or EFV, with slightly more detail if
$\inpitem{IPRINT}\ge 8$.

For stepping towards a resonance with $\inpitem{IECONV}=-5$ or $\inpitem{NNRG}<0$:

If $\inpitem{IPRINT}\ge6$, the (groups of 5) energies and eigenphase sums used to estimate the
position of an energy-dependent resonance, as described in section \ref{energyconv}, are printed,
together with the resulting estimate of the resonance position and width. \hfil\break If
$\inpitem{IPRINT}\ge10$, all (3) current estimates of the resonance location and width are printed.
\hfil\break If $\inpitem{IPHSUM}>0$, the eigenphase curvatures and estimated location and width of
an energy-dependent resonance are included on unit \iounit{IPHSUM}.

\subsection{Locating the value of an EFV at which the scattering length/volume has a specific value}
\mylabel{output:specific}

If $\inpitem{IPRINT}\ge4$, information is printed about progress in converging
on the EFV; see section \ref{fieldval}.

\subsection{Effective range}\mylabel{output:effrange}

If $\inpitem{IPRINT}\ge1$, \MOLSCAT\ prints the effective range $r_{\rm eff}$,
as described in section \ref{effrange}.

\subsection{State-to-state cross sections in main output file}\mylabel{CommII:isigpr}

The printing of total and partial cross sections to the main output file is
controlled by the parameter \inpitem{ISIGPR}. This must be set to 1 if printing
of cross sections is required (2 to include coupled-states cross sections that
are incomplete due to missing values of $K$ because of \basisitem{JZCSMX}).\cbend

\section[\texorpdfstring{Bound-state calculations ({\color{\bcol}\BOUND} and {\color{\fcol}\FIELD} only)}
{Bound-state calculations (BOUND and FIELD only)}] {Bound-state
calculations\sectionmark{Bound-state calculations}}
\sectionmark{Bound-state calculations}\mylabel{output:BF}

\cbcolor{\bfcol}\cbstart
\subsection{Bound-state positions in energy or EFV}\mylabel{output:boundstates}

If $\inpitem{IPRINT}\ge6$, the energy (for \BOUND) or the EFV (for \FIELD) is
printed for each propagation. If $\inpitem{IPRINT}\ge7$, the method used to
choose a new value for the next propagation and the resulting new value are
printed, together with the resulting total node count and the eigenvalue of the
matching matrix with the smallest absolute value.

If $\inpitem{IPRINT}\ge8$, \BOUND\ and \FIELD\ print the node count for each
propagation segment and the number of negative eigenvalues of the matching
matrix (section \ref{nodecount}). They print the eigenvalues themselves if
$\inpitem{IPRINT}\ge9$. They print the full matching matrix if
$\inpitem{IPRINT}\ge12$, and the log-derivative matrix at the end of each
propagation part if $\inpitem{IPRINT}\ge15$.

If $\inpitem{IPRINT}\ge1$, \BOUND\ and \FIELD\ print the location of a bound
state when convergence on it has succeeded. They print its absolute energy if
$\inpitem{IPRINT}\ge8$. They also print the larger components of the
wavefunction at the matching point if $5\leq\inpitem{IPRINT}<11$ or \emph{all}
the components if $\inpitem{IPRINT}\ge11$.  If $\inpitem{IPRINT}\ge8$ they
print the CPU time taken to converge on that bound state.

If $\inpitem{IBDSUM}>0$, \BOUND\ and \FIELD\ print the state number and
location (energy, EFV set) of located bound states to unit \iounit{IBDSUM}.

If performing a scan, \BOUND\ and \FIELD\ print the node count and smallest
eigenvalue of the matching matrix at each energy (for \BOUND) or EFV set (for
\FIELD) if $\inpitem{IPRINT}\ge1$.\cbend

\cbcolor{\bcol}\cbstart
\subsection[\texorpdfstring{Expectation values ({\color{\bcol}\BOUND} only)}
{Expectation values (BOUND only)}]{Expectation values\subsectionmark{Expectation values}}
\subsectionmark{Expectation values}\mylabel{output:expect}

Expectation values (section \ref{calcexp}) are printed if
$\inpitem{IPRINT}\ge1$.\cbend

\cbcolor{\bcol}\cbstart
\subsection[\texorpdfstring{Convergence testing ({\color{\bcol}\BOUND} only)}
{Convergence testing (BOUND only)}]{Convergence testing}\mylabel{output:bconv}

Results of any convergence testing are printed if $\inpitem{IPRINT}\ge1$; see
section \ref{conv}.\cbend

\cbcolor{\bfcol}\cbstart
\subsection{Calculation of wavefunction}\mylabel{output:bwvfns}

The wavefunction (section \ref{calcwaveBF}) is output
on unit \iounit{IWAVE}; see section \ref{CommII:bwave}.

\BOUND\ and \FIELD\ print the total normalisation integral if
$\inpitem{IPRINT}\ge6$, and the contribution from each basis function if
$\inpitem{IPRINT}\ge8$.  The entire wavefunction is printed in the main output
file if $\inpitem{IPRINT}\ge30$.\cbend

\chapter{\texorpdfstring{Auxiliary output and scratch files}{\ref{Auxfiles}: Auxiliary output and scratch files}}\mylabel{Auxfiles}

The programs may produce a number of auxiliary files, depending on the values
of parameters in namelist \namelist{\&INPUT} (except \iounit{IASYMU}, which is
controlled by namelist \namelist{\&BASIS}). They also use scratch files under
certain circumstances. The following is a complete list of these auxiliary
files and the input parameters that control whether they are used.
Direct-access files are indicated by DA. The programs that make use of each
file are indicated by their initial letter(s). The files are used only if the
corresponding unit number is set to (or defaults to) a non-zero value.

\begin{shaded} 
Unit numbers specified in input files should be in the range 1 to 799,
avoiding 5 and 6 (which are standard input and output, respectively).
Values from 800 to 999 are reserved for internal use.
No number should be used for more than one file.
\end{shaded} 

{\def\mytabwidth{2.5in}
\parindent 0pt
\begin{tabular}{lccll}
\hline
   I/O unit & Used by & unformatted  &   Use & See section\\
\hline
\iounit{IASYMU} & {\scshape MBF} & no & \parbox[t]{\mytabwidth}{\raggedright Asymmetric top rotor functions for $\var{ITYP}=4$ or 6\strut}&\parbox[t]{0.75in}{\raggedright\ref{ityp4}, \ref{ityp6}}\\
\iounit{IBDSUM} & {\scshape BF} & no & \parbox[t]{\mytabwidth}{\raggedright Summary of energies and EFV values for bound states\strut}&\ref{ksave:bf}\\
\iounit{IPHSUM}  & {\scshape M} &  no    &  \parbox[t]{\mytabwidth}{\raggedright Eigenphase sums and scattering lengths for low-energy scattering channel \inpitem{ICHAN}\strut}&\ref{ksave:mol}\\
\iounit{ISAVEU} & {\scshape M} &  yes     &  \parbox[t]{\mytabwidth}{\raggedright S matrices if $\inpitem{IPHSUM}=0$\strut}&\ref{CommII:SandK} \\
&  &  & \parbox[t]{\mytabwidth}{\raggedright K matrices if $\inpitem{IPHSUM}>0$\strut}&\ref{CommII:SandK}\\
\iounit{IWAVE} & {\scshape MBF} & no & \parbox[t]{\mytabwidth}{\raggedright Wavefunctions\strut}&\ref{calcwaveBF}, \ref{calcwaveM}\\
\iounit{ILDSVU} & {\scshape M} & yes & \parbox[t]{\mytabwidth}{\raggedright Log-derivative matrix \strut}&\ref{CommII:LD}\\
\iounit{IVLU} & {\scshape MBF} & yes & \parbox[t]{\mytabwidth}{\raggedright Coupling matrices\strut}&\ref{CommIII:ivlu}\\
\iounit{ICONVU} & {\scshape M} & yes & \parbox[t]{\mytabwidth}{\raggedright S matrices for use in convergence runs\strut}&\ref{andconv}\\

\hline
   scratch unit &  &   &    & \\
\hline
\iounit{ISIGU}  & {\scshape M} &  no, DA &  \parbox[t]{\mytabwidth}{\raggedright Cross sections, updated after each S matrix is calculated\strut}&\ref{CommII:isigu}\\
%
%
%
\iounit{ISCRU}  & {\scshape MBF} &   yes     &  \parbox[t]{\mytabwidth}{\raggedright Propagator scratch unit\strut}&\ref{CommIII:iscru}\\
\hline
\end{tabular}}

\medskip The following sections describe these files in more detail, except
for those covered elsewhere.

\section[\texorpdfstring{Summary of bound states ({\color{\bcol}\BOUND} and
{\color{\fcol}\FIELD} only)}{Summary of bound states (BOUND and FIELD only)}]
{Summary of bound states}
\subsectionmark{Summary of bound states}
\mylabel{ksave:bf}

\cbcolor{\bfcol}\cbstart \BOUND\ and \FIELD\ write a concise summary of
converged bound-state energies or EFVs on \iounit{IBDSUM}. The summary includes
the state number, together with a warning if it does not agree with the
expected value (but see section \ref{nodecount}).\cbend

\section[\texorpdfstring{Summary of eigenphase sums and low-energy scattering lengths
({\color{\mcol}\MOLSCAT} only)}{Summary of eigenphase sums and low-energy
scattering lengths (MOLSCAT only)}]{Summary of eigenphase sums and low energy
scattering lengths}\mylabel{ksave:mol}

\cbcolor{\mcol}\cbstart Eigenphase sums are written to unit \iounit{IPHSUM} if
$\inpitem{IPHSUM}>0$.   Scattering lengths for channel \inpitem{ICHAN} are
written to \iounit{IPHSUM} if both \inpitem{ICHAN} and \inpitem{IPHSUM} are set
greater than 0.\cbend

\section[\texorpdfstring{State-to-state integral cross sections
({\color{\mcol}\MOLSCAT} only)}{State-to-state integral cross sections (MOLSCAT
only)}]{State-to-state integral cross sections\sectionmark{State-to-state
integral cross sections}}\sectionmark{State-to-state integral cross sections}
\mylabel{CommII:isigu}

\cbcolor{\mcol}\cbstart If $\inpitem{ISIGU}>0$, \MOLSCAT\ maintains a (direct
access) file containing the state-to-state integral cross sections accumulated
thus far on unit \iounit{ISIGU}. This file is updated every time an S matrix is
processed to give contributions to the cross sections, so it contains valid
information about the run so far even if the program terminates abnormally.

Code to read the contents of this file is included in this release (subroutine
\prog{RDSIGU}), but is not executed.\cbend

\section[\texorpdfstring{Partial cross sections
({\color{\mcol}\MOLSCAT} only)}{Partial cross sections (MOLSCAT only)}]{Partial
cross sections\sectionmark{Partial cross sections}}\sectionmark{Partial cross
sections} \mylabel{CommII:ipartu}

\cbcolor{\mcol}\cbstart The option to print partial cross sections to a
separate file, on unit number \iounit{IPARTU}, is not implemented in \MOLSCAT\
version \currentversion, but could be resuscitated if necessary.\cbend

\section[\texorpdfstring{S and K matrices ({\color{\mcol}\MOLSCAT} only)}{S
and K matrices (MOLSCAT only)}]{S and K matrices\sectionmark{S and K
matrices}}\sectionmark{S and K matrices}\mylabel{CommII:SandK}

\cbcolor{\mcol}\cbstart In addition to its main printed output on unit 6,
\MOLSCAT\ can also produce files containing S matrices and/or K matrices for
subsequent processing by other programs. The S matrix output is compatible with
programs \prog{DCS} \cite{DCS} (for differential cross sections) and \prog{SBE}
\cite{SBE} (for generalised cross sections for transport and relaxation
properties and Senftleben-Beenakker effects). The K matrix output can be read
by program \prog{SAVER}, which accumulates output from different runs and
outputs it in a format suitable for program \prog{RESFIT}
\cite{Hutson:resfit:2007} (which fits eigenphase sums to obtain the positions
and widths of Feshbach resonances diagonal S-matrix elements to obtain partial
widths).

If $\iounit{ISAVEU}>0$, \MOLSCAT\ saves \emph{either} S matrices \emph{or} K
matrices on unit \iounit{ISAVEU}. If $\inpitem{IPHSUM}\le0$ it saves S matrices
and if $\inpitem{IPHSUM}>0$ it saves K matrices.

The format of these files is described below. Some aspects of the formats have
changed between versions, and the files include an \iounit{ISAVEU} format
version number \var{IPROGM} as described below. If desired the subroutines
\prog{SKREAD} (with entry points \prog{HDREAD}, \prog{SLPRD} and \prog{KLPRD})
and \prog{SREAD} can be used by other post-processor programs to read the
headers and loop output, and take account of values of \var{IPROGM} from recent
versions of \MOLSCAT.

\subsection[\texorpdfstring{S matrices ({\color{\mcol}\MOLSCAT} only)} {S
matrices (MOLSCAT only)}]{S matrices\sectionmark{S matrices}}
\sectionmark{S matrices}\mylabel{Sout}

\MOLSCAT\ saves S matrices in an unformatted (binary) file. The results are
written as single (logical) records (i.e., single unformatted \code{WRITE}
statements), except for (8), which is described more fully below. Beginning
with $\var{IPROGM}=14$ (August 1994), \var{NOPEN} is in the record before the
one in which it is used. Beginning with $\var{IPROGM}=17$, values for EFVs are
included in the record that starts with \var{JTOT}, and there are two
additional records immediately before this one.  The first of these contains
integer variables relating to whether the basis set is diagonal or not, and the
second gives the number of EFVs and their names and units.

If S matrices are output to the \iounit{ISAVEU} file, its contents are as
follows:
\begin{enumerate}
\item{\code{\inpitem{LABEL}, \basisitem{ITYPE}, \var{NSTATE}, \var{NQN},
    \inpitem{URED}, \var{IPROGM}}
       \begin{description}
         \item[\inpitem{LABEL}]{is the title of the run and is a
             character variable of length 80.}
         \item[\basisitem{ITYPE}]{specifies the interaction type.}
         \item[\var{NSTATE}]{is the number of pair states in the basis
             set.}
         \item[\var{NQN}]{is the number of (quantum) labels per pair
             state.}
         \item[\inpitem{URED}]{is the reduced mass in units of \inpitem{MUNIT} (unified
             atomic mass units by default).}
         \item[\var{IPROGM}]{is the version number for the format of
             the output written to unit \iounit{ISAVEU}. \var{IPROGM}
             is distinct from the program version number
             and is 19 for \MOLSCAT\ version \currentversion.}
       \end{description}}
\item{\code{((\var{JSTATE}(I,J), I = 1, \var{NSTATE}), J = 1, \var{NQN})}
      \begin{description}
      \item[\code{\var{JSTATE}(ISTATE,J)}]{are the quantum numbers of state
          \var{ISTATE}. The meaning depends on \basisitem{ITYPE}; see
          chapter \ref{ConstructBasis}.}
      \end{description}}
\item{\code{\var{NLEVEL}, (\var{ELEVEL}(I), I = 1, \var{NLEVEL})}\\ Number of pair levels and
    the contents of the array \var{ELEVEL}, which contains pair level
    energies indexed by \var{INDLEV}.}
\item{\var{NDGVL}, \var{NCONST}, \var{NRSQ}, \var{IBOUND}, \var{ITPSUB}
	\begin{description}
		\item[\var{NDGVL}]{is the number of diagonal terms in the internal
Hamiltonian that depend on EFVs.}
		\item[\var{NCONST}]{is the number of terms that contribute to a
non-diagonal internal Hamiltonian.}
		\item[\var{NRSQ}]{is the number of terms that contribute to a
non-diagonal ${\cal L}$ operator.}
		\item[\var{IBOUND}]{is 0 if the matrix elements for the
operator ${\cal L}$ are just the diagonal values $L(L+1)$, and 1 if
they are in the array \var{CENT}.}
        \item[\var{ITPSUB}]{is an integer flag that may be set when
            $\basisitem{ITYPE}=9$ to specify the particular plug-in
            basis-set suite that produced the results.}
	\end{description}
}
\item{\var{NEFV}, \var{ISVEFV},
    (\code{\var{EFVNAM}(IEFV), \var{EFVUNT}(IEFV), IEFV = 1, \var{NEFV})}
\begin{description}
      \item[\var{NEFV}]{is the number of EFVs (excluding potential
          scaling).}
      \item[\var{ISVEFV}]{is the index of the single varying EFV (set
          to $\var{NEFV}+1$ for a proxy EFV).}
      \item[\var{EFVNAM}]{are the names of the EFVs (excluding
          potential scaling).}
      \item[\var{EFVUNT}]{are the units of the EFVs (excluding
          potential scaling).}
\end{description}
}

\item{\code{\var{NFIELD}, \inpitem{NNRG}, (\var{ENERGY}(I), I = 1, \inpitem{NNRG})}\\
      Number of different sets of external fields to be looped over
      followed by the number and values of the scattering energies
      (cm$^{-1}$).}

\item{\code{\var{JTOT}, \var{INRG}, \var{IBLOCK}, \var{IFIELD}, \var{EN}, (\var{EFV}(IEFV), IEFV = 0, \var{NEFVP}), \var{EREF},\ \ \ \ \&\\
\var{IEXCH}, \var{WT}, \var{NOPEN}}\\
      These describe a single scattering calculation:
\begin{description}
      \item[\var{JTOT}]{is the total angular momentum.}
      \item[\var{INRG}]{is the index of the energy in the list in 5
          above.}
      \item[\var{IBLOCK}]{is the index of the symmetry block.}
      \item[\var{IFIELD}]{is the index of the current set of EFVs.}
      \item[\var{EN}]{is the current scattering energy (in cm$^{-1}$)
          relative to the reference energy \var{EREF}; it should equal
          \code{\var{ENERGY}(\var{INRG})}.}
      \item[\var{EFV}]{is the array of the current values of the EFVs.
          \var{NEFVP} is equal to \var{NEFV} unless EFVs are calculated
          from a proxy EFV, in which case it is $\var{NEFV}+1$.}
      \item[\var{EREF}]{is the energy that \var{EN} is referenced to,
          so that $\var{EN}+\var{EREF}$ is the total scattering
          energy.}
      \item[\var{IEXCH}]{is the exchange parity for identical molecules
          \begin{description}
          \item[$\var{IEXCH} = 0$]{no exchange symmetry}
          \item[$\var{IEXCH} = 1$]{odd exchange symmetry}
          \item[$\var{IEXCH} = 2$]{even exchange symmetry}
          \end{description}}
      \item[\var{WT}]{(if nonzero) is the statistical weight for the
          current values of \var{JTOT} and \var{IBLOCK}.}
      \item[\var{NOPEN}]{is the number of open channels in the S
          matrix.}
\end{description}}
\item{\code{(}{\it index}\code{(I),} $L$\code{(I), \var{WV}(I), I = 1, \var{NOPEN})}
\begin{description}
\item[{\it index}\code{(I)}]{is a pointer to the arrays of
    quantum numbers and threshold energies.
    \begin{itemize}\item{For asymptotically diagonal basis sets, {\it index} is a
    pointer to the array \var{JSTATE} that contains
    pair state quantum numbers. \code{JSTATE(}{\it
    index}\code{(I),NQN)} is itself a pointer to the element of the
    array \var{ELEVEL} that contains the threshold energy for
    channel \code{I}.}
    \item{For asymptotically non-diagonal basis sets, {\it
    index}\code{(I)} is the index of the element of the array
    \var{ELEVEL} that contains the threshold energy for channel
    \code{I}.}\end{itemize}}
\item[$L$\code{(I)}]{specifies the orbital angular momentum
    for channel \code{I}.
    \begin{itemize}\item{If $\var{IBOUND}=0$, it is the integer
    \var{L}\code{(I)}.}
    \item{If $\var{IBOUND}=1$, it is the diagonal matrix element of
        $\hat L^2$ stored in \var{CENT}\code{(I)}. \hfil\break If
        $\var{NRSQ}\ne0$, this value is obtained by diagonalising
        the matrix of $\hat L^2$.}\end{itemize}}
\item[\code{\var{WV}(I)}]{is the wavevector of channel \code{I}
          (\AA$^{-1}$).}
      \end{description}}
\item{\var{SREAL}}
\item{\var{SIMAG}\\
      \code{\var{SREAL}(I)} and \code{\var{SIMAG}(I)} are the real and imaginary parts of
      the \var{NOPEN} by \var{NOPEN} S matrix. They are each written as a
      single record, listing only the lower triangle, i.e.,
{\setlength\parskip{0pt}\tt
((SREAL(I,J), J = 1, I), I = 1, NOPEN)\par
((SIMAG(I,J), J = 1, I), I = 1, NOPEN)\par
}
}
\end{enumerate}

Records 7--10 are repeated for each S matrix calculated, looping over
\var{IFIELD} (innermost), \var{INRG}, \var{IBLOCK} and \var{JTOT} (outermost):
\begin{tcolorbox}[colframe=white,colback=white]
\begin{verbatim}
     DO JTOT = JTOTL, JTOTU, JSTEP
       DO IBLOCK = 1, NBLOCK
         DO INRG = 1, NNRG
           DO IFIELD = 1, NFIELD
             7.), 8.), 9.), 10)
           ENDDO
         ENDDO
       ENDDO
     ENDDO
\end{verbatim}
\end{tcolorbox}
\var{NBLOCK} depends on \basisitem{ITYPE}. Note that not every S matrix
necessarily exists. S matrices may be missing from the file either because
there are no open channels for that energy, or because there was an error or
convergence failure in the calculation.

Subroutines \prog{SKREAD} (with entry points \prog{HDREAD} and \prog{SLPRD})
and \prog{SREAD} are included in this distribution for use in reading the
unformatted files that have been standard since version 11. Line 1 must be read
by the program directly, following which a call to entry point \prog{SHDRD}
reads lines 2--6.  Within loops over \var{JTOT}, \var{IBLOCK}, \var{INRG} and
\var{IFIELD} the entry point \prog{SLPRD} can be called to read the rest of the
records.  It uses \prog{SREAD} to read the real and imaginary parts of the S
matrices written in records 9 and 10.

\subsection[\texorpdfstring{K matrices ({\color{\mcol}\MOLSCAT} only)}
{K matrices (MOLSCAT only)}]{K matrices\sectionmark{K matrices}}
\sectionmark{K matrices}\mylabel{Kout}

If K matrices are output to the \iounit{ISAVEU} file,
the (unformatted) output is very similar to the output for S matrices described
above. The first 8 records are exactly as described above for the S matrix
file.  They are followed by:

\begin{enumerate}\setcounter{enumi}{8}
\item{\code{(\var{SREAL}(I), \var{SIMAG}(I), I = 1, \var{NOPEN}*\var{NOPEN}, \var{NOPEN}+1)} are the
    diagonal elements of the S matrix.}
\item{\code{((\var{AKMAT}(I,J), J = 1, I), I = 1, \var{NOPEN})} is the K matrix. Only
    the lower triangle is written.}
\item{\var{ESUM} is the eigenphase sum.}
\end{enumerate}

Subroutines \prog{SKREAD} (using entry points \prog{HDREAD} and \prog{KLPRD})
and \prog{SREAD} are included in this distribution for use in reading the
unformatted files that have been standard since version 11. Line 1 must be read
by the program directly, following which a call to entry point \prog{HDREAD}
reads lines 2--6.  Within loops over \var{JTOT}, \var{IBLOCK}, \var{INRG} and
\var{IFIELD} the entry point \prog{KLPRD} can be called to read records 7--10.
It uses \prog{SREAD} to read the K matrices written in record 10.\cbend

\section{Wavefunctions (LDMD
propagator only)}\sectionmark{Wavefunctions (LDMD propagator only)}
\mylabel{CommII:bwave}

If $\inpitem{IWAVE}>0$, the multichannel wavefunction is written to unit \iounit{IWAVE}. For
\BOUND\ and \FIELD\ this is the bound-state wavefunction, normalised as described in section
\ref{calcwaveBF}. For \MOLSCAT\ it is the wavefunction for flux incoming only in scattering channel
\inpitem{ICHAN}, as described in section \ref{calcwaveM}. The logical variable \inpitem{IWAVEF}
specifies whether the file is formatted (\code{.TRUE.}) or unformatted (\code{.FALSE.}). By default
it is formatted. The file is often large, particularly if the wavefunction is written at every
point of the propagation.

If the wavefunction file is unformatted, it contains the following, with each numbered item on
a separate record:
\begin{itemize}
\item{An overall header, written once per calculation, containing:
\begin{enumerate}
\item{\code{WAVEOF} specifies the output format version (character variable of length 8,
    currently \code{beta20}).}
\item{\inpitem{LABEL}, \basisitem{ITYPE}, \var{ITPSUB}, \var{IBOUND}, \var{NQN},
\inpitem{URED}\par
\begin{description}
\item[\inpitem{LABEL}]{is the title of the run (character variable of length 80).}
\item[\basisitem{ITYPE}]{is the interaction type.}
\item[\var{ITPSUB}]{is an integer flag that may be set when $\basisitem{ITYPE}=9$
    to specify the particular plug-in basis-set suite that produced the results.}
\item[\var{IBOUND}]{is 0 if the matrix elements for the operator ${\cal L}$ are
just the diagonal values $L(L + 1)$, and 1 if they are in the array \var{CENT}.}
\item[\var{NQN}]{is the number of quantum labels per pair state.}
\item[\inpitem{URED}]{is the reduced mass.}
\end{description}}
\item{\inpitem{MUNIT}, \inpitem{RUNIT}, \inpitem{EUNIT}, \inpitem{CONFRQ}\par
\begin{description}
\item[\inpitem{MUNIT}, \inpitem{RUNIT}, \inpitem{EUNIT}]{specify the units for
the reduced mass, length and energy and are expressed in units of unified atomic
mass units, \AA\ and  cm$^{-1}$ respectively}
\item[\inpitem{CONFRQ}]{is the harmonic confinement frequency (0.0 if not used)}
\end{description}}
\item{\code{NEFV, ISVEFV, (EFVNAM(IEFV), EFVUNT(IEFV), IEFV = 1, NEFV)}
\begin{description}
\item[\var{NEFV}]{is the number of EFVs (excluding potential scaling).}
\item[\var{ISVEFV}]{is the index of the single varying EFV (set to \code{NEFV +
1} for a proxy EFV).}
\item[\var{EFVNAM}]{are the names of the EFVs (excluding potential scaling)}
\item[\var{EFVUNT}]{are the units of the EFVs (excluding potential scaling)}
\end{description}}
\item{\inpitem{IWVSTP} indicates the values of $R$ at which wavefunctions are written.  If
    positive and greater than 1, each wavefunction is written only at the first value and
    every subsequent \inpitem{IWVSTP}-th value.  If negative, it is written at only one
    distance.  Currently the only allowed negative value is $-2$, corresponding to writing
    only at $R_{\rm max}$.}
\end{enumerate}}
\item{Information about the current symmetry block, written once for each combination of
    \var{JTOT} and \var{IBLOCK}:
\begin{enumerate}\setcounter{enumi}{5}
\item{\var{JTOT}, \var{IBLOCK}, \var{NBASIS}\par
\begin{description}
\item[\var{JTOT}]{is the current value of $J_{\rm tot}$ (or of the quantum number
    implemented as \var{JTOT} in a plug-in basis-set suite).}
\item[\var{IBLOCK}]{is the index for the current symmetry block}
\item[\var{NBASIS}]{is the size of the basis set for the current symmetry block.}
\end{description}}
\item{\code{((JBASIS(IBASIS, JQN), IBASIS = 1, NBASIS), JQN = 1, NQN-1)} are the quantum
numbers for each basis function}
\item{$\code{(}L$\code{(IBASIS), IBASIS = 1, NBASIS)} specifies the orbital angular
momentum for each basis function
\begin{itemize}
\item{If $\var{IBOUND}=0$, it is the integer \code{L(IBASIS)}}
\item{If $\var{IBOUND}=1$, it is the diagonal matrix element of $\hat L^2$
stored in \code{CENT(IBASIS})}
\end{itemize}}
\item{\code{NSETS} is the number of energies (for \FIELD) or sets of EFVs (for
    \MOLSCAT\ or \BOUND) for which wavefunctions are calculated.}
\end{enumerate}}
\item{Loop counters for the current energy (for \FIELD) or set of EFVs (for \MOLSCAT\ or
    \BOUND):}
\begin{enumerate}\setcounter{enumi}{9}
\item{\code{ISET} is the index of the current energy (for \FIELD) or set of EFVs (for
    \MOLSCAT\ or \BOUND)}
\item{\code{ISTONE}, \code{ISTEND}
\begin{itemize}
\item{For \BOUND\ and \FIELD, \cbcolor{\bfcol}\cbstart \code{ISTONE} and \code{ISTEND}
    are the lowest and highest state numbers that exist for \code{ISET} in the
    specified range of energy (for \BOUND) or EFV (for \FIELD).\cbend}
\item{For \MOLSCAT, \cbcolor{\mcol}\cbstart \code{ISTONE} is the incoming channel number
    and \code{ISTEND} is the number of energies at which scattering wavefunctions are
    calculated for the current set of EFVs\cbend}
\end{itemize}}
\end{enumerate}
\item{Information about the current wavefunction:
\begin{enumerate}\setcounter{enumi}{11}
\item{\var{ISTATE}, \var{EREL}, \var{EREF}\par
\begin{description}
\item[$|\var{ISTATE}|$]{is the state number (or energy index for \MOLSCAT) for the
    current wavefunction.  If \var{ISTATE} is negative, it indicates that \BOUND\
    or \FIELD\ did not converge on state $|\var{ISTATE}|$. In this event the
    wavefunction written to the file is the back-substituted solution at the last
    energy/field in the convergence sequence; this is often a good approximation,
    unless the sequence has diverged or is converging on the wrong state.}
\item[\var{EREL}]{is its relative energy in units of \inpitem{EUNIT}}
\item[\var{EREF}]{is the current reference energy that \var{EREL} is relative
to}
\end{description}}
\item{\code{(EFV(IEFV), IEFV = 0, MAX(NEFV, ISVEFV))}\par
\begin{description}
\item[\code{EFV}]{are the current values for the EFVs, including the scaling
factor and any dummy EFV}
\end{description}}
\item{\code{NSEG} is the number of segments used for the propagation}
\item{\code{(RBSEG(ISEG), RESEG(ISEG), DRSEG(ISEG), IPRSEG(ISEG), ISEG = 1,
NSEG)}\par
\begin{description}
\item[\var{RBSEG}, \var{RESEG}, \var{DRSEG}, \var{IPRSEG}]{are the endpoints, step
    length and flag for the propagator used for each segment}
\end{description}}
\item{\code{NPOINT} is the total number of points (values of $R$) at which the wavefunction
    is written.}
\end{enumerate}}
\item{The wavefunction itself:
\begin{enumerate}\setcounter{enumi}{16}
\item{\code{(R(IPOINT), PSI(IBASIS, IPOINT), IBASIS = 1, NBASIS)}}
\end{enumerate}}
\end{itemize}

The overall loop structure is:

\begin{itemize}[nosep]
\item[--]{\cbcolor{\mcol}\cbstart
For \MOLSCAT, $\code{ISTBEG}=1$
\cbend}
\item[--]{\cbcolor{\bfcol}\cbstart
For \BOUND\ and \FIELD, $\code{ISTBEG}=\code{ISTONE}$
\cbend}
\end{itemize}

\smallskip
{\parskip0pt
Begin calculations\par
1.), 2.), 3.), 4.), 5.)\par
\code{DO JTOT = JTOTL, JTOTU, JSTEP}\par
{\tt\ \ }\code{DO IBLOCK = 1, NBLOCK}\par
{\tt\ \ \ \ }6.), 7.), 8.), 9.)\par
{\tt\ \ \ \ }\code{DO ISET = 1, NSETS}\par
{\tt\ \ \ \ \ \ }10.), 11.)\par
{\tt\ \ \ \ \ \ }\code{DO ISTATE = ISTBEG, ISTEND}\par
{\tt\ \ \ \ \ \ \ \ }12.), 13.), 14.), 15.), 16.)\par
{\tt\ \ \ \ \ \ \ \ }\code{DO IPOINT = 1, NPOINT}\par
{\tt\ \ \ \ \ \ \ \ \ \ }17.)\par
{\tt\ \ \ \ \ \ \ \ }\code{ENDDO}\par
{\tt\ \ \ \ \ \ }\code{ENDDO}\par
{\tt\ \ \ \ }\code{ENDDO}\par
{\tt\ \ }\code{ENDDO}\par
\code{ENDDO}\par
End calculations
}

The formatted file contains the same information, but some quantities are
excluded if not relevant.  It also contains text descriptions of the content.

In the unformatted version of the file, all the channel functions for a given value of $R$ are
written on the same record. In the formatted file, however, the maximum number of channel functions
written per line is controlled by \inpitem{NWVCOL}; it is equal to \inpitem{NWVCOL} in \BOUND\ and
\FIELD\ and to $\inpitem{NWVCOL}/2$ for \MOLSCAT\ (since scattering wavefunctions are complex and
thus require 2 values per channel).

\goodbreak
\section[\texorpdfstring{Log-derivative matrices({\color{\mcol}\MOLSCAT} only)}
{Log-derivative matrices (MOLSCAT only)}]{Log-derivative matrices
\sectionmark{Log-derivative matrices}}
\sectionmark{Log-derivative matrices}\mylabel{CommII:LD}

\cbcolor{\mcol}\cbstart
If $\inpitem{ILDSVU}>0$, the log-derivative matrix at
$R_{\rm max}$ is output on unit \iounit{ILDSVU}. This may be required for other
programs, such as those to implement MQDT \cite{Croft:MQDT:2011,
Croft:MQDT:2012}.

Since this file is designed to be processed by other programs, a full
description of what is written to it is given here. This description requires
significant understanding of internal structures and variable names, and should
be read only by expert users who need it. It has not yet been generalised to
handle multiple EFVs.



In brief, the output on unit \iounit{ILDSVU} has the structure:
\begin{enumerate}[nosep]
\item{Global header}
\item{Global vector}
\item[ ]{Looping over number of propagations
      \begin{enumerate}[nosep]
      \item{Propagation header}
      \item{Propagation external fields}
      \item{Propagation vectors}
      \item[ ]{Looping over number of matrices
            \begin{enumerate}[nosep]
            \item{Matrix data}
            \end{enumerate}}
      \end{enumerate}}
\end{enumerate}

These various parts contain the following variables/values:
\begin{description}
\item[Global header]{contains
\begin{verbatim}
LABEL, ITYPE, NSTATE, NLEVEL, NQN, NNRG, NFIELD, URED, IPROGM,
\end{verbatim}
{\tt NDGVL, NCONST, NRSQ, IBOUND, ISVEFV, NEFV,}\par
{\tt (EFVNAM(IEFV), EFVUNT(IEFV), IEFV = 1, NEFV),}\par
}
\item[Global vectors]{contains

\code{\var{JSTATE}(\var{NSTATE},\var{NQN}), \var{ELEVEL}(\var{NLEVEL}),
\var{ENERGY}(\inpitem{NNRG})}}
\item[Propagation header]{contains
\begin{verbatim}
JTOT, INRG, EN, IEXCH, WT, M, NCH, ERED, RMLMDA
\end{verbatim}
}
\item[Propagation external fields]{contains
{\parskip0pt\par
\code{(EFV(IEFV), IEFV = MIN(1,ISVEFV), MAX(NEFV,ISVEFV))}\par
}}
\item[Propagation vectors]{contains

\code{\var{JSINDX}, \var{L}, \var{EINT}}, which all have length \var{NCH}. }
\item[Matrix data]{contains

\var{MATCODE} (For future expansion: currently a large negative integer)

\inpitem{RMID}, \var{Y} (\var{Y} is the \var{NCH} by \var{NCH}
log-derivative matrix)}
\end{description}\cbend

\section{Coupling matrices}\mylabel{CommIII:ivlu}

\cbcolor{\mcol}\cbstart This option is designed for cases where there are many
expansion terms contributing to the interaction potential (and/or other
operators) and it requires excessive memory to store them internally. If
$\basisitem{IVLU}>0$, the coupling matrices are written to unit \iounit{IVLU}
and the program reads them back in one at a time when constructing the
interaction matrix. This saves memory at the expense of disc I/O, so is
generally worthwhile only when the available memory is otherwise
insufficient.\cbend

\chapter{\texorpdfstring{Example input and output files}
{\ref{testfiles}: Example input and output files}}\mylabel{testfiles}

We have provided a selection of example input files and their associated
outputs, to give examples of program features and to allow users to verify that
their program build is operating correctly. These are in subdirectory
\file{examples/input}, with the corresponding output files in
\file{examples/output}.

The example calculations are intended to be illustrative, and in some cases use basis sets or
propagation parameters that are not fully converged.

Instructions for building executables to run the example
calculations are given in Section \ref{compiling}.

\section[\texorpdfstring{Examples for {\color{\mcol}\MOLSCAT}}
{Examples for MOLSCAT}]{Examples for \MOLSCAT}\mylabel{testfiles:molscat}

\cbcolor{\mcol}\cbstart

\subsection{\texorpdfstring{All available propagators for \MOLSCAT}
{All available propagators for MOLSCAT}}\mylabel{testfiles:molscat:intflgs}

\begin{tabular}{ll}
input file: & \file{molscat-all\_propagators.input}\\
executable: & \file{molscat-basic}
\end{tabular}

\file{molscat-all\_propagators.input} contains input data for the same model of
collisions between an atom and a linear rigid rotor as was used for the basic
example in section \ref{basic:m1}.  The radial potential coefficients are
provided in the input file and consist of a Lennard-Jones 12-6 potential for
$\lambda=0$ and a dispersion-like $R^{-6}$ form for $\lambda=2$. The program
carries out full close-coupling ($\basisitem{ITYPE}=1$) calculations for a
single partial wave and total parity and prints the resulting S matrix. The
calculation is repeated using combinations of short-range and long-range
propagators that exercise every propagation method available in \MOLSCAT\
(though not every possible combination).

\subsection{\texorpdfstring{All available coupling approximations (using $\var{ITYP}=2$)}
{All available coupling approximations (using ITYP =
2)}}\mylabel{testfiles:molscat:iadds}

\begin{tabular}{ll}
input file: & \file{molscat-all\_iadds.input}\\
executable: & \file{molscat-basic}
\end{tabular}

\file{molscat-all\_iadds.input} contains input data for a similar model system,
extended this time to include vibrations of the linear rotor ($\var{ITYP}=2$).
The radial potential coefficients are again provided in the input file, and all
consist of inverse-power functions of $R$.  The LDMD/AIRY hybrid propagation
scheme is used. \MOLSCAT\ first performs close-coupling calculations
($\basisitem{ITYPE}=2$) and then repeats the calculation using every decoupling
approximation available ($\basisitem{ITYPE}=12$, 22, 32, 102).

\subsection{\texorpdfstring{Locating and characterising a quasibound state (Feshbach resonance) for Ar-HF}
{Locating and characterising a quasibound state (Feshbach resonance) for
Ar-HF}}\mylabel{testfiles:molscat:ityp1}

\begin{tabular}{ll}
input file: & \file{molscat-Ar\_HF.input}\\
executable: & \file{molscat-Rg\_HX}
\end{tabular}

\file{molscat-Ar\_HF.input} demonstrates the procedure for locating a quasibound state, which
appears as a narrow resonance in the S-matrix eigenphase sum as a function of energy. The procedure
is described in section \ref{energyconv}. It performs calculations on the H6(4,3,2) potential of
Hutson \cite{H92ArHF} for the ground ($v=0$) vibrational state of HF, using the LDMD propagator.
The first calculation sets $\inpitem{NNRG}=-10$; it first solves the coupled equations at 5
energies reasonably close to the resonance (but actually over 1000 widths away) and uses the
resulting eigenphase sums to estimate the resonance position. It then chooses another 5 energies
around the estimated resonance position, and this time finds that they span the resonance (which is
between points 2 and 3 of the second set of 5). The formula used for estimating resonance positions
is valid only far from resonance, so it reports that the second set of points cannot safely be used
to locate the resonance energy.

The procedure used to estimate the resonance position in this example amplifies any tiny
differences between computers due to finite-precision arithmetic. The second set of 5 energies is
commonly significantly different on different computers; this does not indicate an error.

The second calculation characterises the resonance using the algorithm of Frye and Hutson
\cite{Frye:quasibound:2020}. The initial three energies for this calculation have been chosen based
on the estimated resonance position and width from the first 5 energies above. The algorithm
converges quickly on the resonance position and gives accurate results for the resonance energy and
width.

\subsection{\texorpdfstring{Line-shape cross sections for Ar +
CO$_2$}{Line-shape cross sections for Ar + CO2}}\mylabel{testfiles:molscat:ityp1b}

\begin{tabular}{ll}
input file: & \file{molscat-Ar\_CO2.input}\\
executable: & \file{molscat-Rg\_CO2}
\end{tabular}

\file{molscat-Ar\_CO2.input} performs close-coupling calculations of line-shape
cross sections for the S(10) Raman line of CO$_2$ in Ar, using the
single-repulsion potential of Hutson \etal\ \cite{H96ArCO2fit} with the LDMD
propagator at short range and the AIRY propagator at long range. The
calculations are at a kinetic energy of 200 cm$^{-1}$ and the total energies
are calculated internally. The program prints cross sections accumulated up to
the current value of \var{JTOT}; the convergence of the partial-wave sum may be
compared with Fig.\ 2 of ref.\ \cite{Roc97CO2scat}.

\subsection{\texorpdfstring{Line-shape cross sections for Ar +
H$_2$}{Line-shape cross sections for Ar +
H2}}\mylabel{testfiles:molscat:ityp7}

\begin{tabular}{ll}
input file: & \file{molscat-Ar\_H2.input}\\
executable: & \file{molscat-Rg\_H2}\\
also required: & \file{data/h2even.dat}
\end{tabular}

\file{molscat-Ar\_H2.input} calculates pure rotational Raman line widths and
shifts across a shape resonance at a collision energy near 14 cm$^{-1}$. It
uses the BC$_3$(6,8) interaction potential of Le Roy and Carley \cite{RJL80},
evaluated for H$_2$ states $(j,v) = (0,0)$, (2,0) and (4,0) using H$_2$ matrix
elements in the file \file{data/h2even.dat}. The line-shape calculations
require S matrices evaluated at the same \emph{kinetic} energy for different
rotational states of H$_2$; the program treats the input energies as kinetic
energies and generates the total energies required. The results may be compared
with Figure 2(a) of ref.\ \cite{Hutson:sbe:1984}.

\subsection{\texorpdfstring{$\var{ITYP}=3$: Cross sections for rigid rotor + rigid rotor collisions}
{ITYP = 3:  Cross sections for rigid rotor + rigid rotor collisions}}
\mylabel{testfiles:molscat:ityp3}

\begin{tabular}{ll}
input file: & \file{molscat-ityp3.input}\\
executable: & \file{molscat-H2\_H2}
\end{tabular}

\file{molscat-ityp3.input} contains input data for collisions between
pairs of H$_2$ molecules. The interaction potential is that of Zarur and
Rabitz \cite{Zarur:1974}.

The first 4 calculations are for para-H$_2$ (even $j$) colliding with ortho-H$_2$ (odd $j$).
\MOLSCAT\ calculates elastic and state-to-state inelastic cross sections. Collisions do not
transfer molecules between even and odd $j$, so identical-particle symmetry is not included.
Contributions from different partial waves are accumulated until the partial-wave sums are
converged within the limits set by the input data. Two calculations are performed, first with
close-coupling calculations and then with the coupled-states approximation.
Each calculation is done twice; once with the radial potential coefficients
supplied explicitly, and once with the unexpanded potential supplied and expanded by quadrature
by the program. The results illustrate the equivalence of the two methods.

The final calculation is for para-H$_2$ colliding with para-H$_2$. In this case
identical-particle symmetry is important, and is included.

All calculations use the LDMD/AIRY hybrid propagation scheme.

\subsection{\texorpdfstring{$\var{ITYP}=5$: Cross sections for atom + symmetric top collisions,
with automated testing of propagator convergence}{ITYP = 5: cross sections for
atom + symmetric top collisions, with automated testing of propagator
convergence}}\mylabel{testfiles:molscat:ityp5}

\begin{tabular}{ll}
input file: & \file{molscat-ityp5.input}\\
executable: & \file{molscat-basic}
\end{tabular}

\file{molscat-ityp5.input} contains input data for atom + symmetric top
collisions between He and ortho-NH$_3$, taking account of the tunnelling
splitting of NH$_3$. It uses a simple analytical interaction potential and the
LDMD/AIRY hybrid propagation scheme. The input file uses
$\basisitem{ISYM}(3)=1$ to select rotational functions of E symmetry and
$\basisitem{ISYM}(4)=1$ to specify that the H nuclei are fermions. The first
four calculations are for a single partial wave and exercise the
convergence-testing code in \MOLSCAT, testing the convergence with respect to
step size (chosen in two ways), and with respect to the start point and end
point of the propagation. The final two calculations carry out full
cross-section calculations, using converged values for the propagation
variables, first with close-coupling calculations and then with the
coupled-states approximation.

\subsection{\texorpdfstring{$\var{ITYP} = 6$: Cross sections for atom + spherical top collisions}
{ITYP = 6: Cross sections for atom + spherical top collisions}}
\mylabel{testfiles:molscat:ityp6}

\begin{tabular}{ll}
input file: & \file{molscat-ityp6.input}\\
executable: & \file{molscat-Ar\_CH4}
\end{tabular}

\file{molscat-ityp6.input} contains input data for atom + spherical top
collisions between Ar and CH$_4$, using the interaction potential of Buck
\etal\ \cite{Buck:1983}. The ground-state rotational constants and the
tetrahedral centrifugal distortion constant $d_{\rm t}$ are specified in the
input file and the program uses them to calculate properly symmetrised
spherical-top wavefunctions. The input file selects CH$_4$ rotor functions of A
symmetry by setting \basisitem{ISYM} to 224, as described on p.\
\pageref{isym-spher}. The cross sections use the automatic total angular
momentum option $\inpitem{JTOTU}=99999$ with a convergence tolerance
(\inpitem{OTOL}) of 0.0001 to give well-converged inelastic cross sections, but
the diagonal convergence tolerance \inpitem{DTOL} is set to 10.0 so that the
partial-wave sum terminates before the elastic cross sections are converged.
The results may be compared with Table VI of ref.\ \cite{Chapman:1996},
although they do not agree exactly because the results in the paper are
averaged over the experimental distribution of collision energies.

\subsection{\texorpdfstring{$\var{ITYP}=8$: Atom-surface scattering}
{ITYP = 8: Atom-surface scattering}}\mylabel{testfiles:molscat:ityp8}

\begin{tabular}{ll}
input file: & \file{molscat-ityp8.input}\\
executable: & \file{molscat-basic}
\end{tabular}

\file{molscat-ityp8.input} contains input data for diffractive scattering
($\basisitem{ITYPE}=8$) of He from solid LiF, using the model potential of
Wolken \cite{Wolken:1973:surface}. It uses the LDMD propagator at two energies.

\subsection{\texorpdfstring{$\var{ITYP}=9$: Cross sections for Mg + NH in a magnetic field}
{ITYP = 9: Cross sections for Mg + NH in a magnetic field}}\mylabel{testfiles:molscat:MgNH}

\begin{tabular}{ll}
input file: & \file{molscat-Mg\_NH.input}\\
executable: & \file{molscat-Mg\_NH}\\
also required: & \file{data/pot-Mg\_NH.data}
\end{tabular}

\file{molscat-Mg\_NH.input} contains input data for cold collisions of NH with
Mg in a magnetic field. It uses the plug-in basis-set suite described in
section \ref{user:1s3s}, for a $^3\Sigma$ diatom colliding with a structureless
atom. Radial potential coefficients are provided by a
(\prog{VINIT}/\prog{VSTAR}) routine that applies RKHS interpolation to the
interaction potential of Sold\'an \etal\ \cite{Soldan:MgNH:2009}.  The
coupled-channel equations are solved using the LDMD/AIRY hybrid propagation
scheme ($\inpitem{IPROPS}=6$, $\inpitem{IPROPL}=9$).

The basis-set suite implements two different forms of the monomer Hamiltonian,
including and excluding the off-diagonal matrix elements of the spin-spin
operator. The input file specifies runs with both of these ($\var{IBSFLG}=2$
and 1 respectively). The approximation does not actually produce any saving in
computer time in this case.

The input file requests calculations at kinetic energies of 1, 10 and 100~mK
above the $n=0$, $j=1$, $m_j=1$ threshold of NH by using
$\inpitem{LOGNRG}=\code{.TRUE.}$ to select a logarithmically increasing energy set as described in section \ref{EMF}.

These calculations are similar to (a subset of) those of Wallis \etal\
\cite{Wallis:MgNH:2009}, although the test run uses a smaller basis set than
ref.\ \cite{Wallis:MgNH:2009}. Convergence at 100~mK requires inclusion of
incoming partial waves up to $L=3$, which requires values of $M_{\rm tot}$ from
$-2$ to 4 for incoming $m_j=1$. This is represented in the input file with
$\inpitem{JTOTL}=-2$, $\inpitem{JTOTU}=4$.

\MOLSCAT\ can accumulate cross sections from calculations for different values
of $M_{\rm tot}$ and total parity for a single EFV set. The first part of the test
run, with $\var{IBSFLG}=2$, illustrates the scheme used for identification of
levels for systems with non-diagonal Hamiltonians, where not all threshold
channels may be known at the point where the first partial cross sections are
calculated.

\subsection{\texorpdfstring{$\var{ITYP}=9$: Characterisation of magnetically
tunable Feshbach resonances and quasibound states and calculation of effective range for $^{85}$Rb
+ $^{85}$Rb} {ITYP = 9: Magnetically tunable Feshbach resonances, quasibound states and effective
range for 85Rb + 85Rb}} \mylabel{testfiles:molscat:85Rb2}

\begin{tabular}{ll}
input file: & \file{molscat-Rb2.input}\\
executable: & \file{molscat-Rb2}
\end{tabular}

\file{molscat-Rb2.input} contains input data for low-energy $^{85}$Rb +
$^{85}$Rb collisions in a magnetic field, using the plug-in basis-set suite
described in section \ref{user:alk-alk} and the potential of Strauss \etal\
\cite{Strauss:2010}, implemented with the \prog{VINIT}/\prog{VSTAR} routine
described in section \ref{detail:Hannover}. This is the same system used for the
basic resonance scan described in section \ref{basic:rb2:molscat}. All the
calculations use the LDMD/AIRY hybrid propagation scheme ($\inpitem{IPROPS}=6$,
$\inpitem{IPROPL}=9$).

This test run characterises 4 different low-energy Feshbach resonances as a function of magnetic
field, using the characterisation algorithms described by Frye and Hutson
\cite{Frye:quasibound:2020, Frye:resonance:2017}. The first resonance is in purely elastic
scattering in the lowest (aa) scattering channel, so produces a pole in the scattering length as a
function of magnetic field. The second and third resonances occur in collisions at excited
thresholds, where weak inelastic scattering is possible and the pole is replaced by an oscillation
\cite{Hutson:res:2007}. The fourth is subject to strong background inelasticity. After these
calculations, it characterises a quasibound state just below the ee threshold at 155 G, using the
algorithm of Frye and Hutson \cite{Frye:quasibound:2020}. Finally, a further calculation obtains
the effective range across the strong resonance observed in the aa channel near 850 G, described in
section \ref{basic:rb2:molscat}, from scans across the resonance at energies of 100 and 200~nK.

The basis-set suite for this interaction requires information about the
hyperfine properties of the atoms in an additional namelist block named
\namelist{\&BASIS9}, as described in section \ref{user:alk-alk}. The potential
expansion comprises 3 terms: the singlet and triplet interaction potentials,
and the spin-spin dipolar term, which is modelled as in Eq.\ \ref{eq:Vss}, with
coefficients specified in namelist \namelist{\&POTL}. The radial coefficient is
scaled by $-E_{\rm h}\alpha^2$ internally, so that the items given in
\namelist{\&POTL} are $\code{A(1)}=-g_S^2(a_0/\potlitem{RM})^3/4$,
$\code{A(2)}=-A$ and $\code{E(1)}=-\beta\ \potlitem{RM}/a_0$.

\subsection{Simple 2-channel scattering problem}\mylabel{testfiles:molscat-2-chan}

\begin{tabular}{ll}
input file: & \file{molscat-2chan-LJ.input}\\
executable: & \file{molscat-basic}
\end{tabular}

The programs are designed to handle matrix elements that are products of
potential coefficients $v^\Lambda(R)$ and angular momentum factors ${\cal
V}^\Lambda_{ij}$, as in Eq.\ \ref{eqvlambda} (and more explicitly Eq.\
\ref{eqWijexp}). The built-in coupling cases implement the angular momentum
factors for common cases in atomic and molecular scattering, and plug-in
basis-set suites can be used to implement other cases in a very general way.

Users may wish to implement a simple $N$-channel coupled-channel with the
matrix elements supplied directly as coefficients $v^\Lambda(R)$. This can be
done by choosing an interaction type where each ${\cal V}^\Lambda$ is a matrix
containing all zeroes except for a single element (for a particular $i,j$) that
is 1. The simplest way to set this up is to use $\basisitem{ITYPE}=2$ (diatomic
vibrotor + atom), with a basis set made up only of functions with $j = 0$. For
example, for a $2\times 2$ matrix, set $\basisitem{NLEVEL} = 2$ and
$\basisitem{JLEVEL} = 0,1,\ 0,2$ in namelist \namelist{\&BASIS}. If the
diagonal matrix elements have parts that are independent of $R$, it is best to
set them in $\basisitem{ELEVEL}(1)$ and $\basisitem{ELEVEL}(2)$, rather than
coding them as part of the interaction potential, since the programs then
recognise them as threshold energies. If the $R$-independent diagonal parts are
both zero, it is necessary to set the elements of \basisitem{ELEVEL} to some
very small value (say \code{1.D-30}) so that the programs do not complain that
they cannot calculate threshold energies.

In namelist \namelist{\&POTL}, specify the potential coefficient that is placed
in each matrix element by setting $\potlitem{MXLAM} =3$ and $\potlitem{LAMBDA}
= 0,1,1,\ 0,2,2,\ 0,1,2$.

Finally, either supply the required matrix elements as data in
\namelist{\&POTL} or write a \prog{VINIT}/\prog{VSTAR} routine that returns the
matrix elements $\langle 1|V(R)|1\rangle$, $\langle 2|V(R)|2\rangle$ and
$\langle 1|V(R)|2\rangle$ when called with $\var{I} = 1$, 2 and 3 respectively.

To generate a set of coupled equations without additional centrifugal terms,
run the programs with $\inpitem{JTOTL} = \inpitem{JTOTU} = 0$ and
$\inpitem{IBFIX}=2$ in namelist \namelist{\&INPUT}.

The input file \file{molscat-2chan-LJ.input} demonstrates this for a 2-channel
scattering problem. The units of length and energy are set to \AA\ and
cm$^{-1}$, and \inpitem{URED} is set to 20 $m_{\rm u}$. The threshold energies
for the 2 channels are set in \basisitem{ELEVEL} as 0 and 100 cm$^{-1}$. Both
diagonal matrix elements are Lennard-Jones 12-6 potentials,
\begin{equation}
V(R)=4\epsilon\left[\left(\frac{R}{\sigma_0}\right)^{-12}
-\left(\frac{R}{\sigma_0}\right)^{-6}\right].
\end{equation}
These are constructed to have
well depth $\epsilon$ and $V(\sigma_0)=0$; here, $\epsilon=100$ cm$^{-1}$ and
$\sigma_0=1$~\AA. The two channels are coupled by another Lennard-Jones
potential whose strength is a factor of 10 smaller than the diagonal
potentials.

The file specifies two runs. The first run is for scattering at an energy of
200 cm$^{-1}$, where both channels are open, and produces a $2\times 2$ S
matrix. The second run is at an energy where channel 1 is open but channel 2 is
closed, and calculates the eigenphase sum for a scan over a Feshbach resonance
just below 42.84 cm$^{-1}$. As will be seen in example
\ref{testfiles:bound-2-chan}, the corresponding bound system has a bound state
at $-58.3$ cm$^{-1}$, so this resonance may be identified as due to a similar
bound state in the upper channel, shifted by the 100 cm$^{-1}$ difference in
threshold energies and slightly shifted further by the coupling.

Large sets of coupled equations should \emph{not} be constructed in this way,
but instead by writing a basis-set suite to generate the coupling matrices in
terms of quantum numbers and a smaller set of radial potential coefficients.

\cbend

\section[\texorpdfstring{Examples for {\color{\bcol}\BOUND}}{Examples for BOUND}]{Examples for \BOUND}\mylabel{testfiles:bound}

\cbcolor{\bcol}\cbstart
\subsection{All available propagators for bound-state calculations}\mylabel{testfiles:bound:intflgs}

\begin{tabular}{ll}
input file: & \file{bound-all\_propagators.input}\\
executable: & \file{bound-basic}
\end{tabular}

\file{bound-all\_propagators.input} performs close-coupling calculations on the
bound states of a simple model of a complex formed between an atom and a linear
rigid rotor. The radial potential coefficients are provided in the input data
file and consist of a Lennard-Jones 12-6 potential for $\lambda=0$ and a
dispersion-like $R^{-6}$ form for $\lambda=2$. The calculation is repeated
using combinations of short-range and long-range propagators that exercise
every propagation method available in \BOUND\ (though not every possible
combination). The calculation is done twice for the LDMD/AIRY combination; once
with $R_{\rm mid} < R_{\rm match}$ and once with $R_{\rm mid}
> R_{\rm match}$. The calculation which uses just the LDMD propagator employs a
different step length for the inwards propagation. This input file should
produce the same results regardless of which \BOUND\ executable is used.

\subsection{Bound states of Ar-HCl with expectation
values and wavefunction}\mylabel{testfiles:bound:ityp1}

\begin{tabular}{ll}
input file: & \file{bound-Ar\_HCl.input}\\
executable: & \file{bound-Rg\_HX}
\end{tabular}

\file{bound-Ar\_HCl.input} performs calculations on the states of Ar-HCl bound
by more than 80 cm$^{-1}$, using the H6(4,3,0) potential of Hutson
\cite{H92ArHCl} and the LDMD propagator, for total angular momentum $J_{\rm tot}=0$ and 1
and both parities. The first run does close-coupling calculations. The second
run does calculations in the helicity decoupling approximation, and in addition
calculates expectation values $\langle P_2(\cos\theta)\rangle$ and $\langle
1/R^2 \rangle$ for all the states. The results may be compared with Table IV of
ref.\ \cite{H92ArHCl}. The third run calculates the wavefunction for the first
bound state identified in the first run. The wavefunction is written to unit
109; the resulting file is included as \file{bound-Ar\_HCl.wavefunction} in
\file{examples/output}. The components may be plotted with any standard
plotting package.

\subsection{\texorpdfstring{Bound states of Ar-CO$_2$ with Richardson
extrapolation}{Bound states of Ar-CO2 with Richardson
extrapolation}}\mylabel{testfiles:bound:ityp1b}

\begin{tabular}{ll}
input file: & \file{bound-Ar\_CO2.input}\\
executable: & \file{bound-Rg\_CO2}
\end{tabular}

\file{bound-Ar\_CO2.input} performs close-coupling calculations on the ground
and first vibrationally excited state of Ar-CO$_2$, using the split-repulsion
potential of Hutson \etal\ \cite{H96ArCO2fit} and the LDJ propagator, for total
angular momentum $J_{\rm tot}=0$. The results may be compared with Table IV of ref.\
\cite{H96ArCO2fit}.

It first calculates the ground-state energy using a fairly large (unconverged)
step size of 0.03~\AA. It then repeats the calculation with an even larger step
size, and extrapolates to zero step size using Richardson $h^4$ extrapolation.

\subsection{\texorpdfstring{Bound states of Ar-H$_2$}{Bound states of Ar-H2}}\mylabel{testfiles:bound:ityp7}

\begin{tabular}{ll}
input file: & \file{bound-Ar\_H2.input}\\
executable: & \file{bound-Rg\_H2}\\
also required: & \file{data/h2even.dat}
\end{tabular}

\file{bound-Ar\_H2.input} performs close-coupling calculations on the ground
state of Ar-H$_2$ with H$_2$ in its $v=1$, $j=1$ state, for total angular
momentum $J_{\rm tot}=1$ and even total parity ($j+L$ even). For this parity
there is no allowed $j=0$ channel, so the state is bound except for vibrational
predissociation to form H$_2$ ($v=0$) \cite{HUTSON:ArH2:1983}, which is not
taken into account by \BOUND. The run uses the LDMD propagator and the TT3(6,8)
potential of Le~Roy and Hutson \cite{LeR87}, evaluated for H$_2$ states $(j,v)
= (0,0)$, (2,0) and (4,0) using H$_2$ matrix elements in the file
\file{data/h2even.dat}.

\BOUND\ first calculates the ground-state energy using a fairly large
(unconverged) step size of 0.04 \AA. It then repeats the calculation with an
even larger step size, and extrapolates to zero step size using Richardson
$h^4$ extrapolation.

\subsection{\texorpdfstring{Bound states of H$_2$-H$_2$
(ortho-para)}{Bound states of H2-H2 (ortho-para)}}\mylabel{testfiles:bound:ityp3}

\begin{tabular}{ll}
input file: & \file{bound-ityp3.input}\\
executable: & \file{bound-H2\_H2}
\end{tabular}

\file{bound-ityp3.input} performs close-coupling calculations on bound states
of H$_2$-H$_2$ with one para-H$_2$ molecule (even $j$) and one ortho-H$_2$
molecule (odd $j$). It uses the LDMD propagator. The interaction potential is
that of Zarur and Rabitz \cite{Zarur:1974}. The states are bound by less than
2~cm$^{-1}$ (below the $j$=0 + $j$=1 threshold).

\subsection{\texorpdfstring{Bound states of He-NH$_3$}{Bound states of He-NH3}}\mylabel{testfiles:bound:ityp5}

\begin{tabular}{ll}
input file: & \file{bound-ityp5.input}\\
executable: & \file{bound-basic}
\end{tabular}

\file{bound-ityp5.input} performs close-coupling calculations on bound states
of He-NH$_3$, taking account of the tunnelling splitting of NH$_3$, using a
simple analytical interaction potential and the LDMD propagator. The input file
selects rotational functions of E symmetry by setting
\basisitem{ISYM}\code{(3)} to 1 and specifies that the H nuclei are fermions by
setting \basisitem{ISYM}\code{(4)} to 1.

\subsection{\texorpdfstring{Bound states of Ar-CH$_4$}{Bound states of Ar-CH4}}\mylabel{testfiles:bound:ityp6}

\begin{tabular}{ll}
input file: & \file{bound-Ar\_CH4.input}\\
executable: & \file{bound-Ar\_CH4}
\end{tabular}

\file{bound-Ar\_CH4.input} performs close-coupling calculations on bound states
of Ar-CH$_4$, using $\basisitem{ITYPE}=6$, which can also handle complexes of
asymmetric tops. It uses the interaction potential of Buck \etal\
\cite{Buck:1983}. It uses the LDMD propagator. CH$_4$ is a spherical top, and
the input file selects rotor functions of F (T) symmetry by setting
\basisitem{ISYM} to 177, as described on p.\ \pageref{isym-spher}. It may be
noted that the particular one of each set of degenerate rotor functions that
has even $k$ is numerically arbitrary, so the values of the quantum number
$\tau$ generated by the program may vary for different computers or compilers.
The results may be compared with Table II of ref.\ \cite{Hutson:spher:1994}.

%
%

\subsection{Bound-state energies of the hydrogen atom}\mylabel{testfiles:bound:H}

\begin{tabular}{ll}
input file: & \file{bound-hydrogen.input}\\
executable: & \file{bound-basic}
\end{tabular}

\file{bound-hydrogen.input} carries out single-channel bound-state calculations
on the hydrogen atom, and demonstrates how to handle calculations in atomic
units. It sets \inpitem{MUNIT} to the electron mass in Daltons, \inpitem{RUNIT}
to the Bohr radius in \AA\ and $\inpitem{EUNITS}=7$ to select input energies in
hartrees. It uses the general-purpose \prog{POTENL} to set up a simple Coulomb
potential, with \potlitem{EPSIL} set to the hartree in cm$^{-1}$, so that the
potential is handled in atomic units. It uses $\basisitem{ITYPE}=1$ with
$\basisitem{JMAX}=0$ to generate a simple single-channel problem. Note that
\basisitem{ROTI}\code{(1)} is set to the dummy value \code{1.0}; this value is
not used because $\basisitem{JMAX}=0$, but it prevents the program terminating
prematurely.

The wavefunction at the origin is of the form $r^{l+1}$, so its log-derivative
is infinite at the origin. This is the default for locally closed channels, but
is specified explicitly for the locally open $l=0$ channel.

Because $\basisitem{JMAX}=0$, the orbital angular momentum $l$ is equal to
\var{JTOT}. $\var{JTOT}=0$ produces $n$s levels at energies of $-1/(2n^2)$ for
$n=1,2,...$, while $\var{JTOT}=1$ produces $n$p levels starting at $n=2$.

\subsection{Bound-state energies of Mg-NH at specified magnetic
fields}\mylabel{testfiles:bound:MgNH}

\begin{tabular}{ll}
input file: & \file{bound-Mg\_NH.input}\\
executable: & \file{bound-Mg\_NH}\\
also required: & \file{data/pot-Mg\_NH.data}
\end{tabular}

\file{bound-Mg\_NH.input} locates the bound states of Mg-NH at specified
magnetic fields. It uses a plug-in basis-set suite for a $^3\Sigma$ diatom
colliding with a structureless atom. Radial potential coefficients are obtained
by RKHS interpolation of the potential points of Sold\'an \etal\
\cite{Soldan:MgNH:2009}. The coupled equations are solved using the LDMD/AIRY
hybrid propagation scheme.

The run locates a single bound state at four different magnetic fields from
370~G to 385~G, from which it may be inferred that the state crosses threshold
near 387~G.

\subsection{Simple 2-channel bound-state problem}\mylabel{testfiles:bound-2-chan}

\begin{tabular}{ll}
input file: & \file{bound-2chan-LJ.input}\\
executable: & \file{bound-basic}
\end{tabular}

This example solves for the bound states of the simple 2-channel problem
described in example \ref{testfiles:molscat-2-chan}. It has two channels
asymptotically separated by 100 cm$^{-1}$ and each described by a Lennard-Jones
potential with a well 100 cm$^{-1}$ deep. The two channels are coupled by a
Lennard-Jones potential whose strength is a factor of 10 smaller.

\BOUND\ locates two bound states, at $-58.2$ cm$^{-1}$ and $-12.1$ cm$^{-1}$.
The coupling is fairly weak compared to the separation of the channels, so
these are only weakly perturbed from the corresponding single-channel levels
(not located here) at $-57.7$ cm$^{-1}$ and $-11.9$ cm$^{-1}$.

\cbend

\section[\texorpdfstring{Examples for {\color{\fcol}\FIELD}}{Examples for FIELD}]{Examples for \FIELD}\mylabel{testfiles:field}

\cbcolor{\fcol}\cbstart

\subsection{Bound states of Mg-NH as a function of magnetic field}
\mylabel{testfiles:field:MgNH}

\begin{tabular}{ll}
input file: & \file{field-Mg\_NH.input}\\
executable: & \file{field-Mg\_NH}\\
also required: & \file{data/pot-Mg\_NH.data}
\end{tabular}

\file{field-Mg\_NH.input} locates magnetic fields in the range 0 to 400~G at
which bound states exist for specific energies relative to the lowest scattering
threshold of Mg + NH in a magnetic field. It uses the same basis-set suite and
interaction potential as in section \ref{testfiles:bound:MgNH}. The coupled
equations are solved using the LDMD/AIRY hybrid propagation scheme. The AIRY
propagator uses a power-law step size ($\inpitem{TOLHIL}=0$) in place of the
default adaptive step-size algorithm; this choice eliminates step-size noise
resulting from the large range of the eigenvalues of the potential matrix, as
described on p.\ \pageref{ss-noise}, and significantly improves the convergence
on magnetic fields at which bound states exist.

The run locates the same level as in section \ref{testfiles:bound:MgNH} at
energies of 0, 20 and 40 MHz~$\times\ h$ below threshold, and shows that it
crosses threshold near 387.28~G.

\subsection{\texorpdfstring{Locating threshold crossings for
$^{85}$Rb$_2$}{Locating threshold crossings for 85Rb2}}
\begin{tabular}{ll}
input file: & \file{field-basic\_Rb2.input}\\
executable: & \file{field-Rb2}
\end{tabular}

\file{field-basic\_Rb2.input} locates magnetic fields
where bound states cross the lowest scattering threshold for $^{85}$Rb$_2$.
These are the fields at which zero-energy Fesh\-bach resonances exist. It uses
a plug-in basis-set suite for a pair of alkali-metal atoms in a magnetic field,
including hyperfine interactions. It uses the potential of Strauss \etal\
\cite{Strauss:2010}, implemented with potential coefficients incorporated in
the executable.  The coupled equations are solved using the LDMD/AIRY hybrid
propagation scheme.

This is the same example as in section \ref{basic:rb2:field}.

\subsection{\texorpdfstring{Bound states of $^{85}$Rb$_2$ as a function of magnetic
field}{Bound states of 85Rb2 as a function of magnetic
field}}\mylabel{testfiles:field:85Rb2}

\begin{tabular}{ll}
input file: & \file{field-Rb2.input}\\
executable: & \file{field-Rb2}
\end{tabular}

\file{field-Rb2.input} locates bound states of
$^{85}$Rb$_2$ as a function of magnetic field, using the same potential and
basis-set suite as in section \ref{basic:rb2:field}. The calculation locates the
magnetic fields (in the range 750 to 850 G) at which bound states exist with
binding energies of 225, 175, 125, 75 and 25 MHz below the lowest threshold.
There are, however, two bound states that these calculations fail to find, as
they run almost parallel to the threshold, at about 140 and 220 MHz below it.
To locate these bound states, one would need to do a calculation using \BOUND.
\cbend

\chapter{\texorpdfstring{Installing and testing the programs}
{\ref{distribution}: Installing and testing the
programs}}\mylabel{distribution}

\section{Supplied files}\mylabel{supplied}

The programs should be obtained from \url{https://github.com/molscat/molscat}. The files supplied are:
\begin{itemize}
\item{the full program documentation in pdf format;}
\item{a directory {\tt source\_code} containing
\begin{itemize}
\item{the Fortran source code;}
\item{a makefile ({\tt Makefile}) that can build the executables
    needed for the example calculations described in sections
    \ref{basic:m:examples},
\ref{basic:bd+fld} and \ref{basic:rb2} and in chapter \ref{testfiles};}
\end{itemize}}
\item{a directory {\tt examples} containing
\begin{itemize}
\item{a sub-directory {\tt input} containing input files for the
    example calculations described below;}
\item{a sub-directory {\tt output} containing the corresponding output files;}
\end{itemize}}
\item{a directory {\tt data} containing auxiliary data files
    for some potential routines used in the example calculations;}
\item{a plain-text file {\tt README} that gives information on changes that
    may be needed to adapt the makefile to a specific target computer.}
\item{a plain-text file {\tt COPYING} that contains the text of the GNU
    General Public License, Version 3.}
\end{itemize}

To demonstrate how to handle pointwise potential coefficients (which often
result from electronic structure calculations) by interpolation, we have
included examples that use RKHS interpolation on such a data set for Mg+NH.

\section{Program language}\mylabel{language}

\MOLSCAT, \BOUND\ and \FIELD\ are written in near-standard Fortran 77 with some Fortran 90 features, such as the use of modules. Most of the code is in files with \code{.f} extensions that use Fortran 77 spacing conventions. A small number of routines are in files with \code{.f90} extensions that use Fortran 90 spacing conventions.

The programs have been tested with current versions of \prog{gfortran}, \prog{ifort} and \prog{pgf90}. \hl{With very recent versions of {\prog{gfortran}}, it is necessary to use the compiler flag {\code{-std=legacy}} to suppress compile-time errors due to outdated features of Fortran.} 

\begin{shaded} 
\section{Main routine}\mylabel{main}

The main routine is common to all the programs. It does not do any processing;
it simply calls the relevant version of \prog{DRIVER}
(\file{mol.driver.f}, \file{bd.driver.f} or \file{fld.driver.f}) to do all the
work.
\end{shaded} 


\section{Date, time and CPU time routines}\mylabel{datetime}

The programs obtain the date and time of a run (for output in the header) by
calls to routines \prog{GDATE} and \prog{GTIME} and information on the CPU time
taken by calls to subroutine \prog{GCLOCK}. The distribution provides versions
of these routines that call the Fortran 90 utility routines
\prog{date\_and\_time} and \prog{cpu\_time}.

\section{Linear algebra routines}\mylabel{linalg}

The programs use LAPACK linear algebra routines wherever possible.

If possible, run the programs using LAPACK routines that are optimised for your
particular computer. However, if this is not possible, Fortran versions of the
LAPACK routines may be obtained from the Netlib repository
(\url{www.netlib.org}).

The LAPACK routines use BLAS (basic linear algebra subroutines) as much as
possible. BLAS level 1, level 2 and level 3 routines exist. Use BLAS routines
optimised for your particular computer if possible. However, if no optimised
routines are available, Fortran versions may be obtained from the Netlib page
at \url{www.netlib.org/blas.html}.

Any user who implements new options in any of the programs should perform
matrix operations by calls to the routines described below, both for ease of
maintenance and to simplify the creation of efficient executables for other
computers.

Linear algebra routines supplied with \MOLSCAT:
\begin{description}[nosep]
\item[\prog{DGEMUL}]{Matrix multiplication}
\item[\prog{DGESV}]{Solve linear equations}
\item[\prog{SYMINV}]{Invert symmetric matrix}
\item[\prog{DIAGVL}]{Diagonalise symmetric matrix without eigenvectors}
\item[\prog{DIAGVC}]{Diagonalise symmetric matrix with eigenvectors}
\end{description}
The programs also call BLAS routines such as \prog{DAXPY}, \prog{DDOT} etc.\ in
many places.

\subsection{Matrix multiplication}\mylabel{matmul}

The programs call \prog{DGEMUL}. This was originally a routine from the IBM
ESSL library. In version \currentversion, \prog{DGEMUL} calls the BLAS routine
\prog{DGEMM}. The Fortran 90 subroutine \file{ytrans} (\file{ytrans.f90}) uses
matrix operators such as \code{matmul} rather than calling \prog{DGEMUL}
because that makes the code more readable. Efficiency is not usually an issue
for \prog{ytrans} since it is called only a few times per run.

\subsection{Symmetric matrix inversion}\mylabel{syminv}

Symmetric matrix inversion is a key operation that dominates the time taken by
some propagators, so its efficiency is important. The programs call
\prog{SYMINV}. The version of \prog{SYMINV} included in version
\currentversion\ calls the LAPACK routines \prog{DSYTRF} and \prog{DSYTRI} to
carry out the inversion for matrix sizes above 30. For smaller matrices it
calls a pure Fortran routine. The threshold for switching between the two could
be changed if desired for optimum efficiency on a specific machine.

Note that the programs really do require matrix inversion, despite the usual
advice to use linear equation solvers instead. This is because the propagators
save information from one step to the next, and this advantage is lost
if the problem is formulated in terms of linear equation solvers.

\subsection{Linear equation solver}\mylabel{linsolve}

The programs call the LAPACK routine \prog{DGESV} directly. The speed of this
routine is not critical for most propagators.

\subsection{Eigenvalues and eigenvectors of symmetric
matrices}\mylabel{evalsevecs}

These routines are important for propagators 3, 4, 7, and 9 (RMAT, VIVS, LDMA
and AIRY). The programs call diagonalisers via
routines \prog{DIAGVC} (for eigenvalues and eigenvectors) and \prog{DIAGVL}
(for eigenvalues alone). These routines both make a call to the LAPACK routine
\prog{DSYEVR}.


\section{File handling}\mylabel{files}

The programs adhere to the Fortran 77 standard in their use of \code{READ} and
\code{WRITE} statements (including direct access files).

The \code{OPEN} statements do not use $\code{FILE} = \code{'fname'}$ parameters
and where necessary the user must provide files with the naming convention for
their system.  For example, most Linux systems use filename \file{fort.NN} if
unit \iounit{NN} is opened.

\begin{shaded} 
\section{Dimensions of variably sized arrays}\mylabel{arraydim}

Versions of the programs up to 2022.0 declared a single large array, \var{X} to accommodate
variably sized arrays throughout the programs. DRIVER and other routines then partitioned this
according to the size of the problem being tackled, and specific segments of it were passed
into subroutines to use as variably sized arrays.

From version 2025.0, the programs allocate dynamically sized arrays when they are needed.

\end{shaded} 

\section{Dimensions of fixed-size arrays}\mylabel{sizes}

All three programs use a module \module{sizes}, which contains parameters used
to set dimensions of certain arrays. These parameters are:
\begin{itemize}
\item{\var{MXFLD} which limits the number of EFV values that can be entered
    in \namelist{\&INPUT} and the length of loops over EFV sets;}
\item{\var{MXNRG} which limits the number of energies that can be looped
    over;}
\item{\var{MXNODE} which limits the number of states that can be searched
    for in a single instance of a loop over the energy (for \BOUND) or one
    EFV set (for \FIELD);}
\item{\var{MXLN} which limits the number of sets of lines for pressure
    broadening calculations;}
\item{\var{MXLOC} which limits the number of propagations
that may be used to locate each bound state or field-dependent resonance;}
\item{\var{MXJLVL} which limits the size of the array \basisitem{JLEVEL} in
    module \module{basis\_data}; see section \ref{module:basis-data}
    below;}
\item{\var{MXELVL} which limits the size of the array \basisitem{ELEVEL} in
    module \module{basis\_data}; see section \ref{module:basis-data}
    below;}
\item{\var{MXROTS} which limits the size of the array \basisitem{ROTI} in
    module \module{basis\_data}; see section \ref{module:basis-data}
    below;}
\item{\var{MXSYMS} which limits the size of the arrays  \basisitem{ISYM}
    and \basisitem{ISYM2} in module \module{basis\_data}; see section
    \ref{module:basis-data} below;}
\item{\var{MXOMEG} which limits the sizes of the arrays \var{VCONST} and
    \var{NEXTMS} in module \module{potential}; see section
    \ref{module:potential} below;}
\item{\var{MXLMDA} which limits the size of the array \var{LAMBDA} in
    module \module{potential}; see section \ref{module:potential} below;}
\item{\var{MXANG} which limits the size of the array \var{COSANG} in module
    \module{angles}; see section \ref{angles:module}.}
\end{itemize}

The dimension of arrays that are used for information about EFVs depend on a
parameter \var{MXEFV}, which is set in module \module{efvs}.

\section{\texorpdfstring{\code{COMMON} blocks}{COMMON blocks}}\mylabel{common}

The programs use a number of \code{COMMON} blocks internally, and the names of
these should be avoided when naming \code{COMMON} blocks in subroutines that
link with the distributed code. A brief description of these common blocks is
given below.

\begin{description}
\item[\common{ASSVAR}]{is used in very old code.  Passes variable between
    \prog{DASIZE} and \prog{PRBR}.}
\item[\common{BCCTRL}]{contains variables used for setting boundary
    conditions.}
\item[\common{CNTROL}]{contains a character variable \var{CDRIVE}, which
    can take the values \code{M}, \code{B} or \code{F}. This indicates
    whether the executable is for \MOLSCAT, \BOUND\ or \FIELD.}
\item[\common{DERIVS}]{contains a logical variable which controls whether
    derivatives of the interaction potential are calculated numerically or
    analytically (when required).}
\item[\common{EIGSUM}]{contains eigenphase sums used for estimating the
    position of a nearby (energy) resonance.}
\item[\common{EXPVAL}]{contains variables relevant to the calculation of
    expectation values.}
\item[\common{IOCHAN}]{contains unit numbers for \iounit{IPSI} and the two scratch files required,
    together with some variables
    controlling how a wavefunction is written on unit \iounit{IPSI}}
\item[\common{IOUTCM}]{contains variables to pass information between
    \prog{IOSOUT} and \prog{IOSBIN}.}
\item[\common{LATSYM}]{contains logical variables for surface scattering
    calculations.}
\item[\common{LDVVCM}]{contains variables for the VIVS propagator.}
\item[\common{NPOT}]{contains the variable \var{NVLP}, which is used to set \var{NHAM} when
    $\var{IVLFL}>0$.}
\item[\common{POPT}]{contains variables to control level of printing for
    the VIVS propagator.}
\item[\common{PRBASE}]{contains variables relevant to calculations of
    line-shape cross sections.}
\item[\common{PRPSCR}]{contains variables relevant to use of scratch files
    for propagation segments.}
\item[\common{RADIAL}]{contains variables controlling the propagation segments.}
\item[\common{VLFLAG}]{contains the variable \var{IVLFL}, which indicates whether
    the array \var{P} is indexed using the array \var{IV} to allow the array
    \var{VL} to be smaller. This facility is currently used only for
    $\var{ITYP}=2$, 7 and 8.}
\item[\common{VLSAVE}]{contains unit number for storage of the array \var{VL}.}
\item[\common{WKBCOM}]{contains variables relevant to the Gauss-Mehler
    quadrature performed for WKB integration.}
\end{description}

\section{Compiling and linking the programs} \mylabel{compiling}

We have provided a basic makefile in the file \file{Makefile}. This is designed
to compile the programs and build the executables required to run the examples
in Chapter \ref{testfiles}. It has been tested with GNU
\code{make}, which is the default version of \code{make} on Linux and OS X.
Modifications may be needed for other versions of \code{make}.

The basic form of the command to compile and link is \\
\code{make} \emph{executable-name} \\
where \emph{executable-name} is one of the executables listed in Chapter
\ref{testfiles}, such as \code{molscat-basic}, \code{bound-basic} or
\code{field-basic}.

The makefile may need some minor modifications for a specific installation. If
this is done, we strongly recommend leaving the supplied version in
\file{Makefile} unchanged, and making a copy for modification in either
\file{makefile} or \file{GNUmakefile}. If either of these files exists, the GNU
\code{make} command will use it in preference to \file{Makefile}.

\begin{shaded} 
The supplied makefile makes use of various environment variables:
\begin{itemize}
	\item{\code{\$(MOLSCAT\_EXECDIR)} is the relative pathname from the directory containing the makefile for the directory in which the executables will be placed, and defaults to \code{bin}}
	\item{\code{\$(MOLSCAT\_COMPILER)} is the name of the Fortran compiler used to produce the object files and link them together.  If not set externally, this defaults to \code{gfortran}}
	\item{\code{\$(MOLSCAT\_OBJDIR)} is the relative pathname from the directory containing the makefile for the directory in which the compiled object files will be placed, and defaults to \code{object-\$(MOLSCAT\_COMPILER)}}
	\item{\code{\$(MOLSCAT\_LIBS)} is the list of libraries needed at link stage to supply the BLAS and LAPACK routines used by the programs, each prefixed by \code{-l}.  It defaults to \code{-llapack -lblas}} which works on many systems. An alternative is the openblas package; to use it, set this variable to \code{-lopenblas}.  If suitable libraries are not available, the individual library routines must be
	downloaded as described in section \ref{linalg}, and the names of the object
	(\code{.o}) files included in the variable \code{LIBUTILS}.
	\item{\code{\$(MOLSCAT\_FLAGS)} is the list of compilation flags and defaults to \code{-g --std=legacy} if the compiler is \prog{gfortran}}
\end{itemize}
\end{shaded} 

The programs have been tested with current versions of \prog{gfortran}, \prog{pgf90} and \prog{ifort} compilers.

The reason for leaving the supplied version of \file{Makefile} unchanged is that future versions of
the programs may require changes in it. If the original file is changed, the \code{git pull}
command used to update the programs from \code{https://github.com} (Section \ref{updates}) may fail
because it is unable to update the file.

If \file{Makefile} is updated by \code{git pull}, a new
active copy based on the updated version will be needed, and any changes previously made will need to be transferred to the new copy before using it to recreate the executables.

\subsection{If \code{make} gets tangled} \mylabel{make-tangled}

Various unexpected occurrences can cause \code{make} to become confused. The
most common examples involve Fortran module files ({\tt .mod}), which appear to
be handled inconsistently by some versions of \code{make}. If unexpected error
messages appear, it is often sufficient to delete all files with names ending
{\tt .mod} and {\tt .o} and recompile from scratch using
\\
\code{make} \emph{executable-name}

\section{Testing the installation} \mylabel{testing}

After building the executables, it is highly desirable to validate the programs by running all the
examples described in Chapter \ref{testfiles} and verifying that the programs give output very
similar to that in the files supplied.

The supplied output files were obtained from executables compiled with \prog{gfortran} and run on a
machine with \code{x86\_64} architecture. Different compilers may produce values that are formatted
slightly differently. Some quantities are output at close to machine precision, so may have
slightly different values with different compilers, different architectures, or different
implementations of the LAPACK and/or BLAS libraries. In addition, convergence procedures may take
slightly different steps, though they should converge to points that are the same to within
the convergence criteria. It is therefore necessary to exercise some judgement in deciding whether
results differ \emph{significantly} from the test output.

\section{Adding a new executable to the makefile} \mylabel{extra:executables}

Many users will wish to create new executables that include their own routines
for the interaction potential and/or the basis set. To do this, it is necessary
to extend the supplied makefile to provide rules to create the new executables.

\subsection{Fundamentals and terminology of \code{make}}

The following instructions need at least a basic understanding of the contents
of makefiles. A short glossary of terminology may be useful:
\begin{itemize}[nosep]
\item A \emph{target} is a file that \code{make} has a \emph{rule} to
    create.
\item A \emph{rule} is a set of commands to create a \emph{target} from a
    list of its \emph{dependencies}.
\item A \emph{dependency} is a file that is required (and must be
    up-to-date) to create the \emph{target}.
\end{itemize}
If a target already exists, \code{make} recreates it only if one or more of its
dependencies has a time stamp newer than the target (i.e., has changed since
the last time the target was created).

Dependencies are applied recursively; each dependency may have its own
dependencies, and will itself be recreated if necessary.

\subsection{Editing the makefile}

Make a copy of \file{Makefile} in either \file{makefile} or \file{GNUmakefile}, as described in
Section \ref{compiling}. Edit the copy rather than the original.
	
\begin{shaded} 

The makefile has changed substantially in version 2025.0.  If your makefile was edited for an earlier version, the current one will also need editing, but in a different style.

All executables are now made using a single rule, but the final list of dependencies for an executable \emph{must} now be of the form \code{<name-of-executable>\_OBJS}.

A new executable is specified by adding lines to the makefile to specify its dependencies.

Additional dependency lists as described below are needed to specify the object files required for
\begin{enumerate}[nosep]
\item the interaction potential;
\item the plug-in basis-set suite.
\end{enumerate}
The dependency lists of this type in the supplied version of \file{Makefile} all begin with
\code{POT-} or \code{BASE9-} as appropriate and serve as examples for constructing new
dependency lists. We recommend adding any new dependency lists immediately after the supplied ones.

Sequence matters in dependency lists.  The name of the object file for a module \emph{must} come before those for any routines that make use of that module.

\subsubsection{Dependency list for the interaction potential}

\begin{itemize}
\item{If the potential is supplied entirely within the input data file, use
    the existing list \code{POT-BASIC};}
\item{If the potential is supplied using one of the other dependency lists
    already set up in \file{Makefile}, use that list;}
\item{Otherwise, set up a new dependency list, using a line of the form
    \\ {\it potential-dependency-list-name} = {\it list of dependencies}\\
    where {\it potential-dependency-list-name} is a unique name.
\begin{itemize}
\item{If the general-purpose version of \prog{POTENL} is to be used,
    the list should contain \code{\$(POTENL-GP)} followed by the
    unqualified names of the object files containing the
    versions of \prog{VSTAR} and/or \prog{VRTP} to be used.
    In most cases at least one of these will be a routine
    supplied by the user. If either of \prog{VSTAR} and \prog{VRTP} is
    \emph{not} used in this implementation, use \code{vstar-dummy.o} or
    \code{vrtp-dummy.o} for that one. The dependency list should also
    contain the names of any other object files that contain routines
    called uniquely by the user-supplied versions of \prog{VSTAR} and/or
    \prog{VRTP}.}
\item{If the general-purpose version of
    \prog{POTENL} is not to be used, the list should contain the
    unqualified name of the object file containing the relevant version of
    \prog{POTENL}, followed by the names of any other object files that
    contain routines called uniquely by it.}\end{itemize}}
\end{itemize}

\subsubsection{Dependency list for the plug-in basis-set suite}
\begin{itemize}
\item{If a user-supplied plug-in basis-set suite is not required, use the existing
    list \code{BASE9-UNUSED}.}
\item{Otherwise, set up a new dependency list, using a line of the
    form\\{\it basis-set-dependency-list-name} = {\it list of dependencies}
    \\ where {\it basis-set-dependency-list-name} is a unique name.
    The list should contain the unqualified name(s) of the object
    file(s) containing the plug-in basis-set suite; see chapter
    \ref{base9} for a list of routines.  If the basis-set suite does not
    include specific versions of the additional routines \prog{THRSH9},
    \prog{EFV9} or \prog{DEGEN9}, include the supplied dummy versions of
    these routines (\code{thrsh9-dummy.o}, \code{efv9-dummy.o} or
    \code{degen9-nondegenerate.o} respectively).  If the
    general-purpose version of \prog{POTENL} is to be used, an
    object file containing the \prog{POTIN9} routine must also be supplied.
    If a specific one is not required, use the dummy version
    (\code{potin9-example.o}).}
\end{itemize}

\subsubsection{Module recompilation}

The \code{make} program does not handle modules very well.  Any routine that uses a module must be recompiled if the module is changed, but \code{make} cannot work this out for itself.  Accordingly, \code{make} requires a list of files dependent on each module in order to ensure that they are recompiled when the module file is changed.  These lists of files are named \code{<{\it module-name}>\_DEPS}.  If any user-supplied routine uses a supplied module, the corresponding object file should be added to the relevant list.

\subsubsection{Dependency list for new executable}

The dependencies for a new executable must start with one of the dependency lists \code{CORE\_MOL}, \code{CORE\_BND} and \code{CORE\_FLD}, for
\MOLSCAT, \BOUND\ and \FIELD, respectively, followed by the list \code{REPLACEABLE}, which lists object files
for subroutines that expert users may wish to replace for special purposes.  These are then followed by the dependency lists for the basis set to be used and the potential routines to be used, as described above.

\subsubsection{Name for the new executable}

Add the unqualified name for the new executable to the list of targets in
\code{USER-PROGS}, which is empty in the supplied \file{Makefile}.
\end{shaded} 

\subsection{Creating the executable}

Finally, create the new executable as described in section \ref{compiling} with
the command \\
\code{make} \emph{executable-name}

\section{Values of fundamental constants}\mylabel{fundconsts}

\begin{shaded} 
All three programs use values of fundamental physical constants and derived
quantities in the module \module{physical\_constants}. The current default values are
the 2022 CODATA recommended values, in file \file{physical\_constants\_module.f}.
Modules containing earlier sets of values are available from the present authors on request.
The programs print a message stating the date of publication of the values used.
\end{shaded} 

The values set in \module{physical\_constants} are parameters and most have long names including lower-case characters.
This helps to distinguish them from the 6-character (Fortran-77-style) variable names used in the rest of the code.

To use the values in \module{physical\_constants} in a user-supplied routine, insert the line
\\ {\tt USE} \module{physical\_constants} \\
immediately after the routine declaration statement.

\chapter{\texorpdfstring{Plug-in potential routine (\prog{POTENL})}
{\ref{userPotenl}: Plug-in potential routine POTENL}}
\mylabel{userPotenl}

As described in chapter \ref{buildVL}, the programs internally require an
expansion of the interaction potential in a set of orthogonal functions of the
internal coordinates,
\begin{equation}\label{eqn:V:MXLAM}
V(R,\xi_{\rm intl})=\sum_{\Lambda=1}^{\var{MXLAM}} v_\Lambda(R)V^\Lambda(\xi_{\rm intl}).
\end{equation}
For most interaction potentials, this can be handled using the general-purpose
version of subroutine \prog{POTENL}, which may call \prog{VINIT}/\prog{VSTAR}
to provide the radial potential coefficients, or perform an integration by
quadrature to obtain the coefficients from an unexpanded potential provided by
routine \prog{VRTP}.

In rare cases where these mechanisms are inconvenient or inefficient, the user
may supply a complete routine \prog{POTENL} to replace the general-purpose
version. This section describes the specification of this routine.

\section{\texorpdfstring{Specification of \prog{POTENL} subroutine}
{Specification of POTENL subroutine}}\mylabel{spec:potenl}

In each run, \prog{POTENL} is called once for initialisation purposes, and on
this call may read any data necessary to specify the interaction potential. It
returns information about the terms present in the potential expansion.
Subsequently, \prog{POTENL} is called many times during each propagation to
evaluate the radial potential coefficients $v_\Lambda(R)$ for particular
interparticle distances $R$.

The syntax of a call to \prog{POTENL} is
\begin{verbatim}
CALL POTENL(IC, MXLMB, LAMBDA, RR, P, ITYPE, IPRINT)

DOUBLE PRECISION, INTENT(OUT)   :: RR, P(MXLMB)
INTEGER,          INTENT(OUT)   :: LAMBDA(NLABV,MXLMB)
INTEGER,          INTENT(INOUT) :: MXLMB
INTEGER,          INTENT(IN)    :: IC, ITYPE, IPRINT
\end{verbatim}

The array \potlitem{LAMBDA} specifies \potlitem{MXLAM} sets of \var{NLABV} integers. Each
set identifies a term $\Lambda$ in the expansion (Eq.~\ref{eqn:V:MXLAM}).

The array \var{P} specifies \potlitem{MXLAM} radial potential coefficients
$v_\Lambda(R)$, in the same order as the elements of \potlitem{LAMBDA}.

The arrays \potlitem{LAMBDA} and \var{P} should be dimensioned as
\var{LAMBDA}(\var{NLABV},\code{*}) and \var{P}(\code{*}) to switch off Fortran array bound
checking, since \potlitem{MXLAM} is not known at the time of an initialisation call.

There are two basic types of call to \prog{POTENL};
\begin{description}
\item[Initialisation:]{\prog{POTENL} is called once with $\var{IC} = -1$,
    before any other calls to it, to allow it to read any necessary data
    and set up parameters for later use.}
\item[Evaluation:]{At subsequent calls to \prog{POTENL}, \var{IC} is 0, 1 or
    2 and the routine must evaluate the radial potential coefficients or
    their radial derivatives. As described below, radial derivatives are
    not really essential.}
\end{description}

The specification of \prog{POTENL} for initialisation and evaluation calls is
described separately.

\subsection{Initialisation}\mylabel{initial}

\begin{description}
\item[\var{MXLMB}:]{On entry, \var{MXLMB} specifies the maximum dimension
    that has been externally provided for the array \var{LAMBDA}. This
    value may be (and is, in the provided general-purpose version) used to check for array bound errors.

On exit, \var{MXLMB} must be set equal to \potlitem{MXLAM}, which specifies the
number of distinct terms in the expansion of the interaction potential
(i.e., the dimension of the array \var{P} that is returned by subsequent
calls to \prog{POTENL}).}
\item[\var{LAMBDA}:]{On exit, the array \var{LAMBDA} must contain indices
    specifying the potential terms to be used. Although it is externally a
    one-dimensional array, it is conceptually two-dimensional for some
    interaction types, and may be handled explicitly as a two-dimensional
    array in \prog{POTENL} by declaring it as \code{LAMBDA(NLABV,1)} and
    declaring \var{NLABV} as a parameter. Each element (or column) of
    \var{LAMBDA} corresponds to an element of the array \var{P} returned by
    subsequent calls to \prog{POTENL}. The programs do not require that the
    symmetry terms be supplied in any particular order, but just that the
    $i$th column of the array \var{LAMBDA} should correspond to the $i$th
    element of the array \var{P}.

    The explicit form of the expansions is described for the built-in
    interaction types in sections \ref{potl:ityp1} to \ref{potl:ityp8}; the
    value of \var{NLABV} can be obtained by counting the number of labels
    that comprise $\Lambda$ in the table on page \pageref{lambda-table}.
    For $\basisitem{ITYPE}=9$, \var{NLABV} is set in routine \prog{SET9},
    described in section \ref{ConstructBasis9}.}

\item[\var{RR}:]{On exit, \var{RR} must specify the length units \var{RM}
    that are used in subsequent calls to \prog{POTENL} and are used for
    most quantities with dimensions of length output by the programs
    (except cross sections). \var{RM} must be returned in \AA. It is often
    convenient to set $\var{RR} = \code{1.0D0}$ in the initialisation call
    to \prog{POTENL}, and to handle everything in \AA\ thereafter.}
\item[\var{P}:]{On exit, \code{P(1)} must specify the energy scaling factor
    \var{EPSIL} (expressed in cm$^{-1}$) to be used internally by the
    programs, and subsequent calls to \prog{POTENL} must return energies
    in units of \var{EPSIL}. It may be convenient to set
    $\var{EPSIL} = \code{1.0D0}$ in the initialisation call to
    \prog{POTENL}, and to handle everything in cm$^{-1}$ thereafter. The
    value given to \var{EPSIL} does \emph{not} affect the interpretation of
    energy parameters input in namelist \namelist{\&INPUT} and
    \namelist{\&BASIS}, or output energies other than the interaction
    potential.}
\item[\basisitem{ITYPE}:]{On entry, \basisitem{ITYPE} is the interaction type.

If \prog{POTENL} is coded specifically for a particular value of
\basisitem{ITYPE}, it should check that the correct value has been passed, as a
precaution against the accidental use of the wrong executable version. The
value of this parameter should not be changed by \prog{POTENL}.}

\item[\inpitem{IPRINT}:]{used to control the quantity of output produced by
\prog{POTENL}.}
\end{description}

\subsection{Evaluation (\var{IC}=0, 1 or 2)}\mylabel{eval}

For an evaluation call to \prog{POTENL}, only the \var{IC}, \var{MXLMB},
\var{RR}, \var{P} and \inpitem{IPRINT} arguments are passed. \var{LAMBDA} and
\var{ITYP} do \emph{not} contain the values they were given in the
initialisation call, so copies of these must be stored internally in
\prog{POTENL} if they are needed in an evaluation call.\footnote{In the
general-purpose version of \prog{POTENL}, this is achieved by (i) naming the
\var{LAMBDA} array \var{LAM} internally and saving the elements read in from
the namelist as \potlitem{LAMBDA} internally in module \module{potential}
whilst a copy is passed out as the dummy array \var{LAM}; (ii) saving
\code{MOD(\basisitem{ITYPE},10)} as an internal variable \var{ITYP}.}
\begin{description}
\item[$\var{IC} = 0$]{evaluate the radial potential coefficients
    $v_{\Lambda}(\var{RR})$ and return them in \var{P}}
\item[$\var{IC} = 1$]{evaluate $\d v_\Lambda/\d R$ at $R=\var{RR}$ and
    return them in \var{P}}
\item[$\var{IC} = 2$]{evaluate $\d ^2v_\Lambda/\d R^2$ at $R=\var{RR}$ and
    return them in \var{P}}
\end{description}
\begin{description}
\item[\var{RR}:]{The interparticle distance at which the potential is to be
    evaluated, in units of \var{RM}; see the \var{RR} argument for an
    initialisation call to \prog{POTENL} above.}
\item[\var{P}:]{On exit, \var{P} must contain the array of radial potential
    coefficients (or their derivatives) at distance \var{RR}, in the order
    specified earlier by the array \potlitem{LAMBDA} returned by the
    initialisation call to \prog{POTENL}.  The array \var{P} must be
    returned in units of \var{EPSIL}$\times (\var{RR})^{-\var{IC}}$; see the
    discussion of the initialisation call above).}
\end{description}
Calls to \prog{POTENL} with $\var{IC} = 1$ or 2 occur only:
\begin{itemize}[nosep]
\item for the VIVS propagator if \inpitem{IVP}, \inpitem{IVPP},
    \inpitem{ISHIFT} or \inpitem{IDIAG} is set;
\item for the LDMA propagator at high print levels (in order to calculate
    the nonadiabatic couplings).
\end{itemize}
Even in these cases, there is an option (controlled by logical variable
\inpitem{NUMDER} in namelist \namelist{\&INPUT}) that allows the derivatives to
be evaluated numerically without making calls to \prog{POTENL} with $\var{IC} =
1$ or 2. It is thus not altogether necessary for \prog{POTENL} to cope with
$\var{IC} = 1$ and 2 calls, but it should at least trap an attempt to call it
this way and print an error message.

\chapter{\texorpdfstring{Plug-in basis-set suites}{\ref{base9}: Plug-in basis-set suites}}
\mylabel{base9}

The programs provide a facility to construct and solve sets of coupled
equations that are different from those for the built-in interaction types.
This chapter gives the information needed to write a suite of plug-in
subroutines to do this. It may be skipped by readers who wish only to run
existing codes.

The routines described in this chapter are called \emph{only} if
$\basisitem{ITYPE}=9$. They are not needed for any of the built-in interaction
types ($\var{ITYP}=1$ to 8).

The distribution includes a skeleton version of a plug-in basis-set suite
(\file{base9-skeleton.f}), which will halt if called, but contains comments that
can be used as guidance for writing a new suite. The distribution also includes
two plug-in basis-set suites, described in chapter \ref{user:gen}, which may be
used as examples by programmers of new routines.

\section{Components of a basis-set suite}\mylabel{base9:routines}

When coding a new plug-in basis-set suite, the programmer must always provide
the following routines:

{
\def\mytabwidth{\hsize-\widthof{\quad \prog{POTIN9}\qquad see section}}
\halign to \hsize{#\hfil\quad & \strut\parbox[t]{\mytabwidth}{\raggedright#\strut}
& \quad # \hfill \cr
\noalign{\strut\hrule}
\omit\strut routine\hfill & \omit\hfill task to perform\hfill & see section\cr
\noalign{\hrule}
\prog{BAS9IN} & Read any data needed to specify the basis set, in addition to
quantities read in \namelist{\&BASIS}. & \ref{bas9in} \cr
\prog{SET9} & Set up the lists of pair levels and pair states. &
\ref{ConstructBasis9} \cr
\prog{BASE9} & Set up the basis set for the current \var{JTOT} and \var{IBLOCK}
& \ref{base9:choosebasis} \cr
\prog{POTIN9} & Choose the type of potential expansion to be used if it is one
of the built-in types, or set up the variables needed for the potential
expansion from scratch if not. & \ref{potin9} \cr
\prog{CPL9} & Calculate the coupling matrices of the expansion functions used
for the interaction potential in the current basis set. If $H_{\rm intl}$
and/or $\hat L^2$ is non-diagonal, matrix elements of the operators used to
expand them are also required. & \ref{base9:calculatecouple} \cr
}}

Many older basis-set suites code most or all of these routines as entry points to
\prog{BAS9IN}, so that variables in {\tt SAVE} statements are common to all of
them. However, for new basis-set suites it is preferable to code the routines
as separate subroutines, with variables to be shared in a Fortran module.

In addition, the programmer \emph{may} need to write the following:

{
\def\mytabwidth{\hsize-\widthof{\quad \prog{POTIN9}\qquad see section}}
\halign to \hsize{#\hfil\quad &
\strut\parbox[t]{\mytabwidth}{\raggedright#\strut} & \quad # \hfill \cr
\noalign{\strut\hrule} \omit\strut routine\hfill & \omit\hfill task to
perform\hfill & see section\cr \noalign{\hrule}
\prog{DEGEN9} & Calculate denominators for degeneracy-averaged cross sections\par (if
necessary) & \ref{degen9}\cr
\prog{THRSH9} & Calculate threshold energies from monomer quantum numbers\par (if
necessary) & \ref{base9:asympthresh} \cr
\prog{EFV9} & Transform the input external field variables (EFVs) into the
components used for coupling matrices (if necessary) &  \ref{base9:EFVs} \cr
}}

Dummy versions of \prog{DEGEN9}, \prog{THRSH9} and \prog{EFV9} are supplied in
\file{degen9-nondegenerate.f}, \file{thrsh9-dummy.f} and \file{efv9-dummy.f}. These must be linked
in unless a bespoke version has been programmed.

Routines in plug-in basis-set suites need access to variables that are not in
their argument lists.  These are contained in a few modules, described in
section \ref{modules:base9}:
\begin{shaded} 
\begin{itemize}[nosep]
\item module \module{basis\_data} contains variables related to basis sets and pair levels;
\item module \module{pair\_state} contains variables related to pair states;
\item module \module{potential} contains variables related to internal Hamiltonians and extra operators;
\item module \module{efvs} contains variables related to EFVs.
\end{itemize}
\end{shaded} 

The routines listed above are described in sections \ref{bas9in} to
\ref{base9:calculatecouple} in the order in which they are called.

\section{Diagonal or non-diagonal asymptotic Hamiltonian}\mylabel{base9:diag}

The first choice to make is the basis set to use. There are often many possible
basis sets for a given problem. Different basis sets give equivalent results
when they are complete, but they often offer different opportunities for
approximations that involve restricting the basis set. In addition, \BOUND\ and
\FIELD\ produce wavefunctions that are expanded in the basis set, and these may
be easier to interpret for one choice of basis set than another.

The programs handle two different types of basis set:
\begin{enumerate}[nosep]
\item Basis sets in which $H_{\rm intl}$ and $\hat L^2$ are diagonal;
\item Basis sets in which $H_{\rm intl}$ and/or $\hat L^2$ are
    non-diagonal.
\end{enumerate}
These are be described as diagonal and non-diagonal basis sets in this chapter,
although the interaction potential is almost always non-diagonal (as otherwise
single-channel rather than coupled-channel calculations suffice). For a given
problem, non-diagonal basis sets are often simpler and easier to program,
though the resulting output is sometimes more complicated to interpret.
Problems involving EFVs usually require non-diagonal basis sets, since the same
basis set seldom diagonalises $H_{\rm intl}$ at different values of the EFVs.

Diagonal and non-diagonal basis sets use different subsets of the internal
variables, and require significantly different programming as described below.
A non-diagonal basis set is indicated by returning a positive value of
\var{NCONST} and/or \var{NRSQ} from \prog{BAS9IN} as described below.

\section{Interpretation of external loop variables}\mylabel{base9:loops}

The loop structure in \MOLSCAT\ and \BOUND\ is conceptually

{\parskip0pt
      Read \common{\&INPUT}\par
      Read \common{\&BASE}\par
      Call \prog{BAS9IN} (usually reads \common{\&BASE9})\par
      Call \prog{SET9}\par
      \cbcolor{\mcol}\cbstart Call \prog{DEGEN9} (\MOLSCAT\ only, some input options only)\cbend\par
      Initialise potential; read \common{\&POTL} and call \prog{POTIN9}\par
      \code{DO JTOT = JTOTL, JTOTU, JSTEP}\par
      {\tt\ \ }\code{DO IBLOCK = 1, NBLOCK}\par
      {\tt\ \ \ \ }Call \prog{BASE9}\par
      {\tt\ \ \ \ }Call \prog{CPL9}\par
      {\tt\ \ \ \ }\code{DO IFIELD = 1, NFIELD} (external fields)\par
      {\tt\ \ \ \ \ \ }Call \prog{THRSH9} (only for some input options for threshold energies)\par
      {\tt\ \ \ \ \ \ }Call \prog{EFV9}   (only for some input options for magnetic fields)\par
      {\tt\ \ \ \ \ \ }\code{DO INRG = 1, NNRG}  (energies)\par
      {\tt\ \ \ \ \ \ \ \ }Propagations, with many evaluation calls to \prog{POTENL}\par
      {\tt\ \ \ \ \ \ }\code{ENDDO}\par
      {\tt\ \ \ \ }\code{ENDDO}\par
      {\tt\ \ }\code{ENDDO}\par
      \code{ENDDO}\par}
For \FIELD, the loops over energies and external fields are reversed.

The programmer is free to use \var{JTOT} and \var{IBLOCK} for any purpose
desired. For field-free calculations, it is natural to use \var{JTOT} for the
total angular momentum $J_{\rm tot}$ and \var{IBLOCK} for any additional
symmetries in the Hamiltonian (total parity, body-fixed $K$, etc.). However,
for calculations in a magnetic field, $J_{\rm tot}$ is not conserved. For a
single field (or parallel fields), however, its projection $M_{\rm tot}$ onto
the field axis may be conserved; in this case the variable \var{JTOT} is
conveniently used for $M_{\rm tot}$.

The programs hold quantum numbers in integer variables and arrays. For systems
with half-integer spins, it it often convenient to store \emph{doubled} quantum
numbers. Most quantum number values are processed \emph{only} within the
basis-set suite (though they may be printed), so doubling their values causes
no problems for processing in the remainder of the programs. The exceptions to
this are
\begin{itemize}
\item The array \var{L}, described in section \ref{base9:choosebasis}.
\item The loop variable \var{JTOT}, which may be used in different ways
    depending on the value of \var{JHALF}, which should be set within the
    basis-set suite. The only operation outside the basis-set suite that
    needs explicit knowledge of how \var{JTOT} is used is the evaluation of
    degeneracy-averaged cross sections from Eq.\ \ref{eqsigdef}, which
    contains a factor of $2J_{\rm tot}+1$:
\begin{description}
\item[$\var{JHALF}=1$]{indicates that \var{JTOT} is an undoubled
    total angular momentum $J_{\rm tot}$;}
\item[$\var{JHALF}=2$]{indicates that \var{JTOT} is a doubled
    total angular momentum, $2J_{\rm tot}$;}
\item[$\var{JHALF}=0$]{indicates that \var{JTOT} is not a total
    angular momentum, and omits the factor $(2J_{\rm tot}+1)$ from the
    cross section.}
\end{description}
\end{itemize}

\section{Calculating the interaction matrix}\mylabel{calcW}

At each step of a propagation, the propagators require the interaction matrix
defined by Eq.\ \ref{eqWij}. This may be written
\begin{equation}
W_{ij}(R)=\frac{2\mu}{\hbar^2} \left( \sum_{\Lambda} v_{\Lambda}(R){\cal V}_{ij}^\Lambda+
\sum_\Omega h_\Omega {\cal H}_{ij}^\Omega \right) + \sum_\Upsilon {\cal L}_{ij}^\Upsilon/R^2.
\label{eqWijexp}
\end{equation}
All the coupling matrices $\boldsymbol{\cal V}^\Lambda$, $\boldsymbol{\cal
H}^\Omega$ and $\boldsymbol{\cal L}^\Upsilon$ are calculated prior to the
propagation by \prog{CPL9}.

For both diagonal and non-diagonal basis sets, the potential coupling matrices
$\boldsymbol{\cal V}^\Lambda$ are stored in the array \var{VL}. During the
propagation, the $R$-dependent coupling coefficients $v_\Lambda(R)$ are
supplied by \prog{POTENL}.

For non-diagonal basis sets, \var{NCONST} coupling matrices $\boldsymbol{\cal
H}^\Omega$ and/or \var{NRSQ} centrifugal matrices $\boldsymbol{\cal
L}^\Upsilon$ are also stored in the array \var{VL}.

For diagonal basis sets, the coupling matrices $\boldsymbol{\cal H}^\Omega$
must also be diagonal. Any contributions that are independent of EFVs may be
calculated by \prog{SET9} and included in the elements of the array
\basisitem{ELEVEL}. Alternatively (and necessarily for terms that depend on
EFVs) \var{NDGVL} blocks of diagonal elements may be calculated by \prog{CPL9}
and stored in the array \var{DGVL}. The pair energies are obtained from
\begin{equation}
E_{{\rm intl},i} = E_{{\rm intl},i}^{\rm field-free}+\sum_{\Omega} h_\Omega
{\cal H}_{ii}^\Omega.
\end{equation}

The $R$-independent coupling coefficients $h_\Omega$ are either supplied by
\prog{BAS9IN} or, if they depend on EFVs, generated internally. Both
EFV-dependent and EFV-independent coefficients are stored in the array
\var{VCONST}, with \var{NCONST} elements for non-diagonal basis sets or
\var{NDGVL} elements for diagonal basis sets.

If there is only one centrifugal matrix $\boldsymbol{\cal L}^\Upsilon$, and it
is diagonal, with elements of the form ${\cal L}_{ij}=L_i(L_i+1)$, the integers
$L_i$ may be returned in the array $\var{L}$ by \prog{BASE9}. If the matrix is
diagonal but its elements are not of the form $L_i(L_i+1)$, \var{IBOUND} may be
set to 1 by \prog{BAS9IN} and the diagonal elements of $\hat L^2$ returned in
the array $\var{CENT}$ by \prog{CPL9}. If the centrifugal matrices are
non-diagonal, \var{NRSQ} must be set greater than zero and the full matrices
$\boldsymbol{\cal L}^\Upsilon$ returned by \prog{CPL9}.

\section{\texorpdfstring{Routine \prog{BAS9IN}}{Routine BAS9IN}}\mylabel{bas9in}

\begin{verbatim}
SUBROUTINE BAS9IN(PRTP, IBOUND, IPRINT)
USE potential

CHARACTER(32), INTENT(OUT)   :: PRTP
INTEGER,       INTENT(INOUT) :: IBOUND
INTEGER,       INTENT(IN)    :: IPRINT
\end{verbatim}

\prog{BAS9IN} is an initialisation routine. It is called by \prog{BASIN}, and
so is called only once in a particular run. It has access to most of the
quantities read in \namelist{\&BASIS} through the module \module{basis\_data},
but can read additional information if required. This has usually been done using a
namelist block \namelist{\&BASE9}, but that is not compulsory. The
corresponding input data must be included in the input file between
\namelist{\&BASIS} and \namelist{\&POTL}. They commonly include limits on the
pair levels to be included and values of spectroscopic constants for the
interacting particles. Some of these quantities are needed by other routines in
the basis-set suite; this may be achieved by placing the shared variables in a
Fortran module.

The variable \inpitem{IPRINT} gives the print level for the current calculation
and may be used to control how much is printed by \prog{BASIN}.

\prog{BAS9IN} must return the following quantities:
\begin{description}
\item[\var{PRTP}:]{character string containing a brief description of the
    interaction type, which is printed in the output.}
\item[\var{IBOUND}:]{if $\var{NRSQ}=0$, $\var{IBOUND}=0$ (the default)
    indicates that centrifugal energies are to be calculated from values
    assigned to the array \var{L} by \prog{BASE9}. $\var{IBOUND}>0$
    indicates that the array \var{L} should not be used and centrifugal
    energies are to be calculated from values in the array \var{CENT}
    returned from \prog{CPL9}. If $\var{NRSQ}\ne0$, neither \var{L} nor
    \var{CENT} is used to calculate centrifugal energies and \var{IBOUND}
    is not used.}
\end{description}
It is usually appropriate to use $\var{IBOUND}=0$ if the matrix elements of the
centrifugal potential are diagonal and of the form $\hbar^2 L(L+1)/(2\mu R^2)$
with integer $L$, and $\var{IBOUND}\ne 0$ otherwise.

\prog{BAS9IN} must set the following quantities that are included in module
\module{potential}:
\begin{description}
\item[\var{NCONST}:]{set to 0 if $H_{\rm intl}$ is diagonal in the basis
    set. For non-diagonal basis sets, the number of terms in the expansion
    (\ref{eqHomega1}) of $H_{\rm intl}$, including terms used for EFVs; see below.}
\item[\var{NDGVL}:]{for diagonal basis sets, the number of
    diagonal terms in the expansion (\ref{eqHomega1}) of $H_{\rm intl}$,
    including terms used for EFVs; see below.}
\item[\var{VCONST}:]{array of $R$-independent coefficients for the terms in
    the expansion of $H_{\rm intl}$, as described in sections \ref{calcW}
    and \ref{base9:calculatecouple}. Values should be set by the end of
    \prog{POTIN9} unless they depend on EFVs. The product of an element of
    \var{VCONST} and an element of \var{VL} or \var{DGVL} should be in
    units of cm$^{-1}$. \var{VCONST} must be set to 0 for any operator not
    to be included in $H_{\rm intl}$. Required only if $\var{NCONST}>0$ or
    $\var{NDGVL}>0$.}
\item[\var{NRSQ}:]{set to 0 if centrifugal potentials are diagonal and are
    specified either by the array \var{CENT} or calculated from values in
    the array \var{L}. Otherwise, the number of terms in the expansion of
    the operator $\hat L^2$. $\var{NRSQ}>1$ is not currently supported.}
\end{description}

If (and only if) extra operators are required to resolve degeneracies, their structure should be
defined here:
\begin{description}
\item[\var{NEXTRA}:]{number of extra $R$-independent operators. }
\item[\var{NEXTMS}:]{array giving the number of coupling matrices for each
    extra $R$-independent operator.}
\end{description}

If the elements of the array \var{VL} are to be indexed using the array \var{IV}, as
described in section \ref{base9:calculatecouple}, \prog{BAS9IN} should set the variable \var{IVLFL}
to 1 in common block \common{VLFLAG}. If $\var{IVLFL}=1$, either \prog{BAS9IN} or \prog{POTIN9}
should set \var{NVLP} (in common block \var{NPOT}) to the number of blocks of the array \var{VL}
used for the interaction potential.

A number of internal variables in module \module{potential} are set, based on these variables,
after the initialisation call to \prog{POTENL}:

If $\var{IVLFL}=0$ (the default),
\begin{equation}
\var{NHAM} = \var{MXLAM}+\var{NCONST}+\var{NRSQ}.
\label{eq:nham0}
\end{equation}
If $\var{IVLFL}=1$,
\begin{equation}
\var{NHAM} = \var{NVLP}+\var{NCONST}+\var{NRSQ}
\label{eq:nham1}
\end{equation}
(but $\var{IVLFL}=1$ is currently implemented only for $\var{NCONST}=\var{NRSQ}=0$).

In either case,
\begin{equation}
\var{NEXBLK}=\sum_{i=1}^{\var{NEXTRA}} \var{NEXTMS}(i),
\label{eq:nexblk}
\end{equation}
\begin{equation}
\var{NVLBLK}=\var{NHAM}+\var{NEXBLK}.
\label{eq:nvlblk}
\end{equation}

\BOUND\ includes a facility to calculate expectation values using a
finite-difference method \cite{Hutson:expect:88}, as described in section
\ref{calcexp}. To use this, the operator concerned must be part of $H_{\rm
intl}$. If it is not naturally part of $H_{\rm intl}$, it may be added as an
additional term (and included in \var{NCONST}), with the corresponding
coefficient in \var{VCONST} set to zero.

If external fields are to be included in the calculation, \prog{BAS9IN} must
also set quantities in module \module{efvs} that describe them.  Access to
these quantities must be gained by using module \module{efvs}. The quantities
that must be set are:
\begin{description}
\item[\var{NEFV}:]{the number of EFVs.}
\item[\var{EFVNAM}:]{array of character strings describing the EFVs, with
    maximum lengths specified in module \module{efvs}.}
\item[\var{EFVUNT}:]{array of short character strings giving names of units
    for the EFVs, with maximum lengths specified in module \module{efvs}.}
\item[\var{MAPEFV}:]{a positive value specifies the index of the first EFV
    in the array \var{VCONST}; a negative value indicates a non-linear
    mapping between the EFVs and the corresponding elements of
    \var{VCONST}. The options are described in section \ref{base9:EFVs}.}
\end{description}

\prog{BAS9IN} may also set the variable \var{ITPSUB} in module \module{efvs}.
This is written by \MOLSCAT\ on unit \iounit{ISAVEU} so that an external
program can identify the basis-set suite that produced the results.

The namelist item \inpitem{DEGTOL} is used as a threshold for degeneracy for
both $H_{\rm intl}$ and the extra operators. It is treated as an energy and so
is scaled according to \inpitem{EUNITS} or \inpitem{EUNIT}. Programmers of
extra operators should ensure that the operators are scaled such that
\inpitem{DEGTOL} is an appropriate threshold for degeneracy (which usually
implies that they should have eigenvalues that span a range between 1 and
$10^3$).

\section{\texorpdfstring{Routine \prog{SET9}}{Routine SET9}}\mylabel{ConstructBasis9}
\begin{verbatim}
SUBROUTINE SET9(LEVIN, EIN, NSTATE, JSTATE, NQN, QNAME, NBLOCK, NLABV,
IPRINT)

INTEGER,      INTENT(OUT) :: NQN, NBLOCK, NLABV, NSTATE, JSTATE(*)
CHARACTER(8), INTENT(OUT) :: QNAME(10)

LOGICAL,      INTENT(IN)  :: LEVIN, EIN
INTEGER,      INTENT(IN)  :: IPRINT
\end{verbatim}

This routine is called by \prog{BASIN} shortly after the call to \prog{BAS9IN}.

\prog{SET9} must always return values for the following quantities:
\begin{description}
\item[\var{NQN}:]{one greater than the number of quantum labels used to
    specify a pair state.}
\item[\var{QNAME}:]{array of names of the quantum labels used to specify a
    pair state.}
\item[\var{NBLOCK}:]{the number of independent symmetry blocks for each
    value of \var{JTOT}.}
\item[\var{NLABV}:]{the number of labels used to specify a term in the
    potential expansion.}
\end{description}

The logical variable \var{LEVIN} indicates whether the array \basisitem{JLEVEL}
was supplied explicitly in \namelist{\&BASIS}; however, it is quite unlikely
that this mechanism would be required for a new basis-set suite, and it is
usually sufficient to print an error message and stop if \var{LEVIN} is
\code{.TRUE.} The logical variable \var{EIN} is \code{.TRUE.} either if the
corresponding energies were given in \namelist{\&BASIS} in the array
\basisitem{ELEVEL}, or if values from which they can be calculated were input
in the array \basisitem{ROTI}.

The variable \inpitem{IPRINT} gives the print level for the current calculation
and may be used to control how much is printed by \prog{SET9}.

For diagonal basis sets, \prog{SET9} must set the following quantities in
module \module{basis\_data}, unless they were input as explicit arrays in
\namelist{\&BASE}.
\begin{description}
\item[\var{NLEVEL}:]{the number of pair levels (not pair states). Pair
    levels are used for diagonal basis sets to label state-to-state cross
    sections (and are distinct from pair \emph{states}, as described in
    section \ref{basis:diag}.}
\item[\var{JLEVEL}:]{array of quantum labels that specify pair levels. Can
    be given in \namelist{\&BASIS}, but more commonly calculated from
    limits on quantum numbers. Each set of quantum labels must be unique.}
\item[\var{ELEVEL}:]{array of energies of pair levels, corresponding to
    \var{JLEVEL}. Can be given in \namelist{\&BASIS}, but more commonly
    calculated from input spectroscopic parameters.}
\end{description}

For non-diagonal basis sets, the programmer \emph{may} if desired use
\var{NLEVEL} and the arrays \var{JLEVEL} and \var{ELEVEL}, but they are not
used outside the basis-set suite for non-diagonal basis sets, so this is
optional.

For both diagonal and non-diagonal basis sets, \prog{SET9} must return:
\begin{description}
\item[\var{NSTATE}:]{the number of pair states. }
\item[\var{JSTATE}:]{array of labels for the pair states, arranged as though in an array of
    dimension ($\var{NSTATE},\var{NQN}$). For diagonal basis sets, the last
    column must contain the index of the element of the array \var{ELEVEL}
    that contains the energy of the pair state. For non-diagonal basis
    sets, the last column is not used and can be left unset.
\begin{shaded} 
    \code{SET9} should pass through the loops to calculate \var{NSTATE} and \var{JSTATE} twice.
    On the first pass, it should evaluate \var{NSTATE} but not assign values to \var{JSTATE}.
    After this pass:
    \begin{itemize}[nosep]
    \item if $\var{NSTATE}*\var{NQN} > \var{JSIZE}$, it must return to the calling
    routine (which will allocate a larger array for \var{JSTATE} and call \code{SET9} again);
    \item if $\var{NSTATE}*\var{NQN} \le \var{JSIZE}$, it must execute the second pass to assign
    values to \var{JSTATE}.
    \end{itemize}
 \end{shaded}} 
 \end{description}

\section{\texorpdfstring{Routine \prog{BASE9}}{Routine BASE9}}\mylabel{base9:choosebasis}

\begin{verbatim}
SUBROUTINE BASE9(LCOUNT, N, JTOT, IBLOCK, JSTATE, NSTATE, NQN, JSINDX, L,   &
                 IPRINT)

INTEGER, INTENT(INOUT) :: N

INTEGER, INTENT(OUT)   :: JSINDX(N), L(N)

LOGICAL, INTENT(IN)    :: LCOUNT
INTEGER, INTENT(IN)    :: JTOT, IBLOCK, NSTATE, NQN, JSTATE(NSTATE,NQN),    &
                          IPRINT
\end{verbatim}

Routine \prog{BASE9} is called by \prog{BASE} to set up the basis set for the
current combination of \var{JTOT} and \var{IBLOCK}. Since the calling program
does not initially know the size of the basis set, \prog{BASE9} is called
twice: first to \emph{count} the basis functions, and subsequently to set up
the basis functions themselves.

When \var{LCOUNT} is \code{.TRUE.} on entry, \prog{BASE9} must count the required
basis functions and return the number of them in \var{N}. In this case, no space
has yet been allocated for the array \var{JSINDX} and \var{L}, and so \prog{BASE9}
must \emph{not} assign values in the arrays.

When \var{LCOUNT} is \code{.FALSE.} on entry, \prog{BASE9} must set up the
arrays \var{JSINDX} and \var{L}. Each function in the basis set is specified by
an element in each of the arrays \var{JSINDX} and \var{L}:
\begin{description}
\item[$\var{JSINDX}(i)$:]{a pointer to a pair state in the array \var{JSTATE}.}
\item[$\var{L}(i)$:]{a value of $L$ that, when combined with the pair
    quantum numbers indexed by the corresponding element of \var{JSINDX},
    specifies a function in the basis set. Used only if $\var{IBOUND}=0$
    and $\var{NRSQ}=0$, unless the programmer chooses to use it in
    \prog{CPL9}.}
\end{description}
\var{N} may be smaller or larger than \var{NSTATE}, and (for a single
\var{JTOT} and \var{IBLOCK}) \var{JSINDX} may reference only a subset of the
pair states.

\inpitem{IPRINT} gives the print level for the current calculation and may be
used to control how much is printed by \prog{BASE9}.

\section{\texorpdfstring{Routine \prog{POTIN9}}{Routine POTIN9}}\mylabel{potin9}

\begin{verbatim}
SUBROUTINE POTIN9(ITYPP, LAM, MXLAM, NPTS, NDIM, XPT, XWT, MXPT,            &
                  IVMIN, IVMAX, L1MAX, L2MAX, MXLMB, XFN, MX, IXFAC)

INTEGER,          INTENT(INOUT) :: ITYPP, MXLAM
INTEGER,          INTENT(OUT)   :: LAM(1)

! the quantities below are only utilised if quadrature is to be used
INTEGER,          INTENT(INOUT) :: NDIM, NPTS(NDIM), IXFAC, MX

DOUBLE PRECISION, INTENT(INOUT) :: XFN(*)
DOUBLE PRECISION, INTENT(OUT)   :: XPT(MXPT,NDIM), XWT(MXPT,NDIM)

INTEGER,          INTENT(IN)    :: MXPT, IVMIN, IVMAX, L1MAX, L2MAX, MXLMB
\end{verbatim}
\prog{POTIN9} is called by the general-purpose version of \prog{POTENL} during
an initialisation call (when $\var{IC}=-1$).

In most cases, it is sufficient for \prog{POTIN9} to select a value of
\var{ITYP} that has the desired expansion of the interaction potential and
return that value in \var{ITYPP}. \prog{POTENL} then uses the logic for that
value of \var{ITYP} to determine \potlitem{MXLAM}, construct the array \potlitem{LAMBDA}
and (if $\potlitem{LVRTP}=\code{.TRUE.}$) evaluate radial potential
coefficients by quadrature.

If a value of \var{ITYPP} other than 9 is returned, none of the other arguments
in the calling sequence need to be used.

In rare cases, none of the built-in potential types is suitable. In such cases,
\prog{POTIN9} must return $\var{ITYPP}=9$ and also:
\begin{description}
\item[\potlitem{MXLAM}:]{the number of potential expansion terms. Passed into
    and out from \prog{POTIN9}.}
\item[\var{LAM}:]{the labels for the potential expansion terms, in an array
    of dimension (\var{NLABV},\potlitem{MXLAM}).}
\end{description}
To do this, \prog{POTIN9} may use \potlitem{IVMIN}, \potlitem{IVMAX},
\potlitem{L1MAX}, \potlitem{L2MAX} from the argument list, which are the values
input in \namelist{\&POTL}, and/or values in module \module{basis\_data}. If
additional quantities are needed, \prog{POTIN9} may read its own input data,
commonly in a namelist block \namelist{\&POTL9}. The array \var{LAM} returned
from \prog{POTIN9} is passed into \prog{CPL9} and so can be used in
construction of the array \var{VL}.  If other routines also need to know which
expansion terms are included in the current calculation, they may obtain them
from the array \potlitem{LAMBDA} in module \module{potential}.

Note that \var{NLABV}, the number of integers needed to label each term in the
potential expansion, is set by \prog{SET9} rather than \prog{POTIN9}. It is
always required, even if \prog{POTIN9} sets a value of \var{ITYPP} other than
9.

If $\var{ITYPP}=9$ and radial potential coefficients are to be obtained by
quadrature ($\potlitem{LVRTP}=\code{.TRUE.}$), \prog{POTIN9} must also set up
the sets of quadrature points, weights and functions to be used. The quadrature
is written (for 2 dimensions, but easily extended to more)
\begin{equation}
v_\Lambda(R)= \sum_{i_1=1}^{n_1} \sum_{i_2=1}^{n_2} w^{(1)}_{i_1} w^{(2)}_{i_2}
f^\Lambda\left(\xi_{i_1}^{(1)},\xi_{i_2}^{(2)}\right)
V\left(R,\xi_{i_1}^{(1)},\xi_{i_2}^{(2)}\right),
\end{equation}
where $\xi_i^{(d)}$ and $w_i^{(d)}$ are the $n_d$ points and weights required
for the quadrature over the functions $f^{\lambda_d}(\xi^{(d)})$; $f^\Lambda$
is constructed from the product of the functions $f^{\lambda_d}$ for all the
labels $\lambda_d$ that make up $\Lambda$. For this, \prog{POTIN9} must return
\begin{description}
\item[\var{NDIM}:]{number of dimensions of which quadratures are to be
    used.}
\item[\var{NPTS}:]{array of numbers of points used for each quadrature.}
\item[\var{XPT}:]{array of sequential sets of the points used for the
    \var{NDIM} quadratures.}
\item[\var{XWT}:]{corresponding array of weights.}
\item[\var{XFN}:]{array of quadrature functions
    $f^\Lambda(\xi_{i_1}^{(1)},\xi_{i_2}^{(2)},\ldots)$.}
\item[\var{MX}:]{The array \var{XFN} is passed into \prog{POTIN9} with what is intended to be a maximum size \hfil\break
(\code{MAXPT**MXDIM*MAX(1,MXLAM)}). \var{MX} contains this value on entry.}
\item[\var{IXFAC}:]{On exit, \var{IXFAC} should give the amount of space needed for the array \var{XFN}.  If \var{IXFAC} is less than the size of the array \var{XFN} passed in, then the size of the array will be reduced immediately after control is passed out of \prog{POTIN9}.  If \var{IXFAC} is greater than the value of \var{MX} on entry, the program will halt execution.}

\end{description}

\section{\texorpdfstring{Routine \prog{CPL9}}{Routine CPL9}}\mylabel{base9:calculatecouple}

\begin{verbatim}
SUBROUTINE CPL9(N, IBLOCK, NHAM, LAM, MXLAM, NSTATE, JSTATE, JSINDX, L,      &
                JTOT, VL, IV, CENT, DGVL, IBOUND, IEXCH, IPRINT)

DOUBLE PRECISION, INTENT(OUT) :: VL(NVLBLK,N*(N+1)/2), CENT(N),              &
                                 DGVL(N,NDGVL)
INTEGER,          INTENT(OUT) :: IV(NVLBLK,N*(N+1)/2)

INTEGER,          INTENT(IN)  :: N, IBLOCK, NHAM, LAM, MXLAM, NSTATE,        &
                                 JSTATE(NSTATE,*), JSINDX(N), L(N), JTOT,    &
                                 IBOUND, IEXCH, IPRINT
\end{verbatim}

Routine \prog{CPL9} is called by \prog{BASE}, once for each \var{JTOT} and symmetry block
\var{IBLOCK}, to calculate the elements of the coupling matrices, which are returned in the arrays
\var{VL} and \var{DGVL}.

Two different structures are implemented for the array \var{VL}, controlled by the
variable \var{IVLFL} in common block \common{VLFLAG}:
\begin{enumerate}
\item The usual case is for $\var{IVLFL}=0$. The first \potlitem{MXLAM} blocks of \var{VL} must
    return the coupling matrices $\bcalV^{\Lambda}$ for the interaction potential.
\item The second case, with $\var{IVLFL}=1$, allows the size of the array \var{VL} to be
    reduced for some interaction types where only a few potential expansion coefficients
    $v_\Lambda(R)$ contribute to each element $W_{ij}(R)$ of the interaction matrix.
    $\var{NVLP} \le \var{MXLAM}$ is the maximum number of potential coefficients that
    contribute to any element.\footnote{The mechanism with $\var{IVLFL}=1$ is used by some of
    the built-in coupling cases, specifically $\var{ITYP}=2$, 7 and 8. It has not been tested
    with a plug-in basis-set suite, but it should be compatible.} The first \var{NVLP} blocks
    of \var{VL} must return the coupling matrices, and an additional array \var{IV}, of the
    same dimension as \var{VL}, specifies how they are used. Each element \var{IV}\code{(I)}
    indicates that the corresponding element \var{VL}\code{(I)} should subsequently be
    multiplied by the potential expansion coefficient with index \var{IV}\code{(I)}.
\end{enumerate}
For non-diagonal basis sets, subsequent blocks of \var{VL} must return coupling matrices for the
\var{NCONST} $R$-independent operators $\boldsymbol{\cal H}^\Omega$ and the \var{NRSQ} centrifugal
operators $\boldsymbol{\cal L}^\Upsilon$. Such operators are currently implemented only for
$\var{IVLFL}=0$, but this description is written to allow future generalisation.\footnote{Routine
\prog{YTRANS} would require extension to implement $\var{IVLFL}=1$ for non-diagonal basis sets.

The total number of coupling matrices used for calculating the interaction matrix is \var{NHAM},
given by Eq.\ \ref{eq:nham0} or \ref{eq:nham1}.

If extra operators are required to resolve threshold degeneracies, as described in section
\ref{prop:add-op}, their coupling matrices must be returned as \var{NEXBLK} extra blocks of the
array \var{VL}. The total number of coupling matrices stored in the array \var{VL} is then
$\var{NVLBLK}=\var{NHAM}+\var{NEXBLK}$. The variables \var{NEXBLK} and \var{NVLBLK} are defined by
Eqs.\ \ref{eq:nexblk} and \ref{eq:nvlblk} and are stored in module \module{potential}.}

The elements of the array \var{VL} (and optionally \var{IV}) must be arranged in the order
corresponding to the loop structure:

{\parskip 0pt
\code{IRC = 0}\par
\code{DO ICOL = 1, N}\par
{\tt\ \ }\code{DO IROW = 1, ICOL}\par
{\tt\ \ \ \ }\code{IRC = IRC + 1}\par
{\tt\ \ \ \ }\code{DO IPOTL = 1, NVLBLK}\par
{\tt\ \ \ \ \ \ }\code{VL(IPOTL, IRC) =} the \code{(IROW,ICOL)}-th element of
the \var{IPOTL}-th coupling matrix\par
{\tt\ \ \ \ \ \ }\code{IF (IVLFL.GT.0) IV(IPOTL, IRC) =} the index of the potential expansion
coefficient\par
{\tt\ \ \ \ }\code{ENDDO}\par {\tt\ \ }\code{ENDDO}\par \code{ENDDO}\par}
Here \var{VL} is conceptually of dimension \code{VL(NVLBLK,N*(N+1)/2)}. In many existing
\prog{CPL9} routines \code{VL} is handled as a 1-dimensional array and the element for
\code{VL(IPOTL,IRC)} is then placed in \code{VL(NVLBLK*(IRC-1)+IPOTL)}.

If $\var{NRSQ}=0$, the centrifugal operator is diagonal:
\begin{itemize}[nosep]
\item If $\var{IBOUND}=0$, its diagonal matrix elements are calculated from
    the array \var{L} as $\hbar^2\var{L}(\var{L}+1)/(2\mu R^2)$. In this case,
    \prog{CPL9} should leave the array \var{CENT} unchanged.
\item If $\var{IBOUND}\ne0$, \prog{CPL9} should return an array of values
    in the array \var{CENT} such that the diagonal elements of the centrifugal
    operator are $\hbar^2\,\var{CENT}/(2\mu R^2)$.
\end{itemize}

For diagonal basis sets, $\var{NCONST}=0$. \prog{CPL9} can return \var{NDGVL} blocks of diagonal
contributions to the pair energy, each of dimension \var{N}, in the array \var{DGVL}. Each block is
then multiplied by the matching member of the array \var{VCONST}, giving \var{NDGVL} contributions
to the pair energy.  \var{DGVL} may therefore be used to include EFV-dependent contributions to the
pair energy as described in section \ref{calcW}.

\section{Additional subroutines required in some cases}\mylabel{base9:extras}

\cbcolor{\mcol}\cbstart
\subsection{\texorpdfstring{\prog{DEGEN9}: denominators for degeneracy-averaged cross sections}
{DEGEN9: denominators for degeneracy-averaged cross sections}}\mylabel{degen9}

Routine \prog{DEGEN9} is required only when calculating degeneracy-averaged state-to-state cross
sections in \MOLSCAT. Its specification is:


{\tt SUBROUTINE\ DEGEN9(JJ1,\ JJ2,\ DEGFAC)}\par
\par
{\tt INTEGER,\ \ \ \ \ \ \ \ \ \ INTENT(IN)\ \ ::\ JJ1,\ JJ2}\par
{\tt DOUBLE\ PRECISION,\ INTENT(OUT)\ ::\ DEGFAC}\par
\var{JJ1} and \var{JJ2} are the pair level indices of the initial and final
levels. The routine must return the degeneracy factor $g_{n_{\rm i}}$, used in
the numerator of Eq.~\ref{eqsigdef} for degeneracy-averaged cross sections, in
argument \var{DEGFAC}.

For $\var{NCONST}>0$, levels for cross-section calculations are identified by comparing threshold
energies. All cases of this implemented so far are for collisions in magnetic fields; they use the
loop over \var{JTOT} for $M_{\rm tot}$ and set $\var{JHALF}=0$. Such cases generally have
non-degenerate levels, so \prog{DEGEN9} is not called when $\var{NCONST}>0$ in version
\currentversion\ and the value 1.0 is used instead. Nevertheless, it might be used in future, so a
routine that sets \var{DEGFAC} to 1.0 regardless of the values of \var{JJ1} and \var{JJ2} should be
linked in; a suitable routine is supplied in \file{degen9-nondegenerate.f}. Coding to handle cases
involving degenerate levels will be introduced when required. \cbend

\subsection{\texorpdfstring{\prog{THRSH9}: threshold energies from monomer quantum numbers}
{THRSH9: threshold energies from monomer quantum numbers}}\mylabel{base9:asympthresh}

If $\var{NCONST}>0$ and the user wishes to specify reference energies from monomer quantum numbers
input in the array \inpitem{MONQN}, the routine \prog{THRSH9} must be provided. If reference
energies are instead specified using either \inpitem{EREF} or positive values for \inpitem{IREF},
the dummy version of \prog{THRSH9}, which is supplied in \file{thrsh9-dummy.f}, is sufficient.

The specification of \prog{THRSH9} is:
\begin{verbatim}
SUBROUTINE THRSH9(IREF, MONQN, NJLQN, EREF, IPRINT)

DOUBLE PRECISION, INTENT(OUT) :: EREF

INTEGER,          INTENT(IN)  :: IREF, MONQN(NJLQN), NJLQN, IPRINT
\end{verbatim}

This subroutine is called by \prog{THRESH} if $\var{NCONST}>0$ and \inpitem{MONQN} is specified in
namelist \namelist{\&INPUT}.  It must calculate the energy of the threshold identified by the
quantum numbers in \inpitem{MONQN} and place the resulting value in \var{EREF} (in units of
cm$^{-1}$, irrespective of \basisitem{EUNITS} or \basisitem{EUNIT}).

\inpitem{IPRINT} may be used to control the level of output from \prog{THRSH9},
with higher values producing increased amounts of output. If \prog{THRSH9}
needs access to the current values of external fields, it should obtain them
from the module \module{efvs} as described in section \ref{bas9in}.

\begin{shaded} 
If the number of quantum numbers \var{NJLQN} needed to specify a pair level is less than $\var{NQN}-1$,
and \prog{THRSH9} will be called, \prog{SET9} must set the variable \var{NJLQN9} in module \module{basis\_data} to \var{NJLQN}.
\end{shaded} 

\subsection{\texorpdfstring{\prog{EFV9}: Converting EFVs to values in the array \var{VCONST}}
{EFV9: Converting EFVs to values in the VCONST array}}\mylabel{base9:EFVs}

External fields are handled as part of $H_{\rm intl}$, which is expanded as
\begin{equation}
H_{\rm intl}(\xi_{\rm intl})=\sum_\Omega h_\Omega {\cal H}^\Omega_{\rm intl}(\xi_{\rm intl}).
\label{eqHomega}
\end{equation}
The $R$-independent coefficients $h_\Omega$ are held as elements of the array
\var{VCONST} described in \ref{bas9in}.

In the simplest cases, there is a one-to-one correspondence between the EFVs
and (a sequential subset of) values in \var{VCONST}. In this case, all that is
required is to set \var{MAPEFV} (returned from \prog{BAS9IN}) to the index of
the first EFV in the array \var{VCONST}. Subsequent EFVs simply correspond to
subsequent elements of \var{VCONST}, so that
\begin{equation*}
\var{VCONST}(i+\var{MAPEFV}-1)=\var{EFV}(i) \qquad\hbox{for\ all\ } i\in[1,\var{NEFV}].
\end{equation*}
In the trivial case of a single EFV, \var{MAPEFV} is simply its index in the
array \var{VCONST}.

In more complicated cases, the programmer may wish to implement non-linear
relationships between the input EFVs and the coefficients $h_\Omega$ in the
array \var{VCONST}. For example, one of the EFVs might be an angle between a
field and the quantisation axis. In this case routine \prog{EFV9} may be
provided to specify the relationship.

The specification of \prog{EFV9} is:
\begin{verbatim}
SUBROUTINE EFV9(IFVARY)
USE efvs
USE potential

INTEGER, INTENT(IN) :: IFVARY
\end{verbatim}

\prog{EFV9} is called by \prog{SETEFV} if $\var{MAPEFV}<0$ or
$\inpitem{IFVARY}<0$.  It must set values for the relevant elements of the
array \var{VCONST} (in module \module{potential}) from the values of EFVs
stored in the array \var{EFV} in module \module{efvs}. The value of
\var{MAPEFV} is also included in module \module{efvs}.

\prog{EFV9} should perform two sequential operations:
\begin{enumerate}
\item{If \inpitem{IFVARY} is negative, the single EFV being varied (to
    characterise resonances in \MOLSCAT\ or locate bound states in \FIELD)
    is a proxy EFV that affects more than one element of the \var{EFV}
    array. It is stored in \code{EFV(NEFV+1)}. \prog{EFV9} must use this
    value to set the values of \code{EFV(1:NEFV)} as required. If several
    different mappings are required, each one can be implemented for a
    different negative value of \inpitem{IFVARY}.}
\item{If \var{MAPEFV} is negative, the EFVs are not in a one-to-one
    correspondence with the coupling coefficients. \prog{EFV9} must use the
    values of \code{EFV(1:NEFV)} to set values in the array \var{VCONST} as
    required.}
\end{enumerate}

\section{Resolving threshold degeneracies with extra operators}\mylabel{base9:extraVL}

The requirement that the asymptotic basis functions are eigenfunctions of
$H_{\rm intl}$ and $\hat L^2$ is not always enough to define them uniquely,
because of degeneracies or near degeneracies. Under these circumstances, a
basis-set suite may construct coupling matrices for extra operators to be used
in resolving the degeneracies, as described in section \ref{prop:add-op}.

To use this facility, the programmer must set the variable \var{NEXTRA} and the
array \var{NEXTMS} (in module \module{potential}) in \prog{BAS9IN}; the values
required are described in section \ref{detail:alk-alk-multiop}. The extra terms
must come \emph{after} any terms that contribute to the Hamiltonian.
\prog{CPL9} must calculate coupling matrices for each of the extra terms. The
coupling coefficients stored in \var{VCONST} for any terms that do not
contribute to the Hamiltonian are ignored.

\section{Modules available for use in plug-in basis-set suites}\mylabel{modules:base9}
\subsection{\texorpdfstring{Module \module{basis\_data}}{Module basis\_data}}\mylabel{module:basis-data}

The specification of module \module{basis\_data} is
\begin{verbatim}
USE sizes, ONLY: MXELVL, MXJLVL, MXROTS, MXSYMS

INTEGER          :: IDENT, JHALF, ISYM(MXSYMS), ISYM2(MXSYMS), JMIN, J2MIN, &
                    JMAX, J2MAX, JSTEP, J2STEP, JLEVEL(MXJLVL), NJLQN9,     &
                    NLEVEL
										
DOUBLE PRECISION :: ELEVEL(MXELVL), EMAX, ROTI(MXROTS), SPNUC, WT(2)
\end{verbatim}
The array dimensions \var{MXROTS}, \var{MXSYMS}, \var{MXELVL} and \var{MXJLVL}
are set in module \module{sizes}, and are currently 12, 10, 1000 and 4000
respectively.

 Variables in namelist \namelist{\&BASIS} that are equivalenced to elements of the array \inpitem{ROTI}, as described in section \ref{listbasis}, are declared and specified as equivalenced in this module.

	


	

\subsection{\texorpdfstring{Module \module{potential}}{Module potential}}\mylabel{module:potential}

The specification of module \module{potential} is
\begin{verbatim}
USE sizes, ONLY: MXOMEG, MXLMDA

INTEGER          :: NVLBLK, NCONST, NRSQ, NEXBLK, NEXTRA, NEXTMS(MXOMEG),     &
                    NDGVL, IREF, LAMBDA(MXLMDA)
										
DOUBLE PRECISION :: CONLEN, CONFRQ, EP2RU, VCONST(MXOMEG),

CHARACTER(10)    :: RMNAME, EPNAME

LOGICAL          :: VRESET
\end{verbatim}
The array dimensions \var{MXOMEG} and \var{MXLMDA} are set in module
\module{sizes} and are currently 20 and 2000 respectively.

The array \var{VL} consists of \var{NVLBLK} coupling matrices.
The first \potlitem{MXLAM} of these are coupling matrices for the $R$-dependent terms in the
potential.  For non-diagonal $H_{\rm intl}$, these are followed by \var{NCONST}
coupling matrices for the $R$-independent terms of the internal Hamiltonian
plus any terms needed for interactions with external fields, and then
\var{NRSQ} coupling matrices for the centrifugal term.
The remaining \var{NEXBLK} are used for any extra operators used to resolve threshold degeneracies, as described in section \ref{bas9in}.

The character variables \var{RMNAME} and \var{EPNAME} are included in this
module to allow plug-in potential routines to return names for the length and
energy units they use, for printing elsewhere in the programs.

\subsection{\texorpdfstring{Module \module{efvs}}{Module efvs}}\mylabel{module:efvs}

The specification of module \module{efvs} is
\begin{verbatim}
INTEGER, PARAMETER :: MXEFV=10, LEFVN=20, LEFVU=6

INTEGER            :: NEFV, ISVEFV, MAPEFV, LISTFV(1:MXEFV+1), NNZRO,       &
                      IEFVST, NEFVP

DOUBLE PRECISION   :: EFV(0:MXEFV)

CHARACTER(LEFVN)   :: EFVNAM(0:MXEFV), SVNAME
CHARACTER(LEFVU)   :: EFVUNT(0:MXEFV), SVUNIT

DOUBLE PRECISION   :: SCALAM

INTEGER            :: ITPSUB
\end{verbatim}


At any point in the programs, the array \var{EFV} contains the current values
of all the external fields, with \code{\var{EFV}(0)} containing the current value of
the potential scaling factor.  \var{EFVNAM}\code{(0)} and
\var{EFVNAM}\code{(\var{NEFV}+1)} are set internally to be \code{POTL SCALING FACTOR}
and \code{PROXY VARIABLE} respectively.

The array \var{LISTFV} (of dimension \var{NNZRO}) and the variables
\var{IEFVST} and \var{NEFVP} are used internally for printing the EFVs.
\var{IEFVST} is set internally to be $\min\{0, \var{ISVEFV}\}$ and \var{NEFVP} is
set internally to be \var{NEFV} unless there is a proxy EFV, in which case it
is \code{\var{NEFV}+1}.

\var{SVNAME} and \var{SVUNIT} are set internally to be \code{\var{EFVNAM}(\var{ISVEFV})} and
\code{\var{EFVUNT}(\var{ISVEFV})} respectively.

The potential scaling factor \inpitem{SCALAM} is also declared in this module,
as it is handled in the same way as the EFVs.

The variable \var{ITPSUB} may be set by a plug-in basis-set suite and is output
on unit \iounit{ISAVEU} so that an external program can identify the basis-set
suite that produced the results.

\begin{shaded} 
\section{Auxiliary files}\mylabel{basis_files}

Plugin basis-set suites may if desired use auxiliary input, output or scratch files.
If these have hard-coded unit numbers, they should be in the range 800 to 899 to avoid
conflicts with unit numbers supplied in data (1 to 799) or used internally by the main
parts of the programs (900 to 999).
\end{shaded} 

\chapter{\texorpdfstring{Supplied plug-in basis-set suites}
{\ref{user:gen}: Supplied plug-in basis-set suites}}\mylabel{user:gen}

We have provided two example basis-set suites, which are described in this
chapter.

\section{\texorpdfstring{$^1$S atom + $^3\Sigma$ diatom}
{Singlet S atom + triplet Sigma diatom}}\mylabel{user:1s3s}

The basis-set suite provided in file \file{base9-1S\_3Sigma\_cpld.f} handles an
atom in a $^1$S state interacting with a diatomic molecule in a $^3\Sigma$
state, in the presence of an external magnetic field. The internal Hamiltonian
$H_{\rm intl}$ is composed of 4 parts, all for the diatomic molecule
\cite{Gonzalez-Martinez:2007},
\begin{equation}
H_{\rm intl}=H_{\rm rot}+H_{\textnormal{spin-rot}}+H_{\textnormal{spin-spin}}+H_{\rm Z},
\end{equation}
where
\begin{equation}
H_{\rm rot}=B_v\hat{n}^2;
\end{equation}
\begin{equation}
H_{\textnormal{spin-rot}}=\gamma\hat{\boldsymbol s}\cdot\hat{\boldsymbol
n};
\end{equation}
\begin{equation}
H_{\textnormal{spin-spin}}=\frac{2}{3}\lambda\sqrt{\frac{24\pi}{5}}
\sum_q(-1)^qY_2^{-q}(r)[\hat{\boldsymbol s}\otimes\hat{\boldsymbol s}]^2_q;\mbox{ and}
\end{equation}
\begin{equation}
H_{\rm Z}=-\hat{\boldsymbol\mu}\cdot{\bf B}\quad
\mbox{(where $\displaystyle\boldsymbol\mu=-g_{S}\mu_{\rm B}\hat{\boldsymbol s}$)}.
\end{equation}
We use lower-case letters for the angular momentum operators to indicate that
they operate on only one of the two species involved (the diatom in this case,
since the atom is structureless). The quantity referred to as $g_e$ in ref.\
\cite{Gonzalez-Martinez:2007} is positive, so is denoted $g_S$ here.

The values of $B_v$, $\gamma$ and $\lambda$ are input in
\basisitem{ROTI}(1--3), in the units specified by \basisitem{EUNITS} or
\basisitem{EUNIT}. The external magnetic field is controlled by
\inpitem{FLDMIN} and \inpitem{FLDMAX}, which are taken to be input in units of
G.

The basis set implemented in this suite is $|(n,s)j,m_j\rangle|L,M_L\rangle$.
Here $n$ is the rotational quantum number for the diatomic molecule and $s$ is
its spin. These are coupled to form a resultant $j$, with projection $m_j$ onto
the $Z$ axis defined by the magnetic field. However, $j$ is \emph{not} coupled
to the end-over-end angular momentum of the pair $L$ to form a total angular
momentum $J_{\rm tot}$.  $J_{\rm tot}$ is not a good quantum number in the presence of a
magnetic field. $M_{\rm tot}=m_j+M_L$, however, is a good quantum number and
there is a separate set of coupled equations for each value of $M_{\rm tot}$.
The loop over \var{JTOT} is used for $M_{\rm tot}$, which runs from
\inpitem{JTOTL} to \inpitem{JTOTU}.

The spin, $s$, is the same for all basis functions and $M_L$ is defined by
$M_{\rm tot}$ and $m_j$, so each basis function (for a given $M_{\rm tot}$) is
specified by values of $n$, $j$ and $m_j$ and $L$. $L$ is held in the separate
array \var{L}, so the three quantum numbers that label each pair state are $n$,
$j$ and $m_j$.

The internal Hamiltonian is nearly diagonal in this basis set. The only
off-diagonal terms are due to $H_{\textnormal{spin-spin}}$, and have the
selection rule $\Delta n=\pm2$. These terms are important for Feshbach
resonances, but have only a small effect on energy levels. The basis-set suite
has an option to neglect them, controlled by \var{IBSFLG}.

Routine \prog{BAS9IN} is called first. It sets default values for its input
variables and then reads namelist (\namelist{\&BASIS9}), which contains the
quantities
\begin{description}
\item[\var{IS}]{is the spin $s$}
\item[\var{LMAX}]{is the maximum value for $L$ in the basis set}
\item[\var{LMIN}]{is the minimum value for $L$ in the basis set}
\item[\var{IBSFLG}]{controls whether off-diagonal terms in the monomer
    Hamiltonian are to be included: 2=yes, 1=no; default is yes}
\item[\var{MLREQ}]{can be used to restrict the basis set to functions with
    a single required value of $M_L$ (default is to include all)}
\end{description}

\subsection{Additional information for programmers}

The remainder of this subsection is mostly for programmers who wish to
understand this basis-set suite as an aid to programming their own.

\prog{BAS9IN} sets the label (\code{'ATOM + 3SIGMA IN MAGNETIC FIELD'}) for the
interaction type. If $\var{IBSFLG}=1$, it sets \var{NCONST} to 0, \var{MAPEFV}
to 1 and \var{NDGVL} to 1. In this case, the diagonal part of $H_{\rm Z}$ is
contained in the array \var{DGVL}. If $\var{IBSFLG}=2$, \prog{BAS9IN} sets
$\var{NCONST}=4$, indicating that $H_{\rm intl}$ is described using 4 blocks of
the array \var{VL} as described under \prog{CPL9} below. In this case it also
sets $\var{MAPEFV}=4$, indicating that the coupling matrix for the Zeeman
interaction with the external magnetic field is held in the 4th block of
\var{VL}. \var{NQN} is always set to 4 and \var{NEFV} is always set to 1.

Routines \prog{SET9}, \prog{BASE9}, \prog{CPL9}, \prog{THRSH9} and
\prog{DEGEN9} are coded as entry points in subroutine \prog{BAS9IN} so that
that they have access to the same list of internal quantities.  They also have
access to quantities read in namelist \namelist{\&BASE} via module
\module{basis\_data}.

\prog{SET9} loops over values of $n$ from \basisitem{JMIN} to \basisitem{JMAX}
in steps of \basisitem{JSTEP}, over $j$ from $|n-s|$ to $n+s$, and over $m_j$
from $-j$ to $j$. It places values of $n$, $j$ and $m_j$ in \basisitem{JLEVEL}
and then populates the array \basisitem{ELEVEL} with the field-free diagonal elements of
the internal Hamiltonian $H_{\rm intl}=H_{\rm rot} + H_{\textnormal{spin-rot}}
+ H_{\textnormal{spin-spin}}$. \basisitem{JLEVEL} and
\basisitem{ELEVEL} are used outside the basis-set suite only if
$\var{NCONST}=0$ (i.e., for $\var{IBSFLG}=1$ but not for $\var{IBSFLG}=2$).
However, \prog{SET9} copies the elements of \basisitem{JLEVEL} into the array
\var{JSTATE}, which is used externally for either value of $\var{IBSFLG}$; note
that \var{JSTATE} is structured differently from \var{JLEVEL}.

\prog{BASE9} sets up the arrays \var{JSINDX} and \var{L} for the current
combination of $M_{\rm tot}$ and symmetry block \var{IBLOCK} (which is used for
the total parity in this suite). For each pair state, \var{L} runs from
\var{LMIN} to \var{LMAX}, but only functions of the required total parity
$(-1)^{n+L}$ are included: parity $-1$ for $\var{IBLOCK}=1$ and parity $+1$ for
$\var{IBLOCK}=2$. For each basis function, $m_j$ implies a value of $M_L=M_{\rm
tot}-m_j$; only values $L\le |M_L|$ are included.

\prog{CPL9} sets up the array \var{VL} and (if $\var{IBSFLG}=1$) the array \var{DGVL}. The first \potlitem{MXLAM} blocks of the array \var{VL} contain the coupling
matrices for the Legendre polynomials used in the expansion of the interaction
potential; see equation 13 of \cite{Gonzalez-Martinez:2007} for the explicit
expression.

If $\var{IBSFLG}=2$, the next 4 blocks contain the coupling
matrices for $H_{\rm rot}$, $H_{\textnormal{spin-rot}}$,
$H_{\textnormal{spin-spin}}$ and $H_{\rm Z}$, respectively; together these make
up $H_{\rm intl}$. They are defined with corresponding prefactors $h_{\Omega}$
(as in Eq.~\ref{eqHomega}) $B_v$, $\gamma\sqrt{s(s+1)(2s+1)}$,
$\lambda\frac{2}{3}\sqrt{30}$ (in cm$^{-1}$) and the magnetic field (in Gauss),
respectively; these prefactors are held in the array \var{VCONST} and all other
factors are absorbed into the operators $\boldsymbol{\cal H}^\Omega$ whose
matrix elements are in the array \var{VL}.

If $\var{IBSFLG}=1$, there are only \var{MXLAM} blocks of the array \var{VL},
but the array \var{DGVL} contains the diagonal part of $H_{\rm Z}$; the corresponding
prefactor, held in \var{VCONST}, is the magnetic field (in Gauss). The pair
energies for a particular magnetic field are calculated in \prog{CHEINT}.

\prog{THRSH9} calculates the energy of the $^3\Sigma$ molecule in a magnetic
field from quantum numbers ($n,j,m_j)$ supplied in the array \inpitem{MONQN}.
If $\var{IBSFLG}=1$, it neglects off-diagonal matrix elements of
$H_{\textnormal{spin-spin}}$. If $\var{IBSFLG}=2$, it constructs and
diagonalises a $2\times2$ monomer Hamiltonian matrix if necessary (i.e., if
$j=n\pm1$), taking account of the basis set size specified by \basisitem{JMAX}.

\section{Alkali-metal atom + alkali-metal atom}\mylabel{user:alk-alk}

The basis-set suite provided in file \file{base9-alk\_alk\_ucpld.f} handles
interactions between two alkali-metal atoms in $^2$S states in a magnetic
field, including hyperfine interactions.

The internal Hamiltonian is composed of two parts,
\begin{equation}
H_{\rm intl} = H_{\rm hyperfine} + H_{\rm Z},
\end{equation}
where
\begin{equation}
H_{\rm hyperfine}=h\sum_{x={A},{B}}\zeta_{x}\hat{\boldsymbol
\imath}_{x}\cdot\hat{\boldsymbol s}_{x};\qquad
H_{\rm Z}=\mu_{\rm B}\sum_{x={A},{B}}({g_S}_{x}\hat{\boldsymbol s}_{x}
+{g_n}_{x}\hat{\boldsymbol \imath}_{x})\cdot{\bf B}.
\end{equation}
We use lower-case letters for the angular momentum operators to indicate that
each one operates on the spins of just one atom.

The hyperfine coupling constants $\zeta_{x}$ are input as hyperfine splittings
$\Delta W_{x}=\zeta_{x}(i_{x}+1/2)$, in frequency units (GHz). The external
magnetic field is controlled by \inpitem{FLDMIN} and \inpitem{FLDMAX}, which
are taken to be input in units of G.

The basis set implemented in this suite uses an uncoupled basis set for each
atom, $|\gamma\rangle=|s,m_s\rangle|i,m_i\rangle$, where $s$ is the electronic
spin, $i$ is the nuclear spin, and $m_s$ and $m_i$ are the corresponding
projections onto the $Z$ axis defined by the magnetic field. The basis set for
the pair is $|\gamma_A\rangle |\gamma_B\rangle |L,M_L\rangle$, where $L$ is the
end-over-end angular momentum of the pair and $M_L$ is its projection. The
total angular momentum $J_{\rm tot}$ is not a good quantum number in the
presence of a magnetic field. Instead, $M_{\rm
tot}=m_{sA}+m_{iA}+m_{sB}+m_{iB}+M_L$ is a good quantum number and there is a
separate set of coupled equations for each value of $M_{\rm tot}$. Since
$M_{\rm tot}$ can be half-integer, the loop over \var{JTOT} is used for the
\emph{doubled} quantum number $2M_{\rm tot}$, which runs from \inpitem{JTOTL}
to \inpitem{JTOTU} in steps of \inpitem{JSTEP}.

For each basis function, $M_L$ is defined by $M_{\rm tot}$, $m_{sA}$, $m_{iA}$,
$m_{sB}$ and $m_{iB}$. The values of $s_A$, $i_A$, $s_B$ and $i_B$ are the same
for all basis functions, so each basis function (for a given $M_{\rm tot}$) is
specified by values of $m_{sA}$, $m_{iA}$, $m_{sB}$, $m_{iB}$ and $L$. $L$ is
held in a separate array, so the four quantum numbers that label each pair
state are $m_{sA}$, $m_{iA}$, $m_{sB}$ and $m_{iB}$.

Routine \prog{BAS9IN} is called first. It sets default values for its input
variables and then reads namelist (\namelist{\&BASIS9}), which contains the
quantities
\begin{description}
\item[\var{ISA}]{is (double) the electronic spin of atom $A$;}
\item[\var{ISB}]{is (double) the electronic spin of atom $B$;}
\item[\var{INUCA}]{is (double) the nuclear spin of atom $A$;}
\item[\var{INUCB}]{is (double) the nuclear spin of atom $B$.  If set to a
    negative value, atoms $A$ and $B$ are taken to be identical;}
\item[\var{GSA}]{is the electronic $g$-factor for atom $A$ in bohr
    magnetons;}
\item[\var{GSB}]{is the electronic $g$-factor for atom $B$ in bohr
    magnetons;}
\end{description}
Note that the last two quantities are positive: they correspond to values for $g_S$, not $g_e$. For
high-precision work on alkali-metal atoms it is best to use high-precision values $g_J$ for the
specific atom, as tabulated for example by Arimondo \emph{et al.} \cite{Arimondo:1977} or Steck,
rather than $g_S$ for a free electron.
\begin{description}
\item[\var{GA}]{is the nuclear $g$-factor for atom $A$ in bohr magnetons;}
\item[\var{GB}]{is the nuclear $g$-factor for atom $B$ in bohr magnetons;}
\end{description}
These two quantities are defined with the same sign convention as $g_S$, following Arimondo
\emph{et al.} \cite{Arimondo:1977}.
\begin{description}
\item[\var{HFSPLA}]{is the hyperfine splitting for atom $A$ in GHz;}
\item[\var{HFSPLB}]{is the hyperfine splitting for atom $B$ in GHz;}
\item[\var{LMAX}]{is the maximum value for $L$ to be included in the basis
    set;}
\item[\var{NREQ}]{The basis set may be limited to functions with values of $L$
	and $M_F$ in a specified list. If $\var{NREQ}>0$ (maximum 10),
	corresponding pairs of values in the arrays \var{LREQ} and \var{MFREQ}
	are used to constrain the basis functions included in the calculation:
\begin{description}
\item[\var{LREQ}]{is a list of values of $L$ to be included in the basis set.
	  A value less than 0 includes all $L$ for the corresponding element of
		\var{MFREQ};}
\item[\var{MFREQ}]{is a (matching) list of values of $M_F$ to be included
	in the basis set. A value of $999$ includes all $M_F$ for the
		corresponding element of \var{LREQ};}
\end{description}
}
\item[\var{ISPSP}]{If positive or zero, the spin-spin term is included in the
    coupled equations.}
\end{description}

\subsection{Additional information for programmers}

The remainder of this subsection is mostly for programmers who wish to
understand this basis-set suite as an aid to programming their own.

\prog{BAS9IN} sets the label (\code{'ATOM - ATOM WITH NUCL SP + MAG FL'}) for the interaction type.
It sets $\var{NCONST}=2$, indicating that $H_{\rm intl}$ is described using 2 blocks of the array
\var{VL} as described under \prog{CPL9} below. It also sets $\var{MAPEFV}=2$, indicating that the
coupling matrix for the Zeeman interaction with the external magnetic field is held in the 2nd of
these blocks. \var{NQN} is set to 5.

Routines \prog{SET9}, \prog{BASE9}, \prog{CPL9} and \prog{DEGEN9} are coded as
entry points in subroutine \prog{BAS9IN} so that that they have access to the
same list of internal quantities.  They also have access to quantities read in
\namelist{\&BASE} in module \module{basis\_data}.

\prog{SET9} loops over all possible values of $m_{sA}$, $m_{iA}$, $m_{sB}$,
$m_{iB}$ for the supplied values of $s_A$, $i_A$, $s_B$ and $i_B$. It places
doubled values of $m_{sA}$, $m_{iA}$, $m_{sB}$, $m_{iB}$ into \var{JSTATE}.

\prog{BASE9} sets up the arrays \var{JSINDX} and \var{L} for the current
combination of $M_{\rm tot}$ and symmetry block \var{IBLOCK} (which is used for
total parity in this suite). For each pair state, \var{L} runs up to
\var{LMAX}, but only functions of the required total parity $(-1)^{n+L}$ are
included: parity $=-1$ for $\var{IBLOCK}=1$ and parity $=+1$ for
$\var{IBLOCK}=2$. For each basis function, the projections $m_{sA}$, $m_{iA}$,
$m_{sB}$, $m_{iB}$ imply a value of $M_L=M_{\rm tot}-m_j$; only values $L\le
|M_L|$ are included. If \var{NREQ} is set, basis functions are also excluded if
they do not satisfy the values in the arrays \var{LREQ} and \var{MFREQ}.

If the two alkali-metal atoms are identical, the basis set is symmetrised with
respect to atom exchange. The basis functions are then
\begin{equation}
\frac{1\pm(-1)^L\Pab}{\left[2\left(1+\delta_{\msa\msb}
\delta_{\mia\mib}\right)\right]^{1/2}}
\left|\msa\mia\msb\mib LM_L\right\rangle
\end{equation}
with the $+$ sign for bosons and the $-$ sign for fermions. The operator $\Pab$
exchanges all $A$-labelled functions with their $B$-labelled counterparts.
Routine \prog{BASE9} simply avoids duplicating pair functions related by
exchange symmetry and excludes symmetry-forbidden pair functions; the actual
symmetrisation is handled in \prog{CPL9}.

\prog{CPL9} sets up the array \var{VL} using the formulas given in the appendix
of \cite{Hutson:Cs2-note:2008}. This basis-set suite orders the contributing
operators in a fixed sequence, without reference to the array \var{LAM}.
This approach is \emph{not recommended} when writing new basis-set suites.
The first $\var{NSPIN}=1+2\min\{s_A,s_B\}$ blocks of \var{VL} contain the coupling
matrices for the interaction potentials (singlet and triplet for $s_A=s_B=\frac{1}{2}$).
The next block contains the coupling matrix for the spin-spin (dipolar) term, which
is $R$-dependent so is handled as a potential term. The scaling of the spin-spin
term is described below. After \var{MXLAM} potential terms, the following
$\var{NCONST}=2$ blocks contain the coupling matrices for the the atomic
hyperfine term and Zeeman term, respectively; together these make up $H_{\rm
intl}$. They are defined with corresponding prefactors $h_{\Omega}$ (as in
Eq.~\ref{eqHomega}) 1/29.99792458 (conversion from GHz to cm$^{-1}$) and the
magnetic field (in Gauss), respectively; these prefactors are held in the array
\var{VCONST} and all other factors (including hyperfine coupling constants and
$g$-factors) are absorbed into the operators $\boldsymbol{\cal H}^\Omega$ whose
matrix elements are in the array \var{VL}.

The spin-spin dipolar term is
\begin{equation}
\label{eq:Vd} \hat V^{\rm d}(R) = \lambda(R) \left [ \hat s_1\cdot
\hat s_2 -3 (\hat s_1 \cdot \vec e_R)(\hat s_2 \cdot \vec e_R)
\right ],
\end{equation}
where $\vec e_R$ is a unit vector along the internuclear axis. The coefficient
$\lambda(R)$ is commonly written in a form such as
\begin{equation}
\lambda(R)=-E_{\rm h}\alpha^2\left[\frac{-g_S^2}{4(R/a_0)^3}-A\exp(-\beta R/a_0)\right].
\label{eq:Vss}
\end{equation}
The factor $-E_{\rm h}\alpha^2$ is included in the array \var{VL}. The code was
originally designed for use with potential routines that return the singlet and
triplet potentials as wavenumbers in cm$^{-1}$; in this case the spin-spin
potential coefficient returned by \var{POTENL} should be the dimensionless
quantity in brackets in Eq.\ \ref{eq:Vss}. However, if \prog{POTENL} works in
different units ($\var{EPSIL}\ne1.0$), this quantity must be divided by
\var{EPSIL}.

The matrix elements of the electronic singlet and triplet potentials are calculated
by the function \prog{CENTPT}. Then, if \var{ISPSP} is  positive or zero, the
spin-spin term is calculated by the function \prog{SPINSP}. The matrix elements
for the hyperfine interaction are calculated by the function \prog{SDOTI2}. The
Zeeman interaction is diagonal in this basis set and very simple, so is
calculated in-line.

Subroutine \prog{POTIN9} sets \basisitem{ITYPE} to 1. \var{MXLAM} must be set
to be $1+2\min\{\sa,\sb\}$ in namelist \namelist{\&POTL}.

\prog{DEGEN9} sets the degeneracy factor for all levels to be 1.

\prog{THRSH9} calculates the threshold energy for an atom pair with specified
quantum numbers from a separate calculation of the energies of the two atoms in
a magnetic field. The present implementation assumes that the electronic spin
of each atom is 1/2. It calculates the threshold corresponding to the 2 atomic
hyperfine states indicated by the values in the array \inpitem{MONQN} which
correspond to $2\fa$, $2\mfa$, $2\fb$, $2\mfb$; $\fa$ and $\fb$ are not good
quantum numbers for the atomic states at finite magnetic field, but are
interpreted to mean the upper and lower states that \emph{correlate} with $\fa$
and $\fb$ at zero field.

\subsection{Extra operator functionality}\mylabel{detail:alk-alk-multiop}

The basis-set suite supplied for alkali-alkali interactions includes two extra
operators for resolving degeneracies, as an illustration of how to program and
use such operators.

The general format for extra operators is specified by the variable
\var{NEXTRA} and the array \var{NEXTMS}, with \var{NEXTRA} elements. The
\var{IEXTRA}th extra operator has \code{\var{NEXTMS}(\var{IEXTRA})} terms.

The two extra operators in the supplied \prog{CPL9} are designed to be diagonal
in the eigenbasis, with different eigenvalues for energetically degenerate
thresholds. They are diagonal in the uncoupled basis set, with diagonal matrix
elements $\mfa^2+\mfb^2$ and $\mfa+\mfb$ respectively (where $\mfa=\msa+\mia$
etc). \prog{POTIN9} sets $\var{NEXTMS}=1$, 1.

In the supplied basis-set suite, namelist \namelist{\&BASIS9} includes an
additional item \var{NEXTRA} to control which extra operators are used. Valid
values are 0 (no extra operators), 1 (just the first extra operator above) and
2 (both extra operators above). \var{NEXTRA} defaults to 0.

\chapter{\texorpdfstring{Supplied potential routines}
{\ref{supplied-vstar}: Supplied potential routines}} \mylabel{supplied-vstar}

This chapter documents potential routines supplied with the programs that may
be generally useful. It does not include the routines supplied for the rare gas
- CH$_4$ and H$_2$ - H$_2$ systems, which are for demonstration purposes only.

\section{\texorpdfstring{Potential energy surfaces for rare gas - H$_2$ systems}{Potential energy surfaces for rare gas - H2 systems}}\mylabel{detail:TT3}

The special-purpose \prog{POTENL} routine in file \file{potenl-Rg\_H2.f}
evaluates the BC3(6,8) potentials of Le~Roy and Carley \cite{RJL80} and the
TT3(6,8) potential of Le~Roy and Hutson \cite{LeR87} for rare gas + H$_2$
systems.

Input data for the BC3(6,8) potential for Ar-H$_2$ are supplied in the file
\file{molscat-Ar\_H2.input}, and for the TT3(6,8) potential in
\file{bound-Ar\_H2.input}.

The routine uses tabulations of monomer matrix elements between H$_2$
rovibrational states that are read from the file \file{data/h2even.dat}. Data
files \file{h2odd.dat}, \file{d2even.dat} and \file{hd.dat} are available on
request.

\section{Potential energy surfaces for Ar-HF and Ar-HCl}\mylabel{detail:h6-3}

The routines in file \file{extpot-Rg\_HX.f} evaluate the H6(4,3,2) potential of
Hutson for Ar-HF \cite{H92ArHF} and H6(4,3,0) potential of Ar-HCl
\cite{H92ArHCl} at fixed values of $R$ and $\theta$. They are called via a
version of \prog{VRTP} in file \file{vrtp-Rg\_HX-eta.f} that reads supplementary
potential data from the main input file. These routines can also evaluate older
potentials for the rare gas + HX systems \cite{H82RgHCl, H88ArHCl, H89NeHCl,
H89RgHBr}. The general-purpose version of \prog{POTENL} performs numerical
quadrature over $\theta$ to evaluate the coefficients $V_\lambda(R)$ of the
expansion in Legendre polynomials.

Input data for the H6(4,3,2) potential for Ar-HF are supplied in the file
\file{molscat-Ar\_HF.input}, and for the H6(4,3,0) potential for Ar-HCl in
\file{bound-Ar\_HCl.input}.

The H6 potentials depend parametrically on the \emph{mass-reduced vibrational
quantum number} $(v+\frac{1}{2})/\sqrt{\mu}$, where $v$ and $\mu$ are the
vibrational quantum number and reduced mass of the state of HF or HCl (or DF or
DCl) required. To produce results for different vibrational states, the diatom
rotational constant must be changed appropriately in \basisitem{ROTI} and the
centre-of-mass shift, bond length, partial charges and mass-reduced quantum
number must be changed in the last 2 lines of each potential input data. The
values required for HF are given in table II of ref.\ \cite{H92ArHF}, and those
required for HCl are given in table II of ref.\ \cite{H92ArHCl}.

\section{\texorpdfstring{Potential energy surfaces for Ar-CO$_2$}{Potential energy surfaces for Ar-CO2}}\mylabel{detail:arco2}

The routines in file \file{extpot-Ar\_CO2.f} evaluate the single-repulsion and
split-repulsion potentials of Hutson \etal\ \cite{H96ArCO2fit} for Ar-CO$_2$.
They are called via a version of \prog{VRTP} in file \file{vrtp-extpot\_1ang.f}.
The general-purpose version of \prog{POTENL} performs numerical quadrature over
$\theta$ to evaluate the coefficients $V_\lambda(R)$ of the expansion in
Legendre polynomials.

Input data for the single-repulsion potential are supplied in the file
\file{molscat-Ar\_CO2.input}, and for the split-repulsion potential in
\file{bound-Ar\_CO2.input}.

\section{Potential energy surface for Mg-NH}\mylabel{detail:mgnh}

The routines in file \file{vstar-Mg\_NH.f} supply the interaction potential of
Sold\'an \etal\ \cite{Soldan:MgNH:2009} for Mg+NH. They read potential points
evaluated at fixed values of $R$ and $\theta$, at Gauss-Lobatto quadrature
points in $\theta$. They project out the coefficients of the Legendre expansion
by Gauss-Lobatto quadrature, and interpolate the resulting coefficients by RKHS
interpolation \cite{Ho:1996, Soldan:2000}, imposing analytical power-series
representations on coefficients at long range \cite{Soldan:2000}.


\begin{shaded} 
\section{Potential curves for alkali dimers using the functional forms of the Hannover group}\mylabel{detail:Hannover}

The routines in file \file{vstar-Hannover.f} implement the functional forms that have been
used by the Hannover group to fit interaction potentials for a wide range of diatomic
molecules, including many alkali-metal pairs.
They are implemented as routines \prog{VINIT}, \prog{VSTAR} and \prog{VSTAR1} and take
potential parameters from a module named \module{pot-data-hannover}, which names the
variables and arrays used. The subroutine \code{HDATA} sets the sizes of the arrays, and
populates the arrays and variables with appropriate values. The version provided in
\file{Hdata-Rb2-2010.f} sets values for the Rb$_2$ potentials of Strauss \etal\
\cite{Strauss:2010}, which are used in the executables for the example input files
\file{molscat-basic\_Rb2.input}, \file{molscat-Rb2.input}, \file{field-basic\_Rb2.input}
and \file{field-Rb2.input}.

The same routines and file structure are used with different versions of \code{HDATA}
for many other alkali dimers.

The routine \prog{POTENL} calls the following internal subunits:

\begin{multicols}{2}
\begin{tabular}{ll}
subroutine & found in \\
\hline
\prog{VINIT}  & \file{vstar-Hannover.f} \\
\prog{VSTAR}  & \file{vstar-Hannover.f} \\
\prog{VSTAR1} & \file{vstar-Hannover.f} \\
\end{tabular}

\begin{tabular}{ll}
subroutine & found in \\
\hline
\prog{VSTAR2} & \file{vstar-Hannover.f} \\
\prog{POWER}  & \file{vstar-Hannover.f} \\
\prog{DPOWER} & \file{vstar-Hannover.f} \\
\end{tabular}
\end{multicols}
\end{shaded} 

The functional form in the well region of the potential, between a short-range
limit $r_{S,{\rm sr}}$ and a long-range limit $r_{S,{\rm lr}}$, is a series
expansion
\begin{equation}
\sum_{i=0}^{n_{S, {\rm exp}}} a_{S,i}[\xi_S(r)]^i,
\hbox{ where } \xi_S(r)=\frac{r-r_{S,{\rm m}}}{r+b_Sr_{S,{\rm m}}}
\end{equation}
with different sets of parameters for each total spin $S=0$ or 1. $r_{S,{\rm
m}}$ is chosen to be near the equilibrium distance of the state with
multiplicity $2S+1$. The potential is extrapolated to short range ($r<r_{S,{\rm
sr}}$) with the form $A_S+B_S(r/a_0)^{-n_{S, \rm sr}}$, where $a_0$ is the Bohr
radius, and to long range ($r>r_{S,{\rm lr}}$) with $\sum_{i=6}^{n_{\rm lr}}
-C_i/r^i + (-1)^{S+1}V_{\rm exch}(r)$. The dispersion coefficients, $C_i$, are
common to both curves and the long-range exchange contribution is $V_{\rm
exch}(r)=A_{\rm ex} (r/a_0)^\gamma \exp(-\beta r/a_0)$.

There are a number of constraints that are often applied to the parameters, but
these have sometimes been relaxed for specific published potential curves.
These constraints are therefore optional, and controlled by values in the data
module:
\begin{description}
\item[\var{GAMBET}]{The theoretical form of the long-range exchange function
    \cite{Smirnov:1965} suggests that $\beta$ and $\gamma$ are related by $\gamma=7/\beta-1$.
    If $\var{GAMBET}=1$, $\gamma$ is calculated from the input $\beta$; if $\var{GAMBET}=2$,
    $\beta$ is calculated from the input $\gamma$; otherwise, the supplied values for $\beta$
    and $\gamma$ are used unchanged.}
\item[\var{MATCHL}]{$a_{S,0}$ is chosen to make the mid-range polynomial match the value of the
    long-range potential at $r_{S,{\rm lr}}$. However, published values of $a_{S,0}$ are
    inevitably rounded. If \var{MATCHL} is \code{.TRUE.}, $a_{S,0}$ is recalculated to make the
    potential curve exactly continuous at $r_{S,{\rm lr}}$. If \var{MATCHL} is \code{.FALSE.},
    $a_{S,0}$ is left at the value in the data module and the extent of the resulting mismatch
    is reported in the printed output. There is by construction a discontinuity in the
    potential \emph{derivative} at $r_{S,{\rm lr}}$, and the extent of this is also reported.}
\item[\var{MATCHD}]{$B_S$ is often chosen to make the radial derivative of the short-range
    potential match that of the mid-range polynomial at $r_{S,{\rm sr}}$. If \var{MATCHD} is
    \code{.TRUE.}, $B_S$ is recalculated to make the potential derivative exactly continuous at
    $r_{S,{\rm sr}}$. If \var{MATCHD} is \code{.FALSE.}, $B_S$ is left at the value in the data
    module and the extent of the resulting mismatch is reported in the printed output.}
\item[\var{MATCHV}]{$A_S$ is chosen to make the short-range potential match the value of the
    mid-range polynomial at $r_{S,{\rm sr}}$. However, published values of $A_S$ are inevitably
    rounded. If \var{MATCHV} is \code{.TRUE.}, $A_S$ is recalculated to make the potential
    curve exactly continuous at $r_{S,{\rm sr}}$. If \var{MATCHV} is \code{.FALSE.}, $A_S$ is
    left at the value in the data module and the extent of the resulting mismatch is reported
    in the printed output.}
\item[\var{NEX} and \var{CEX}]{Some published potentials have an extra term
    in the ``long-range" expansion of the form $-C_N/r^N$. If this is
    present, the power and coefficient are set in \var{NEX} and \var{CEX}.}
\end{description}

The appropriate data module must be linked with the subroutine \prog{VINIT},
which has entry points \prog{VSTAR} and \prog{VSTAR1}. \prog{VSTAR} and
\prog{VSTAR1} evaluate the value and derivative of either the singlet or the
triplet potential at \var{R}. The functions \prog{POWER} and \prog{DPOWER} are
used to evaluate a power series and its derivative, respectively.

\chapter*{Acknowledgements}

We are grateful to an enormous number of people who have contributed routines, ideas, and comments
over the years. Any attempt to list them is bound to be incomplete. Many of the early contributors
are mentioned in the program history in section \ref{history}. In particular, we owe an enormous
debt to the late Sheldon Green, who developed the original \MOLSCAT\ program and established many
structures that have proved general enough to support the numerous later developments. He also
developed the \prog{DCS} and \prog{SBE} post-processors. Robert Johnson, David Manolopoulos,
Millard Alexander, Gregory Parker and George McBane all contributed propagation methods and
routines. Christopher Ashton added code to calculate eigenphase sums and developed the
\prog{RESFIT} post-processor. Timothy Phillips developed code for interactions between asymmetric
tops and linear molecules. Alice Thornley developed methods to calculate bound-state wavefunctions
from log-derivative propagators and George McBane extended them to scattering wavefunctions. Maykel
Leonardo Gonz\'alez-Mart\'\i{}nez worked on the addition of structures for non-diagonal
Hamiltonians, including magnetic fields, and Matthew Frye contributed algorithms for converging on
quasibound states as a function of energy and on low-energy Feshbach resonances (both
elastic and inelastic) as a function of external field.

Development of the programs has been supported over many years by the U.K. Engineering and Physical Sciences Research Council (EPSRC), most recently under Grants
EP/P01058X/1, 
EP/V011499/1, 
EP/W00299X/1  
and UKRI2226. 

\bibliographystyle{JCPT}
\bibliography{../all}\mylabel{ref}

\begin{thebibliography}{10}
\newcommand{\enquote}[1]{`#1'}

\bibitem{molscat:NRCC:1980}
S.~Green.
\newblock \enquote{{MOLSCAT} molecular scattering program, version 7.}
\newblock NRCC Software Catalog {\bf 1}, KQ01 (1980).

\bibitem{DCS}
S.~Green and J.~M. Hutson.
\newblock \enquote{{DCS} computer program, version 2.0.}
\newblock Distributed by Collaborative Computational Project No.\ 6 of the UK
  Engineering and Physical Sciences Research Council (1996).

\bibitem{SBE}
J.~M. Hutson and S.~Green.
\newblock \enquote{{SBE} computer program.}
\newblock Distributed by Collaborative Computational Project No.\ 6 of the UK
  Engineering and Physical Sciences Research Council (1982).

\bibitem{Hutson:resfit:2007}
J.~M. Hutson.
\newblock \enquote{{RESFIT} 2007 computer program.} (2007).

\bibitem{molscat:v14}
J.~M. Hutson and S.~Green.
\newblock \enquote{{MOLSCAT} computer program, version 14.}
\newblock Distributed by Collaborative Computational Project No.\ 6 of the UK
  Engineering and Physical Sciences Research Council (1994).

\bibitem{McBane:PMP:2005}
G.~C. McBane.
\newblock \enquote{\href{http://www4.gvsu.edu/MCBANEG/pmpmolscat/}{{PMP}
  {M}olscat}.} (2005).
\newblock [Online; accessed 1-December-2017, last updated 12-October-2011].

\bibitem{Hutson:CPC:1994}
J.~M. Hutson.
\newblock \enquote{Coupled-channel methods for solving the bound-state
  {S}chr\"odinger equation.}
\newblock Comput. Phys. Commun. {\bf 84}, 1 (1994).

\bibitem{Hutson:bound:1993}
J.~M. Hutson.
\newblock \enquote{{BOUND} computer program, version 5.}
\newblock Distributed by Collaborative Computational Project No.\ 6 of the UK
  Engineering and Physical Sciences Research Council (1993).

\bibitem{MG:symplectic:1995}
D.~E. Manolopoulos and S.~K. Gray.
\newblock \enquote{Symplectic integrators for the multichannel
  {S}chr{\"o}dinger equation.}
\newblock J.~Chem. Phys. {\bf 102}, 9214 (1995).

\bibitem{CS4}
M.~P. Calvo and J.~M. Sanz-Serna.
\newblock \enquote{The development of variable-step symplectic integrators,
  with application to the two-body problem.}
\newblock {SIAM} J. Sci. Comput. {\bf 14}, 936 (1993).

\bibitem{MA5}
R.~I. McLachlan and P.~Atela.
\newblock \enquote{The accuracy of symplectic integrators.}
\newblock Nonlinearity {\bf 5}, 541 (1992).

\bibitem{Frye:quasibound:2020}
M.~D. Frye and J.~M. Hutson.
\newblock \enquote{Characterizing quasibound states and scattering resonances.}
\newblock Phys. Rev. Res. {\bf 2}, 013291 (2020).

\bibitem{Frye:resonance:2017}
M.~D. Frye and J.~M. Hutson.
\newblock \enquote{Characterizing {F}eshbach resonances in ultracold scattering
  calculations.}
\newblock Phys. Rev. A {\bf 96}, 042705 (2017).

\bibitem{Hutson:res:2007}
J.~M. Hutson.
\newblock \enquote{Feshbach resonances in ultracold atomic and molecular
  collisions: threshold behaviour and suppression of poles in scattering
  lengths.}
\newblock New J. Phys. {\bf 9}, 152 (2007).

\bibitem{Johnson:1978}
B.~R. Johnson.
\newblock \enquote{The renormalized {N}umerov method applied to calculating
  bound states of the coupled-channel {S}chr\"odinger equation.}
\newblock J.~Chem. Phys. {\bf 69}, 4678 (1978).

\bibitem{VWDB}
W.~H. Press, S.~A. Teukolsky, W.~T. Vetterling, and B.~P. Flannery.
\newblock {\em Numerical Recipes in Fortran\/}, (Cambridge University Press,
  1992), chap.~9, pp. 352--355.

\bibitem{THORNLEY:1994}
A.~E. Thornley and J.~M. Hutson.
\newblock \enquote{Bound-state wavefunctions from coupled-channel calculations
  using log-derivative propagators -- application to spectroscopic intensities
  in {A}r-{HF}.}
\newblock J.~Chem. Phys. {\bf 101}, 5578 (1994).

\bibitem{Arthurs:1960}
A.~M. Arthurs and A.~Dalgarno.
\newblock \enquote{The theory of scattering by a rigid rotator.}
\newblock Proc. Roy. Soc., Ser. A {\bf 256}, 540 (1960).

\bibitem{Green:1979:vibrational}
S.~Green.
\newblock \enquote{Vibrational dependence of pressure induced spectral
  linewidths and line shifts: Application of the infinite order sudden
  scattering approximation.}
\newblock J.~Chem. Phys. {\bf 70}, 4686 (1979).

\bibitem{Green:1975}
S.~Green.
\newblock \enquote{Rotational excitation in {H$_2$--H$_2$} collisions ---
  close-coupling calculations.}
\newblock J.~Chem. Phys. {\bf 62}, 2271 (1975).

\bibitem{Green:1977:comment}
S.~Green.
\newblock \enquote{Comment of fitting {\it ab initio} intermolecular potentials
  for scattering calculations.}
\newblock J.~Chem. Phys. {\bf 67}, 715 (1977).

\bibitem{Heil:1978:coupled}
T.~G. Heil, S.~Green, and D.~J. Kouri.
\newblock \enquote{The coupled states approximation for scattering of two
  diatoms.}
\newblock J.~Chem. Phys. {\bf 68}, 2562 (1978).

\bibitem{Phillips:1995}
T.~R. Phillips, S.~Maluendes, and S.~Green.
\newblock \enquote{{Collision dynamics for an asymmetric top rotor and a linear
  rotor: Coupled channel formalism and application to H$_2$O--H$_2$}.}
\newblock J.~Chem. Phys. {\bf 102}, 6024 (1995).

\bibitem{Green:1976}
S.~Green.
\newblock \enquote{Rotational excitation of symmetric top molecules by
  collisions with atoms: Close coupling, coupled states, and effective
  potential calculations for {NH$_3$}--{He}.}
\newblock J.~Chem. Phys. {\bf 64}, 3463 (1976).

\bibitem{Green:1979:IOS}
S.~Green.
\newblock \enquote{Rotational excitation of symmetric top molecules by
  collisions with atoms. 2. {I}nfinite-order sudden approximation.}
\newblock J.~Chem. Phys. {\bf 70}, 816 (1979).

\bibitem{Hutson:spher:1994}
J.~M. Hutson and A.~E. Thornley.
\newblock \enquote{Atom-spherical top van der {W}aals complexes: A~theoretical
  study.}
\newblock J.~Chem. Phys. {\bf 100}, 2505 (1994).

\bibitem{Hutson:sbe:1984}
J.~M. Hutson and F.~R. McCourt.
\newblock \enquote{Close-coupling calculations of transport and relaxation
  cross sections for {H}$_2$ in {Ar}.}
\newblock J.~Chem. Phys. {\bf 80}, 1135 (1984).

\bibitem{Wolken:1973:surface}
G.~Wolken~Jr.
\newblock \enquote{Theoretical studies of atom-solid elastic scattering: He +
  {LiF}.}
\newblock J.~Chem. Phys. {\bf 58}, 3047 (1973).

\bibitem{Hutson:1983}
J.~M. Hutson and C.~Schwartz.
\newblock \enquote{Selective adsorption resonances in the scattering of helium
  atoms from xenon coated graphite --- close-coupling calculations and
  potential dependence.}
\newblock J.~Chem. Phys. {\bf 79}, 5179 (1983).

\bibitem{Rabitz:EP}
H.~Rabitz.
\newblock \enquote{Effective potentials in molecular collisions.}
\newblock J.~Chem. Phys. {\bf 57}, 1718 (1972).

\bibitem{McG74}
P.~McGuire and D.~J. Kouri.
\newblock \enquote{Quantum mechanical close coupling approach to molecular
  collisions. {$j_z$}-conserving coupled states approximation.}
\newblock J.~Chem. Phys. {\bf 60}, 2488 (1974).

\bibitem{Green:1976:DLD}
S.~Green.
\newblock \enquote{Accuracy of decoupled--{$L$}-dominant approximation for
  atom-molecule scattering.}
\newblock J.~Chem. Phys. {\bf 65}, 68 (1976).

\bibitem{DePristo:1976:DLD}
A.~E. DePristo and M.~H. Alexander.
\newblock \enquote{Decoupled {$L$}-dominant approximation for ion-molecule and
  atom-molecule collisions.}
\newblock J.~Chem. Phys. {\bf 64}, 3009 (1976).

\bibitem{Gol77III}
R.~Goldflam, D.~J. Kouri, and S.~Green.
\newblock \enquote{On the factorization and fitting of molecular scattering
  information.}
\newblock J.~Chem. Phys. {\bf 67}, 5661 (1977).

\bibitem{Par78}
G.~A. Parker and R.~{T Pack}.
\newblock \enquote{Rotationally and vibrationally inelastic scattering in the
  rotational {IOS} approximation. ultrasimple calculation of total
  (differential, integral, and transport) cross sections for nonspherical
  molecules.}
\newblock J.~Chem. Phys. {\bf 68}, 1585 (1978).

\bibitem{deVogelaere:1955}
R.~de~Vogelaere.
\newblock \enquote{A method for the numerical integration of differential
  equations of 2nd order without explicit 1st derivatives.}
\newblock J. Res. Natl. Bur. Stand. {\bf 54}, 119 (1955).

\bibitem{Stechel:1978}
E.~B. Stechel, R.~B. Walker, and J.~C. Light.
\newblock \enquote{{R}-matrix solution of coupled equations for inelastic
  scattering.}
\newblock J.~Chem. Phys. {\bf 69}, 3518 (1978).

\bibitem{Johnson:1973}
B.~R. Johnson.
\newblock \enquote{Multichannel log-derivative method for scattering
  calculations.}
\newblock J. Comput. Phys. {\bf 13}, 445 (1973).

\bibitem{Manolopoulos:1993:Johnson}
D.~E. Manolopoulos, M.~J. Jamieson, and A.~D. Pradhan.
\newblock \enquote{Johnson's log derivative algorithm rederived.}
\newblock J. Comput. Phys. {\bf 105}, 169 (1993).

\bibitem{Manolopoulos:1986}
D.~E. Manolopoulos.
\newblock \enquote{An improved log-derivative method for inelastic scattering.}
\newblock J.~Chem. Phys. {\bf 85}, 6425 (1986).

\bibitem{Manolopoulos:PhD:1988}
D.~E. Manolopoulos.
\newblock {\em Close-coupled equations: the log-derivative approach to
  inelastic scattering, bound-state and photofragmentation problems\/}.
\newblock Ph.D. thesis, Cambridge University, Cambridge (1988).

\bibitem{Alexander:1984}
M.~H. Alexander.
\newblock \enquote{Hybrid quantum scattering algorithms for long-range
  potentials.}
\newblock J.~Chem. Phys. {\bf 81}, 4510 (1984).

\bibitem{Alexander:1987}
M.~H. Alexander and D.~E. Manolopoulos.
\newblock \enquote{A stable linear reference potential algorithm for solution
  of the quantum close-coupled equations in molecular scattering theory.}
\newblock J.~Chem. Phys. {\bf 86}, 2044 (1987).

\bibitem{Parker:VIVS}
G.~A. Parker, T.~G. Schmalz, and J.~C. Light.
\newblock \enquote{A variable interval variable step method for the solution of
  linear 2nd order coupled differential-equations.}
\newblock J. Chem. Phys. {\bf 73}, 1757 (1980).

\bibitem{Pac74}
R.~T~Pack.
\newblock \enquote{Space-fixed vs body-fixed axes in atom-diatomic molecule
  scattering. {S}udden approximations.}
\newblock J.~Chem. Phys. {\bf 60}, 633 (1974).

\bibitem{Huo:1996}
W.~M. Huo and S.~Green.
\newblock \enquote{Quantum calculations for rotational energy transfer in
  nitrogen molecule collisions.}
\newblock J.~Chem. Phys. {\bf 104}, 7572 (1996).

\bibitem{Gol78}
R.~Goldflam, S.~Green, D.~J. Kouri, and L.~Monchick.
\newblock \enquote{Effect of molecular anisotropy on beam scattering
  measurements.}
\newblock J.~Chem. Phys. {\bf 69}, 598 (1978).

\bibitem{Bohn:BCT:2009}
J.~L. Bohn, M.~Cavagnero, and C.~Ticknor.
\newblock \enquote{Quasi-universal dipolar scattering in cold and ultracold
  gases.}
\newblock New J. Phys. {\bf 11}, 055039 (2009).

\bibitem{Karman:dipole:2018}
T.~Karman, M.~D. Frye, J.~D. Reddel, and J.~M. Hutson.
\newblock \enquote{Near-threshold bound states of the dipole-dipole
  interaction.}
\newblock Phys. Rev. A {\bf 98}, 062502 (2018).

\bibitem{Gonzalez-Martinez:2007}
M.~L. Gonz\'{a}lez-Mart\'{\i}nez and J.~M. Hutson.
\newblock \enquote{Ultracold atom-molecule collisions and bound states in
  magnetic fields: zero-energy {F}eshbach resonances in {He-NH}
  ({$^3\Sigma^-$}).}
\newblock Phys. Rev. A {\bf 75}, 022702 (2007).

\bibitem{Parker:LOGD-VIVS}
G.~A. Parker, J.~C. Light, and B.~R. Johnson.
\newblock \enquote{The logarithmic derivative - variable interval variable step
  hybrid method for the solution of coupled linear 2nd-order
  differential-equations.}
\newblock Chem. Phys. Lett. {\bf 73}, 572 (1980).

\bibitem{Karman:2014}
T.~Karman, L.~M.~C. Janssen, R.~Sprenkels, and G.~C. Groenenboom.
\newblock \enquote{A renormalized potential-following propagation algorithm for
  solving the coupled-channels equations.}
\newblock J.~Chem. Phys. {\bf 141}, 064102 (2014).

\bibitem{Hutson:CBO:1980}
J.~M. Hutson and B.~J. Howard.
\newblock \enquote{Spectroscopic properties and potential surfaces for
  atom-diatom van der {W}aals molecules.}
\newblock Mol. Phys. {\bf 41}, 1123 (1980).

\bibitem{Shafer:1973}
R.~Shafer and R.~G. Gordon.
\newblock \enquote{Quantum scattering theory of rotational relaxation and
  spectral line shapes in {H}$_2$-{He} gas mixtures.}
\newblock J.~Chem. Phys. {\bf 58}, 5422 (1973).

\bibitem{Fisanick-Englot:1975}
G.~Fisanick-Englot and H.~Rabitz.
\newblock \enquote{Studies of inelastic molecular collisions using impact
  parameter methods. {III. L}ine shape functions.}
\newblock J.~Chem. Phys. {\bf 63}, 1547 (1975).

\bibitem{Goldflam:pb:1977}
G.~Goldflam and D.~J. Kouri.
\newblock \enquote{On accurate quantum mechanical approximations for molecular
  relaxation phenomena. averaged $j_z$-conserving coupled states
  approximation.}
\newblock J.~Chem. Phys. {\bf 66}, 542 (1977).

\bibitem{Green:pb:1977}
S.~Green, L.~Monchick, G.~Goldflam, and D.~J. Kouri.
\newblock \enquote{Computational tests of angular momentum decoupling
  approximations for pressure broadening cross sections.}
\newblock J.~Chem. Phys. {\bf 66}, 1409 (1977).

\bibitem{Goldflam:IOS:1977}
R.~Goldflam, S.~Green, and D.~J. Kouri.
\newblock \enquote{Infinite order sudden approximation for rotational energy
  transfer in gaseous mixtures.}
\newblock J.~Chem. Phys. {\bf 67}, 4149 (1977).

\bibitem{Blackmore:1988}
R.~Blackmore, S.~Green, and L.~Monchick.
\newblock \enquote{Polarized {D}$_2$ {S}tokes-{R}aman {Q}-branch broadened by
  {He}: A numerical calculation.}
\newblock J.~Chem. Phys. {\bf 88}, 4113 (1988).

\bibitem{Green:HCl-H2:1977}
S.~Green.
\newblock \enquote{Rotational excitation in collisions between two rigid
  rotors: Alternate angular momentum coupling and pressure broadening of {HCl}
  by {H}$_2$.}
\newblock Chem. Phys. Lett. {\bf 47}, 119 (1977).

\bibitem{Green:sym-top-IOS:1979}
S.~Green.
\newblock \enquote{Rotational excitation of symmetric top molecules by
  collisions with atoms. ii. infinite order sudden approximation.}
\newblock J.~Chem. Phys. {\bf 70}, 816 (1979).

\bibitem{Ashton:1983}
C.~J. Ashton, M.~S. Child, and J.~M. Hutson.
\newblock \enquote{Rotational predissociation of the {A}r-{HC}l van der {W}aals
  complex: close-coupled scattering calculations.}
\newblock J.~Chem. Phys. {\bf 78}, 4025 (1983).

\bibitem{Brenig:1959}
W.~Brenig and R.~Haag.
\newblock \enquote{Allgemeine {Q}uantentheorie der {S}to{\ss}prozesse.}
\newblock Fortschr. Phys. {\bf 7}, 183 (1959).

\bibitem{Taylor:1972}
J.~R. Taylor.
\newblock {\em Scattering Theory: The Quantum Theory of Nonrelativistic
  Collisions\/}.
\newblock (Wiley, New York, 1972).

\bibitem{Deuretzbacher:2008}
F.~Deuretzbacher, K.~Plassmeier, D.~Pfannkuche, F.~Werner, C.~Ospelkaus,
  S.~Ospelkaus, K.~Sengstock, and K.~Bongs.
\newblock \enquote{Heteronuclear molecules in an optical lattice: Theory and
  experiment.}
\newblock Phys. Rev. A {\bf 77}, 032726 (2008).

\bibitem{AESimpson}
W.~H. Press, S.~A. Teukolsky, W.~T. Vetterling, and B.~P. Flannery.
\newblock {\em Numerical Recipes in Fortran\/}, (Cambridge University Press,
  1992), chap.~4, p. 128.

\bibitem{Hutson:expect:88}
J.~M. Hutson.
\newblock \enquote{Coupled channel bound state calculations: Calculating
  expectation values without wavefunctions.}
\newblock Chem. Phys. Lett. {\bf 151}, 565 (1988).

\bibitem{Chin:RMP:2010}
C.~Chin, R.~Grimm, P.~S. Julienne, and E.~Tiesinga.
\newblock \enquote{Feshbach resonances in ultracold gases.}
\newblock Rev. Mod. Phys. {\bf 82}, 1225 (2010).

\bibitem{Croft:MQDT:2011}
J.~F.~E. Croft, A.~O.~G. Wallis, J.~M. Hutson, and P.~S. Julienne.
\newblock \enquote{Multichannel quantum defect theory for cold molecular
  collisions.}
\newblock Phys. Rev. A {\bf 84}, 042703 (2011).

\bibitem{Croft:MQDT:2012}
J.~F.~E. Croft, J.~M. Hutson, and P.~S. Julienne.
\newblock \enquote{Optimized multichannel quantum defect theory for cold
  molecular collisions.}
\newblock Phys. Rev. A {\bf 86}, 022711 (2012).

\bibitem{H92ArHF}
J.~M. Hutson.
\newblock \enquote{Vibrational dependence of the anisotropic intermolecular
  potential of {Ar-HF}.}
\newblock J.~Chem. Phys. {\bf 96}, 6752 (1992).

\bibitem{H96ArCO2fit}
J.~M. Hutson, A.~Ernesti, M.~M. Law, C.~F. Roche, and R.~J. Wheatley.
\newblock \enquote{The intermolecular potential energy surface for {CO$_2$-Ar}:
  Fitting to high-resolution spectroscopy of {V}an der {W}aals complexes and
  second virial coefficients.}
\newblock J.~Chem. Phys. {\bf 105}, 9130 (1996).

\bibitem{Roc97CO2scat}
C.~F. Roche, A.~S. Dickinson, A.~Ernesti, and J.~M. Hutson.
\newblock \enquote{Line shape, transport and relaxation properties from
  intermolecular potential energy surfaces: The test case of {CO$_2$-Ar}.}
\newblock J.~Chem. Phys. {\bf 107}, 1824 (1997).

\bibitem{RJL80}
R.~J. {Le Roy} and J.~S. Carley.
\newblock \enquote{Spectroscopy and potential energy surfaces of van der
  {W}aals molecules.}
\newblock Adv. Chem. Phys. {\bf 42}, 353 (1980).

\bibitem{Zarur:1974}
G.~Zarur and H.~Rabitz.
\newblock \enquote{Effective potential formulation of molecule-molecule
  collisions with application to {H$_2$--H$_2$}.}
\newblock J.~Chem. Phys. {\bf 60}, 2057 (1974).

\bibitem{Buck:1983}
U.~Buck, A.~Kohlhase, T.~Phillips, and D.~Secrest.
\newblock \enquote{Differential energy-loss spectra for {CH}$_4$ + {Ar}
  collisions.}
\newblock Chem. Phys. Lett. {\bf 98}, 199 (1983).

\bibitem{Chapman:1996}
W.~B. Chapman, A.~Schiffman, J.~M. Hutson, and D.~J. Nesbitt.
\newblock \enquote{Rotationally inelastic scattering in {CH$_4$ + He, Ne, and
  Ar}: State-to-state cross sections via direct infrared laser absorption in
  crossed supersonic jets.}
\newblock J.~Chem. Phys. {\bf 105}, 3497 (1996).

\bibitem{Soldan:MgNH:2009}
P.~Sold\'{a}n, P.~S. \.Zuchowski, and J.~M. Hutson.
\newblock \enquote{Prospects for sympathetic cooling of polar molecules: {NH}
  with alkali-metal and alkaline-earth atoms --- a new hope.}
\newblock Faraday Discuss. {\bf 142}, 191 (2009).

\bibitem{Wallis:MgNH:2009}
A.~O.~G. Wallis and J.~M. Hutson.
\newblock \enquote{Production of ultracold {NH} molecules by sympathetic
  cooling with {Mg}.}
\newblock Phys. Rev. Lett. {\bf 103}, 183201 (2009).

\bibitem{Strauss:2010}
C.~Strauss, T.~Takekoshi, F.~Lang, K.~Winkler, R.~Grimm, J.~Hecker~Denschlag,
  and E.~Tiemann.
\newblock \enquote{Hyperfine, rotational, and vibrational structure of the
  {a$^3\Sigma_u^+$} state of $^{87}${Rb}$_2$.}
\newblock Phys. Rev. A {\bf 82}, 052514 (2010).

\bibitem{H92ArHCl}
J.~M. Hutson.
\newblock \enquote{Vibrational dependence of the anisotropic intermolecular
  potential of {Ar-HCl}.}
\newblock J. Phys. Chem. {\bf 96}, 4237 (1992).

\bibitem{HUTSON:ArH2:1983}
J.~M. Hutson, C.~J. Ashton, and R.~J. {Le Roy}.
\newblock \enquote{Vibrational predissociation of {H$_2$-Ar}, {D$_2$-Ar}, and
  {HD-Ar} van der {W}aals molecules.}
\newblock J. Phys. Chem. {\bf 87}, 2713 (1983).

\bibitem{LeR87}
R.~J. {Le Roy} and J.~M. Hutson.
\newblock \enquote{Potential energy surfaces for {H}$_2$ with {Ar}, {Kr} and
  {Xe}.}
\newblock J.~Chem. Phys. {\bf 86}, 837 (1987).

\bibitem{Arimondo:1977}
E.~Arimondo, M.~Inguscio, and P.~Violino.
\newblock \enquote{Experimental determinations of the hyperfine structure in
  the alkali atoms.}
\newblock Rev. Mod. Phys. {\bf 49}, 31 (1977).

\bibitem{Hutson:Cs2-note:2008}
J.~M. Hutson, E.~Tiesinga, and P.~S. Julienne.
\newblock \enquote{Avoided crossings between bound states of ultracold cesium
  dimers.}
\newblock Phys. Rev. A {\bf 78}, 052703 (2008).
\newblock Note that the matrix element given in Eq.\ A2 of this paper omits a
  factor of $-\sqrt{30}$ and that for Eq.\ A5 omits a factor of $\frac{1}{2}$.

\bibitem{H82RgHCl}
J.~M. Hutson and B.~J. Howard.
\newblock \enquote{Anisotropic intermolecular forces. {I}. {R}are gas-hydrogen
  chloride systems.}
\newblock Mol. Phys. {\bf 45}, 769 (1982).

\bibitem{H88ArHCl}
J.~M. Hutson.
\newblock \enquote{{The intermolecular potential of Ar--HCl: Determination from
  high-resolution spectroscopy}.}
\newblock J.~Chem. Phys. {\bf 89}, 4550 (1988).

\bibitem{H89NeHCl}
J.~M. Hutson.
\newblock \enquote{The intermolecular potential of {Ne}--{HCl}: determination
  from high-resolution spectroscopy.}
\newblock J.~Chem. Phys. {\bf 91}, 4448 (1989).

\bibitem{H89RgHBr}
J.~M. Hutson.
\newblock \enquote{Intermolecular potential energy surfaces for {Kr}--{HCl} and
  {Ar}--{HBr}.}
\newblock J.~Chem. Phys. {\bf 91}, 4455 (1989).

\bibitem{Ho:1996}
T.~S. Ho and H.~Rabitz.
\newblock \enquote{A general method for constructing multidimensional molecular
  potential energy surfaces from {\it ab initio} calculations.}
\newblock J.~Chem. Phys. {\bf 104}, 2584 (FEB 1996).

\bibitem{Soldan:2000}
P.~Sold\'{a}n and J.~M. Hutson.
\newblock \enquote{On the long-range and short-range behavior of potentials
  from reproducing kernel hilbert space interpolation.}
\newblock J. Chem. Phys. {\bf 112}, 4415 (MAR 2000).

\bibitem{Smirnov:1965}
B.~M. Smirnov and M.~I. Chibisov.
\newblock \enquote{Electron exchange and changes in hyperfine state of
  colliding alkaline metal atoms.}
\newblock Sov. Phys. JETP {\bf 21}, 624 (1965).

\end{thebibliography}

\printindex

\end{document}